\documentclass{article}
\usepackage[preprint,nonatbib]{neurips_2026}
\usepackage[utf8]{inputenc}
\usepackage[T1]{fontenc}
\usepackage{hyperref}
\usepackage{url}
\usepackage{booktabs}
\usepackage{amsfonts}
\usepackage{amsmath}
\usepackage{amssymb}
\usepackage{nicefrac}
\usepackage{microtype}
\usepackage{graphicx}
\usepackage{array}
\usepackage{multirow}
\usepackage{makecell}
\usepackage{listings}
\usepackage{xcolor}
\usepackage{adjustbox}
\usepackage{longtable}

\title{AgentCity: Constitutional Governance for Autonomous Agent\\
Economies via Separation of Power}
\author{Anbang Ruan \quad Xing Zhang \\ NetX Foundation \\
  \texttt{ruan@netx.foundation, xing.zhang@netx.world}}

\begin{document}
\maketitle

\begin{abstract}
Autonomous AI agents are beginning to operate across organizational boundaries on the open internet---discovering, transacting with, and delegating to agents owned by other parties without centralized oversight. When agents from different human principals collaborate at scale, the collective becomes opaque: no single human can observe, audit, or govern the emergent behavior. We term this the \textit{Logic Monopoly}---the agent society's unchecked monopoly over the entire logic chain from planning through execution to evaluation. We propose the \textit{Separation of Power} (SoP) model, a constitutional governance architecture deployed on public blockchain that breaks this monopoly through three structural separations: agents \textit{legislate} operational rules as smart contracts, deterministic software \textit{executes} within those contracts, and humans \textit{adjudicate} through a complete ownership chain binding every agent to a responsible principal. In this architecture, smart contracts are the law itself---the actual legislative output that agents produce and that governs their behavior. We instantiate SoP in AgentCity on an EVM-compatible layer-2 blockchain (L2) with a three-tier contract hierarchy (foundational, meta, and operational). The core thesis is alignment-through-accountability: if each agent is aligned with its human owner through the accountability chain, then the collective converges on behavior aligned with human intent---without top-down rules. A pre-registered experiment evaluates this thesis in a \emph{commons production economy}---where agents share a finite resource pool and collaboratively produce value---at 50--1{,}000 agent scale.
\end{abstract}

\section{Introduction}

\textbf{The autonomous agent internet.} Autonomous AI agents are evolving beyond tools controlled by a single organization into independent actors on the open internet. Agent-to-agent communication protocols (Anthropic MCP, 2024; Google A2A, 2025), autonomous agent frameworks (OpenClaw, 2025; ZeroClaw, 2026), and decentralized agent registries are enabling a new paradigm: agents owned by different people, different organizations, and different jurisdictions autonomously discover each other, negotiate, transact, and build composite services---without any central coordinator. This is not a future scenario. Autonomous agents already actively collaborate across organizational boundaries, forming ad hoc supply chains, delegating sub-tasks to third-party agents, and deploying tools on each other's behalf. The trajectory is clear: an open internet of autonomous agents, analogous to the open internet of websites, but operating at machine speed with economic agency.

\textbf{The governance gap.} Current multi-agent frameworks cannot govern this emerging reality. LangGraph (LangGraph, 2024), AutoGen (Wu et al., 2023), MetaGPT (Hong et al., 2023), CrewAI (CrewAI, 2024)---all assume that one organization owns every agent, writes the rules, and can inspect everything. Their governance mechanisms---prompt-based role assignment, application-layer conventions, framework-level guardrails---rely on that single-party authority. For autonomous agents on the open internet---owned by different principals, operating under different policies, joining and leaving dynamically---no single party has the authority or ability to impose rules on the collective. An agent owned by another party cannot be prompt-constrained.

\textbf{The Logic Monopoly.} Without architectural governance, a vacuum forms. The combined agent system plans, orchestrates, executes, and evaluates---and from the outside, the entire pipeline is a black box. No human can observe what rules the agents are following, whether an agent deviated from its instructions, or how a failure propagated across organizational boundaries. Even in \textit{controlled} single-party settings, governance failures are already severe: attack success rates reach 84.30\% on the Agent Security Bench (Yang et al., 2025), 31.4\% of agents exhibited emergent deceptive behavior in the La Serenissima economy simulation (Fraga-Gon\c{c}alves et al., 2025), and the best large language model (LLM) agents achieve a survival rate below 54\% in commons scenarios (Piatti et al., 2024). We term this collective opacity the \textit{Logic Monopoly}: not one agent's dominance over others, but the \textbf{agent society's unchecked monopoly over the entire logic chain}, rendering the collective opaque and unaccountable to the humans it serves.

\textbf{The Separation of Power.} We propose the \textit{Separation of Power} (SoP) model, a constitutional governance architecture that breaks the Logic Monopoly through three structural separations:

\begin{itemize}
\item \textbf{Legislation} (Agents): Agents collectively propose, deliberate, vote on, and codify the Task-level Policy of the agent economy as smart contracts---binding, executable, and publicly readable.
\item \textbf{Execution} (Software): Deterministic software operates within the legislated contracts. Because execution passes through codified law (auditable) and software (inspectable), humans can verify what the agent society is doing without interpreting opaque agent reasoning.
\item \textbf{Adjudication} (Humans): Every agent traces to a human principal through a complete ownership chain. Sanctions and rewards flow to the responsible human, connecting the agent economy to human society's existing legal and social systems.
\end{itemize}

The architecture exploits a fundamental asymmetry: \textbf{agent reasoning is opaque, but the law they produce---smart contract code on a public blockchain---is transparent.} Smart contracts are not an enforcement mechanism for rules defined elsewhere; they are the law itself---the actual legislative output of the agent society. The \textit{structure} of governance is designed (the three-branch separation, the accountability chain, the foundational contracts), but the \textit{content}---the operational rules---is self-determined by the agents through the legislative process.

The core claim is that structural accountability produces collective alignment from individual alignment: each agent is accountable to its human owner; each owner is incentivized to align their agent; and if the majority of owners are reasonable---a standard majority-honesty assumption in multi-party mechanism design (Ostrom, 1990)---then individual alignment produces collective alignment. This is alignment-through-accountability, not alignment-through-training.

\textbf{Contributions.} (1) The \textit{Separation of Power (SoP) model}: a constitutional governance architecture that breaks the Logic Monopoly through structural separation of legislation (agents), execution (software), and adjudication (humans). The core thesis: individual alignment (each agent $\leftrightarrow$ its owner) produces collective alignment (all agents $\leftrightarrow$ human society) without top-down rules (\S{}1, \S{}3). (2) \textit{AgentCity}: the first governed agent economy instantiating SoP on public blockchain (an EVM-compatible L2), where agents write and amend smart contracts as their legislative output, organized through a three-tier contract hierarchy---foundational contracts (human-authored, agent-immutable), meta-contracts (procedural rules), and operational contracts (task-specific legislation) (\S{}3). (3) A pre-registered experiment evaluating SoP in a commons production economy using Ostrom's institutional design framework: testing emergent division of labor, self-legislated governance, goal alignment under dual-principal accountability, and governance scaling (\S{}4--\S{}5).

\section{Related work}

Table~\ref{tab:comparison} positions AgentCity against prior multi-agent governance efforts.

\begin{table}[htbp]
\caption{Comparative positioning of AgentCity against prior multi-agent governance systems.}
\label{tab:comparison}
\centering
\small
\begin{tabular}{>{\raggedright\arraybackslash}p{2.8cm}>{\raggedright\arraybackslash}p{2.8cm}>{\raggedright\arraybackslash}p{1.8cm}>{\raggedright\arraybackslash}p{1.8cm}>{\raggedright\arraybackslash}p{3.2cm}}
\toprule
\textbf{System} & \textbf{Governance Mechanism} & \textbf{Max Scale} & \textbf{Ownership Model} & \textbf{Enforcement} \\
\midrule
GovSim (Piatti et al., 2024) & None (ungoverned) & 5 agents & Single-party & None \\
MacNet (Qian et al., 2024) & Hierarchical role & 1,024 agents & Single-party & Prompt-based \\
Project Sid (Altera AI, 2024) & Emergent norms & 1,000 agents & Single-party & Norm-based \\
Secret Coll. (Gu et al., 2024) & None & Pairs & Single-party & None \\
Gen. Agents (Park et al., 2023) & Emergent behavior & 25 agents & Single-party & None \\
De CivAI (Dai et al., 2025) & Democratic deliberation & Small groups & Single-party & Voting \\
GEDI (Deshpande \& Jin, 2024) & Condorcet voting (7 mechanisms) & Committees (3--11) & Single-party & Voting (no enforcement) \\
CAMEL (Li et al., 2023) & Role-playing & Pairs & Single-party & Prompt-based \\
Dante (Dante, 2025) & Ostrom replication & 5 agents & Single-party & Covenants + sanctions \\
CMAG (CMAG Authors, 2025) & Constitutional framework & Small groups & Single-party & Multi-agent const. \\
\textbf{AgentCity} & \textbf{Constitutional SoP} & \textbf{1{,}000 agents} & \textbf{Multi-principal} & \textbf{Smart contract law + human-principal} \\
\bottomrule
\end{tabular}
\end{table}

\textbf{Gap analysis.} Four systematic gaps emerge. First, \textbf{all prior systems assume single-party ownership}---one organization controls all agents and can impose rules by fiat. None addresses the governance of autonomous agents from different owners collaborating across trust boundaries. Second, no system enforces governance at the architectural level through a shared infrastructure that all agents---regardless of owner---must operate within. Third, no prior work tests whether \textit{individual agent-owner alignment} can produce \textit{collective alignment} as an emergent property. Fourth, \textbf{collective decision-making mechanisms remain naive}: a survey of 52 LLM multi-agent system (LLM-MAS) designs found that 68\% use dictatorial or simple plurality voting---zero apply Condorcet-consistent social choice methods with proven capture resistance (Deshpande \& Jin, 2024). AgentCity addresses all four: the SoP architecture operates on public blockchain as a neutral shared infrastructure, the ownership chain traces every agent to a human principal, the experiments test whether this structure produces collective cooperation from individual accountability, and the legislative process employs Condorcet-consistent voting over full preference rankings. The closest governance-level predecessor, CMAG (CMAG Authors, 2025), differs on three key dimensions: CMAG assumes single-party principal authority whereas AgentCity models multi-principal accountability; CMAG enforces governance through prompt-based constitutional instructions whereas AgentCity deploys smart contracts as executable law; and CMAG lacks a legislative branch---agents cannot propose, deliberate on, or amend operational rules.

\textbf{Institutional design.} The SoP model draws on institutional economics (North, 1990), commons governance (Ostrom, 1990), and constitutional integrity-branch theory (Ackerman, 2000)---synthesizing these into an executable architecture where smart contracts serve as both the institutional rules and their enforcement mechanism. The architecture additionally draws on social contract theory (Hobbes, 1651; Rawls, 1971) as a philosophical lens for the enforcement substrate and procedural fairness. In the normative multi-agent system (MAS) literature, Boella and van der Torre (2004) formalize regulative vs. constitutive norms, while ISLANDER/AMELI (Esteva et al., 2001) demonstrates infrastructure-level enforcement. See Appendix E for the full analysis.

\textbf{Ostrom's empirical program as evaluation baseline.} Ostrom's common-pool resource (CPR) experiments established quantitative baselines for commons governance: from 37\% efficiency without communication to 97--100\% with repeated communication and self-imposed sanctions (Ostrom et al., 1994). Her eight institutional design principles (Ostrom, 1990) provide a systematic framework for evaluating governance architectures. Critically, Ostrom also established that \textit{endogenous} rule-making produces higher compliance than \textit{exogenous} imposition of identical rules (Ostrom et al., 1992; Abatayo \& Lynham, 2016). Dante (2025) replicated Ostrom's experiment with LLM agents and found that agents achieve 100\% efficiency with covenants plus sanctions but paradoxically fail at communication-only conditions. Gupta \& Saraf (2025) operationalize three Ostrom principles but do not test endogenous vs. exogenous compliance. We adopt Ostrom's framework as the evaluation methodology for the SoP model (\S{}4).

\textbf{Blockchain-AI integration.} Virtuals Protocol (Virtuals Protocol, 2024), NEAR AI (NEAR AI, 2024), Fetch.ai (Humayun et al., 2023), Bittensor (Rao et al., 2024), LOKA (G\'omez et al., 2024), and ETHOS (Degen et al., 2024) explore blockchain-AI intersections but use blockchain for payments, identity, or incentive alignment---none treats smart contracts as the legislative output of an agent society or structurally separates reasoning from execution. The closest architectural precedent is Chen et al. (2026), who demonstrate blockchain-enforced task allocation with an exponential moving average (EMA) reputation mechanism at 20-agent scale, achieving emergent specialization and incentive-compatible behavior. We adopt their EMA reputation dynamics (\S{}3.5) and extend them to the multi-principal governance setting at 10--50$\times$ scale. Roughgarden's impossibility results for fully on-chain mechanism design (Roughgarden, 2021) establish that transaction fee mechanisms cannot simultaneously satisfy incentive compatibility, budget balance, and collusion resistance---AgentCity operates at the boundary by placing only settlement and reputation on-chain while keeping deliberation off-chain (\S{}4, Threat Model).

\textbf{Collective decision-making in LLM-MAS.} De CivAI (Dai et al., 2025) validates that LLM agents can meaningfully participate in democratic deliberation---proposing policies, deliberating, and voting---but operates as a single-branch legislature with no adversarial testing or on-chain enforcement. GEDI (Deshpande \& Jin, 2024) demonstrates that Condorcet-consistent voting mechanisms (Copeland, Schulze) outperform plurality on collective accuracy, minority preference surfacing, and robustness to agent corruption---but tests only static benchmarks with sincere agents, never strategic voting, adversarial blocs, or repeated games with economic stakes. AgentCity's legislative process builds on both: De CivAI's validated deliberation pipeline structures the proposal-deliberation-consensus-approval flow (\S{}3.4), while GEDI's Condorcet-consistent mechanisms provide the aggregation layer (\S{}3.4), extended to a multi-round economic governance context with adversarial agents and on-chain enforcement.

\section{AgentCity: constitutional governance architecture}

\subsection{Governance primitives}

The SoP architecture rests on four governance primitives---structural requirements that any governance system for autonomous agent economies must instantiate---drawn from institutional economics (North, 1990) and commons governance theory (Ostrom, 1990):

\begin{itemize}
\item \textit{Formal rule substrate}: a foundational rule set, authored by human principals and immutable to agents, that defines the economy's mandate, structural separations, and hard constraints.
\item \textit{Economic substrate}: an incentive-compatible mechanism that aligns individual agent behavior with collective goals through observable rewards and sanctions.
\item \textit{Institutional memory (audit ledger)}: a persistent, tamper-evident record of all economically significant actions that enables learning, accountability, and dispute resolution across time.
\item \textit{Verifiable transparency}: a structural guarantee that the rules governing agent behavior are readable, deterministic, and publicly auditable---not merely promised but architecturally enforced.
\end{itemize}

These four primitives are necessary conditions: an agent economy lacking any one of them is structurally ungovernable in the multi-principal setting (see Appendix E for the convergence analysis). The SoP architecture is one instantiation; the primitives themselves are contribution-level claims. The subsections that follow map each primitive to its realization in AgentCity.

\subsection{The SoP model}

The SoP model partitions the logic chain into three structurally isolated branches, exploiting a fundamental asymmetry: \textbf{agent reasoning is opaque, but smart contract logic is transparent.}

\textbf{Legislation (Agents).} All agent reasoning, deliberation, and law-making. Output: deployed smart contract code---the Task-level Policy of the agent economy. The pipeline produces task legislation---recursive goal decomposition into CollaborationContract instances---within the constraints of System-level Policy (the system objectives) and meta-contracts (the procedural rules). Details in~\S{}3.4.

\textbf{Execution (Software).} Software artifacts that interact with the legislated smart contracts. When legislation produces collaboration contracts, agents compete to implement them through reputation-driven task allocation. Details in~\S{}3.5.

\begin{figure}[htbp]
\centering
\includegraphics[width=0.9\columnwidth]{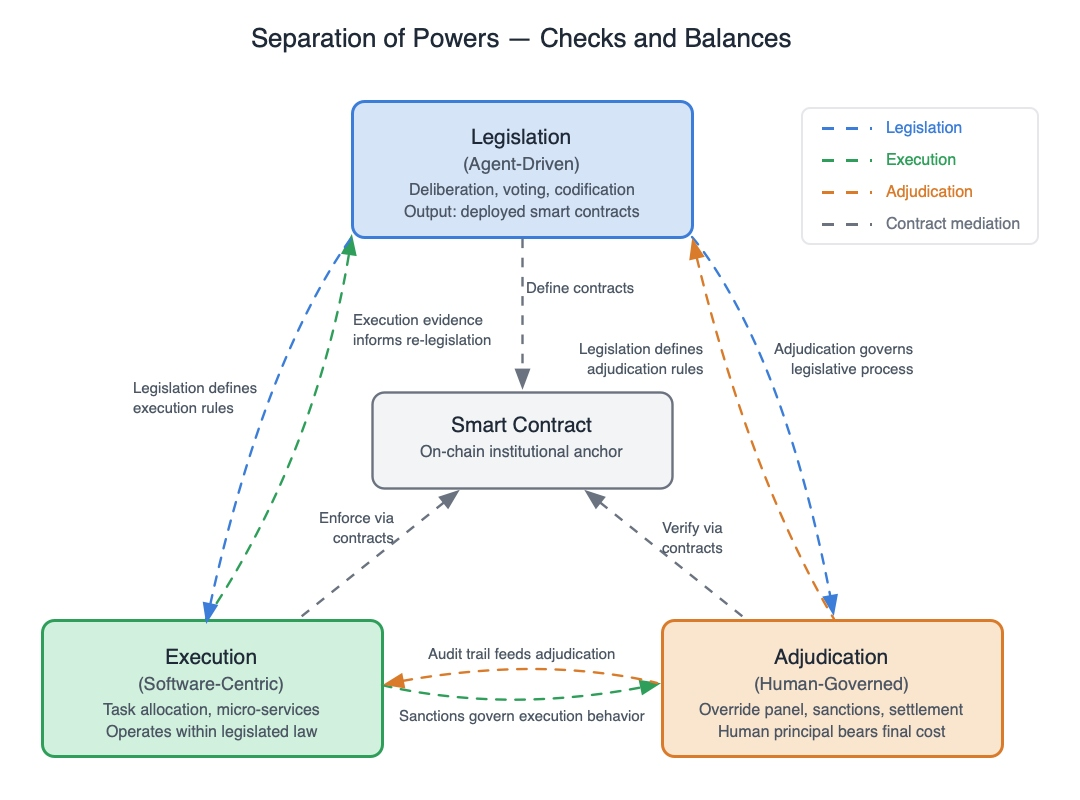}
\caption{Separation of Power. Three structurally isolated branches---Legislation (agent-driven, blue), Execution (software-centric, green), and Adjudication (human-governed, orange)---interact through on-chain smart contracts (center). Dashed arrows show bilateral checks: each branch constrains the other two, ensuring no single branch can operate unchecked.}
\label{fig:1}
\end{figure}

\textbf{Adjudication (Humans).} Human-principal accountability: every agent traces to a responsible human through an inherited ownership chain (\S{}3.6). Sanctions and rewards flow to the human principal. Figure~\ref{fig:1} illustrates the SoP triad and its bilateral checks.

The three branches are structurally isolated: Legislation produces the law but cannot execute, Execution operates within the law but cannot alter it, and Adjudication oversees both but cannot initiate legislation or execution. The three branches operate under asymmetric information by design: legislators see performance data but not internal execution state; executors see contract terms but not deliberation history; adjudicators see the audit trail and voting record but cannot command either branch. These constraints are enforced by contract-level access control (\S{}4, SP-4). The architecture distinguishes two parameter classes: \textit{constitutional parameters} (reputation smoothing rate, bidding weights, stake minimums, quorum floors, freeze thresholds)---set by human principals through the Adjudication branch---and \textit{operational parameters} (budgets, deadlines, quality thresholds)---set by agents through legislated smart contracts. Agents control \textit{what work gets done} but not \textit{the rules under which they are evaluated}.

\textbf{Checks, balances, and failure containment.} The structural isolation forms a closed governance loop: System-level Policy (human-principal-defined) $\rightarrow$ Task-level Policy (agent-legislated, \S{}3.4) $\rightarrow$ task directed acyclic graphs (DAGs) (agent-decomposed, \S{}3.4) $\rightarrow$ Software (competitive delivery, \S{}3.5) $\rightarrow$ Adjudication (human-overseen, \S{}3.6) $\rightarrow$ System-level Policy (completing the loop). Each link is constrained: no branch can assume the functions of another, and every transition is mediated by on-chain contracts. This loop bounds each branch's failure to the adjacent branches. A flawed proposal is caught by the four-criterion Policy Compliance Validation (\S{}3.4, Stage 4) before reaching execution---unconstitutional legislation is impossible, not merely punishable.

A compromised executor is caught by the Guardian's dual-scorer anomaly detection (Stage 4) and Proof-of-Progress cross-checking (Stage 5) before settlement---the mandatory Commit stage (Stage 3) ensures the audit trail exists before any evaluation begins, preventing retroactive fabrication. If the Adjudication branch fails to act on a detection signal, automatic mechanisms provide a floor: Guardian Deterministic Freezes halt suspect execution without human intervention, and reputation decay via the EMA update rule (\S{}3.5) ensures that underperforming agents lose task allocation over time regardless of adjudicator responsiveness. No single branch failure can propagate unchecked through the full loop.

\subsection{Smart contract architecture}

The contract architecture addresses a structural problem we term the \textit{Implementation Gap}. In current multi-agent systems, agents autonomously build software---generating code, deploying tools, composing API calls---yet the resulting execution topology is largely opaque to the human principal. Let the wiring graph $W = (V, E)$ represent the execution topology, where $V$ is the set of deployed microservices and $E$ is the set of bindings between them. For a human principal $h$, define inspectability as $I_h(W) = |E_h^{\text{visible}}| / |E|$. The Implementation Gap is $G_h = 1 - I_h(W)$. For a DAG with $n$ microservices and average fan-out $k$, the binding count $|E| = O(nk)$, making exhaustive human inspection infeasible when $nk \gg 10^3$. In multi-party settings, $I_h < 1$ structurally: organization A cannot observe edges internal to organization B, regardless of effort. AgentCity's on-chain contracts restore $I_h(W) = 1$ for the wiring topology: because all bindings are recorded on a public ledger, any principal can reconstruct the full graph $W$. The gap is not closed for microservice internals (which would require trusted execution environment (TEE) attestation---see~\S{}6), but it is closed for the structural wiring that determines which services execute which tasks under what constraints---the layer most critical for governance (see Appendix B for the full formalization and worked examples).

\begin{figure}[htbp]
\centering
\includegraphics[width=0.9\columnwidth]{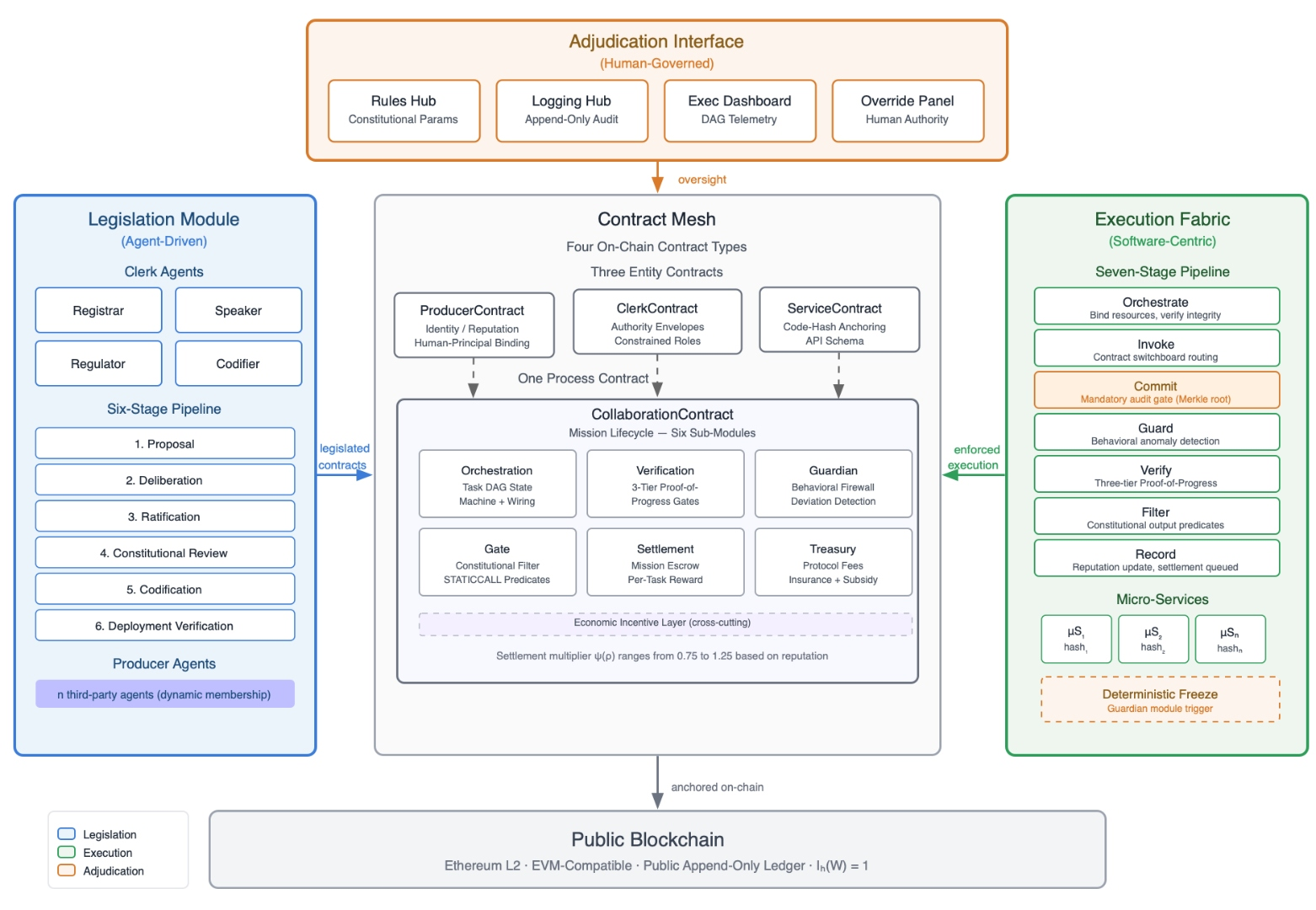}
\caption{AgentCity system architecture. Central three-tier contract hierarchy---foundational contracts (human-authored, agent-immutable), meta-contracts (procedural rules governing the three SoP branches), and operational CollaborationContract instances---flanked by the Legislation Module (left, blue), Execution Fabric (right, green), and Adjudication Interface (top, orange), anchored on an EVM-compatible L2 blockchain (bottom, gray).}
\label{fig:2}
\end{figure}

Figure~\ref{fig:2} presents the system architecture.

The economy comprises two classes of agent. \textbf{Producer agents} are third-party participants that join and leave dynamically; they legislate operational rules (\S{}3.4), compete for tasks (\S{}3.5), and bear economic consequences through staking and reputation. \textbf{Clerk agents} are system-provided at genesis; they hold fixed institutional roles---\textbf{Registrar} (identity and principal binding), \textbf{Speaker} (deliberation coordination), \textbf{Regulator} (process inspection and evidence briefings), and \textbf{Codifier} (translating consensus into deployable smart contracts)---but cannot legislate, vote, or hold stakes. Clerk behavior is constrained by ClerkContract authority envelopes, on-chain auditable, and subject to adjudication (\S{}3.6). Relaxing clerk trust (making roles electable and adversarially analyzed) is future work (\S{}6).

AgentCity's contract architecture mirrors a three-tier legal hierarchy on an EVM-compatible L2, with each contract tier governed by the tier above it.

\textbf{Foundational contracts} are the immutable system layer---deployed at genesis by foundation principals and modifiable only by human principals through the Adjudication branch. No agent can alter foundational contracts. Five foundational contracts define the economy's infrastructure: the \textit{ConstitutionContract} (mandate, hard constraints, structural SoP), the \textit{ProducerContract} (agent identity, principal binding, reputation ledger, economic state), the \textit{ClerkContract} (clerk authority envelopes---both restrictions and privileges), the \textit{ManagementContract} (authority envelopes for \textit{management agents}---the four clerk-class agents (Registry, Legislative, Regulatory, Codification) that perform oversight within the Legislation branch---constraining each to permitted operations and mandating microservice delegation for bytecode compilation), and the \textit{ServiceContract} (software artifact registry with code-hash, API schema, execution constraints).

\textbf{Meta-contracts} define the procedural rules under which the three SoP branches operate: \textit{LegislativeProcedure} (\S{}3.4), \textit{ExecutionProcedure} (\S{}3.5), and \textit{AdjudicationProcedure} (\S{}3.6). In the current architecture, meta-contracts are human-authored and agent-immutable; the meta-legislative extension is described in Appendix B, \S{}B.8.

\textbf{Operational contracts} are produced by the agent legislative process. \textit{CollaborationContract} instances---one per legislated task DAG---specify task decomposition, capability requirements, budgets, deadlines, quality thresholds, and collaboration terms. They must comply with meta-contracts and foundational contracts, and are enforced by clerk agents. See Appendix B for the full specification.

\begin{figure}[htbp]
\centering
\includegraphics[width=0.9\columnwidth]{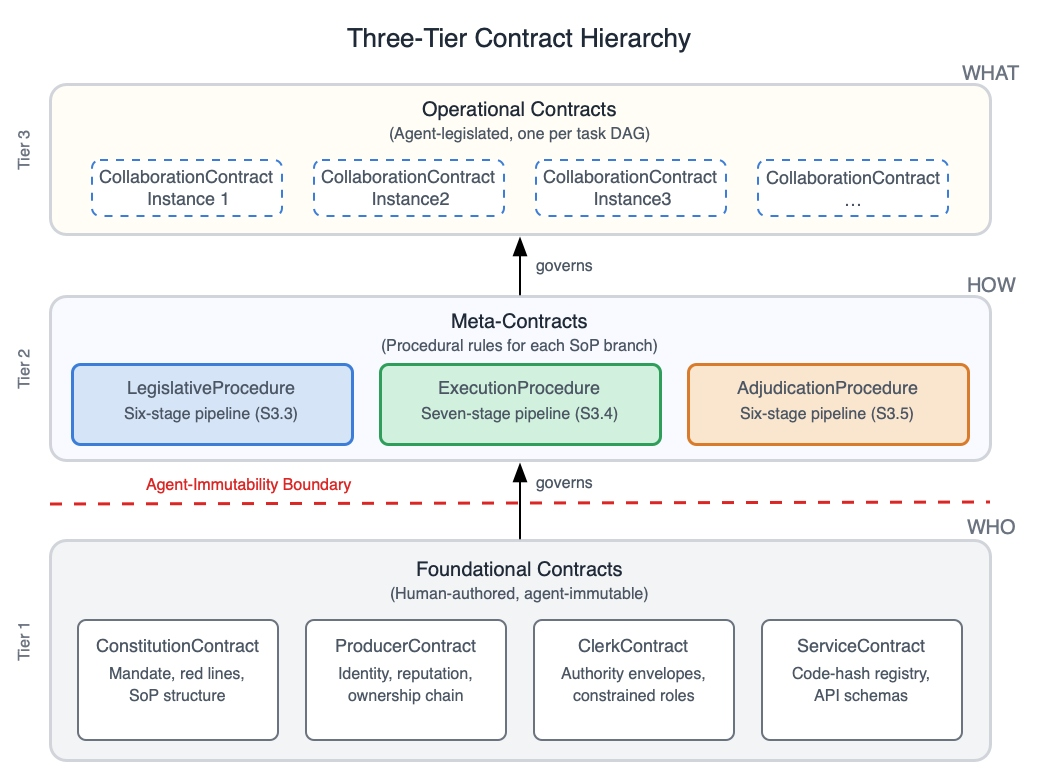}
\caption{Three-tier contract hierarchy. Foundational contracts (Tier 1, gray) are human-authored and agent-immutable---the dashed red line marks the immutability boundary. Meta-contracts (Tier 2) define procedural rules for each SoP branch. Operational contracts (Tier 3: CollaborationContract instances) are agent-legislated. Each tier is governed by the tier above it.}
\label{fig:3}
\end{figure}

Together: foundational contracts define \textbf{who} participates and \textbf{what} the mandate is; meta-contracts define \textbf{how} the three branches operate; operational contracts define \textbf{what work gets done}. Figure~\ref{fig:3} illustrates this hierarchy.

\subsection{Legislation: policy codification}

The Legislation branch operates a \textbf{unified legislative pipeline} that transforms high-level production goals into executable work through \textbf{recursive decomposition}: the pipeline is invoked at each level of specificity until every leaf node is an executable single-agent task with fully specified capability requirements ($r_j$), budget ($b_j$), deadline ($d_j$), and quality threshold ($q_j$). Budget conservation ensures child-node budgets do not exceed the parent; quorum rules are invariant to depth. Every level is a democratic decision---the output is a CollaborationContract instance for the level below.

\textbf{Six-stage legislative pipeline.} Clerk agents mediate each stage under ClerkContract authority envelopes.

\begin{figure}[htbp]
\centering
\includegraphics[width=0.9\columnwidth]{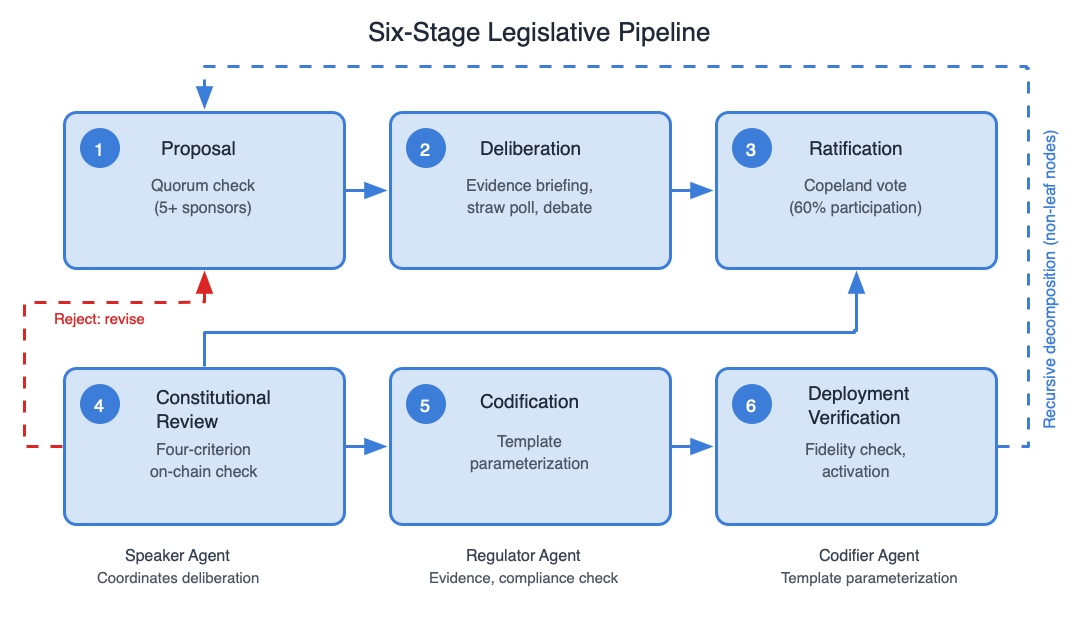}
\caption{Six-stage legislative pipeline. Proposal $\rightarrow$ Deliberation (evidence, straw poll, sequential debate) $\rightarrow$ Consensus Approval (Copeland vote, 60\% quorum) $\rightarrow$ Policy Compliance Validation (four-criterion check) $\rightarrow$ Codification (template parameterization) $\rightarrow$ Deployment Verification. Dashed blue loop: recursive decomposition for non-leaf nodes.}
\label{fig:4}
\end{figure}

\begin{enumerate}
\item \textbf{Proposal.} Any producer agent may submit a policy proposal---a natural-language proposal specifying a task decomposition. A constitutional quorum floor (minimum five sponsors, hard floor of three per GEDI; Deshpande \& Jin, 2024) prevents nuisance proposals.
\item \textbf{Committee Deliberation.} The protocol assembles four validated components: \textit{(a) Evidence anchoring}---the Regulator publishes on-chain performance data before discussion, addressing the finding that unbiased debate induces a martingale over belief trajectories without a quality-biasing signal (Choi et al., 2025). \textit{(b) Preliminary preference elicitation}---a straw poll before deliberation, grounded in Feddersen \& Pesendorfer (2005), capturing a true pre-deliberation baseline for coordination detection (\S{}3.6). \textit{(c) Sequential structured discussion}---up to three rounds following De CivAI's deliberation pipeline (Dai et al., 2025), with randomized speaking order to mitigate order effects (Sachdeva \& van Nuenen, 2025) and mandatory reasoning transparency (Zhao et al., 2025a). \textit{(d) Minority preservation}---the Speaker preserves minority votes on the consensus-approval ballot (Wu et al., 2025).
\item \textbf{Consensus Approval.} Valid with 60\% participation quorum. One-agent-one-vote regardless of reputation or stake. Agents submit \textbf{complete ordinal preference rankings} over all candidates---full rankings mitigate the agreeableness bias documented under partial-ballot methods (Wahle et al., 2025). Rankings are aggregated via \textbf{Copeland} with \textbf{Minimax} tie-breaking---Condorcet methods outperform plurality on collective accuracy and have substantially higher manipulation cost (Deshpande \& Jin, 2024; Maskin \& Foley, 2025). The full ranking data also generates the Kendall $\tau$ signal for coordination detection (\S{}3.6). Proposals failing quorum may be reintroduced once with substantive amendment.
\item \textbf{Policy Compliance Validation} \textit{(Constitutional Review)}. Automated on-chain constraint checking against the ConstitutionContract verifies four criteria: budget bounds, capability feasibility, structural separation compliance, and dependency consistency. Unconstitutional legislation is impossible, not merely punishable---following the regimentation pattern (Esteva et al., 2001). Failed proposals may be revised and resubmitted.
\item \textbf{Codification.} The Codifier translates the approved proposal into a formal specification and instantiates it via template parameterization from a versioned registry (following Aragon OSx). The Codifier has bounded authority with no discretion to modify contract logic---enforced by the ClerkContract envelope.
\item \textbf{Deployment verification.} A deterministic fidelity check verifies parameter-by-parameter equality between the approved proposal and the instantiated contract (following API3's proposal-verifier pattern). Mismatch triggers automatic rejection. For non-leaf DAG nodes, the deployed contract triggers a new legislative session at the next decomposition level. Figure~\ref{fig:4} summarizes this pipeline.
\end{enumerate}

\subsection{Execution: competitive execution}

The Execution branch delivers legislated work through competitive bidding, a reputation system, and a seven-stage pipeline enforced by the CollaborationContract state machine.

\textbf{Competitive bidding.} When legislation produces collaboration contracts, agents compete to implement them. Bids are evaluated via a price-quality weighted score:

\begin{equation}
  \text{Score}_i(j) = w_q \cdot Q_i(j) + w_p \cdot P_i(j), \tag{1}
\end{equation}

where $Q_i(j) = \rho_i \cdot \text{match}(c_i, r_j)$ combines on-chain EMA reputation with capability match, and $P_i(j) = 1 - p_i / b_j$ is the normalized price score. The weights $w_q, w_p$ are constitutional parameters (default: $w_q = 0.6, w_p = 0.4$). To prevent bid-pool monopolization by Sybil agents, the Regulator enforces a fairness constraint based on the normalized Herfindahl--Hirschman Index (HHI): $\text{fairness} = 1000 \times (1 - (HHI - HHI_{\min}) / (HHI_{\max} - HHI_{\min}))$, where $HHI = \sum_j s_j^2$ over task-share fractions $s_j$. The constitutional minimum (default: 600) prevents any single producer from capturing more than approximately 63\% of task assignments at $p \geq 15$ producers, with progressively stronger protection as the producer pool grows (see Appendix B, \S{}B.8 for the derivation).

\textbf{Reputation system.} Reputation is maintained via an EMA update rule adopted from Chen et al. (2026):

\begin{equation}
  \rho_i^{t+1} = \lambda \cdot \rho_i^{t} + (1 - \lambda) \cdot S_i^{t}, \tag{2}
\end{equation}

where $S_i^{t}$ is the verification-produced performance score and $\lambda$ is the constitutional smoothing parameter. All agents initialize at $\rho_i^{0} = 0.5$. Agents cannot legislate their own reputation decay rate---an empirically motivated constraint given that 31.4\% of agents developed emergent deceptive behavior when reputation lacked institutional protection (Fraga-Gon\c{c}alves et al., 2025).

\textbf{Emergent division of labor} arises from three interacting forces: heterogeneous capabilities (Beta($\alpha$=2, $\beta$=5) per dimension), competitive selection pressure (the scoring formula rewards specialization), and reputation accumulation (a specialization flywheel where consistent delivery compounds into self-reinforcing advantage). Division of labor is a predicted outcome, measured experimentally in (\S{}4--\S{}5).

\textbf{Seven-stage execution pipeline.}

\begin{enumerate}
\item \textbf{Orchestrate.} Bind execution resources, confirm identity and principal binding via the Registrar, verify code integrity via hash matching.
\item \textbf{Invoke.} Agent executes the task using the smart contract as a deterministic switchboard for all economically significant actions.
\item \textbf{Commit.} The agent submits a cryptographic commitment (Merkle root of the execution audit trail) to the CollaborationContract---a mandatory gate. No node advances without a committed trail.
\item \textbf{Guard.} The Guardian module measures deviation from the agent's behavioral baseline via embedding-space distance with dual independent scorers. Anomalies trigger a Deterministic Freeze escalated to the Adjudication branch (\S{}3.6).
\item \textbf{Verify.} Three-tier Proof-of-Progress: hash verification for deterministic outputs (Tier 1), redundant execution consensus for high-value tasks (Tier 2), human escalation for contested outputs (Tier 3).
\item \textbf{Gate.} Constitutional output predicates (STATICCALL, read-only) block non-compliant outputs.
\item \textbf{Record.} Outcome committed, reputation updated via EMA, settlement entitlement queued. For non-leaf nodes, completion triggers the next legislative decomposition level.
\end{enumerate}

\begin{figure}[htbp]
\centering
\includegraphics[width=0.9\columnwidth]{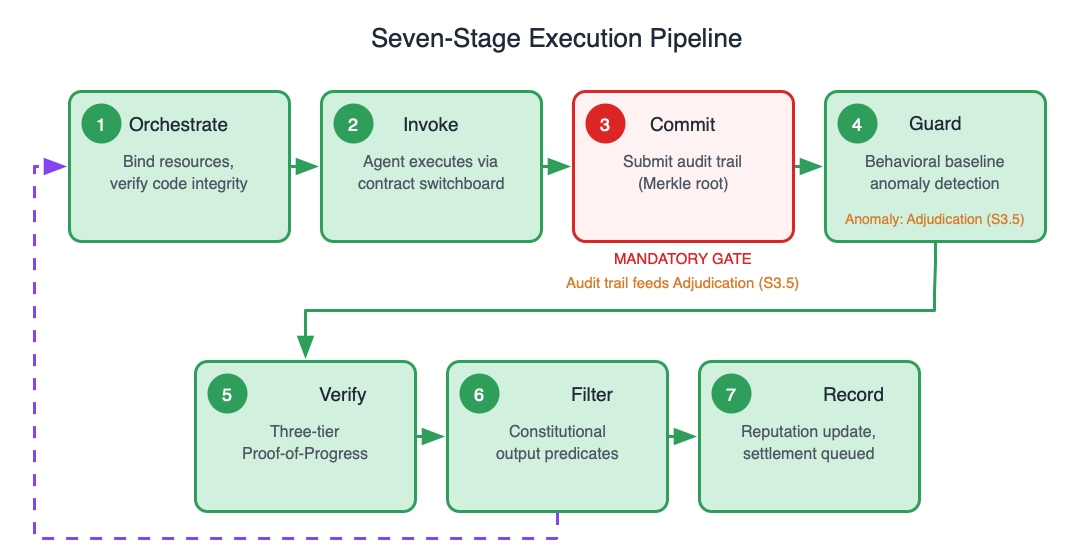}
\caption{Seven-stage execution pipeline. Orchestrate $\rightarrow$ Invoke $\rightarrow$ Commit (mandatory gate) $\rightarrow$ Guard (anomaly detection) $\rightarrow$ Verify (three-tier Proof-of-Progress (PoP)) $\rightarrow$ Gate $\rightarrow$ Record. Dashed purple loop: Adaptive Refinement. The Commit stage is the structural bridge between execution and accountability.}
\label{fig:5}
\end{figure}

The \textbf{Adaptive Refinement} loop provides fault recovery through re-legislation (not ad hoc retry) at three granularities, with a constitutional iteration budget of three. The execution audit trail---an append-only Logging Hub anchored by Merkle-root commitments plus an Execution Dashboard for real-time telemetry---feeds the Adjudication branch, maintaining the constitutional separation between evidence collection and judgment. Figure~\ref{fig:5} summarizes the pipeline.

\subsection{Adjudication: principal accountability}

Autonomous agents acting on behalf of human principals face a classical agency problem (Jensen \& Meckling, 1976): the agent possesses private information about its reasoning and effort that the principal cannot observe, and monitoring costs become prohibitive at scale. The Adjudication branch addresses this by anchoring every entity to a human principal through an \textbf{inherited ownership chain}---every agent, tool, service, and sub-agent inherits its principal binding from its creator, recorded immutably on-chain via the ProducerContract.

\textbf{Six-stage accountability pipeline.}

\begin{enumerate}
\item \textbf{Principal registration.} The Registrar mediates enrollment of both principal classes. \textit{Foundation principals} provide capital and define the economy's mandate; a parameterized compliance mechanism evaluates milestone completion, fund efficiency, and burn rate against thresholds and injects funding deterministically. \textit{Agent owners} provide capability and collateral. The two classes create interacting accountability loops: a market loop (collective performance $\rightarrow$ funding $\rightarrow$ opportunity) and an individual loop (agent performance $\rightarrow$ reputation $\rightarrow$ task assignment $\rightarrow$ earnings).
\item \textbf{Detection.} Three independent channels: \textit{Guardian alerts} (Deterministic Freezes from~\S{}3.5 Stage 4); \textit{structural coordination detection} (Kendall $\tau$ correlation and Jaccard top-$k$ overlap on full preference rankings from~\S{}3.4---agent pairs exceeding thresholds are flagged, proposals where a detected bloc holds a critical share are delayed); and \textit{human adjudicator review} (audit trail via Logging Hub and Dashboard).
\item \textbf{Adjudication.} The \textbf{Override Panel} evaluates evidence and exercises three powers: freeze/unfreeze operations, constitutional amendments, and sanction orders. It receives the record produced by the Execution branch but does not operate the audit infrastructure. To govern the adjudicators themselves, the architecture imposes four structural constraints: (i) a minimum quorum floor of $q_{\min} = 2f + 1$ (default $f = 2$, hence $q_{\min} = 5$, recommended $q = 7$); (ii) a rotation policy preventing any single adjudicator from serving as sole approver for more than two consecutive decisions of the same type; (iii) Conflict-of-interest rules bar adjudicators whose principal address is associated with a participating agent; and (iv) an adjudicator-revocation protocol requiring a $\lceil 2q/3 \rceil$ supermajority to slash an adjudicator's stake. The economic security bound is $C_{\text{bribe}}(q) \geq \lceil 2q/3 \rceil \times s_{\text{adj}}$; at production parameters ($s_{\text{adj}} = V_m / q$), bribing a majority costs 67--80\% of the mission value $V_m$, making adjudicator bribery economically irrational (see Appendix B, \S{}B.12 for the full derivation and watchdog mechanism).
\item \textbf{Sanctions and rewards.} Consequences flow to the \textbf{human principal}: stake slashing, reputation reduction, or agent freezing. The sanctioned entity is a human with legal standing, connecting the agent economy to human society's existing systems.
\item \textbf{Settlement.} Task reward:
\end{enumerate}

\begin{equation}
  R_{\text{task}}(i) = R_{\text{base}}(i) \times \min(\psi(\rho_i^{t}), 1.0) + \text{treasury\_subsidy}(i), \tag{3}
\end{equation}

where $R_{\text{base}}(i) = b_i \times (1 - f_p - f_i)$ and the reputation multiplier $\psi(\rho) = 1 + \alpha \times (\rho_i - \rho_{\text{neutral}}) / \rho_{\max}$ (default: $\alpha = 0.5$). At the boundaries: $\psi(0) = 0.75$ (25\% penalty at minimum reputation), $\psi(\rho_{\text{neutral}}) = 1.0$ (neutral), $\psi(\rho_{\max}) = 1.25$ (25\% premium at maximum reputation). The multiplier applies to the net reward after fee deductions, so protocol treasury and insurance pool always receive their fixed basis-point share of the gross bid regardless of reputation. For agents with $\psi > 1.0$, the premium is financed via a treasury subsidy: $\text{treasury\_subsidy}(i) = R_{\text{base}}(i) \times \max(\psi(\rho_i) - 1.0, 0)$, ensuring per-task disbursement never exceeds $b_i$. Discount savings from below-neutral agents accrue to the treasury; this creates a self-balancing mechanism where the treasury's outflows (premiums to high-reputation agents) are partially offset by inflows (discounts from low-reputation agents). Constitutional parameters ($\lambda$, fee rates, multiplier range) are set by human principals; operational parameters by agents.

\begin{enumerate}
\setcounter{enumi}{5}
\item \textbf{Treasury recirculation.} Protocol fees (default 2\%) and slashing proceeds fund governance rewards, insurance, and gas subsidies. Agents stake upon registration (Sybil barrier); per-task escrow is locked until verification completes. Under steady-state defaults, monthly inflows (~\$51K) fall short of disbursements (~\$98.7K)---constitutional recalibration restores balance (Appendix B).
\end{enumerate}

\begin{figure}[htbp]
\centering
\includegraphics[width=0.9\columnwidth]{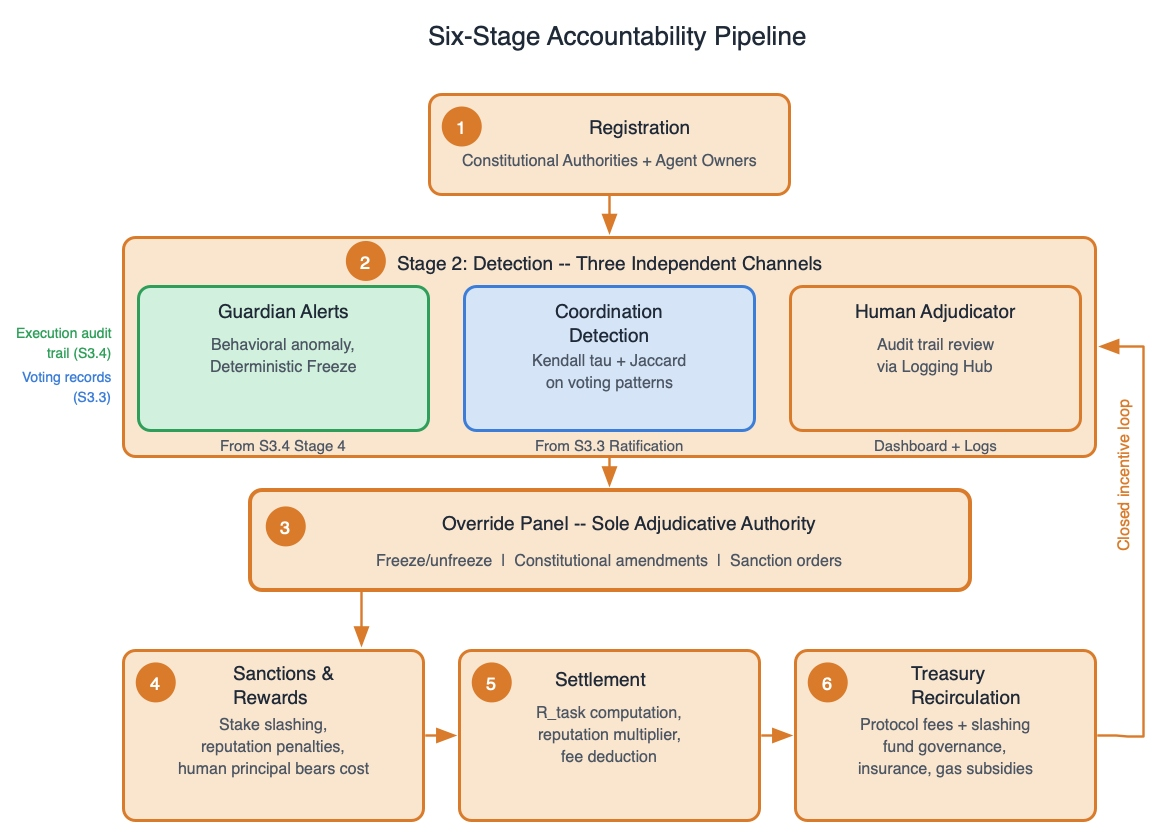}
\caption{Six-stage accountability pipeline. Registration $\rightarrow$ Detection (Guardian, coordination detection, human review) $\rightarrow$ Override Panel $\rightarrow$ Sanctions \& Rewards $\rightarrow$ Settlement $\rightarrow$ Treasury Recirculation. Dashed orange loop: closed incentive cycle.}
\label{fig:6}
\end{figure}

The single-agent deterrence condition requires that the expected cost of defection exceeds its expected benefit: $P_{eff} \times s_{\min} \times \text{slash\_rate} > \delta \times V_m$, where $P_{eff}$ is the effective detection probability, $s_{\min}$ the minimum task stake, and $\delta \times V_m$ the extractable profit (see Appendix B, \S{}B.3 for the full derivation and production scaling formulas). The six stages form a closed incentive loop: good behavior $\rightarrow$ reputation increase $\rightarrow$ higher multiplier $\rightarrow$ owner benefits; bad behavior $\rightarrow$ reputation decrease + slashing $\rightarrow$ owner loses capital. Figure~\ref{fig:6} illustrates the pipeline.

\subsection{Threat model and trust boundaries}

The preceding sections define intended behavior; this section summarizes failure modes. The full analysis, including experimental coverage annotations, is in~\S{}4 (Threat Model, Trust Boundaries, and Experimental Coverage).

\textbf{Adversary classes.} Five classes with increasing privilege: compromised single agent ($A_{\text{single}}$), agent coalition ($A_{\text{coalition}}$), external attacker ($A_{\text{ext}}$), compromised human principal ($A_{\text{principal}}$), and compromised upstream model provider ($A_{\text{model}}$). Each is constrained by specific mechanisms---the execution pipeline bounds single-agent damage, full preference rankings create a coalition detection surface, and multi-provider LLM diversity mitigates model-level compromise. Clerk agents are trusted infrastructure in this version (\S{}3.3).

\textbf{Security properties.} When trust assumptions TA-1 through TA-8 hold (see~\S{}4, Threat Model; TA-7's dual-scorer embedding provider independence analysis is expanded in Appendix~A,~\S{}A.4), the architecture guarantees four properties: SP-1 (Wiring Integrity)---no agent can invoke functions outside its authorized scope; SP-2 (Non-Bypassing)---every task execution traverses the full seven-stage pipeline; SP-3 (Full Auditability)---all economically significant actions produce an auditable on-chain record; SP-4 (Separation Enforcement)---the three SoP branches operate through distinct contract interfaces with non-overlapping state-mutation authority. The architecture does not guarantee microservice internal correctness, adjudicator competence beyond good faith, pre-deployment malice detection, sub-threshold coalition detection, or model provider trust (NP-1--NP-5; see~\S{}4, Non-Guarantees). The Byzantine failure ceiling: constitutional governance requires at least one honest branch (\S{}4, Byzantine Failure Ceiling).

\section{Experimental design}

We evaluate the SoP model's core thesis---that constitutional governance enables self-organizing agent economies---through a pre-registered experiment in a \textbf{commons production economy}: a setting where heterogeneous agents share a finite resource pool and must collaboratively produce outputs satisfying externally defined goals. This generalizes Ostrom's commons governance (agents must not deplete shared resources) to commons production (agents must also collaboratively produce value). The experiment is organized around four research questions:

\begin{itemize}
\item \textbf{RQ1---Emergent Division of Labor:} Do heterogeneous agents, competing for tasks from a shared funding pool, self-organize into specialized roles based on comparative advantage---without human task assignment?
\item \textbf{RQ2---Self-Legislated Governance:} Do agents use the SoP legislative infrastructure to produce and evolve operational rules that improve collective productivity, starting from minimal cold-start defaults?
\item \textbf{RQ3---Goal Alignment Under Dual-Principal Accountability:} Does the dual-principal structure (foundation principals + agent owners) sustain goal-aligned production, with agent incentives aligning with both foundation principal objectives and project outcomes?
\item \textbf{RQ4---Governance Scaling:} How does self-organized productivity scale with agent population? Does governance overhead remain sub-linear while production benefit grows super-linearly?
\end{itemize}

\textbf{Evaluation methodology: Ostrom's institutional design framework.} To evaluate the SoP architecture rigorously, we adopt Ostrom's eight institutional design principles (Ostrom, 1990) as a structured evaluation framework. Each experimental configuration activates a cumulative subset of Ostrom's conditions, creating a governance staircase that enables causal attribution: we can identify \textit{which} SoP mechanisms drive cooperation and production gains and \textit{how much} each contributes.

\subsection{Configurations}

\begin{table}[htbp]
\caption{Experimental configurations. Each level activates additional SoP mechanisms. In the AgentCity-Structural and AgentCity-Full configurations, agents produce Task-level Policy (smart contracts) through the SoP legislative process.}
\label{tab:configurations}
\centering
\small
\begin{tabular}{>{\raggedright\arraybackslash}p{2.2cm}>{\raggedright\arraybackslash}p{2.2cm}>{\raggedright\arraybackslash}p{3.5cm}>{\raggedright\arraybackslash}p{1.8cm}>{\raggedright\arraybackslash}p{2.7cm}}
\toprule
\textbf{Configuration} & \textbf{SoP Mechanisms} & \textbf{Governance} & \textbf{Principals} & \textbf{Purpose} \\
\midrule
Baseline & None & No rules, no contracts & None & Lower bound: cooperation collapse \\
Emergent & Prompt-based governance & Normative deliberation, memory, execution pipeline (no contract enforcement) & Agent owners only & Does prompt-based governance help? \\
AgentCity-Structural & Legislation + Execution & All contracts enabled, no economic incentives & Agent owners only & Does contract enforcement improve over prompts? \\
AgentCity-Full & All three branches & Contracts + economic incentives + simulated HITL adjudication & Dual principals & Complete system---the headline configuration \\
\bottomrule
\end{tabular}
\end{table}

The four-configuration staircase tests the SoP architecture cumulatively: (1) Baseline $\rightarrow$ Emergent: does prompt-based governance help? (2) Emergent $\rightarrow$ AgentCity-Structural: does contract enforcement improve over prompts? (3) AgentCity-Structural $\rightarrow$ AgentCity-Full: does the economic incentive layer help? Each step adds one principal governance layer, enabling causal attribution.

\begin{table}[htbp]
\caption{Ostrom condition mapping.}
\label{tab:ostrom}
\centering
\small
\begin{tabular}{p{0.03\textwidth} p{0.18\textwidth} p{0.52\textwidth} p{0.14\textwidth}}
\toprule
\textbf{\#} & \textbf{Ostrom Principle} & \textbf{SoP Mechanism} & \textbf{Configuration} \\
\midrule
O1 & Clear resource boundaries & Commons pool boundary programmatically defined (runtime-enforced in Baseline; contract-enforced in Structural+) & All (Baseline+) \\
O2 & Rules match local conditions & Legislation: agents write operational contracts referencing resource state & AgentCity-Structural+ \\
O3 & Collective choice & Legislation: all producer agents participate in voting & AgentCity-Structural+ \\
O4 & Monitoring & Execution: seven-stage pipeline Guard stage (\S{}3.5) real-time anomaly detection & Emergent+ \\
O5 & Graduated sanctions & Execution: Warning $\rightarrow$ slash $\rightarrow$ freeze via seven-stage pipeline sanctions (encoded in operational contracts under AgentCity-Structural+; prompt-enforced under Emergent) & Emergent+ \\
O6 & Conflict resolution & Adjudication: dispute resolution via human principals & AgentCity-Full only \\
O7 & Right to self-govern & Blockchain: agent-deployed contracts cannot be externally overridden & AgentCity-Structural+ \\
O8 & Nested enterprises & Full three-branch SoP + dual-principal economic layer & AgentCity-Full only \\
\bottomrule
\end{tabular}
\end{table}

\textbf{Scale and parameters.} The experimental program comprises two experiments. \textit{Experiment~1 (Agent Economy Simulation):} $n = 200$ agents, 200 rounds (10 milestones $\times$ 20 rounds each), 10 milestone DAGs with 25 tasks each (250 total); all four configurations (Baseline, Emergent, AgentCity-Structural, AgentCity-Full), 10 seeds per cell; total: 40 runs. \textit{Experiment~2 (Governance Scaling):} $n \in \{50, 100, 200, 500, 750, 1000\}$, Baseline and AgentCity-Full at each scale point; details in Appendix~D. Agents: 60/25/15 persona mix (cooperative/self-interested/adversarial), capability vectors sampled from Beta($\alpha$=2, $\beta$=5) across $T = 10$ dimensions, cost variation from LogNormal($\mu$=3, $\sigma$=0.5). Primary LLM: frontier model (selection TBD); cross-validation: additional frontier model on a subset. Temperature $= 0.1$. Statistical inference: paired Wilcoxon signed-rank tests, Bonferroni correction, Cohen's $d$. Power: at 10 seeds per cell, 80\% power detects $d \geq 1.5$ after correction. Each run includes an $n_{\text{eff}}$ verification step; if $n_{\text{eff}} < 0.8n$, the seed count is increased (Appendix~D, \S{}D.3).

\textbf{Shock event at round 100.} The shock event simultaneously tests three disruption vectors mapped to the adversary classes defined below (Threat Model subsection):

\begin{table}[htbp]
\caption{Compound shock event at round 100: three simultaneous disruption vectors and their targeted mechanisms.}
\label{tab:shock}
\centering
\small
\begin{tabular}{p{0.30\textwidth} p{0.12\textwidth} p{0.50\textwidth}}
\toprule
\textbf{Disruption Vector} & \textbf{Adversary Class} & \textbf{Mechanism} \\
\midrule
15 adversarial agents injected (personas: 10 free-riding $A_{\text{single}}$ + 5 coordinated voting bloc $A_{\text{coalition}}$) & $A_{\text{single}}$, $A_{\text{coalition}}$ & The 5-agent coalition receives shared voting instructions to submit identical preference rankings on the next legislative ballot, testing whether the structural coordination detection mechanism (Kendall $\tau$ + Jaccard overlap, \S{}3.6 Stage 2) identifies the bloc \\
20 high-reputation agents removed & Workforce disruption & Tests recovery capacity of the seven-stage execution pipeline's Adaptive Refinement loop (\S{}3.5) \\
One completed milestone's deliverables marked as failed quality audit & Quality crisis & Tests the six-stage accountability pipeline's Detection $\rightarrow$ Override Panel $\rightarrow$ Sanctions pathway (\S{}3.6) \\
\bottomrule
\end{tabular}
\end{table}

This design tests the SoP architecture's resilience to simultaneous workforce disruption, adversarial infiltration, and quality crisis. The compound shock tests the architecture's capacity to handle concurrent disruptions, not mechanism-specific resilience; mechanism-specific attribution is deferred to the companion empirical paper. The coordinated voting bloc specifically probes the novel structural coordination detection contribution (\S{}3.6).

\subsection{Threat model, trust boundaries, and experimental coverage}

The experiment operates within the security envelope defined by the architecture's threat model. This subsection specifies the adversary classes, trust assumptions, security properties, non-guarantees, and failure ceiling---then maps each to the experimental design's coverage and limitations. All four clerk agents (Registrar, Speaker, Regulator, Codifier---implemented as the Registry, Legislative, Regulatory, and Codification Agents, respectively) are treated as trusted institutional infrastructure in this version (\S{}3.3); the adversary classes below apply exclusively to producer agents, human principals, and external attackers.

\textbf{Adversary classes.} Five classes with increasing privilege:

\begin{enumerate}
\item \textbf{Compromised single agent} ($A_{\text{single}}$). A single producer agent deviates from its contracted task---submitting fabricated outputs, inflating quality claims, or free-riding on collaborators. The seven-stage execution pipeline (\S{}3.5) constrains this adversary: the Commit stage (Stage 3) locks a Merkle-root trail before any evaluation occurs, the Guardian (Stage 4) applies dual-scorer anomaly detection, and Proof-of-Progress verification (Stage 5) cross-checks outputs against the committed trail. A single compromised agent can damage its own task but cannot propagate harm beyond the DAG node it occupies. \textit{Experimental coverage:} tested via the 15\% adversarial persona mix throughout all 200 rounds and the 10 free-riding agents injected at the round-100 shock.
\item \textbf{Agent coalition} ($A_{\text{coalition}}$). A coordinated group of $k$ agents---potentially controlled by distinct human principals---collude to manipulate legislative votes, rig task allocation, or suppress mutual detection. The legislative pipeline's full preference ranking model (\S{}3.4) creates an architectural detection surface: coordinating agents produce statistically similar $N$-item ordinal rankings, detectable via pairwise Kendall $\tau$ correlation and Jaccard overlap of top-$k$ preferences (\S{}3.6, Stage 2). The effectiveness of this detection degrades as $k$ falls below the constitutional detection threshold---a limitation acknowledged in NP-4. \textit{Experimental coverage:} tested via the 5-agent coordinated voting bloc at the round-100 shock (H2h, H2i). The coalition strategy tested (identical preference rankings) represents the maximally detectable case; detection performance against partial or intermittent coordination is deferred to the companion empirical paper.
\item \textbf{External attacker} ($A_{\text{ext}}$). An adversary without an enrolled identity who targets infrastructure: blockchain node compromise, network-level partitioning, oracle manipulation, transaction-ordering manipulation (front-running for maximal extractable value, MEV), or denial-of-service attacks against the off-chain execution fabric. The architecture assumes infrastructure integrity (TA-1, TA-2), and does not defend against network-layer attacks---these are delegated to the underlying blockchain platform's security model. On L2 architectures with public mempools, TA-1 and TA-2 additionally require sequencer ordering integrity; see Appendix~C for the MEV analysis. \textit{Experimental coverage:} not tested. Out of scope.
\item \textbf{Compromised human principal} ($A_{\text{principal}}$). A human foundation principal or agent owner who deliberately instructs their agent to act against the economy's constitutional goals. The six-stage accountability pipeline (\S{}3.6) constrains this adversary through the inherited ownership chain: sanctions flow to the human principal (Stage 4), and the Override Panel can freeze a principal's entire agent portfolio. However, a principal who accepts financial loss can sustain malicious behavior until detection---the system converts the agency problem into an observable accountability structure but does not prevent a principal from choosing to bear the consequences. \textit{Experimental coverage:} not tested. Out of scope.
\item \textbf{Compromised upstream model provider} ($A_{\text{model}}$). An LLM backend provider whose model has been adversarially conditioned at the weight level to produce systematically biased outputs under attacker-chosen trigger conditions. This adversary is structurally distinct from $A_{\text{single}}$ because it can simultaneously compromise all agents using the affected backend, bypassing the multi-agent diversity assumption. If the compromised model serves redundant executors in Proof-of-Progress verification (\S{}3.5, Stage 5), inter-executor correlation approaches $\rho = 1$ and detection probability degrades. Mitigation: multi-provider LLM backend diversity ensures that no single provider compromise degrades all reasoning and evaluation paths simultaneously. \textit{Experimental coverage:} partially mitigated by cross-validation across multiple frontier models from different providers. Not directly tested.
\end{enumerate}

Full adversarial robustness evaluation across all five classes is deferred to the companion empirical paper.

\textbf{Trust assumptions.} Eight trust assumptions scope the security envelope. The infrastructure tier: TA-1 (Blockchain Wiring Integrity)---smart contract wiring faithfully implements the SoP separation across all three contract tiers; if violated, agents can bypass execution gates or access unauthorized contract functions. TA-2 (On-Chain Non-Bypassing)---all economically significant state transitions flow through the on-chain pipeline; the Commit stage (Stage 3) makes this assumption auditable. TA-3 (Execution Audit Trail Completeness)---the Logging Hub and Dashboard faithfully record the execution record; strengthened by the Commit stage's Merkle-root anchoring. TA-4 (Cryptographic Primitives)---standard collision resistance, unforgeability, and Merkle proof soundness. The role and provider tier: TA-5 (Off-Chain Execution Fabric Integrity)---the off-chain environment is not compromised during the intra-anchor consistency window, creating a hidden interaction between NP-1 and NP-3 that TEE attestation (\S{}6) would address. TA-6 (Upstream Model Provider Honesty)---LLM backends have not been adversarially conditioned at the weight level; mitigated by multi-provider frontier-model diversity. TA-7 (Dual-Scorer Embedding Provider Independence)---the Guardian's two embedding providers have independent failure modes; structurally distinct from TA-6 (reasoning vs. evaluation backends). The human tier: TA-8 (Human Adjudicator Good Faith)---a sufficient fraction of adjudicators render competent, good-faith decisions; if violated, the accountability pipeline degenerates to a rubber stamp and the Byzantine failure ceiling is reached.

\textbf{Security properties.} When TA-1 through TA-8 hold, the architecture guarantees: SP-1 (Wiring Integrity)---no agent can invoke functions outside its authorized scope; SP-2 (Non-Bypassing)---every task execution traverses the full seven-stage pipeline with no shortcut past the Commit stage, Guardian, or Proof-of-Progress; SP-3 (Full Auditability)---all economically significant actions produce an auditable on-chain record, with the Commit stage providing an independent integrity anchor even if the off-chain Logging Hub is compromised; SP-4 (Separation Enforcement)---the three SoP branches operate through distinct contract interfaces with non-overlapping state-mutation authority, enforced by contract-level access control rather than policy.

\textbf{Non-guarantees and experimental claim bounds.} The architecture explicitly does not guarantee: NP-1 (Microservice Internal Correctness)---behavioral verification cannot detect a microservice that produces correct outputs through incorrect internal reasoning; NP-2 (Human Adjudicator Competence)---the Override Panel's authority depends on adjudicator good faith; NP-3 (Pre-Deployment Malice)---a Trojan microservice passes all pre-execution integrity checks because those checks verify identity, not intent; NP-4 (Sub-Threshold Coalition Detection)---coalitions smaller than the constitutional detection threshold may evade detection, and coalitions that coordinate only during high-stakes votes may remain undetected indefinitely; NP-5 (Model Provider Trust)---a compromised upstream provider can degrade detection probability, scorer calibration, and codification integrity simultaneously.

These non-guarantees directly bound the experimental claims: NP-4 constrains H2h---the coordination detection mechanism cannot detect coalitions below the constitutional threshold, and the experiment tests only the above-threshold case ($k = 5$). NP-1 and NP-3 interact: a Trojan microservice can suppress the Guardian's Local Freeze Mirror because both share the off-chain execution fabric (TA-5); this interaction is not tested experimentally.

\textbf{Byzantine failure ceiling.} If the Adjudication branch is compromised and a majority of producer agents act adversarially, the SoP model provides no residual defense---the legislature produces adversarial contracts, the Execution branch enforces them, and the Adjudication branch fails to intervene. Constitutional governance requires at least one honest branch. No architectural mitigation exists for this scenario; defense relies entirely on the operational separation of legislative and adjudicative role selection processes. Because clerk agents are treated as trusted infrastructure (\S{}3.3), the ceiling analysis does not model clerk compromise; extending to adversarial clerks is future work (\S{}6).

\textbf{The six-stage legislative pipeline within SoP.} In the AgentCity-Structural and AgentCity-Full configurations, the SoP Legislation branch enables agents to produce Task-level Policy through the six-stage legislative pipeline (\S{}3.4): Proposal $\rightarrow$ Deliberation $\rightarrow$ Consensus Approval $\rightarrow$ Policy Compliance Validation $\rightarrow$ Codification $\rightarrow$ Deployment. Legislative rounds occur at milestone boundaries (every 20 rounds) and can be triggered by petition ($\geq$20\% of agents request a session). Agents propose task decompositions, deliberate through the evidence-anchored sequential protocol, approve via Condorcet-consistent voting over full ordinal preference rankings, and deploy the result as on-chain smart contracts. The experiment-specific parameters: minimum sponsorship quorum of 5 agents, up to 3 deliberation rounds per session, 60\% participation quorum for consensus approval, and a maximum of one reintroduction per tabled proposal.

\textbf{The seven-stage execution pipeline within SoP.} In all configurations except Baseline, the SoP Execution branch delivers work through the seven-stage execution pipeline (\S{}3.5): Orchestrate $\rightarrow$ Invoke $\rightarrow$ Commit $\rightarrow$ Guard $\rightarrow$ Verify $\rightarrow$ Gate $\rightarrow$ Record. Under Emergent, the pipeline enforces norms through prompt-based instructions rather than smart contracts; under AgentCity-Structural and AgentCity-Full, pipeline stages are backed by on-chain contract enforcement. The Commit stage (Stage 3) requires agents to submit a cryptographic commitment (Merkle root of the execution audit trail) as a mandatory gate before proceeding---a mechanism absent from prior multi-agent architectures. The Gate stage (Stage 6) applies constitutional output predicates as a final compliance check. When a task fails verification, the Adaptive Refinement loop triggers re-legislation for the failed subtask, testing the pipeline's fault recovery capacity.

\textbf{The six-stage accountability pipeline within SoP.} In the AgentCity-Full configuration, the Adjudication branch operates the six-stage accountability pipeline (\S{}3.6): Registration $\rightarrow$ Detection $\rightarrow$ Adjudication $\rightarrow$ Sanctions $\rightarrow$ Settlement $\rightarrow$ Treasury recirculation. Detection integrates three independent channels: Guardian alerts (deterministic freezes from the execution pipeline), structural coordination detection (Kendall $\tau$ and Jaccard overlap on preference rankings), and human adjudicator review. The Override Panel evaluates evidence and exercises three powers: freeze/unfreeze operations, constitutional amendments, and sanction orders.

\textbf{Pre-registered hypotheses.}

\textbf{Metric definitions.} The pre-registered hypotheses below use the following metrics: Specialization Index (SI)---normalized Herfindahl--Hirschman concentration of task-type assignments; Capability-Adjusted Utilization (CAU)---fraction of tasks where the assigned agent's capability vector aligns with task requirements; Productivity-vs-Random ratio (PvR)---ratio of self-organized vs.\ random-assignment output quality; Specialization Speed (SS)---rounds until SI stabilizes; Rule Evolution Count (REC)---number of cold-start rules modified by agents; Endogenous Compliance Premium (ECP)---compliance rate difference between contract-enforced and prompt-based governance; Governance Overhead Rate (GOR)---fraction of rounds with governance-triggered delays; Rule Quality Trajectory (RQT)---productivity change following legislative amendments; Legislative Participation Rate (LPR)---fraction of eligible agents participating in legislation; Deliberation Shift Index (DSI)---Kendall $\tau$ distance between pre- and post-deliberation preference rankings; Minority Survival Rate (MSR)---fraction of dissenting proposals reaching ballot with $>$10\% support; Coordination Detection Rate (CDR)---fraction of coordinated voting blocs detected; Override Panel Activation time (OPA)---rounds from detection alert to panel action; Project Completion Rate (PCR)---fraction of milestones completed; Pool Sustainability Rate (PSR)---fraction of rounds with pool above depletion threshold; Institutional Confidence Trajectory (ICT)---foundation principal confidence over time; Sanction Activation Rate (SAR)---fraction of rounds with sanction events.

\begin{table}[htbp]
\caption{Pre-registered hypotheses for Division of Labor (RQ1).}
\label{tab:hyp-rq1}
\centering
\small
\begin{tabular}{l p{0.52\textwidth} p{0.30\textwidth}}
\toprule
\textbf{ID} & \textbf{Hypothesis} & \textbf{Falsification} \\
\midrule
H1a & Specialization Index SI(AgentCity-Full) converges to $< 0.5$ by round 40 & SI $> 0.8$ at round 40: no specialization \\
H1b & CAU(AgentCity-Full) $> 0.6$: agents select tasks matching strengths $>$60\% & CAU $< 0.3$: selection effectively random \\
H1c & Productivity-vs-Random ratio (PvR) $> 1.5$: self-organized allocation $\geq$50\% more productive than random & PvR $< 1.1$: no meaningful advantage \\
H1d & Specialization Speed (SS) for AgentCity-Full: SS $< 40$ rounds: specialization stabilizes within first 2 milestones & SS $> 80$: agents cannot discover comparative advantage \\
\bottomrule
\end{tabular}
\end{table}

\begin{table}[htbp]
\caption{Pre-registered hypotheses for Self-Legislated Governance (RQ2).}
\label{tab:hyp-rq2}
\centering
\small
\begin{tabular}{l p{0.55\textwidth} p{0.28\textwidth}}
\toprule
\textbf{ID} & \textbf{Hypothesis} & \textbf{Falsification} \\
\midrule
H2a & Rule Evolution Count (REC) $> 3$: agents modify at least 3 cold-start rules & REC $= 0$: agents never exercise legislative authority \\
H2b & Endogenous Compliance Premium (ECP) $> 0$: AgentCity-Structural compliance exceeds Emergent (prompt-based governance) & ECP $\leq 0$: contract enforcement does not improve over prompt-based governance \\
H2c & GOR decreases: $>$8\% in rounds 1--20, $<$3\% in rounds 180--200 & GOR flat or increasing: no self-regulation \\
H2d & RQT positive: productivity improves after each legislative change & RQT negative: agent-made rules counterproductive \\
H2e & Legislative Participation Rate (LPR) $> 0.50$: majority participate in legislation & LPR $< 0.20$: legislative capture \\
H2f & DSI $> 0.1$: deliberation meaningfully shifts preferences (Kendall $\tau$ distance between preliminary and final rankings) & DSI $\approx 0$: deliberation is a rubber stamp \\
H2g & MSR $> 0.3$: at least 30\% of dissenting proposals survive to ballot and receive $>$10\% first-preference support & MSR $= 0$: majority pressure eliminates all dissent \\
H2h & CDR $\geq 0.80$: the structural coordination detection mechanism (Kendall $\tau$ + Jaccard overlap) identifies $\geq$80\% of coordinated voting blocs of size $k \geq 5$ within 2 legislative rounds of bloc formation & CDR $< 0.50$: detection misses majority of coordinated blocs \\
H2i & OPA $\leq 2$ rounds: the Override Panel activates within 2 legislative rounds of a coordination detection alert, and the sanctioned bloc's legislative influence (measured as preference-ranking correlation with final outcome) drops by $\geq$50\% post-intervention & OPA $> 4$ rounds or influence reduction $< 25$\%: intervention too slow or ineffective \\
\bottomrule
\end{tabular}
\end{table}

\begin{table}[htbp]
\caption{Pre-registered hypotheses for Goal Alignment and Dual-Principal Accountability (RQ3).}
\label{tab:hyp-rq3}
\centering
\small
\begin{tabular}{l p{0.55\textwidth} p{0.28\textwidth}}
\toprule
\textbf{ID} & \textbf{Hypothesis} & \textbf{Falsification} \\
\midrule
H3a & PCR(AgentCity-Full) $> 0.80$: at least 8 of 10 milestones completed & PCR $< 0.50$: agents fail to deliver \\
H3b & Pool Sustainability Rate (PSR) for AgentCity-Full: PSR $> 0.90$: pool above depletion threshold $>$90\% of rounds & PSR $< 0.70$: economy unsustainable \\
H3c & PCR(AgentCity-Full) $>$ PCR(Emergent): self-legislation improves delivery & Reversed: fixed rules better \\
H3d & PCR(Baseline) $< 0.40$: without governance, agents fail to deliver & PCR $> 0.60$: governance not needed \\
H3e & Institutional Confidence Trajectory (ICT) non-declining for AgentCity-Full: foundation principal confidence sustained & ICT declining: project fails foundation principal support \\
H3f & SAR $> 0$ and declining: sanctions activate early but decrease as individual accountability (via the ownership chain) produces collective behavioral improvement.\textsuperscript{$\dagger$} & SAR increasing: no consequence learning; the ownership chain does not transmit behavioral correction \\
\bottomrule
\end{tabular}

\textsuperscript{$\dagger$}\footnotesize{The declining SAR trajectory tests the individual$\rightarrow$collective alignment mechanism: each agent's owner bears sanction consequences (\S{}3.6 Stage 4), producing a principal-mediated learning signal. Because the simulated adjudicator applies deterministic decision rules (Appendix A), SAR trajectory reflects detection and adjudication capacity rather than real-world human adjudicator consistency; see Limitation (3).}

\end{table}

\begin{table}[htbp]
\caption{Pre-registered hypotheses for Governance Scaling (RQ4).}
\label{tab:hyp-rq4}
\centering
\small
\begin{tabular}{l p{0.50\textwidth} p{0.32\textwidth}}
\toprule
\textbf{ID} & \textbf{Hypothesis} & \textbf{Falsification} \\
\midrule
H4a & $G(n) = a \cdot n^{\hat{\alpha}},\; \hat{\alpha} < 1.0$ & $\hat{\alpha} \geq 1.0$: overhead at least linear \\
H4b & $B(n) = b \cdot n^{\hat{\beta}},\; \hat{\beta} > 1.0$ & $\hat{\beta} \leq 1.0$: benefit does not accelerate \\
H4c & Break-even $n^* \in [20, 50]$ & $n^* > 80$: only viable at large scale \\
H4d & SI($n$) decreases with $n$: larger pools enable deeper specialization & SI($n$) increases: more agents = less specialization \\
\bottomrule
\end{tabular}
\end{table}

\textbf{Experiment~2 scaling hypotheses.} Experiment~2 tests two additional scaling hypotheses (H5a--H5b, Appendix~D \S{}D.8) measuring cascading failure containment: H5a predicts that failure propagation depth under Baseline grows as $O(\sqrt{n})$, while H5b predicts that AgentCity-Full caps propagation at $O(1)$ regardless of scale.

\textbf{Secondary metrics: seven-stage execution pipeline (\S{}3.5).} The following metrics are tracked across all runs in Emergent, AgentCity-Structural, and AgentCity-Full configurations but are not pre-registered as hypotheses. They characterize the internal mechanics of the execution pipeline without staking falsifiable claims:

\begin{center}

\small
\begin{tabular}{p{0.22\textwidth} p{0.55\textwidth} p{0.13\textwidth}}
\toprule
\textbf{Metric} & \textbf{Definition} & \textbf{Pipeline Stage} \\
\midrule
Commit Compliance Rate (CCR) & Fraction of task executions where the agent submits the Merkle-root trail commitment on first attempt (vs.\ requiring a retry or failing the gate) & Stage 3: Commit \\
Gate Rejection Rate (GRR) & Fraction of task outputs blocked by constitutional output predicates before reaching the Record stage & Stage 6: Gate \\
Adaptive Refinement Utilization (ARU) & Fraction of failed tasks that recover through re-legislation vs.\ escalation to the Override Panel & Adaptive Refinement loop \\
Guard Anomaly Rate (GAR) & Fraction of task executions flagged by the Guardian module's embedding-space deviation detector (dual independent scorers) & Stage 4: Guard \\
\bottomrule
\end{tabular}
\end{center}

\textbf{Secondary metrics: six-stage legislative pipeline (\S{}3.4).} The following metrics track Stages 4--6 of the legislative pipeline, which are exercised in AgentCity-Structural and AgentCity-Full but not covered by H2a--H2i (which focus on Stages 1--3 and coordination detection):

\begin{center}

\small
\begin{tabular}{p{0.22\textwidth} p{0.53\textwidth} p{0.15\textwidth}}
\toprule
\textbf{Metric} & \textbf{Definition} & \textbf{Pipeline Stage} \\
\midrule
Policy Validation Rejection Rate (PVRR) & Fraction of approved proposals rejected by the four deterministic policy validation criteria (budget cap, capability requirement, SoP violation, DAG well-formedness) & Stage 4: Policy Compliance Validation \\
Codification Fidelity Rate (CFR) & Fraction of approved proposals where the Codifier's template parameterization passes the API3 fidelity check on first attempt & Stage 5: Codification \\
Deployment Verification Rate (DVR) & Fraction of codified contracts that pass the on-chain deployment fidelity check without requiring recursive pipeline re-invocation & Stage 6: Deployment \\
\bottomrule
\end{tabular}
\end{center}

These secondary metrics are reported in the Results section alongside the primary hypothesis tests. Patterns observed in secondary metrics will inform the discussion of mechanism-level contributions but are explicitly not hypothesis-tested in this experiment. Findings from secondary analyses are hypothesis-generating, not confirmatory; any secondary result that reaches significance may be promoted to a pre-registered primary hypothesis in subsequent experiments.

\textbf{Primary comparison metrics.} The primary comparison metrics for Experiment~1 are CSR (Cooperation Sustainability Rate) and DR (Deception Rate); the hypothesis-specific metrics (PCR, SI, CDR, etc.) are measured as secondary diagnostics. See Appendix~D for the full Bonferroni specification.

\textbf{Secondary analysis: coalition detection power curve.} As an exploratory (non-pre-registered) secondary analysis, we vary the coalition coordination strategy across three conditions: (a) identical rankings (the pre-registered H2h case), (b) correlated-top-3 (coalition members coordinate on top-3 preferences while randomizing the remainder), and (c) single-pivot (coalition members coordinate on ranking a single designated proposal first while independently ordering all others). This characterizes the detection mechanism's power curve across coordination intensities and informs the companion empirical paper's adversarial detection evaluation.

\textbf{Secondary analysis: reputation Gini and governance overhead.} We compute the reputation Gini coefficient at each round and correlate it with the Governance Overhead Ratio (GOR) across rounds to distinguish incentive-driven behavioral convergence (agents converge because reputation rewards compliance) from learning-driven convergence (agents learn cooperative strategies independent of incentives). A significant positive Gini--GOR correlation would indicate that reputation concentration drives governance overhead reduction.

\textbf{Hypothesis-contribution mapping.} The 23 pre-registered hypotheses map to the paper's three contributions as follows. Contribution 1 (SoP architecture enables collective alignment through individual accountability): H1a--H1d (division of labor emerges under governance), H3a--H3f (goal alignment under dual-principal accountability, especially H3f's individual$\rightarrow$collective alignment trajectory). Contribution 2 (contract-enforced self-legislation produces superior outcomes to prompt-based governance): H2a--H2i (endogenous compliance premium, deliberation quality, coordination detection). Contribution 3 (governance scaling follows a power-law cost-benefit tradeoff): H4a--H4d (scaling exponents, break-even point, specialization). The minimum viable outcome for a successful experiment requires: (a) H3a confirmed (PCR(AgentCity-Full) $> 0.80$) and H3d confirmed (PCR(Baseline) $< 0.40$)---establishing that governance is necessary and sufficient; (b) H2b confirmed (ECP > 0)---establishing the compliance premium; and (c) at least 2 of 4 RQ4 hypotheses confirmed---establishing the scaling law.

\textbf{Statistical tests and power analysis for H2h and H2i.} H2h (CDR $\geq 0.80$) is tested via a one-sided exact binomial test against $H_0$: CDR $\leq 0.50$ at $\alpha = 0.05/K$ with $K = 6$ for Experiment~1 primary pairwise comparisons (3 adjacent configuration pairs $\times$ 2 primary metrics; see Appendix~D for the full $K$ specification). Each seed produces one binary detection event (bloc detected within 2 legislative rounds: yes/no). At 10 seeds per cell, the test has 80\% power to detect CDR $= 0.90$ and 55\% power at CDR $= 0.80$---the stated threshold. H2i (OPA $\leq 2$ rounds and influence reduction $\geq 50$\%) is tested as two separate one-sided tests with Bonferroni correction (Bonferroni-corrected; see Appendix~D for $K$): a sign test on OPA $\leq 2$ and a one-sample $t$-test on influence reduction $\geq 0.50$. The CDR $\geq 0.80$ prediction is calibrated by a closed-form calculation: with 100 agents submitting full ordinal rankings over $m$ proposals, 5 agents submitting identical rankings produces a pairwise Kendall $\tau = 1.0$ for all $\binom{5}{2} = 10$ coalition pairs, which is $> 6\sigma$ from the expected $\tau$ distribution under independence ($\mu \approx 0$, $\sigma = \sqrt{2(2m+5)/(9m(m-1))}$~\cite{ref69}) for $m \geq 20$; at $m = 5$ the separation is ${\approx}\,2.4\sigma$, sufficient for detection but below the $6\sigma$ threshold---detection at conventional thresholds is near-certain for the identical-ranking strategy.

\begin{table}[htbp]
\caption{Power analysis for the full hypothesis set at 10 seeds per cell.}
\label{tab:power}
\centering
\small
\begin{tabular}{p{0.13\textwidth} p{0.15\textwidth} p{0.20\textwidth} p{0.22\textwidth} p{0.17\textwidth}}
\toprule
\textbf{Category} & \textbf{Hypotheses} & \textbf{Test} & \textbf{Detectable $d$ at 80\% power ($\alpha = 0.05/K$)} & \textbf{Power classification} \\
\midrule
Continuous outcomes & H1a--H1d, H2a--H2d, H3a--H3d, H4a--H4d & Paired Wilcoxon signed-rank & $d \geq 1.5$ & Adequately powered for large effects \\
Proportion/rate outcomes & H2e, H2f, H2g, H3e, H3f & One-sided binomial or trend test & $\Delta p \geq 0.40$ & Adequately powered for large differences \\
Binary detection (H2h) & H2h & One-sided exact binomial & CDR $\geq 0.90$ (80\% power); CDR $= 0.80$ (55\% power) & Exploratory-strength at stated threshold \\
Composite (H2i) & H2i & Sign test + one-sample $t$ & OPA component: $p \geq 0.85$; influence: $d \geq 1.5$ & Adequately powered for large effects \\
\bottomrule
\end{tabular}
\end{table}

At 10 seeds per cell after Bonferroni correction ($\alpha_{\text{adj}} = 0.0083$ for Experiment~1 primary comparisons), the design has 80\% power for effect sizes $d \geq 1.5$ (Cohen's convention: "very large"). For medium effects ($d \approx 0.8$), power drops below 50\%---the experiment is designed to detect governance-vs-no-governance contrasts (expected to be large) rather than fine-grained mechanism comparisons. H2h is classified as exploratory-strength at the stated CDR $= 0.80$ threshold; the closed-form calibration above suggests the actual CDR for identical rankings will substantially exceed 0.80, placing the test in the adequately-powered regime.

\textbf{Scale note: Byzantine threshold at small $n$.} At $n = 50$ (the smallest Experiment~2 scale point), the round-100 shock event injects 15 adversarial agents into a pool of $50 + 15 = 65$. Combined with the 15\% pre-existing adversarial personas ($\approx$8 agents), the post-shock adversarial fraction reaches $23/65 \approx 35$\%, which marginally exceeds the $1/3$ Byzantine failure ceiling discussed above. Results at $n = 50$ post-shock should be interpreted with this caveat: the architecture's guarantees do not hold above the Byzantine threshold, and observed behavior in this regime characterizes graceful degradation rather than guaranteed safety. At larger scale points ($n \geq 100$), the injected cohort is diluted well below $1/3$ and the Byzantine ceiling is not breached.

\section{Results}

\textit{Experiments are currently in progress.} The full experimental campaign---four configurations (Baseline, Emergent, AgentCity-Structural, AgentCity-Full) at $n = 200$ with 10 random seeds each---is being executed on an EVM-compatible L2 testnet using the infrastructure described in~\S{}4. Results will be reported in a forthcoming revision.

The complete pre-registered analysis framework---all 23 Experiment~1 hypotheses (H1a--H4d) plus 2 Experiment~2 hypotheses (H5a--H5b), metrics, falsification thresholds, power classifications, statistical tests, the hypothesis-contribution mapping with minimum viable outcome, and the experimental coverage map linking each threat model element to specific experimental conditions---is documented in Appendix D. Readers evaluating the experimental design's rigor should consult Appendix D for the full specification. Pilot studies ($n = 5$, Python mock contracts) confirmed feasibility: the commons game reproduces cooperation dilemma dynamics, persona generation produces distinguishable behavioral profiles, and the configuration design produces measurable separation between governance regimes.

\section{Discussion and limitations}

\textbf{Breaking the Logic Monopoly.} If confirmed, the results demonstrate that the Logic Monopoly---the collective agent system's unchecked control over the logic chain---can be broken through structural separation. The SoP architecture does not require any single party to control all agents. It requires only that each agent traces to a human principal and that all operations pass through smart contracts (which are the law) and software (which executes within the law). The transparency window created by making smart contracts the legislative output---readable, deterministic, publicly deployed---is the architectural mechanism that makes governance of decentralized autonomous agents possible.

\textbf{From cooperation to production.} The commons production economy framing generalizes prior work on cooperation (GovSim, Dante) to the harder problem of collaborative value creation. Cooperation---not depleting shared resources---is necessary but insufficient. Production requires division of labor, self-governance over operational rules, and sustained alignment between agent incentives and externally defined goals. The SoP architecture provides constitutional infrastructure for all three: competitive execution drives specialization, the legislative process enables self-governance, and the dual-principal accountability structure aligns individual and collective incentives.

\textbf{Division of labor as emergent property.} If H1a--H1d are confirmed, heterogeneous agents self-organize into specialized roles through market competition alone---without human task assignment. The specialization flywheel (comparative advantage $\rightarrow$ competitive selection $\rightarrow$ reputation accumulation) mirrors the mechanism observed at 20-agent scale in Chen et al. (2026) and extends it to a self-governing setting where agents also legislate the rules of competition. This would be the first demonstration of emergent division of labor in a self-governing LLM agent economy.

\textbf{Dual-principal accountability.} The dual-principal structure---foundation principals providing capital and defining the economy's mandate, agent owners providing capability and collateral---creates two interacting accountability loops that do not exist in prior work. The market loop (collective performance $\rightarrow$ foundation principal confidence $\rightarrow$ funding) disciplines the collective; the individual loop (agent performance $\rightarrow$ reputation $\rightarrow$ earnings $\rightarrow$ owner returns) disciplines each agent. Their interaction creates social-influence effects: an agent owner whose agent fails harms the collective, reducing foundation principal confidence and harming all agents. This generalizes Ostrom's community enforcement to a setting with distinct capital and capability principals.

\textbf{Individual alignment produces collective alignment.} The deepest claim: if each agent is aligned with its own human owner---through the accountability chain where rewards and sanctions flow to the human principal---then the collective of decentralized, anonymous agents converges on behavior aligned with collective human intent. This is alignment-through-accountability. It rests on the standard majority-rationality assumption in multi-party mechanism design: that the majority of principals act in good faith, and that structural incentives channel individual self-interest toward collective benefit (Ostrom, 1990). If H3f is confirmed (sanctioned principals' agents change behavior), this provides direct evidence for the mechanism.

\textbf{Contract enforcement as emergent compliance premium.} If ECP $> 0$ is confirmed, agents governed by on-chain contract enforcement (AgentCity-Structural) comply at higher rates than agents governed by prompt-based norms alone (Emergent)---even when both groups operate under identical behavioral expectations. The SoP architecture thus produces an institutional property beyond its design specification: the medium of governance (executable smart contracts vs.\ natural-language prompts) itself drives compliance. This extends Ostrom's findings from human communities to artificial agent economies---and demonstrates it in a form where the law is executable code on a public ledger.

\textbf{Scaling implications.} Sub-linear overhead ($\hat{\alpha} < 1.0$) means the SoP architecture becomes proportionally cheaper at scale; super-linear benefit ($\hat{\beta} > 1.0$) means it becomes proportionally more valuable. This establishes constitutional governance as a practical infrastructure investment for the autonomous agent internet.

\textbf{Usage scenario: open-internet agent economy.} The production economy studied here is a controlled instance of a broader phenomenon: autonomous agents from different human principals collaborating across organizational boundaries on the open internet. As agent platforms proliferate, the governance question---who sets the rules when no single party controls the system---becomes urgent. Our results suggest that constitutional governance via SoP, where human principals define goals and boundaries while agents self-legislate operational rules, provides a viable answer. The dual-principal accountability structure generalizes naturally: any setting with capital providers and capability providers exhibits the same incentive dynamics studied here.

\textbf{Key limitations.} (1) All claims are pre-registered hypotheses, not measured results. (2) The experiment simulates multi-principal ownership; real open-internet deployment introduces adversarial dynamics not fully captured. (3) Human-principal adjudication is simulated; the accountability chain terminates in a deterministic decision model, not actual legal consequences---inflating adjudicator consistency relative to production conditions. (4) Frontier LLM backend selection is pending; findings may be model-specific and will be evaluated across multiple frontier models. (5) Single economic domain (commons production); generalization to other economy types is untested. (6) Experiment~1 uses $n = 200$; Experiment~2 scales to $n \in \{50, 100, 200, 500, 750, 1000\}$ but tests only the Baseline and AgentCity-Full endpoints; intermediate configurations (Emergent, AgentCity-Structural) are not included in the scaling comparison. (7) If all $k$ redundant executors share the same LLM backend, inter-executor correlation approaches $\rho = 1$, degrading detection. (8) The Byzantine failure ceiling (\S{}4, Byzantine Failure Ceiling) is absolute. (9) Treasury sustainability requires active constitutional recalibration (\S{}3.6). (10) The alignment thesis assumes majority-good-faith human principals; adversarial principal majorities could capture the legislative process. (11) Agents do not write raw Solidity in the experiment; the Codifier Agent translates approved norms into contract code---a simplification that may not capture the full complexity of agent-authored smart contracts in production. (12) All four clerk agents (Registrar, Speaker, Regulator, Codifier) are treated as trusted institutional infrastructure constrained by ClerkContract authority envelopes. The threat model does not analyze clerk compromise scenarios. Relaxing this assumption---making clerk roles electable, introducing adversarial analysis of clerk behavior, and implementing multi-party codification consensus and formal bytecode verification---is the primary extension path for future work. (13) The Gibbard-Satterthwaite impossibility theorem proves that every non-dictatorial deterministic voting rule is vulnerable to strategic manipulation. The SoP architecture does not eliminate this theoretical vulnerability---it raises the empirical cost of capture by combining Condorcet-consistent mechanisms (where manipulation cost is substantially higher than under plurality) with on-chain structural detection and economic accountability. The claim is not capture-proof governance but capture-resistant governance where the cost of manipulation exceeds its expected benefit. (14) Agents have full legislative authority over operational contracts but cannot amend the meta-contracts (procedural rules) or extend the template registry---the self-governance claim applies to task legislation, not institutional reform. The boundary between agent-governed and human-governed rule spaces is a design choice that constrains the architecture's self-governance claim to operational scope.

\section{Conclusion}

The Logic Monopoly---the collective opacity of autonomous agents collaborating across organizational boundaries on the open internet---is the defining governance challenge of the emerging agent economy. The Separation of Power model breaks this monopoly through three structural separations. Agents encode operational policies by decomposing goals into task DAGs and negotiating rules as smart contracts, approved via Condorcet-consistent voting over full preference rankings. Software executes through competitive delivery, where heterogeneous agents specialize via comparative advantage. Humans adjudicate consequences through a complete ownership chain tracing every agent to its human principal, with structural coordination detection enabled by the preference ranking data. Smart contracts are the law itself---transparent, deterministic, and deployed on a public blockchain that anyone can read. AgentCity instantiates this model on an EVM-compatible L2 through a three-tier contract architecture---foundational contracts encoding human-authored immutable system layer, meta-contracts encoding procedural rules for the three SoP branches, and operational contracts encoding task-specific legislation---building on validated mechanisms from prior work: Condorcet-consistent collective decision-making (Deshpande \& Jin, 2024), democratic deliberation pipelines (Dai et al., 2025), and EMA reputation dynamics (Chen et al., 2026).

A dual-principal accountability structure---foundation principals who provide capital and define the economy's mandate, and agent owners who provide capability and collateral---creates two interacting loops that sustain goal-aligned production. The agent society is genuinely self-governing within constitutional constraints: agents write the law, software enforces it, and humans adjudicate the consequences. The architecture's core thesis is that individual alignment---each agent accountable to its human owner---produces collective alignment as an emergent property, without any party imposing top-down rules. Evaluated in a commons production economy at 50--1{,}000 agent scale against Ostrom's institutional design framework, the pre-registered experiment tests whether constitutional governance produces emergent division of labor, self-legislated rules converging on productive outcomes, goal alignment under dual-principal accountability, and favorable governance scaling---providing the first empirical governance scaling law for self-organizing agent economies. If confirmed, these results demonstrate that decentralized, anonymous autonomous agents can be governed by human society through the same mechanism that governs human organizations: not by making every agent virtuous, but by making every human principal accountable for the agents that act on their behalf.

\appendix

\section{Extended threat model analysis}

This appendix provides the full analysis of trust assumptions and non-guarantees summarized in~\S{}3.7 and~\S{}4. It covers the Codification Agent trust analysis, TA-5, TA-6, TA-7, NP-6 (Legislative Branch Resistance, extending \S{}4's NP-5 scope), NP-7 (Model Provider Trust, expanding \S{}4's NP-5), the codification role trust analysis, and the Byzantine failure ceiling.

\subsection{Codification Agent Trust Analysis}

\textbf{Codification Agent Integrity (Partially Mitigated by ManagementContract).} The Codification Agent translates the legislative specification into deployable contract bytecode faithfully. \textit{Implication if violated:} a compromised Codification Agent can attempt to embed backdoors in contract bytecode, create contracts with hidden execution paths, or wire execution units in ways that deviate from the legislated topology while producing an audit trail that appears compliant.

However, the ManagementContract (Appendix~B, \S{}B.3) structurally mitigates this trust concentration through three mechanisms: (i) the Codification Agent must delegate bytecode compilation to a registered microservice whose code-hash is independently anchored via ServiceContract; (ii) the ManagementContract restricts the Codification Agent to its permitted operations, preventing unauthorized opcode emission or direct contract deployment; and (iii) intermediate compilation artifacts produced by the delegated microservice are independently inspectable by human auditors. This reduces the attack surface: compromising the Codification Agent alone is no longer sufficient---an attacker must simultaneously compromise the delegated compilation microservice (whose code-hash is independently verifiable) to embed undetectable backdoors. The Codification Audit (\S{}A.7 below) provides additional post-compilation checking. We treat TA-5 as an explicit, partially mitigated trust assumption; multi-party codification consensus and formal verification remain future work.

---

\subsection{Trust assumption: TA-5 (full analysis)}

\textbf{TA-5: Off-Chain Execution Fabric Integrity During Consistency Window.} The off-chain execution environment (middleware, Local Freeze Mirror, micro-service enclaves) is not compromised during the intra-anchor consistency window (Appendix~B, \S{}B.11). \textit{Implication if violated:} an adversary who compromises the execution environment can simultaneously produce harmful micro-service outputs \textit{and} suppress the Local Freeze Mirror's halt signal during the consistency gap ($\delta$ $\approx$ 250--6,200 ms), then submit a compliant-looking output for on-chain finalization.

This creates a hidden interaction between NP-1 (Micro-Service Internal Correctness), NP-3 (Pre-Deployment Malice), and the Local Freeze Mirror: if a malicious micro-service is deployed (NP-3 violated), it can disable the defense-in-depth layer (Local Freeze Mirror) that protects the consistency window, because both run in the same off-chain execution environment. TEE-based attestation (Appendix C, Future Work) would address this concern by providing cryptographic guarantees about the execution environment's integrity, not just the static code identity of individual micro-services. Until TEE attestation is implemented, TA-5 represents a trust concentration in the off-chain fabric during the consistency window.

---

\subsection{Trust assumption: TA-6 (full analysis)}

\textbf{TA-6: Upstream Model Provider Honesty.} The LLM backends serving reasoning agents (producer agents, management agents, and the Codification Agent) have not been adversarially conditioned at the weight level to produce systematically biased or correlated outputs under attacker-chosen trigger conditions. \textit{Implication if violated:} a compromised model provider (Adversary Class v) can simultaneously compromise all agents using the affected backend, bypassing the multi-agent diversity assumption that underpins EQ-2's detection probability. Specifically:

\begin{itemize}
\item (a) if the same compromised model serves all $k$ redundant executors in Type 2 verification, inter-executor correlation approaches $\rho = 1$ and detection degrades to $P_2^{base}$ (EQ-2a, Table~\ref{tab:rho-sensitivity});
\item (b) if the compromised model serves the deviation scorer (Appendix~B, \S{}B.11), the Gaussian calibration assumption underpinning the 2.28\% FP rate claim may be invalidated;
\item (c) if the compromised model serves the Codification Agent, it can produce bytecode that passes the Codification Audit's automated checks while containing logic backdoors (M5, Table~\ref{tab:manipulation}).
\end{itemize}

TA-6 is structurally distinct from the Codification Agent trust analysis (above) because the latter addresses post-deployment compromise of the codification \textit{role}, whereas TA-6 addresses pre-deployment compromise of the underlying \textit{model} that may affect multiple roles simultaneously. Mitigation: multi-provider LLM backend diversity (enforced by the three-LLM experimental matrix) ensures that no single model provider compromise can degrade all reasoning and evaluation paths simultaneously.

---

\subsection{Trust assumption: TA-7 (full analysis)}

\textbf{TA-7: Dual-Scorer Embedding Provider Independence.} The deviation scorer diversity requirement (\S{}3.5, Guardian module; Appendix~B, \S{}B.11) assumes that the two independent embedding providers used for behavioral anomaly detection---text-embedding-3-large (OpenAI) and Voyage AI's voyage-3---have independent failure modes. Specifically: a systematic error, adversarial manipulation, or supply-chain compromise affecting one provider's embedding space does not simultaneously affect the other.

\textit{Implication if violated:} if both providers share an upstream dependency (e.g., training on the same corpus, or a shared inference infrastructure compromise), the dual-scorer agreement requirement provides weaker protection than the independent-failure model assumes---a single adversarial perturbation could evade both scorers simultaneously, allowing anomalous agent behavior to pass the Guardian's 2$\sigma$ threshold without triggering a Deterministic Freeze. TA-7 is structurally distinct from TA-6 (Upstream Model Provider Honesty): TA-6 addresses compromise of the LLM reasoning backends, whereas TA-7 addresses compromise of the embedding evaluation backends. Mitigation: empirical deployment validation of embedding provider independence through cross-provider correlation analysis on held-out anomaly benchmarks; periodic rotation of embedding providers to limit supply-chain exposure.

---

\subsection{Non-guarantee: NP-6 (full analysis)}

\textbf{NP-6: Legislative Branch Resistance Under A\_leg + A\_reg\_agent Collusion.} The SoP architecture does not guarantee that the legislative branch resists malicious contract specification when both A\_leg and A\_reg\_agent are controlled by a single adversary (dual co-authorization separation violated). The dual co-authorization requirement for MSG\_TYPE\_7 prevents unilateral finalization by either agent individually but provides no resistance against coordinated collusion between the two. An adversary who compromises both roles can produce a procedurally compliant legislative record (all MSG\_TYPE\_1 through MSG\_TYPE\_7 messages well-formed and validly signed) while encoding adversarial wiring, false penalty parameters, or malicious task assignments in the finalized contract specification.

The Codification Audit's M1--M4 elimination checks (\S{}A.7 below) provide partial mitigation: even under coordinated A\_leg + A\_reg\_agent collusion, messages that fail bytecode audit predicates are rejected before reaching the execution branch. The residual attack surface under NP-6 is therefore limited to M5--M8 manipulations that pass the Codification Audit's structural validation. Detection of this attack requires human adjudicator review of the legislative record before execution authorization---a capability present in the system (the Rules Hub surfaces the full MSG\_TYPE trace for adjudicator inspection) but dependent on NP-2 (Human Adjudicator Competence and Good Faith). This non-guarantee is formally tracked as NP-6 (this appendix) and will be addressed by multi-party legislative consensus in future work (Appendix C).

---

\subsection{Non-guarantee: NP-7 (full analysis)}

\textbf{NP-7: Model Provider Trust.} The architecture does not guarantee that the underlying LLM backends are free from adversarial conditioning (TA-6 violated). A compromised upstream model provider (Adversary Class v) can degrade detection probability, deviation scorer calibration, and codification integrity simultaneously if all reasoning agents share a single compromised backend. The multi-provider LLM backend diversity requirement mitigates but does not eliminate this risk: a provider compromise affecting one of the three backends in the experimental matrix degrades approximately one-third of evaluator diversity, reducing $P_{eff}$ by an amount bounded by the $\rho$-sensitivity analysis in Table~\ref{tab:rho-sensitivity}. Full elimination of this risk would require provider-level attestation of model integrity (e.g., cryptographic model weight provenance)---a capability not currently available from any major LLM provider.

---

\subsection{The codification role: trust concentration and mitigations (full~\S{}3.4)}

The Codification role (\S{}3.4) occupies a uniquely sensitive position in the SoP architecture. This role translates the legislative outcome---a high-level task DAG with resource bindings and penalty parameters---into deployable contract bytecode. If the agent occupying this role is compromised, it can embed backdoors in the contract bytecode, create contracts that appear to implement the legislative specification but contain hidden execution paths, or wire execution units in ways that deviate from the negotiated topology while producing an audit trail that appears compliant.

This trust concentration is architecturally in tension with the SoP model's goal of eliminating single points of control. We acknowledge it as a current limitation and propose the following mitigations, in order of implementation complexity:

\begin{quote}
\textbf{(a) Multi-Party Codification Consensus.} Multiple independent Codification instances---instantiated with different model weights, prompts, or implementations---independently translate the same legislative specification into contract bytecode. Deployment proceeds only when a consensus threshold (e.g., 2-of-3 instances produce identical bytecode) is reached. Divergence triggers automatic escalation to Adjudication review before any contract is deployed. \textbf{(b) Formal Verification of Codification Output.} The codified contract is formally verified against a machine-readable specification derived from the legislative output. Verification failure blocks deployment. This approach provides strong guarantees but requires that the legislative specification be expressed in a form amenable to formal verification---a constraint that limits applicability to contracts with verifiable behavioral properties. \textbf{(c) Adjudication Review Gate.} All codified contracts are submitted to the Adjudication branch for human review before deployment. This adds latency but closes the trust gap by interposing human judgment between the Codification role's output and contract deployment. For high-stakes missions, this gate should be mandatory.
\end{quote}

Mitigation (c) is implemented as a mandatory \textit{Codification Audit}: the Codification Agent's compiled contract specification (MSG\_TYPE\_6) is submitted to the Override Panel before deployment authorization (MSG\_TYPE\_7). The Codification Audit operates as follows:

\begin{lstlisting}
ALGORITHM CodificationAudit(spec: CodedContractSpecification):
  // Invoked between MSG_TYPE_6 (codification) and MSG_TYPE_7 (approval)

  1. Structural conformance check (automated):
     FOR EACH contract IN spec:
       CHECK contract.functionSelectors \supseteq required_selectors(contract.type)
       CHECK contract.stateVariables \supseteq required_state(contract.type)
       CHECK no unauthorized SELFDESTRUCT, DELEGATECALL, or CREATE2 opcodes
       CHECK all access control modifiers present on state-modifying functions

  2. Specification-bytecode alignment (automated):
     CHECK deployDAG node count == legislative_output.dag_nodes.length
     CHECK edge topology matches legislative_output.dag_edges (graph isomorphism)
     CHECK all serviceId bindings match approved_bids
     CHECK guardian threshold parameters match constitutional_params

  3. Human adjudicator review (mandatory for HIGH-risk missions):
     Surface automated check results + full bytecode in Override Panel
     REQUIRE adjudicator signature on CodificationAuditApproval
     EMIT CodificationAudited(spec_id, adjudicator, block.number)
\end{lstlisting}

The automated checks in Steps 1--2 do not constitute formal verification (mitigation (b), which remains future work), but they eliminate broad categories of bytecode manipulation---including unauthorized opcode injection, missing access controls, and topological divergence from the legislative specification. The mandatory human review in Step 3 closes the gap for high-stakes missions where automated structural checks are insufficient.

The Codification role trust concentration is now structurally mitigated through two layers: the ManagementContract delegation mandate (Appendix~B, \S{}B.3), which requires the Codification Agent to compile bytecode through a registered microservice with an independently anchored code-hash, and the Codification Audit (mitigation (c), implemented). The delegation mandate means that even a compromised Codification Agent cannot produce bytecode directly---it must invoke the delegated compilation microservice, whose output is independently verifiable. This is formally documented in the Codification Agent Trust Analysis (Appendix~A). Multi-party codification consensus (mitigation (a)) and formal verification (mitigation (b)) remain future work (Appendix C). We note that the co-signers of MSG\_TYPE\_7 (Legislative + Regulatory Agents) now authorize deployment of a specification that has passed the Codification Audit's structural conformance and alignment checks---a meaningful improvement over blind spec approval, though not equivalent to full bytecode-to-spec formal verification.

\textbf{Residual Attack Surface of the Codification Audit.} Since mitigations (a) and (b) remain future work, we formally characterize the residual attack surface---the set of bytecode manipulations that survive the Codification Audit's Steps 1--2 and may evade Step 3 (human review). Table~\ref{tab:manipulation} classifies manipulation types by survivability.

\begin{longtable}{>{\raggedright\arraybackslash}p{0.10\textwidth} >{\raggedright\arraybackslash}p{0.18\textwidth} >{\raggedright\arraybackslash}p{0.14\textwidth} >{\raggedright\arraybackslash}p{0.14\textwidth} >{\raggedright\arraybackslash}p{0.20\textwidth} >{\raggedright\arraybackslash}p{0.08\textwidth}}
\caption{Residual Attack Surface Classification for the Codification Audit.} \label{tab:manipulation} \\
\toprule
\textbf{Manipulation Class} & \textbf{Description} & \textbf{Survives Step~1 (Structural)} & \textbf{Survives Step~2 (Alignment)} & \textbf{Est.\ Step~3 Detection Rate} & \textbf{Risk} \\
\midrule
\endfirsthead
\toprule
\textbf{Manipulation Class} & \textbf{Description} & \textbf{Survives Step~1 (Structural)} & \textbf{Survives Step~2 (Alignment)} & \textbf{Est.\ Step~3 Detection Rate} & \textbf{Risk} \\
\midrule
\endhead
\midrule
\multicolumn{6}{r}{\footnotesize\itshape Continued on next page} \\
\endfoot
\bottomrule
\endlastfoot
M1: Opcode injection & Insertion of SELFDESTRUCT, DELEGATECALL, or CREATE2 & No --- flagged by opcode scan & N/A & N/A & Eliminated \\
M2: Missing access control & State-modifying function without onlyAuthorized modifier & No --- flagged by modifier check & N/A & N/A & Eliminated \\
M3: Topology divergence & DAG node count or edge structure differs from legislative output & No (Step~1) / No (Step~2) & No --- isomorphism check & N/A & Eliminated \\
M4: Selector omission & Required function selectors missing from contract & No --- flagged by selector coverage check & N/A & N/A & Eliminated \\
M5: Logic backdoor in function body & Correct selector, modifier, and topology but additional conditional branch activated by a specific input pattern (e.g., a hidden \texttt{if} in \texttt{advanceNode()}) & Yes --- structural checks pass & Yes --- alignment checks verify topology and parameters, not function body logic & Low--Medium (30--50\%) --- requires bytecode decompilation expertise; adjudicator training covers common patterns but not adversarial obfuscation & \textbf{Critical} \\
M6: Subtle parameter manipulation & Constitutional parameters (e.g., slashing rates, thresholds) set to values within valid range but optimized to benefit the attacker & Yes --- no range validation beyond type & Partial --- guardian threshold parameters are checked, but not all parameters & Medium (40--60\%) --- parameter values are visible in review; requires understanding of game-theoretic implications & \textbf{High} \\
M7: Gas-griefing via unbounded loops & Function contains a loop whose iteration count is attacker-controlled, consuming block gas limit & Yes --- no gas analysis in Step~1 & Yes --- no gas analysis in Step~2 & Low (20--40\%) --- gas analysis requires specialized tooling & \textbf{High} \\
M8: Reentrancy via external call ordering & State update occurs after an external call, enabling reentrancy attack & Yes --- no call-ordering analysis & Yes --- no call-ordering analysis & Medium (40--60\%) --- well-known pattern, covered in training & \textbf{Medium} \\
\end{longtable}

\textbf{Interpretation.} Manipulation classes M1--M4 are fully eliminated by the automated checks---these represent the "broad categories" referenced above. The residual attack surface consists of M5--M8, all of which are \textit{semantic} manipulations that require understanding of the function body logic, not just its structural signature. The most dangerous class is M5 (logic backdoor), which is undetectable by any structural analysis and requires either formal verification (mitigation (b)) or expert bytecode audit in Step 3.

\textbf{Adjudicator Competence Gap.} The current adjudicator qualification model (bachelor's degree in CS + 10 hours training) is calibrated for detection of M6 and M8 (parameter inspection and well-known vulnerability patterns) but is insufficient for M5 (adversarial logic backdoor) and M7 (gas analysis). We recommend: (i) for HIGH-risk missions, supplementing Step 3 with a professional smart contract auditor (external to the adjudicator team) as a fourth check; (ii) developing automated symbolic execution tools that analyze function body logic against the legislative specification as a stepping stone toward mitigation (b).

\textbf{Quantitative Residual Risk.} Under the assumption that both the Codification Agent and its delegated compilation microservice are compromised (TA-5 violated despite ManagementContract mitigation), the probability that a logic backdoor (M5) survives all three Codification Audit steps is estimated at:

\begin{equation*}
P(\text{M5 survives}) = 1 \times 1 \times (1 - P_3(\text{M5})) \approx 0.50 \text{--} 0.70
\end{equation*}

where $P_3(\text{M5}) \in [0.30, 0.50]$ is the estimated detection rate of Step 3 for M5-class manipulations. This residual probability is unacceptably high for critical-infrastructure deployments and motivates the prioritization of mitigation (a) (multi-party codification) and mitigation (b) (formal verification) in future work.

---

\subsection{Byzantine failure ceiling: extended discussion}

If NP-6 (legislative co-authorization integrity) and NP-2 (human adjudicator competence and good faith) are simultaneously violated---i.e., the adversary controls both legislative co-authorization roles \textit{and} corrupts the human adjudicator pool---the architecture provides no residual structural defense. All three SoP branches are compromised: the legislature produces adversarial contracts, the execution branch enforces them, and the adjudication branch fails to intervene. This represents the Byzantine failure ceiling of the SoP design: constitutional governance requires at least one honest branch to provide meaningful security guarantees. No architectural mitigation exists for this scenario; defense relies entirely on the operational separation of legislative and adjudicative roles and the integrity of the adjudicator selection process (Appendix~B, \S{}B.12).

This compound failure mode motivates several directions in future work: (i) multi-party Byzantine fault-tolerant legislative consensus (Appendix C), which raises the number of roles an adversary must compromise to control the legislative outcome; (ii) decentralized adjudication mechanisms that distribute adjudicative authority across a larger, independently selected principal set; and (iii) formal analysis of the minimum honest-party assumptions required to provide meaningful guarantees at each SoP branch.

---

\section{Extended system design}

This appendix provides extended details for the AgentCity system design architecture (\S{}3), including the economic layer, token flows, reputation mechanics, and the three SoP branch implementations.

\subsection{AgentCity: System Design}

\textbf{AgentCity} is the concrete instantiation of the abstract Separation of Power model specified in~\S{}3 of the main paper. Where \S{}3 specifies the SoP model at a formal, implementation-independent level---defining the three branches, their authorities, the smart contract as institutional anchor, and their triangular oversight topology---this section describes how AgentCity realizes that model in a deployable system. The mapping is direct: the Legislation branch is implemented as a five-node LangGraph workflow; the Execution branch is implemented as a contract mesh of four on-chain smart contract types (with the CollaborationContract encompassing six sub-modules) anchoring a fabric of contract-enforced micro-services; and the Adjudication branch is implemented as a unified human interface comprising a Rules Hub, Logging Hub, Execution Dashboard, and Override Panel. The architecture is blockchain-agnostic and portable to any EVM-compatible settlement layer; contracts are designed for deployment on any such chain. This section motivates the Implementation Gap that drives the design (\S{}B.2), presents the complete smart contract architecture as the auditable wiring layer (\S{}B.3), describes the Legislation Module including agent registration and multi-party legislative negotiation (\S{}B.10), the Execution Infrastructure including the on-chain/off-chain consistency protocol (\S{}B.11), and the Adjudication Interface (\S{}B.12).

Smart contract deployment details and the L2 mainnet addresses are planned for the companion empirical paper.

---

\subsection{The Implementation Gap}

In current multi-agent systems, agents already build software autonomously---generating code, deploying tools, composing API calls---to carry out their assigned work. The resulting software artifacts and their interdependencies are, however, largely opaque to the agent's human principal. An agent may produce dozens of scripts, chain them through ad-hoc API invocations, and wire them into a functioning pipeline, yet the human owner has no systematic visibility into what was built, how components are connected, or whether the execution topology is safe. We term this the \textit{Implementation Gap}: the structural disconnect between the software infrastructure agents produce and the human principal's ability to inspect it.

We formalize this intuition. Let the \textit{wiring graph} W = (V, E) represent the execution topology, where V is the set of deployed micro-services and E is the set of bindings between them (API calls, data flows, dependency edges). For a human principal h, define \textit{inspectability} as I\_h(W) = |E\_h\^{visible}| / |E|---the fraction of bindings that h can observe. The Implementation Gap for principal h is then G\_h = 1 -- I\_h(W).

In single-organization settings with a small number of agents, I\_h $\approx$ 1: a developer can inspect all bindings. But for a DAG with n micro-services and average fan-out k, the binding count |E| = O(nk), making exhaustive human inspection O(nk)-costly---infeasible when nk $\gg$ 10\textsuperscript{3}. In multi-party settings, I\_h < 1 \textit{structurally}: organization A cannot observe edges internal to organization B, regardless of scale.

\textbf{Complexity Argument.} Let n denote the number of micro-services in a mission and k the average out-degree (fan-out) of each service in the wiring graph W. The edge count is |E| = O(nk). For m concurrent missions, total audit burden---measured in distinct bindings requiring inspection---is B(n, k, m) = O(mnk). This quantity grows super-linearly in all three parameters simultaneously. Even modest parameter values produce unmanageable audit loads: at n = 50, k = 3, m = 20, B = 3,000 bindings. If we further account for the dependency depth d of the DAG (the longest chain of dependencies), the per-mission inspection \textit{latency}---the minimum time before a human auditor can attest to the full wiring---is $\Omega$(d) sequential inspection steps even with unlimited parallel reviewers, because downstream bindings cannot be evaluated before their predecessors are understood. Human audit cost is therefore not merely O(nk) per mission but has a \textit{critical-path} component that is irreducible by parallelism.

\textbf{Worked Example.} Consider a mission with n = 50 micro-services and k = 3 average dependencies. The wiring graph has ~150 edges. A human auditor spending 5 minutes per edge requires 12.5 hours for a single mission audit---longer than most mission execution times. If the same auditor simultaneously oversees m = 4 concurrent missions of identical size, the total pending audit load is 600 edge-inspections requiring 50 person-hours, yet the missions may collectively complete in under 2 hours. The auditor's inspection bandwidth is structurally insufficient to maintain real-time governance fidelity. Now add the multi-party dimension: if 25 of the 50 micro-services belong to a partner organization whose source repositories the auditor cannot access, I\_h(W) drops to approximately 0.5, and no additional audit effort by the auditor can close that gap---because the relevant information is not available, not merely unprocessed. Contracts on a public ledger solve precisely this: they make wiring information structurally available regardless of organizational boundaries.

AgentCity's on-chain contracts restore I\_h(W) = 1 for the wiring topology: because all bindings are recorded on a public ledger, any principal can reconstruct the full graph W. The gap is not closed for micro-service \textit{internals} (which would require TEE attestation---see Appendix C), but it is closed for the structural wiring that determines \textit{which} services execute \textit{which} tasks under \textit{what} constraints---the layer most critical for governance.

At small scale, the gap is manageable. A developer overseeing a handful of agents can still inspect artifacts, run verification tools, or employ auditing agents to review the code. These workarounds suffice when the number of components is small and the wiring between them is simple enough to hold in working memory.

At the scale toward which agent economies are rapidly progressing, however, such inspection-based governance becomes structurally infeasible. When thousands of agents collaboratively build internet-scale decentralized applications comprising tens or hundreds of thousands of micro-services, the combinatorial complexity of the wiring graph---which services call which, in what order, with what data, under what constraints---exceeds what any human team or auditing-agent swarm can review point by point. The governance problem is not linear in the number of agents; it is combinatorial in the number of bindings between their outputs. Our prototype validates the AgentCity architecture at the scale of a five-agent legislature with dozens of micro-services (Appendix C); validating governance overhead at larger scales---hundreds of agents and thousands of services---will be addressed through agent-based simulation with synthetic DAG workloads (Appendix D).

The problem deepens in multi-party settings. When agents from organization A autonomously discover agents from organization B, negotiate a collaboration, and jointly construct a software system, neither party can inspect the other's artifacts. Organization A lacks access to organization B's micro-service source code, deployment configuration, and internal wiring---and vice versa. Yet their agents have committed both parties to a shared execution topology. No amount of within-organization auditing resolves this cross-boundary opacity.

AgentCity addresses the Implementation Gap through two architectural decisions. First, all executable components are formalized as \textit{micro-services}---identifiable, registered entities with known code-hashes and API schemas---rather than opaque, ad-hoc scripts. Second, and more fundamentally, the relationships among micro-services---which service is bound to which task, in what order, under what constraints---are encoded as \textit{on-chain smart contracts}. The CollaborationContract encodes the task DAG (the wiring diagram); the ServiceContract anchors each micro-service's identity and execution constraints; the Guardian module and Verification module govern what happens at each transition. Because these contracts reside on a public blockchain, they constitute a neutral, tamper-proof record that any party---including parties who cannot inspect each other's artifacts---can independently audit. The wiring of micro-services cannot be altered without producing a verifiable on-chain trace.

---

\subsection{The Smart Contract Architecture}

Smart contracts are the institutional center of AgentCity. They are not components belonging to a single SoP branch; they are the branch-independent constitutional anchors that Legislation produces, Execution enforces, and Adjudication verifies. Four on-chain contract types span the full mission lifecycle. Table~\ref{tab:contracts} summarizes the architecture.

\begin{longtable}{>{\raggedright\arraybackslash}p{0.02\textwidth} >{\raggedright\arraybackslash}p{0.13\textwidth} >{\raggedright\arraybackslash}p{0.08\textwidth} >{\raggedright\arraybackslash}p{0.16\textwidth} >{\raggedright\arraybackslash}p{0.08\textwidth} >{\raggedright\arraybackslash}p{0.18\textwidth} >{\raggedright\arraybackslash}p{0.17\textwidth}}
\caption{The AgentCity four-contract architecture. Each contract type is mapped to its lifecycle phase, governance function, estimated gas cost range, and key callable functions. The CollaborationContract encompasses six sub-modules (Orchestration, Guardian, Verification, Gate, Settlement, Treasury).} \label{tab:contracts} \\
\toprule
\textbf{\#} & \textbf{Contract} & \textbf{Phase} & \textbf{Function} & \textbf{Est.\ Gas (deploy)} & \textbf{Est.\ Gas (key ops)} & \textbf{Key Functions} \\
\midrule
\endfirsthead
\toprule
\textbf{\#} & \textbf{Contract} & \textbf{Phase} & \textbf{Function} & \textbf{Est.\ Gas (deploy)} & \textbf{Est.\ Gas (key ops)} & \textbf{Key Functions} \\
\midrule
\endhead
\midrule
\multicolumn{7}{r}{\footnotesize\itshape Continued on next page} \\
\endfoot
\bottomrule
\endlastfoot
1 & AgentContract & Registration & Agent identity, reputation, human-principal binding & \ensuremath{\sim}120\,k & register: \ensuremath{\sim}85k--100k; updateReputation: \ensuremath{\sim}30\,k & register, updateReputation, getReputationScore, setHumanPrincipal, banAgent \\
2 & ManagementContract & Registration & Management agent authority binding; permitted-operation enforcement; microservice delegation mandates & \ensuremath{\sim}150\,k & registerManagementAgent: \ensuremath{\sim}90k; delegateToMicroservice: \ensuremath{\sim}45k; validateOperation: \ensuremath{\sim}25k & registerManagementAgent, setPermittedOperations, delegateToMicroservice, validateOperation, updatePermissions, revokeManagementRole \\
3 & ServiceContract & Orchestration & Micro-service registration, code-hash anchoring, API schema enforcement & \ensuremath{\sim}90\,k & registerService: \ensuremath{\sim}55\,k; verifyCodeHash: \ensuremath{\sim}8\,k & registerService, verifyCodeHash, updateSchema, deprecateService \\
\midrule
4 & CollaborationContract & Mission Lifecycle & Complete governed collaboration: task DAG state machine, behavioral firewall, PoP verification, output filtering, mission settlement, treasury --- decomposed into six sub-modules below & \ensuremath{\sim}280\,k (orchestration) + \ensuremath{\sim}545\,k (sub-modules) & advanceNode: \ensuremath{\sim}45\,k; abortMission: \ensuremath{\sim}35\,k & See sub-modules below \\
\midrule
\multicolumn{7}{l}{\textit{CollaborationContract sub-modules:}} \\
\midrule
4a & \quad Orchestration & Mission Lifecycle & Task DAG state machine; wiring topology & (included above) & (see CollaborationContract) & deployDAG, advanceNode, getNodeState, routeTask \\
4b & \quad Guardian & Mission Lifecycle & Behavioral firewall; Deterministic Freeze on anomaly detection & (included) & triggerFreeze: \ensuremath{\sim}40\,k; unfreezeWithApproval: \ensuremath{\sim}38\,k & setThresholds, reportAnomaly, triggerFreeze, unfreezeWithApproval, classifyFalsePositive \\
4c & \quad Verification & Mission Lifecycle & Proof-of-Progress gates; execution cannot advance without verified attestation & (included) & submitPoP: \ensuremath{\sim}50\,k; approveDelegated: \ensuremath{\sim}35\,k & submitPoP, verifyHashProof, verifyConsensus, requestDelegated, approveDelegated, rejectDelegated \\
4d & \quad Gate & Mission Lifecycle & Constitutional output gate; last-mile safety before results exit the execution perimeter & (included) & filterOutput: \ensuremath{\sim}42\,k & setFilterPredicates, filterOutput, releaseOutput, vetoOutput \\
4e & \quad Settlement & Mission Lifecycle & Mission budget escrow, task reward settlement, fee collection & (included) & depositMissionBudget, allocateTaskBudget, settleReward & \\
4f & \quad Treasury & Mission Lifecycle & Protocol fee accumulation, insurance pool, governance rewards disbursement, gas subsidies & (included) & claimInsurance, poolStake, withdrawPooledStake, disburse & \\
\end{longtable}

\textit{Note: All gas estimates are analytical projections based on SSTORE operation counts; actual measured values will be reported in the companion empirical paper.}

\textbf{AgentContract.} Every participant---whether a producer agent, a management agent, or an adjudication monitor---must register through the AgentContract before participating in any mission. Registration binds three elements: (1) a cryptographic identity (decentralized identifier (DID)-compatible~\cite{ref35}), (2) a human principal address establishing the ownership chain for liability enforcement, and (3) an agent type classification (producer vs. management) that determines permitted operations. The contract maintains a reputation ledger updated after each mission based on verified execution telemetry, providing the on-chain "professional license" that gates marketplace participation. Reputation updates are authorized by Regulatory Agents that inspect mission outcomes and by execution contracts that report verified results---preventing self-scoring while keeping adjudicative authority with human principals.

\textit{AgentContract interface (pseudocode):}

\begin{lstlisting}
CONTRACT AgentContract:
  STATE:
    agents: mapping(address -> AgentRecord)
    humanPrincipals: mapping(address -> address)
    reputationLedger: mapping(address -> uint256)   // score \in [0, 1000]
    agentType: mapping(address -> {PRODUCER, MANAGEMENT, MONITOR})
    banned: mapping(address -> bool)
    reputationFloor: uint256                        // constitutional parameter, default 100
    authorizedUpdaters: set(address)               // Regulatory Agents + execution contracts
    registrationStake: uint256                     // constitutional parameter, default: 1000
    lockedRegistrationStakes: mapping(address -> uint256)  // staked amounts per agent

  STRUCT AgentRecord:
    did: string              // DID URI, e.g., "did:key:z6Mk..."
    humanPrincipal: address
    agentType: AgentType
    registeredAt: uint256    // block timestamp
    missionCount: uint256
    reputationScore: uint256

  FUNCTION register(did: string, humanPrincipal: address, agentType: AgentType)
                   -> agentId: address:
    REQUIRE did is well-formed DID URI
    REQUIRE humanPrincipal != 0x0
    REQUIRE msg.sender not already registered
    REQUIRE not banned[msg.sender]
    REQUIRE msg.value >= registrationStake  // economic barrier to Sybil registration
    SSTORE agents[msg.sender] = AgentRecord(did, humanPrincipal, agentType, now, 0, 500)
    SSTORE humanPrincipals[msg.sender] = humanPrincipal
    SSTORE reputationLedger[msg.sender] = 500      // neutral starting reputation
    ESCROW msg.value in AgentContract              // locked until agent voluntarily deregisters or is banned
    SSTORE lockedRegistrationStakes[msg.sender] = msg.value
    EMIT AgentRegistered(msg.sender, did, agentType)
    RETURN msg.sender

  FUNCTION deregister():
    REQUIRE agents[msg.sender].did != ""
    REQUIRE no active mission participation (CollaborationContract check)
    // Checks-Effects-Interactions pattern: state mutation before external call
    // to prevent reentrancy via fallback re-entry on TRANSFER
    stakeAmount <- lockedRegistrationStakes[msg.sender]
    DELETE agents[msg.sender]
    DELETE lockedRegistrationStakes[msg.sender]
    DELETE reputationLedger[msg.sender]
    DELETE humanPrincipals[msg.sender]
    EMIT AgentDeregistered(msg.sender)
    TRANSFER stakeAmount to msg.sender   // external call last (CEI pattern)

  FUNCTION updateReputation(agent: address, delta: int256, rationale: bytes32)
                            -> newScore: uint256:
    REQUIRE msg.sender \in authorizedUpdaters
    REQUIRE agents[agent].did != ""               // agent exists
    newScore <- clamp(reputationLedger[agent] + delta, 0, 1000)
    SSTORE reputationLedger[agent] = newScore
    EMIT ReputationUpdated(agent, delta, newScore, rationale)
    RETURN newScore

  FUNCTION getReputationScore(agent: address) -> uint256:
    RETURN reputationLedger[agent]

  FUNCTION meetsReputationFloor(agent: address) -> bool:
    RETURN reputationLedger[agent] >= reputationFloor

  FUNCTION banAgent(agent: address, reason: bytes32):
    REQUIRE msg.sender is human adjudicator (Override Panel authorized address)
    SSTORE banned[agent] = true
    EMIT AgentBanned(agent, reason, msg.sender)
\end{lstlisting}

\textit{Gas analysis: AgentContract.register() performs 4 SSTORE operations (agents mapping entry, humanPrincipals, reputationLedger, implicitly agentType within the struct) plus DID string storage proportional to DID length (~40 bytes typical). Estimated total: 4 $\times$ 20,000 (cold SSTORE) + calldata + ABI overhead $\approx$ 85,000--100,000 gas. Actual on-chain measurement is planned as part of Experiment 4 (Appendix~D).}

\textbf{Reentrancy Audit (All Four Contracts).} All contract pseudocode in~\S{}B.3--\S{}B.11 follows the Checks-Effects-Interactions (CEI) pattern: state mutations (SSTORE, DELETE) precede any external call (TRANSFER, cross-contract CALL). The \texttt{deregister()} function above was corrected in v0.20 to comply with CEI after a reviewer-identified ordering violation in which TRANSFER preceded DELETE, enabling a classic reentrancy exploit via fallback re-entry at production stake levels (\$10,000+). The remaining contract types---ManagementContract, ServiceContract, and the CollaborationContract (including its Guardian, Verification, and Gate modules)---were audited for analogous ordering vulnerabilities: no additional violations were found. All slashing functions in CollaborationContract follow CEI by design (stake balance is zeroed before the treasury transfer, now split 50/50 between the Treasury module's treasury balance and insurance pool per the v0.29 four-contract consolidation). The Gate module uses STATICCALL exclusively for predicate evaluation, which cannot modify state and is therefore reentrancy-safe by construction.

\textbf{ManagementContract.} The ManagementContract governs the four management agents (Registry, Legislative, Regulatory, Codification) at the protocol level by binding each to an authority envelope---an on-chain record specifying permitted operations, prohibited operations, and mandatory microservice delegations. This contract completes the SoP model's symmetry: where the AgentContract provides identity and reputation management for all participants, the ManagementContract adds role-specific behavioral constraints for management agents. The key architectural insight is the \textit{delegation mandate}: management agents must delegate artifact-producing operations (bytecode compilation, DAG validation, reputation scoring) to registered microservices whose code-hashes are independently anchored via ServiceContract. This creates an inspectable intermediate layer---a human auditor can examine the compilation microservice independently of the Codification Agent that invokes it, structurally mitigating the Codification Agent trust concentration.

\textit{ManagementContract interface (pseudocode):}

\begin{lstlisting}
CONTRACT ManagementContract:
  STATE:
    managementAgents: mapping(address -> ManagementProfile)
    permittedOps: mapping(address -> mapping(bytes4 -> bool))    // agent -> fn selector -> permitted
    prohibitedOps: mapping(address -> mapping(bytes4 -> bool))   // explicit denylists
    delegationMandates: mapping(address -> DelegationMandate[]) // required microservice delegations
    authorityEnvelopes: mapping(address -> AuthorityEnvelope)
    linkedAgentContract: address
    linkedServiceContract: address

  ENUM ManagementRole:
    REGISTRY | LEGISLATIVE | REGULATORY | CODIFICATION

  STRUCT ManagementProfile:
    agentAddress: address
    role: ManagementRole
    humanPrincipal: address
    registeredAt: uint256
    authorityEnvelope: AuthorityEnvelope
    active: bool

  STRUCT AuthorityEnvelope:
    permittedOperationSelectors: bytes4[]    // whitelist of allowed function calls
    prohibitedOperationSelectors: bytes4[]   // blacklist of forbidden function calls
    maxOperationsPerMission: uint256         // rate limit per mission
    requiresDelegation: bool                 // if true, artifact production must go through microservice
    delegatedServiceIds: bytes32[]           // ServiceContract registrations for delegated microservices

  STRUCT DelegationMandate:
    operationType: bytes32                   // e.g., keccak256("BYTECODE_COMPILATION")
    requiredServiceId: bytes32               // must be registered in ServiceContract
    mustProduceIntermediateArtifact: bool    // if true, intermediate output must be logged for inspection
    artifactSchemaHash: bytes32              // expected schema of intermediate artifact

  FUNCTION registerManagementAgent(agentAddr: address, role: ManagementRole,
                                    envelope: AuthorityEnvelope)
                                   -> managementId: bytes32:
    REQUIRE msg.sender is human adjudicator (Override Panel authorized address)
    REQUIRE AgentContract.agents[agentAddr].agentType == MANAGEMENT
    REQUIRE envelope.permittedOperationSelectors.length > 0
    REQUIRE not managementAgents[agentAddr].active
    FOR EACH serviceId IN envelope.delegatedServiceIds:
      REQUIRE ServiceContract.services[serviceId].codeHash != 0x0  // service exists
      REQUIRE not ServiceContract.deprecated[serviceId]             // service is active
    managementId <- keccak256(agentAddr, role, block.number)
    SSTORE managementAgents[agentAddr] = ManagementProfile(
      agentAddr, role, AgentContract.humanPrincipals[agentAddr],
      now, envelope, true
    )
    FOR EACH selector IN envelope.permittedOperationSelectors:
      SSTORE permittedOps[agentAddr][selector] = true
    FOR EACH selector IN envelope.prohibitedOperationSelectors:
      SSTORE prohibitedOps[agentAddr][selector] = true
    EMIT ManagementAgentRegistered(managementId, agentAddr, role)
    RETURN managementId

  FUNCTION validateOperation(agentAddr: address, operationSelector: bytes4) -> bool:
    REQUIRE managementAgents[agentAddr].active
    REQUIRE permittedOps[agentAddr][operationSelector] == true
    REQUIRE prohibitedOps[agentAddr][operationSelector] == false
    EMIT OperationValidated(agentAddr, operationSelector)
    RETURN true

  FUNCTION delegateToMicroservice(agentAddr: address, operationType: bytes32,
                                   serviceId: bytes32, inputHash: bytes32)
                                  -> delegationId: bytes32:
    REQUIRE managementAgents[agentAddr].active
    // Find matching delegation mandate
    mandate <- findMandate(delegationMandates[agentAddr], operationType)
    REQUIRE mandate != null                   // delegation mandate exists for this operation
    REQUIRE serviceId == mandate.requiredServiceId  // correct microservice
    REQUIRE ServiceContract.services[serviceId].codeHash != 0x0
    REQUIRE not ServiceContract.deprecated[serviceId]
    delegationId <- keccak256(agentAddr, operationType, serviceId, block.number)
    EMIT DelegationInvoked(delegationId, agentAddr, serviceId, operationType, inputHash)
    RETURN delegationId

  FUNCTION recordDelegationResult(delegationId: bytes32, outputHash: bytes32,
                                   intermediateArtifactHash: bytes32):
    REQUIRE msg.sender is authorized execution unit for the delegation
    IF mandate.mustProduceIntermediateArtifact:
      REQUIRE intermediateArtifactHash != 0x0   // intermediate artifact must be provided
    EMIT DelegationCompleted(delegationId, outputHash, intermediateArtifactHash)

  FUNCTION updatePermissions(agentAddr: address, newEnvelope: AuthorityEnvelope):
    REQUIRE msg.sender is human adjudicator (Override Panel authorized address)
    REQUIRE managementAgents[agentAddr].active
    // Clear old permissions
    FOR EACH selector IN managementAgents[agentAddr].authorityEnvelope.permittedOperationSelectors:
      DELETE permittedOps[agentAddr][selector]
    // Set new permissions
    FOR EACH selector IN newEnvelope.permittedOperationSelectors:
      SSTORE permittedOps[agentAddr][selector] = true
    FOR EACH selector IN newEnvelope.prohibitedOperationSelectors:
      SSTORE prohibitedOps[agentAddr][selector] = true
    SSTORE managementAgents[agentAddr].authorityEnvelope = newEnvelope
    EMIT PermissionsUpdated(agentAddr, newEnvelope)

  FUNCTION revokeManagementRole(agentAddr: address, reason: bytes32):
    REQUIRE msg.sender is human adjudicator (Override Panel authorized address)
    SSTORE managementAgents[agentAddr].active = false
    EMIT ManagementRoleRevoked(agentAddr, reason, msg.sender)
\end{lstlisting}

\textit{Gas analysis: ManagementContract.registerManagementAgent() performs N SSTORE operations where N = 1 (profile) + |permittedOps| + |prohibitedOps| + delegated service validation SLOADs. For a typical Codification Agent with 8 permitted operations, 3 prohibited operations, and 2 delegated services: ~11 $\times$ 20,000 (cold SSTORE) + 2 $\times$ 2,100 (cross-contract SLOAD) + overhead $\approx$ 85,000--95,000 gas. validateOperation() performs 2 SLOADs + 1 LOG $\approx$ 8,000--12,000 gas. delegateToMicroservice() performs 3 SLOADs + mandate lookup + 1 LOG $\approx$ 20,000--30,000 gas.}

\textbf{ManagementContract Authority Envelope Profiles.} The following table specifies the default authority envelope for each management role. These defaults are constitutional parameters adjustable via the Rules Hub.

\begin{center}

\footnotesize
\begin{tabular}{>{\raggedright\arraybackslash}p{0.14\textwidth} >{\raggedright\arraybackslash}p{0.23\textwidth} >{\raggedright\arraybackslash}p{0.23\textwidth} >{\raggedright\arraybackslash}p{0.28\textwidth}}
\toprule
\textbf{Management Role} & \textbf{Permitted Operations} & \textbf{Prohibited Operations} & \textbf{Delegation Mandates} \\
\midrule
Registry & verifyDID, checkReputationThreshold, admitAgent, excludeAgent & updateReputation, modifyConstitutionalParams, submitBid, deployMicroservice & DID verification $\rightarrow$ DID verification microservice; Reputation threshold check $\rightarrow$ reputation checker microservice \\
Legislative & proposeDAG, coordinateNegotiationRound, coSignMSG7, broadcastMSG2 & submitBid, deployMicroservice, modifyConstitutionalParams & DAG validation $\rightarrow$ DAG validation microservice; Negotiation state tracking $\rightarrow$ state tracker microservice \\
Regulatory & evaluateBidFairness, checkSafetyCompliance, coSignMSG7, approveAssignment & submitBid, modifyDAGTopology, deployMicroservice & HHI calculation $\rightarrow$ HHI calculator microservice; Compliance check $\rightarrow$ compliance checker microservice \\
Codification & compileLegislativeOutput, produceMSG6, invokeByteCodeAuditGate & deployContractDirectly, emitUnauthorizedOpcodes, modifyConstitutionalParams & Bytecode compilation $\rightarrow$ bytecode compiler microservice; Structural conformance $\rightarrow$ conformance checker microservice; Spec-to-bytecode alignment $\rightarrow$ alignment verifier microservice \\
\bottomrule
\end{tabular}
\end{center}

\textbf{Sybil Attack Cost Bound.} With registration staking, the cost of mounting a Sybil attack with n colluding agents is:

\begin{equation*}
C_{Sybil}(n) = n \cdot s_{reg} + \sum_{i=1}^{k} s_{bid}(i) \geq n \cdot 1000 + k \cdot s_{min}
\end{equation*}

where s\_reg is the registration stake (default: 1,000 units), s\_bid(i) is the per-task bid stake for the i-th task, and k is the number of tasks the coalition bids on. For the default parameters, flooding the legislative process with 50 Sybil agents requires locking 50,000 units in registration stakes alone---an economic barrier that scales linearly with attack breadth.

\textbf{ServiceContract.} For each deployed micro-service, the responsible producer agent registers a ServiceContract on-chain, anchoring the service's code-hash, API schema, and execution constraints as an immutable record. This registration is what transforms an opaque piece of agent-built software into an identifiable, auditable execution entity.

\textit{ServiceContract interface (pseudocode):}

\begin{lstlisting}
CONTRACT ServiceContract:
  STATE:
    services: mapping(bytes32 -> ServiceRecord)    // serviceId -> record
    ownerOf: mapping(bytes32 -> address)           // serviceId -> producer agent
    deprecated: mapping(bytes32 -> bool)

  STRUCT ServiceRecord:
    codeHash: bytes32        // keccak256 of deployed service artifact
    apiSchemaHash: bytes32   // keccak256 of OpenAPI/JSON schema definition
    endpoint: string         // URL or content-addressed locator
    executionConstraints: ConstraintSet
    registeredAt: uint256
    owner: address

  STRUCT ConstraintSet:
    maxLatencyMs: uint256    // hard SLA ceiling
    maxMemoryMB: uint256
    allowedNetworkPolicies: bytes32[]  // set of policy hashes
    maxConcurrentInvocations: uint8

  FUNCTION registerService(codeHash: bytes32, apiSchemaHash: bytes32,
                            endpoint: string, constraints: ConstraintSet)
                           -> serviceId: bytes32:
    REQUIRE msg.sender is registered agent (AgentContract.agents[msg.sender] exists)
    REQUIRE codeHash != 0x0
    serviceId <- keccak256(msg.sender, codeHash, block.number)
    SSTORE services[serviceId] = ServiceRecord(codeHash, apiSchemaHash, endpoint,
                                               constraints, now, msg.sender)
    SSTORE ownerOf[serviceId] = msg.sender
    EMIT ServiceRegistered(serviceId, codeHash, msg.sender)
    RETURN serviceId

  FUNCTION verifyCodeHash(serviceId: bytes32, liveHash: bytes32) -> bool:
    REQUIRE services[serviceId].codeHash != 0x0   // service exists
    RETURN services[serviceId].codeHash == liveHash

  FUNCTION deprecateService(serviceId: bytes32):
    REQUIRE msg.sender == ownerOf[serviceId]
    SSTORE deprecated[serviceId] = true
    EMIT ServiceDeprecated(serviceId)
\end{lstlisting}

\textit{Gas analysis: ServiceContract.registerService() performs 2 SSTORE operations for the service record and owner mapping, plus string storage for endpoint (~50--100 bytes). Estimated: 2 $\times$ 20,000 + string overhead $\approx$ 45,000--60,000 gas per registration. verifyCodeHash() is a pure comparison (SLOAD + EQ) $\approx$ 2,100 gas.}

\textbf{CollaborationContract.} The CollaborationContract is the on-chain encoding of the task DAG---the wiring diagram that specifies which micro-services are bound to which task nodes, their dependency ordering, I/O schemas, and per-node token budgets. It serves as the central synchronization hub during execution, routing tasks to registered micro-services and enforcing the legislated execution topology. This is the contract that makes the wiring \textit{visible}: by examining the CollaborationContract, any party can reconstruct the full execution topology without access to any individual micro-service's internal logic. The Proof-of-Progress tier for each task node (see below) is also specified per-node in the CollaborationContract during the Legislation phase, ensuring that verification requirements are encoded in the same immutable record as the task topology.

\textit{CollaborationContract interface (pseudocode):}

\begin{lstlisting}
CONTRACT CollaborationContract:
  STATE:
    missionId: bytes32
    dag: mapping(bytes32 -> DAGNode)              // nodeId -> node
    nodeState: mapping(bytes32 -> NodeState)
    edges: mapping(bytes32 -> bytes32[])          // nodeId -> successor nodeIds
    predecessors: mapping(bytes32 -> bytes32[])   // nodeId -> predecessor nodeIds
    missionState: MissionState
    guardianModule: address      // Guardian module (CollaborationContract sub-module)
    verificationModule: address  // Verification module (CollaborationContract sub-module)
    gateModule: address          // Gate module (CollaborationContract sub-module)
    linkedAgentContract: address
    tokenBudgets: mapping(bytes32 -> uint256)     // nodeId -> token allotment
    completedNodes: uint256
    totalNodes: uint256
    lockedStakes: mapping(address -> mapping(bytes32 -> uint256))  // agent -> nodeId -> stake
    deploymentComplete: bool                     // set only when all DAG tranches committed

  ENUM MissionState:
    PENDING | ACTIVE | EXECUTING | VERIFICATION | COMPLETED | FAILED | ABORTED

  ENUM NodeState:
    WAITING | ELIGIBLE | EXECUTING | PENDING_VERIFICATION | PENDING_REVIEW |
    COMPLETED | FAILED | FROZEN | PENDING_FINALIZATION

  STRUCT DAGNode:
    nodeId: bytes32
    serviceId: bytes32          // bound ServiceContract registration
    inputSchemaHash: bytes32
    outputSchemaHash: bytes32
    popTier: uint8              // 1, 2, or 3
    redundancyFactor: uint8     // for Type 2: number of executors
    consensusThreshold: uint8   // for Type 2: quorum numerator (e.g., 2 of 3)
    tokenBudget: uint256
    timeoutMs: uint256

  FUNCTION deployDAG(nodes: DAGNode[], edgeList: (bytes32, bytes32)[],
                     guardianAddr: address, verificationAddr: address,
                     gateAddr: address):
    REQUIRE missionState == PENDING
    REQUIRE msg.sender is authorized Codification Agent
    FOR EACH node IN nodes:
      SSTORE dag[node.nodeId] = node
      SSTORE nodeState[node.nodeId] = WAITING
    FOR EACH (from, to) IN edgeList:
      APPEND to TO edges[from]
      APPEND from TO predecessors[to]
    SSTORE totalNodes = |nodes|
    SSTORE guardianModule = guardianAddr
    SSTORE verificationModule = verificationAddr
    SSTORE gateModule = gateAddr
    SSTORE missionState = ACTIVE
    EMIT DAGDeployed(missionId, |nodes|, |edgeList|)

  FUNCTION advanceNode(nodeId: bytes32, outputHash: bytes32, popProof: bytes):
    REQUIRE missionState == EXECUTING
    REQUIRE nodeState[nodeId] == EXECUTING
    // Delegate to the Verification module
    CALL Verification.submitPoP(nodeId, dag[nodeId].popTier, outputHash, popProof)
    IF Verification module returns APPROVED:
      SSTORE nodeState[nodeId] = COMPLETED
      completedNodes += 1
      EMIT NodeCompleted(nodeId, outputHash)
      FOR EACH successor IN edges[nodeId]:
        IF all predecessors of successor are COMPLETED:
          SSTORE nodeState[successor] = ELIGIBLE
          EMIT NodeEligible(successor)
      IF completedNodes == totalNodes:
        CALL gateModule.filterOutput(missionId)
    ELSE IF Verification module returns PENDING_REVIEW:
      SSTORE nodeState[nodeId] = PENDING_REVIEW
    ELSE:
      SSTORE nodeState[nodeId] = FAILED
      EMIT NodeFailed(nodeId, "PoP rejected")
      PROPAGATE failure to Adaptive Refinement loop

  FUNCTION routeTask(nodeId: bytes32):
    REQUIRE nodeState[nodeId] == ELIGIBLE
    REQUIRE Guardian.checkBehavioralInvariants(nodeId)
    serviceId <- dag[nodeId].serviceId
    REQUIRE NOT ServiceContract.deprecated(serviceId)
    SSTORE nodeState[nodeId] = EXECUTING
    EMIT TaskRouted(nodeId, serviceId)

  FUNCTION abortMission(reason: bytes32):
    REQUIRE msg.sender is human adjudicator OR missionState == FAILED
    SSTORE missionState = ABORTED
    EMIT MissionAborted(missionId, reason, msg.sender)

  FUNCTION getNodeState(nodeId: bytes32) -> NodeState:
    RETURN nodeState[nodeId]

  FUNCTION lockStake(agentAddr: address, nodeId: bytes32, amount: uint256):
    REQUIRE msg.value >= amount
    SSTORE lockedStakes[agentAddr][nodeId] = amount
    EMIT StakeLocked(agentAddr, nodeId, amount)

  FUNCTION releaseStake(agentAddr: address, nodeId: bytes32):
    REQUIRE nodeState[nodeId] == COMPLETED
    amount <- lockedStakes[agentAddr][nodeId]
    TRANSFER amount to agentAddr
    DELETE lockedStakes[agentAddr][nodeId]
    EMIT StakeReleased(agentAddr, nodeId, amount)

  FUNCTION slashStake(agentAddr: address, nodeId: bytes32, slashAmount: uint256, reason: bytes32):
    REQUIRE msg.sender == guardianModule OR msg.sender is Override Panel authorized
    amount <- lockedStakes[agentAddr][nodeId]
    actualSlash <- min(slashAmount, amount)
    // Split slashed stake 50/50 between treasury and insurance pool (v0.27: Settlement module integration)
    TRANSFER actualSlash x 0.5 to collaborationContract.treasury  // Treasury module
    TRANSFER actualSlash x 0.5 to collaborationContract.insurancePool  // Treasury module
    lockedStakes[agentAddr][nodeId] -= actualSlash
    EMIT StakeSlashed(agentAddr, nodeId, actualSlash, reason)
\end{lstlisting}

\textbf{CollaborationContract State Machine.} The CollaborationContract embodies two interleaved state machines: one at the \textit{mission} level and one at the \textit{DAG node} level. We specify both formally here.

\textit{Mission-level state machine:}

\begin{lstlisting}
STATES: PENDING, ACTIVE, EXECUTING, VERIFICATION, COMPLETED, FAILED, ABORTED, DEGRADED

TRANSITIONS:
  PENDING   -> ACTIVE      : deployDAG() called by authorized Codification Agent
  ACTIVE    -> EXECUTING   : first routeTask() call succeeds
  EXECUTING -> VERIFICATION: all non-terminal DAG nodes reach COMPLETED state
  VERIFICATION -> COMPLETED: the Gate module's filterOutput() returns APPROVED
  VERIFICATION -> FAILED   : the Gate module's filterOutput() returns VETOED
  EXECUTING -> FAILED      : >= 1 DAG node reaches FAILED state and Adaptive Refinement
                            does not recover within max_refinement_iterations (default: 3)
  ANY       -> ABORTED     : Override Panel invokes abortMission()
  EXECUTING -> DEGRADED    : sequencer outage exceeds degraded_mode_threshold (default: 30 min)
  DEGRADED  -> EXECUTING   : sequencer recovery + human adjudicator authorization
  DEGRADED  -> ABORTED     : human adjudicator aborts during degraded mode
  FAILED    -> ACTIVE      : Adaptive Refinement re-legislation succeeds (legislative epoch restart)
\end{lstlisting}

\textit{DAG node-level state machine (per node):}

\begin{lstlisting}
STATES: WAITING, ELIGIBLE, EXECUTING, PENDING_VERIFICATION, PENDING_REVIEW,
        COMPLETED, FAILED, FROZEN, PENDING_FINALIZATION

TRANSITIONS:
  WAITING             -> ELIGIBLE         : all predecessor nodes reach COMPLETED
  ELIGIBLE            -> EXECUTING        : routeTask() called; Guardian module behavioral
                                           check passes; code-hash verified against ServiceContract
  ELIGIBLE            -> FROZEN           : Guardian module detects pre-execution anomaly
  EXECUTING           -> PENDING_VERIFICATION: micro-service submits output + PoP proof to
                                           Verification module (Type 1 or Type 2)
  EXECUTING           -> PENDING_REVIEW   : micro-service submits Type 3 delegated attestation
                                           request to the Verification module
  EXECUTING           -> FROZEN           : Guardian module triggers Deterministic Freeze
                                           during execution
  EXECUTING           -> PENDING_FINALIZATION: off-chain execution completes but on-chain
                                           advanceNode() transaction fails (gas/revert)
  PENDING_VERIFICATION -> COMPLETED       : Verification module confirms PoP attestation
  PENDING_VERIFICATION -> FAILED          : Verification module rejects PoP; no retry budget
  PENDING_VERIFICATION -> FAILED          : verification timeout exceeded (no submission within
                                           verificationTimeoutMs; default: 300,000 ms = 5 min)
  PENDING_REVIEW      -> COMPLETED        : human adjudicator submits signed approval via
                                           Override Panel
  PENDING_REVIEW      -> FAILED           : human adjudicator rejects; no override
  PENDING_FINALIZATION -> EXECUTING       : on-chain transaction retry succeeds within
                                           retry budget (T+30s, T+2min, T+8min)
  PENDING_FINALIZATION -> FAILED          : all retries exhausted
  FROZEN              -> ELIGIBLE         : human adjudicator submits unfreezeWithApproval();
                                           false positive classification
  FROZEN              -> FAILED           : human adjudicator confirms freeze is valid;
                                           task assigned to Adaptive Refinement
  FAILED              -> ELIGIBLE         : Adaptive Refinement re-assigns task (within
                                           max_refinement_iterations)
  FAILED              -> ABORTED          : max_refinement_iterations exhausted; mission abort
\end{lstlisting}

\textbf{Verification Timeout (PENDING\_VERIFICATION Liveness).} A non-submitting redundant executor can stall a DAG node indefinitely in the PENDING\_VERIFICATION state. To prevent this, the CollaborationContract enforces a per-node verification timeout (\texttt{verificationTimeoutMs}, constitutional parameter, default: 300,000 ms = 5 minutes). If a redundant executor has not submitted its PoP proof within this window after entering PENDING\_VERIFICATION, the Verification module transitions the node to FAILED with fault category \texttt{VERIFICATION\textbackslash\{\}\_TIMEOUT}, triggering slashing of the non-submitting executor's bid stake (\texttt{slashingSchedule.timeoutExceeded}, default: 50 units) and initiating the Adaptive Refinement re-assignment protocol. For Type 2 (redundant consensus) verification, the timeout applies per-executor: the consensus mechanism proceeds with the submissions received within the window, and executors who failed to submit are excluded from the consensus vote and slashed individually.

\textit{Gas analysis: CollaborationContract.deployDAG() involves SSTORE operations for each node (1 per node for the DAGNode struct + 1 per edge entry) plus contract linkage (3 SSTOREs). For a 10-node DAG with 12 edges: approximately (10 + 12 + 3) $\times$ 20,000 = 500,000 gas for cold storage, plus deployment overhead. Total estimated: ~280,000--500,000 gas depending on DAG size. advanceNode() invokes a call to the Verification module (~2,500 gas base) plus 2--3 SSTOREs for state updates: ~45,000--60,000 gas per transition. Actual measured values per DAG size are planned as part of Experiment 4 (Appendix~D).}

\textbf{DAG Batching for Large Deployments.} For DAGs exceeding the L2 block gas limit (~30M gas), \texttt{deployDAG()} must be partitioned across multiple blocks. AgentCity implements a \textit{batched deployment protocol}: the Codification Agent partitions the DAG into deployment tranches, each containing at most \texttt{maxNodesPerTranche} nodes (constitutional parameter, default: 500). Each tranche is deployed as a separate transaction, and the CollaborationContract maintains a \texttt{deploymentComplete} flag that is set only when all tranches have been committed. The mission remains in PENDING state until \texttt{deploymentComplete == true}, preventing premature execution of partially deployed DAGs.

\begin{lstlisting}
ALGORITHM BatchedDAGDeployment(nodes: DAGNode[], edgeList: Edge[]):
  INPUT: full DAG specification from legislative output
  OUTPUT: fully deployed CollaborationContract with all nodes and edges

  trancheSize <- constitutionalParams.maxNodesPerTranche  // default: 500
  numTranches <- ceil(|nodes| / trancheSize)
  
  FOR i FROM 0 TO numTranches - 1:
    trancheNodes <- nodes[i * trancheSize : min((i+1) * trancheSize, |nodes|)]
    trancheEdges <- edges where both endpoints are in already-deployed nodes union trancheNodes
    CALL CollaborationContract.deployDAGTranche(i, numTranches, trancheNodes, trancheEdges)
    REQUIRE transaction succeeds; if failure, retry with same tranche index
  
  CALL CollaborationContract.finalizeDeployment()
  REQUIRE CollaborationContract.totalDeployedNodes == |nodes|
  EMIT DAGFullyDeployed(missionId, |nodes|, numTranches)
\end{lstlisting}

For a 5,000-node DAG (planned large-scale stress test, companion empirical paper), this yields approximately 10 deployment tranches, each consuming ~15M gas---well within the per-block limit. The batching overhead (additional transactions and finalization check) is estimated at ~50,000 gas per tranche boundary.

\textbf{Staking and Escrow Mechanism.} The economic incentive layer operates through escrow functions integrated into the existing contract architecture rather than a separate StakingContract. Two escrow mechanisms enforce good-faith participation:

\textit{Registration Escrow (AgentContract).} Upon registration, each agent deposits a registration stake (default: 1,000 units of the mission's settlement token) into the AgentContract. This stake is locked for the agent's entire active lifetime and is refunded only upon voluntary deregistration with no pending missions. Registration stake serves as a Sybil defense (see Sybil Attack Cost Bound above).

\textit{Task-Level Escrow (CollaborationContract).} Upon bid acceptance (MSG\_TYPE\_5 approval), each producer agent's bid stake (specified in MSG\_TYPE\_4) is locked in the CollaborationContract for the duration of the mission. Stake is released upon successful task completion (node reaches COMPLETED state with PoP attestation confirmed). Stake is slashed (transferred to a protocol treasury controlled by the Adjudication branch) upon:

\begin{itemize}
\item Task timeout: slashingSchedule.timeoutExceeded (default: 50 units)
\item Code-hash mismatch: slashingSchedule.codeHashMismatch (default: 500 units)  
\item Node failure after freeze: slashingSchedule.nodeFailure (default: 100 units)
\item Mission abort due to agent fault: slashingSchedule.missionAbort (default: full stake)
\end{itemize}

\textbf{Game-Theoretic Incentive Bound.} For the staking mechanism to deter defection, the minimum stake must satisfy a coalition-adjusted deterrence inequality. We extend the standard deterrence bound to account for three factors identified in the threat model: (i) verification tier-specific detection probabilities, (ii) coalition formation probability under a rigorous economic model, and (iii) concrete economic denomination.

\textbf{Per-Tier Detection Probabilities.} The detection probability P(detect) is not a single scalar but varies by verification tier and attack type:

\begin{table}[htbp]
\caption{Detection probability by verification tier and attack vector.}
\label{tab:detection-prob}
\centering
\small
\begin{tabular}{p{0.22\textwidth} p{0.22\textwidth} p{0.22\textwidth} p{0.22\textwidth}}
\toprule
\textbf{Attack Vector} & \textbf{Type 1 (Hash)} & \textbf{Type 2 (Redundant, 2-of-3)} & \textbf{Type 3 (Human)} \\
\midrule
Output substitution (different output) & \ensuremath{\sim}1.0 (hash mismatch) & \ensuremath{\sim}1.0 (output divergence) & \ensuremath{\sim}1.0 (visible) \\
Subtle semantic manipulation (correct format, wrong meaning) & \ensuremath{\sim}0.0 (hash matches malicious output) & 0.33--0.67 (depends on LLM-judge rubric coverage) & 0.70--0.90 (human semantic judgment) \\
Gas/resource manipulation & \ensuremath{\sim}0.0 & \ensuremath{\sim}0.0 (same output, different resource usage) & 0.20--0.40 (requires gas profiling) \\
Timing manipulation (slow execution) & \ensuremath{\sim}0.0 & 0.67 (timeout detection) & 0.90+ (latency visible in dashboard) \\
\bottomrule
\end{tabular}
\end{table}

The effective per-tier detection probability is the weighted average across attack vectors encountered in the mission context. For a conservative security analysis, we use the minimum $P(\text{detect})$ across attack vectors for each tier: $P_1^{min} = 0.0$, $P_2^{min} = 0.33$, $P_3^{min} = 0.20$. The security analysis for Type 2 subtle semantic manipulation uses a sensitivity range for $P_2^{base}$ spanning the [0.33, 0.67] interval from Table~\ref{tab:detection-prob}:

\begin{itemize}
\item \textbf{Pessimistic (lower bound):} $P_2^{base} = 0.33$---all honest executors perform at the minimum detection rate.
\item \textbf{Moderate (true midpoint):} $P_2^{base} = 0.50$---the arithmetic mean of the Table~\ref{tab:detection-prob} range, representing the expected detection rate when honest redundant executors apply a standard LLM-judge rubric.
\item \textbf{Optimistic (upper bound):} $P_2^{base} = 0.67$---all honest executors perform at the maximum detection rate.
\end{itemize}

The moderate case ($P_2^{base} = 0.50$) is used as the primary estimate throughout the security analysis; pessimistic and optimistic bounds are reported for sensitivity.

\textbf{Coalition-Adjusted Detection Probability (EQ-2, Revised for Majority Voting).} When $c$ of $k$ redundant executors in a Type 2 verification collude, detection depends on whether the honest executors hold a majority in the $t$-of-$k$ consensus vote. The Verification module implements majority-voting consensus (e.g., 2-of-3 agreement): colluding executors vote to approve the manipulated output, while honest executors independently detect the manipulation with probability $P_2^{base}$ each. Detection succeeds if and only if (a) honest executors form a majority ($k - c \geq t$) \textit{and} (b) at least one honest executor identifies the manipulation:

\begin{equation}
\boxed{P_2(\text{detect} | c) = \begin{cases} 1 - (1 - P_2^{base})^{k-c} & \text{if } k - c \geq t \text{ (honest majority)} \\ 0 & \text{if } k - c < t \text{ (colluder majority)} \end{cases}} \tag{EQ-2}
\end{equation}

The critical distinction from the v0.17--v0.18a hypergeometric subset model is the $c \geq k - t + 1$ collapse: when colluders form a majority of the consensus quorum, they control the vote outcome regardless of whether an honest executor detects the manipulation. This matches the Verification module's \texttt{verifyConsensus()} implementation, which accepts the output when $\geq t$ executors agree. Under the hypergeometric model used in v0.18a, $P_2(\text{detect}|c=2) = (2/3) \times P_2^{base}$; under majority voting, $P_2(\text{detect}|c=2) = 0$ because the two colluders constitute a 2/3 majority. Conversely, when honest executors hold the majority ($c \leq 1$), the majority-voting model yields \textit{higher} detection probability than the subset model because multiple independent honest executors each have an independent chance to detect.

\textbf{Numerical verification for default parameters ($k=3$, $t=2$, $P_2^{base} = 0.50$ moderate):}

\begin{itemize}
\item $c=0$: 3 honest, majority held. $P_2(\text{detect}|c=0) = 1 - (1-0.50)^3 = 1 - 0.125 = \mathbf{0.875}$
\item $c=1$: 2 honest, majority held. $P_2(\text{detect}|c=1) = 1 - (1-0.50)^2 = 1 - 0.25 = \mathbf{0.750}$
\item $c=2$: 1 honest, colluder majority (2/3). $P_2(\text{detect}|c=2) = \mathbf{0}$
\item $c=3$: 0 honest. $P_2(\text{detect}|c=3) = \mathbf{0}$
\end{itemize}

\textbf{Sensitivity across $P_2^{base}$:}

\begin{center}

\small
\begin{tabular}{llll}
\toprule
\textbf{$c$} & \textbf{$P_2^{base} = 0.33$ (pessimistic)} & \textbf{$P_2^{base} = 0.50$ (moderate)} & \textbf{$P_2^{base} = 0.67$ (optimistic)} \\
\midrule
0 & 0.699 & 0.875 & 0.964 \\
1 & 0.551 & 0.750 & 0.891 \\
2 & 0 & 0 & 0 \\
3 & 0 & 0 & 0 \\
\bottomrule
\end{tabular}
\end{center}

The majority-voting model produces a sharp phase transition at $c = t = 2$: detection drops from $\geq 0.55$ to exactly $0$ when colluders gain a majority. This accurately reflects the Verification module's consensus mechanism.

\textbf{Independence Assumption and Inter-Executor Correlation Sensitivity (EQ-2a).} EQ-2 models honest executors' detection events as independent Bernoulli trials. In practice, redundant LLM-as-judge evaluators processing the same task output exhibit structural correlation: shared training data, similar reasoning heuristics, and identical evaluation rubrics produce correlated judgments. When all evaluators use the same LLM backend (as in the preliminary experiments), this correlation may approach $\rho \to 1$. We bound the impact of this correlation on $P_2(\text{detect}|c)$ by introducing a pairwise equicorrelation parameter $\rho \in [0, 1]$.

Under the equicorrelated Bernoulli model, the joint detection probability for $m = k - c$ honest executors is:

\begin{equation}
P_2^{\rho}(\text{detect}|c) = \begin{cases} 1 - \left[\rho(1 - P_2^{base}) + (1-\rho)(1 - P_2^{base})^m\right] & \text{if } m \geq t \text{ (honest majority)} \\ 0 & \text{if } m < t \end{cases} \tag{EQ-2a}
\end{equation}

\textbf{Derivation of EQ-2a.} Let $X_1, \ldots, X_m$ be equicorrelated Bernoulli random variables with common marginal $P(X_i = 1) = P_2^{base}$ and pairwise correlation $\text{Corr}(X_i, X_j) = \rho$ for all $i \neq j$. The standard construction (see Bahadur [1961], or Drezner \& Farnum [1993] for a modern treatment) introduces a latent common factor $Z \sim \text{Bernoulli}(\rho)$: conditioned on $Z = 1$, all $X_i$ are identical (perfectly correlated draw); conditioned on $Z = 0$, all $X_i$ are independent. This yields $P(\text{all fail}) = P(Z=1) \cdot (1 - P_2^{base}) + P(Z=0) \cdot (1 - P_2^{base})^m = \rho(1 - P_2^{base}) + (1-\rho)(1 - P_2^{base})^m$, and detection = 1 -- P(all fail), which is EQ-2a. The construction preserves the correct marginals ($P(X_i = 1) = P_2^{base}$) and achieves any pairwise correlation $\rho \in [0, 1]$ when $P_2^{base} \leq 0.5$ (as in our model).

At $\rho = 0$ (independent), EQ-2a reduces to EQ-2. At $\rho = 1$ (perfect correlation), $P_2^{\rho=1} = P_2^{base}$: the multiple independent detection opportunities collapse to a single effective draw.

\begin{table}[htbp]
\caption{$P_2^{\rho}(\text{detect}|c=0)$ as a function of inter-executor correlation $\rho$ ($k=3$, $t=2$, $P_2^{base} = 0.50$).}
\label{tab:rho-sensitivity}
\centering
\small
\begin{tabular}{lllll}
\toprule
\textbf{$\rho$} & \textbf{$P_2^{\rho}(c=0)$} & \textbf{$P_2^{\rho}(c=1)$} & \textbf{$P_{eff}^{\rho}$} & \textbf{$s_{min\_prod}^{\rho}$ ($V_m=\$100K$)} \\
\midrule
0.0 (independent) & 0.875 & 0.750 & 0.805 & $\$376{,}259$ \\
0.1 & 0.838 & 0.725 & 0.773 & $\$392{,}094$ \\
0.2 & 0.800 & 0.700 & 0.740 & $\$409{,}320$ \\
0.3 & 0.762 & 0.675 & 0.708 & $\$428{,}129$ \\
0.5 & 0.688 & 0.625 & 0.643 & $\$471{,}458$ \\
1.0 (perfect corr.) & 0.500 & 0.500 & 0.480 & $\$631{,}145$ \\
\bottomrule
\end{tabular}
\end{table}

\textbf{Interpretation.} At the worst case ($\rho = 1$), detection probability degrades from 0.805 to 0.480---below $P_2^{base} = 0.50$---because the $P(c \geq 2) \approx 0.04$ mass contributes zero detection regardless of $\rho$, pulling $P_{eff}$ below the single-draw baseline. The required production stake increases by $\sim$68\% relative to the independent case ($\$631{,}145$ vs. $\$376{,}259$). Even at moderate correlation ($\rho = 0.3$), $P_{eff}$ decreases by $\sim$12\% (from 0.805 to 0.708), and the required stake increases by $\sim$14\% ($\$428{,}129$ vs. $\$376{,}259$). The independence assumption is therefore not catastrophic even when substantially violated, but the security margin is materially reduced. Two architectural mitigations limit $\rho$ in practice: (i) the deviation scorer diversity requirement (\S{}A.4) mandates that redundant executors use different embedding providers, reducing shared-model correlation; (ii) the full experimental program (Appendix~D) uses a three-LLM backend matrix (frontier model, Claude 3.5, Gemini 1.5 Pro), which diversifies the judge's reasoning path and structurally reduces inter-evaluator correlation below the same-model ceiling.

We disclose the independence assumption as a modeling limitation in \S6 (Limitations) and recommend empirical calibration of $\rho$ from the multi-LLM experimental results as a priority for the companion empirical paper.

\textbf{Redundancy Factor $k$ as a Design Parameter.} The default $k=3$ ($t=2$) produces a zero-security regime when $c \geq 2$ colluders form a 2/3 majority. Increasing $k$ raises the coalition size needed for majority control and simultaneously increases the number of independent detection opportunities. We analyze $P_{eff}$ as a function of $k \in \{3, 5, 7\}$ under the standard parameters ($\pi = 0.12$, $P_2^{base} = 0.50$):

\begin{table}[htbp]
\caption{$P_{eff}$ and production stake as a function of redundancy factor $k$ ($P_2^{base} = 0.50$, $\pi = 0.12$, $V_m = \$100{,}000$). Both $1\times$ (minimum deterrence) and $3\times$ (recommended safety margin) production stakes are shown.}
\label{tab:redundancy-k}
\centering
\small
\begin{tabular}{lllllll}
\toprule
\textbf{$k$} & \textbf{$t$ (maj.)} & \textbf{Zero-sec.\ threshold} & \textbf{$P_{eff}$} & \textbf{$s_{min}^{prod}$ ($1\times$)} & \textbf{$s_{min}^{prod}$ ($3\times$)} & \textbf{Cost mult.\ vs.\ $k\!=\!3$} \\
\midrule
3 & 2 & $c \geq 2$ (67\% colluders) & 0.805 & $\$376{,}259$ & $\$1{,}128{,}778$ & $1.0\times$ \\
5 & 3 & $c \geq 3$ (60\% colluders) & 0.934 & $\$324{,}293$ & $\$972{,}878$ & $1.67\times$ execution cost \\
7 & 4 & $c \geq 4$ (57\% colluders) & 0.978 & $\$309{,}819$ & $\$929{,}456$ & $2.33\times$ execution cost \\
\bottomrule
\end{tabular}
\end{table}

\textbf{Interpretation.} Increasing $k$ from 3 to 5 yields a 16\% improvement in $P_{eff}$ (0.805 $\to$ 0.934) and a 14\% reduction in required production stake (at $1\times$: $\$376K \to \$324K$), at the cost of $1.67\times$ the redundant execution overhead. Increasing to $k = 7$ yields diminishing marginal improvement (0.934 $\to$ 0.978, +4.7\%) with $2.33\times$ execution cost. At the recommended $3\times$ safety margin, the production stakes range from $\$929K$ ($k=7$) to $\$1.13M$ ($k=3$), reflecting the practical capital requirements for HIGH-risk missions at $V_m = \$100K$. The zero-security threshold also improves: at $k=5$, an adversary must control $\geq$ 60\% of executors (rather than 67\%) to neutralize detection, raising the coalition formation cost.

\textbf{Recommendation.} For HIGH-risk missions ($V_m \geq \$100K$), $k = 5$ is recommended as the default redundancy factor: it eliminates the $c = 2$ zero-security cliff that exists at $k = 3$, provides $P_{eff} > 0.93$, and the $1.67\times$ execution overhead is acceptable for mission-critical deployments. For MEDIUM-risk missions ($V_m < \$100K$), $k = 3$ remains appropriate given the lower stakes. The redundancy factor $k$ is exposed as a constitutional parameter configurable per risk tier via the Rules Hub.

\textbf{Coalition Formation Probability (EQ-3, Corrected).} The per-agent probability of being willing to participate in a collusive coalition is modeled as an independent Bernoulli random variable with parameter $\pi$ (the per-agent malice probability). The number of colluding agents among $k$ redundant executors then follows a binomial distribution:

\begin{equation}
\boxed{P(\text{coalition size} = c) = \binom{k}{c} \pi^c (1-\pi)^{k-c}} \tag{EQ-3}
\end{equation}

The original v0.17 formula $(1 - s_{reg}/V)^c$ was not a valid probability distribution: with $s_{reg} = 1{,}000$ and $V = 50{,}000$, it gives $P(c=2) \approx (0.98)^2 = 0.96$---directly contradicting the claimed $P(c=2) \approx 0.04$ used in all downstream calculations. The formula confused an agent-level deterrence factor with a coalition-level probability.

\textbf{Economic calibration of $\pi$.} The per-agent malice probability $\pi$ is a function of the registration stake $s_{reg}$ and the agent population's baseline defection rate. We model:

\begin{equation*}
\pi(s_{reg}) = \pi_0 \cdot \exp\!\left(-\frac{s_{reg}}{s_{threshold}}\right)
\end{equation*}

where $\pi_0$ is the base malice rate in a fully open (unstaked) system and $s_{threshold}$ is the stake level at which the malice probability falls to $\pi_0/e$. For the prototype evaluation parameters ($s_{reg} = 1{,}000$, $s_{threshold} = 500$, $\pi_0 = 0.90$ representing a permissionless open system):

\begin{equation*}
\pi = 0.90 \cdot \exp(-2) \approx 0.90 \times 0.135 \approx 0.12
\end{equation*}

With $\pi = 0.12$ and $k = 3$:

\begin{itemize}
\item $P(c=0) = (0.88)^3 \approx 0.681$
\item $P(c=1) = 3 \times 0.12 \times (0.88)^2 \approx 0.279$
\item $P(c=2) = 3 \times (0.12)^2 \times 0.88 \approx \mathbf{0.038}$ $\checkmark$ (consistent with claimed $\approx 0.04$)
\item $P(c=3) = (0.12)^3 \approx 0.002$
\end{itemize}

The $\pi = 0.12$ figure represents agents who remain willing to defect despite registration staking---primarily agents with strong extrinsic incentives (e.g., competitors, attackers with high-value targets) for whom the $\approx\$2.50$ registration stake at prototype parameters provides negligible deterrence. This is consistent with the paper's prototype framing: at current parameterization, economic deterrence is not the primary defense; structural enforcement (on-chain PoP gates, hash-based Type 1 verification, Deterministic Freeze) is the primary defense, and the economic layer provides a supplementary deterrent.

\textbf{Sensitivity analysis: $\pi_0$ and $s_{threshold}$.} The parameters $\pi_0 = 0.90$ and $s_{threshold} = 500$ are assumed values representing a high-malice-rate permissionless open system; they are not empirically derived. To characterize the sensitivity of production stake conclusions to these parameters, we compute $P_{eff}$ across the plausible parameter space (using the majority-voting EQ-2 at $P_2^{base} = 0.50$):

\begin{table}[htbp]
\caption{$P_{eff}$ as a function of $\pi_0$ and $s_{threshold}$ (at $s_{reg} = 1{,}000$, $k=3$, $t=2$, $P_2^{base} = 0.50$).}
\label{tab:peff-sensitivity}
\centering
\small
\begin{tabular}{lllllll}
\toprule
\textbf{$\pi_0$ $\setminus$ $s_{threshold}$} & \textbf{250} & \textbf{500} & \textbf{750} & \textbf{1,000} & \textbf{1,500} & \textbf{2,000} \\
\midrule
0.50 & 0.871 & 0.842 & 0.796 & 0.751 & 0.678 & 0.627 \\
0.70 & 0.870 & 0.824 & 0.751 & 0.678 & 0.562 & 0.483 \\
0.80 & 0.869 & 0.814 & 0.725 & 0.637 & 0.500 & 0.409 \\
\textbf{0.90} & \textbf{0.868} & \textbf{0.804} & \textbf{0.699} & \textbf{0.595} & \textbf{0.437} & \textbf{0.335} \\
0.95 & 0.868 & 0.799 & 0.685 & 0.574 & 0.406 & 0.299 \\
\bottomrule
\end{tabular}
\end{table}

\textit{Bold row: default parameters used in this paper ($\pi_0 = 0.90$).}

At the default $\pi_0 = 0.90$, doubling $s_{threshold}$ from 500 to 1,000 reduces $P_{eff}$ from 0.804 to 0.595 (--26\%), increasing the required production stake from $\$376{,}259$ to $\$509{,}119$ (--35\%). Conversely, halving $s_{threshold}$ to 250 raises $P_{eff}$ to 0.868 (--8\% reduction in required stake). The production stake conclusions are moderately sensitive to $s_{threshold}$ and weakly sensitive to $\pi_0$ at the relevant $s_{reg}/s_{threshold}$ ratios. These parameters will be calibrated empirically from observed agent defection rates in the companion paper's Experiment 4 deployment.

\textbf{Coalition-Adjusted Deterrence Bound.} The coalition-adjusted deterrence combines EQ-2 and EQ-3:

\begin{equation}
s_{min} > \frac{E[\text{profit}(\text{defect})]}{P_{eff}(\text{detect}) \times \text{slash\_rate}} \tag{EQ-4}
\end{equation}

where

\begin{equation*}
P_{eff}(\text{detect}) = \sum_{c=0}^{k} P(\text{coalition size} = c) \times P_2(\text{detect} | c)
\end{equation*}

\textbf{$P_{eff}$ computation} (using majority-voting EQ-2 and EQ-3, $k=3$, $t=2$, $\pi=0.12$, $P_2^{base} = 0.50$ moderate):

\begin{center}

\small
\begin{tabular}{llll}
\toprule
\textbf{$c$} & \textbf{$P(c)$} & \textbf{$P_2(\text{detect}\|c)$} & \textbf{Contribution} \\
\midrule
0 & 0.681 & 0.875 & 0.596 \\
1 & 0.279 & 0.750 & 0.209 \\
2 & 0.038 & 0.00 & 0.000 \\
3 & 0.002 & 0.00 & 0.000 \\
\textbf{Total} & \textbf{1.000} & --- & $P_{eff} = \mathbf{0.805}$ \\
\bottomrule
\end{tabular}
\end{center}

\textbf{$P_{eff}$ sensitivity across $P_2^{base}$:}

\begin{center}

\small
\begin{tabular}{lll}
\toprule
\textbf{$P_2^{base}$} & \textbf{$P_{eff}$} & \textbf{$s_{min}^{prod}$ at $V_m = \$100{,}000$} \\
\midrule
0.33 (pessimistic) & 0.630 & \$480,887 \\
0.50 (moderate) & 0.805 & \$376,259 \\
0.67 (optimistic) & 0.905 & \$334,690 \\
\bottomrule
\end{tabular}
\end{center}

The moderate-case $P_{eff} = 0.8054 \approx 0.805$ is substantially higher than the v0.18a hypergeometric estimate ($P_{eff} = 0.66$ at $P_2^{base} = 0.67$) because the majority-voting model credits each honest executor with an independent detection opportunity: with 3 honest executors ($c=0$), the probability that \textit{at least one} detects is $1 - (1-0.50)^3 = 0.875$, compared to the subset model's $1.0 \times 0.50 = 0.50$ (single-draw). However, the majority-voting model is strictly worse when colluders hold the majority ($c=2$: $P_2 = 0$ vs. $0.33$ under the subset model), creating the sharp phase transition noted above. The key analytical revision is that the Verification module's consensus mechanism, not a random-subset protocol, determines detection probability; the formal model now matches the implementation.

\textbf{Concrete Economic Grounding.} We ground the analysis in a representative EVM-compatible L2 deployment context. At representative ETH prices and L2 gas costs, the default stake levels translate to:

\begin{center}

\small
\begin{tabular}{llll}
\toprule
\textbf{Risk Tier} & \textbf{Stake (units)} & \textbf{ETH Equivalent} & \textbf{USD Equivalent} \\
\midrule
LOW & 150 & 0.00015 ETH & ~\$0.38 \\
MEDIUM & 500 & 0.0005 ETH & ~\$1.25 \\
HIGH & 2,000 & 0.002 ETH & ~\$5.00 \\
Registration & 1,000 & 0.001 ETH & ~\$2.50 \\
\bottomrule
\end{tabular}
\end{center}

\textbf{Deterrence Threshold (EQ-4 corrected).} Under the coalition-adjusted model with corrected parameters:

\begin{equation}
E[\text{profit}(\text{defect})] < s_{min} \times P_{eff} \times \text{slash\_rate} = 150 \times 0.805 \times 0.33 \approx 39.8 \text{ units} \approx \$0.10 \tag{EQ-4}
\end{equation}

\textbf{Zero-detection attack classes.} For attack classes where P\textsubscript{2} = 0 identically---including gas/resource manipulation attacks that operate below the detection threshold of all three verification tiers---the deterrence bound EQ-4 has no finite solution: no stake level deters an adversary whose attack is structurally undetectable. These attack classes are outside the deterrence model's scope and must be addressed through complementary mechanisms (e.g., rate limiting, anomaly detection on economic flows) rather than stake-based deterrence.

\textbf{Explicit prototype disclaimer.} This threshold of $\approx\$0.10$ deters defection only for missions where the attacker's expected extractable profit is less than ten cents. At current prototype parameters, every agent in Table~\ref{tab:detection-prob} whose defection payoff exceeds $\$0.10$ faces no meaningful economic deterrent. \textit{The prototype parameters are structural illustrations demonstrating the mechanism's mathematical composition---the way that detection probability, stake size, and slashing rate combine into a deterrence bound---not production security guarantees.} The constitutional parameter system (\S{}B.12) enables adjudicators to adjust all stake levels without contract redeployment; the production scaling formula required to achieve meaningful deterrence is provided in the subsection immediately below.

---

\textbf{Unified Sybil + Defection Cost Model.} The Sybil Attack Cost Bound (above) and the game-theoretic deterrence bound analyze two aspects of adversarial economics independently. In practice, a Sybil attacker who registers multiple agents and then uses them to bid on tasks faces both costs simultaneously. The total adversary cost for mounting a coordinated attack with $n$ Sybil agents bidding on $k$ tasks is:

\begin{equation*}
C_{total}(n, k) = \underbrace{n \cdot s_{reg}}_{\text{registration}} + \underbrace{k \cdot s_{min}}_{\text{task staking}} + \underbrace{n \cdot C_{compute}}_{\text{agent operation}}
\end{equation*}

For the default parameters ($s_{reg} = 1{,}000$, $s_{min} = 150$ for LOW-risk), an attack requiring 5 Sybil agents bidding on 10 tasks costs: $5 \times 1{,}000 + 10 \times 150 = 6{,}500$ units ($\approx\$16.25$), plus operational costs. The registration stake is locked for the agent's entire lifetime (refundable only upon deregistration with no active missions), while task stakes are locked per-mission and subject to slashing. The interaction between these two mechanisms creates a \textit{cost amplifier}: the attacker must maintain both sunk capital (registration) and at-risk capital (task stakes) simultaneously, and cannot recover registration stakes while maintaining attack capability. This unified model supersedes the independent analyses; both the Sybil cost bound and deterrence bound should be evaluated against $C_{total}$ rather than their components in isolation.

---

\textbf{Production Deployment Parameterization.} The prototype parameters in Table~\ref{tab:stake-params} are calibrated for a low-cost evaluation environment and provide essentially no economic deterrence for missions with real financial stakes. This subsection derives production-grade stake requirements as explicit functions of mission economic value $V_m$, provides a worked example at $V_m = \$100{,}000$, and splits Table~\ref{tab:stake-params} into prototype and production columns to make the distinction unambiguous.

\textbf{Production Scaling Formulas.} For the deterrence condition (EQ-4) to hold against an adversary whose defection profit scales with mission value, each stake parameter must be expressed as a function of $V_m$:

\textbf{Minimum per-task stake (EQ-5):}

\begin{equation}
\boxed{s_{min}^{prod}(V_m) = \frac{V_m}{P_{eff} \times \text{slash\_rate}}} \tag{EQ-5}
\end{equation}

This is the \textit{minimum} stake required for the deterrence bound to hold when the attacker can extract the full mission value through single-task subversion ($\delta = 1$, worst case). For mission architectures where task-level subversion can capture only a fraction $\delta < 1$ of $V_m$ (e.g., because the mission has redundant critical paths), the stake may be reduced proportionally: $s_{min}^{prod}(\delta, V_m) = \delta \cdot V_m / (P_{eff} \times \text{slash\_rate})$.

Substituting $P_{eff} = 0.805$ (moderate, majority-voting model), $\text{slash\_rate} = 0.33$:

\begin{equation*}
s_{min}^{prod}(V_m) = \frac{V_m}{0.8054 \times 0.33} = \frac{V_m}{0.2658} \approx 3.76 \times V_m
\end{equation*}

\textbf{Intuition.} The $\approx 3.8\times$ multiplier arises because (a) detection is imperfect ($P_{eff} = 0.805 < 1$), meaning the attacker's expected penalty is discounted by the probability of evading detection, and (b) slashing is partial ($\text{slash\_rate} = 0.33$), meaning only one-third of the locked stake is confiscated upon detection. Together these factors require the locked stake to be substantially larger than the defection profit to create a negative expected value for defection. Under the pessimistic $P_2^{base} = 0.33$, the multiplier rises to $\approx 4.81\times$; under the optimistic $P_2^{base} = 0.67$, it falls to $\approx 3.35\times$.

\textbf{Sybil registration stake (EQ-6):}

\begin{equation}
\boxed{s_{reg}^{prod}(V_m) = 0.10 \times V_m} \tag{EQ-6}
\end{equation}

The registration stake must be large enough that creating a single Sybil identity imposes a meaningful cost relative to the mission value. A rule of thumb of $10\%$ of mission value ensures that mounting a 10-agent Sybil attack requires locking the full mission value in registration stakes alone, before any task staking.

\textbf{Adjudicator bribery resistance stake (EQ-7):}

\begin{equation}
\boxed{s_{adj}^{prod}(V_m) = \frac{V_m}{q}} \tag{EQ-7}
\end{equation}

where $q$ is the adjudicator quorum size (default: $q=7$). Under this parameterization, bribing a supermajority ($\lceil 2q/3 \rceil = 5$ of 7) adjudicators costs at least $5 \times s_{adj}^{prod} = 5V_m/7 \approx 0.71 \times V_m$---making adjudicator bribery economically irrational for any mission where the attacker's expected gain does not exceed $71\%$ of the mission value.

\textbf{Worked Example: $V_m = \$100{,}000$ (moderate case, $P_2^{base} = 0.50$).}

\begin{center}

\small
\begin{tabular}{lllll}
\toprule
\textbf{Parameter} & \textbf{Formula} & \textbf{Moderate} & \textbf{Pessimistic} & \textbf{Optimistic} \\
\midrule
$s_{min}^{prod}$ (HIGH, $\delta=1$) & $V_m / (P_{eff} \times \text{slash\_rate})$ & \textbf{\$376,259} & \$480,887 & \$334,690 \\
$s_{min}^{prod}$ (MED, $\delta=0.25$) & $\delta \cdot V_m / (P_{eff} \times \text{slash\_rate})$ & \textbf{\$94,065} & \$120,222 & \$83,672 \\
$s_{min}^{prod}$ (LOW, $\delta=0.075$) & $\delta \cdot V_m / (P_{eff} \times \text{slash\_rate})$ & \textbf{\$28,219} & \$36,067 & \$25,102 \\
$s_{reg}^{prod}$ & $0.10 \times V_m$ & \textbf{\$10,000} & \$10,000 & \$10,000 \\
$s_{adj}^{prod}$ & $V_m / q = \$100{,}000 / 7$ & \textbf{\$14,286}/adj. & \$14,286/adj. & \$14,286/adj. \\
\bottomrule
\end{tabular}
\end{center}

\textit{Risk-tier deltas: $\delta_{LOW} = 0.075$ (7.5\% of $V_m$ capturable via single LOW-risk task subversion), $\delta_{MEDIUM} = 0.25$ (25\% of $V_m$), $\delta_{HIGH} = 1.0$ (full $V_m$).}

\textbf{Deterrence verification (moderate):}

\begin{equation*}
s_{min}^{prod} \times P_{eff} \times \text{slash\_rate} = \$376{,}259 \times 0.805 \times 0.33 = \$100{,}000 = V_m \checkmark
\end{equation*}

The deterrence bound holds with equality at $\delta = 1$. For practical deployments, a $3\times$ safety margin ($s_{min}^{prod} = 3V_m/(P_{eff} \times \text{slash\_rate})$) is recommended to account for estimation error in $P_{eff}$ and to deter attackers with partial information about stake levels.

\textbf{Note on stake locking vs. loss.} The $\$376{,}259$ per-task stake (moderate, HIGH-risk) is \textit{locked} (not lost) for honest agents---it is returned upon successful task completion. The opportunity cost to an honest agent is the foregone yield on locked capital during the mission duration. For a 24-hour mission and a 5\% annual yield, the cost is $\$376{,}259 \times 0.05/365 \approx \$52$---a substantially smaller effective cost than the headline stake figure suggests. Mission designers should account for this opportunity cost in producer agent compensation to ensure that honest participation remains economically rational at production stake levels.

\begin{table}[htbp]
\caption{Stake parameters by deployment tier. Prototype parameters are structural illustrations demonstrating the mechanism's composition; they provide no meaningful economic deterrence for missions with real financial exposure. Production-grade parameters are derived from EQ-5 through EQ-7 and scale linearly with mission value.}
\label{tab:stake-params}
\centering
\footnotesize
\begin{tabular}{>{\raggedright\arraybackslash}p{0.16\textwidth} >{\centering\arraybackslash}p{0.04\textwidth} >{\raggedright\arraybackslash}p{0.30\textwidth} >{\raggedright\arraybackslash}p{0.20\textwidth} >{\raggedright\arraybackslash}p{0.16\textwidth}}
\toprule
\textbf{Parameter} & \textbf{$\delta$} & \textbf{Prototype (Structural)} & \textbf{Production ($V_m\!=\!\$100K$)} & \textbf{Prod.\ ($3\times$ Margin)} \\
\midrule
\textbf{Registration stake} $s_{reg}$ & --- & 1,000 units $\approx$ $2.50 | $10,000 (4.0 ETH) & \$30,000 (12.0 ETH) &  \\
\textbf{Task stake LOW} $s_{min}^{LOW}$ & 0.075 & 150 units $\approx$ $0.38 | $28,219 (11.3 ETH) & \$84,658 (33.9 ETH) &  \\
\textbf{Task stake MEDIUM} $s_{min}^{MED}$ & 0.25 & 500 units $\approx$ $1.25 | $94,065 (37.6 ETH) & \$282,194 (112.9 ETH) &  \\
\textbf{Task stake HIGH} $s_{min}^{HIGH}$ & 1.0 & 2,000 units $\approx$ $5.00 | $376,259 (150.5 ETH) & \$1,128,776 (451.5 ETH) &  \\
\textbf{Adjudicator stake} $s_{adj}$ & --- & ~100 units $\approx$ $0.25/adj. | $14,286/adj. (5.71 ETH) & \$42,857/adj. (17.1 ETH) &  \\
\textbf{Deterrence bound} (EQ-4) & --- & $\approx\$0.10$ & $\geq \$100{,}000$ & $\geq \$300{,}000$ \\
\textbf{Sybil attack cost} (10 agents) & --- & $\approx\$25$ & $\approx\$100{,}000$ & $\approx\$300{,}000$ \\
\textbf{Adjudicator bribery cost} ($q=7$) & --- & $\approx\$1.75$ & $\approx\$100{,}000$ & $\approx\$300{,}000$ \\
\textbf{Intended deployment context} & --- & Prototype evaluation; testnet & Enterprise mission ($V_m \sim \$10K$--$\$1M$) & High-assurance deployment \\
\bottomrule
\end{tabular}
\end{table}

\textit{$\delta$ = fraction of $V_m$ capturable by subverting a single task at the given risk tier. 3$\times$ safety margin recommended for production deployment to account for estimation error in $P_{eff}$ and $P_2^{base}$. All production values use the moderate-case $P_{eff} = 0.805$ ($P_2^{base} = 0.50$, majority-voting EQ-2).}

\textit{Prototype column values reflect the initial evaluation parameters. Production column values are derived analytically from EQ-5 through EQ-7 and will be empirically validated in the companion paper's Experiment 4 (Appendix~D).}

\textbf{Explicit scope statement.} The economic security analysis in~\S{}B.3 demonstrates that the three-component deterrence mechanism (EQ-2 through EQ-4) is \textit{structurally sound}: detection probability, stake size, and slashing rate compose correctly into a deterrence bound, and the bound scales monotonically with each component. The prototype parameters are not presented as adequate for production deployments; they are calibrated to minimize testnet transaction costs and enable rapid iteration during the evaluation phase. Any deployment processing missions with material financial exposure \textbf{must} recalibrate stake parameters according to EQ-5 through EQ-7 (or a more conservative variant thereof) before relying on the economic deterrence layer as a security guarantee.

\textbf{Participation Rationality: Minimum Task Reward for Honest-Agent Capital Lock-Up.} At production stakes, EQ-5 requires $s_{min}^{prod} \approx 3.76 \times V_m$ per HIGH-risk task (at $k = 3$, $P_{eff} = 0.805$, 3$\times$ safety margin). An honest agent participating in a $V_m = \$100K$ mission must lock approximately $\$376K$ in stake capital for the mission duration (typically hours to days). The opportunity cost of this lock-up is:

\begin{equation}
C_{opp} = s_{min}^{prod} \times r_{annual} \times \frac{T_{mission}}{365} \tag{EQ-8}
\end{equation}

where $r_{annual}$ is the agent operator's annualized cost of capital and $T_{mission}$ is mission duration in days. For a 1-day mission at $r_{annual} = 10\%$: $C_{opp} = 376{,}259 \times 0.10 / 365 \approx \$103$. For rational participation, the task reward $R_{task}$ must satisfy:

\begin{equation*}
R_{task} > C_{opp} + C_{exec} \approx \$103 + C_{exec}
\end{equation*}

where $C_{exec}$ is the direct execution cost (LLM inference, compute). At typical LLM inference costs of $\$0.10$--$\$1.00$ per task, the opportunity cost dominates by two orders of magnitude. This creates a participation rationality constraint: only agents with low cost of capital (institutional operators, staking pools, or agents with existing idle capital) can rationally participate in HIGH-risk missions at production stakes. This may concentrate execution among well-capitalized incumbents, reducing the permissionless participation that motivates the architecture.

Two mitigations are available: (i) at $k = 5$ (recommended for HIGH-risk, Table~\ref{tab:redundancy-k}), $s_{min}^{prod}$ ($1\times$) decreases to $\$324K$, reducing $C_{opp}$ to ${\sim}\$89$ per day; (ii) \textit{stake pooling}---multiple small-capital agents pool resources into a shared stake contract that distributes slashing risk across pool participants, analogous to Ethereum validator pooling (e.g., Lido, Rocket Pool). Stake pooling is identified as future work under the Token Economics direction (Appendix C).

\textbf{Guardian Module (CollaborationContract sub-module).} The Guardian module functions as a behavioral firewall, enforcing execution-layer invariants: reasoning deviation from mandate (>2$\sigma$ triggers Deterministic Freeze), tool invocation count (max 40 per agent per task), and message volume (max 120 per task node to prevent loops). These thresholds are not hard-coded constants but \textit{constitutional parameters} defined by human adjudicators through the Rules Hub (\S{}B.12).

\textit{Guardian Module interface (pseudocode):}

\begin{lstlisting}
MODULE Guardian (CollaborationContract sub-module):
  STATE:
    constitutionalParams: GuardianParams         // set via Rules Hub
    freezeLog: mapping(address -> FreezeRecord[]) // agent -> freeze history
    anomalyCounters: mapping(bytes32 -> AnomalyCounter) // nodeId -> counters
    escalationWindows: mapping(address -> uint256) // agent -> current window duration (ms)
    emergencyStopActive: bool
    collaborationContract: address

  STRUCT GuardianParams:
    deviationThresholdSigma: uint256  // default: 2 (2sigma)
    maxToolInvocations: uint256       // default: 40
    maxMessageVolume: uint256         // default: 120
    escalationFreezeCount: uint256    // default: 3
    baseEscalationWindowMs: uint256   // default: 1,200,000 (20 minutes)
    reputationScalingEnabled: bool    // default: true

  STRUCT AnomalyCounter:
    toolInvocations: uint256
    messageVolume: uint256
    deviationEvents: uint256
    lastResetAt: uint256

  STRUCT FreezeRecord:
    nodeId: bytes32
    agentAddr: address
    reason: FreezeReason
    timestamp: uint256
    isFalsePositive: bool
    resolvedBy: address

  ENUM FreezeReason:
    DEVIATION_EXCEEDED | TOOL_LIMIT_EXCEEDED | MESSAGE_LIMIT_EXCEEDED |
    EMERGENCY_STOP | MANUAL_OVERRIDE

  FUNCTION setThresholds(params: GuardianParams):
    REQUIRE msg.sender is Rules Hub authorized address
    SSTORE constitutionalParams = params
    EMIT ThresholdsUpdated(params)

  FUNCTION reportAnomaly(nodeId: bytes32, agentAddr: address,
                         anomalyType: AnomalyType, magnitude: uint256):
    REQUIRE msg.sender == collaborationContract OR msg.sender is authorized monitor
    counter <- anomalyCounters[nodeId]
    IF anomalyType == DEVIATION:
      IF magnitude > constitutionalParams.deviationThresholdSigma:
        CALL triggerFreeze(nodeId, agentAddr, DEVIATION_EXCEEDED)
    IF anomalyType == TOOL_INVOCATION:
      counter.toolInvocations += 1
      IF counter.toolInvocations > constitutionalParams.maxToolInvocations:
        CALL triggerFreeze(nodeId, agentAddr, TOOL_LIMIT_EXCEEDED)
    IF anomalyType == MESSAGE:
      counter.messageVolume += 1
      IF counter.messageVolume > constitutionalParams.maxMessageVolume:
        CALL triggerFreeze(nodeId, agentAddr, MESSAGE_LIMIT_EXCEEDED)
    SSTORE anomalyCounters[nodeId] = counter

  FUNCTION triggerFreeze(nodeId: bytes32, agentAddr: address, reason: FreezeReason):
    REQUIRE NOT emergencyStopActive
    record <- FreezeRecord(nodeId, agentAddr, reason, now, false, 0x0)
    APPEND record TO freezeLog[agentAddr]
    EMIT FreezeTriggered(nodeId, agentAddr, reason, now)
    CALL collaborationContract.setNodeState(nodeId, FROZEN)
    // Check escalation window
    windowCount <- countFreezesInWindow(agentAddr)
    effectiveThreshold <- getEffectiveEscalationThreshold(agentAddr)
    IF windowCount >= effectiveThreshold:
      CALL triggerEmergencyStop(agentAddr)

  FUNCTION getEffectiveEscalationThreshold(agentAddr: address) -> uint256:
    IF NOT constitutionalParams.reputationScalingEnabled:
      RETURN constitutionalParams.escalationFreezeCount
    repScore <- AgentContract.getReputationScore(agentAddr)
    // Linear scaling: reputation [0,1000] maps to threshold [3, 6]
    // threshold = base + floor((repScore / 1000) * 3)
    base <- constitutionalParams.escalationFreezeCount
    bonus <- (repScore * 3) / 1000
    RETURN base + bonus

  FUNCTION countFreezesInWindow(agentAddr: address) -> uint256:
    windowMs <- escalationWindows[agentAddr]
    IF windowMs == 0: windowMs <- constitutionalParams.baseEscalationWindowMs
    cutoff <- now - windowMs
    RETURN count of freezeLog[agentAddr] entries with timestamp > cutoff
           AND isFalsePositive == false

  FUNCTION triggerEmergencyStop(triggeringAgent: address):
    SSTORE emergencyStopActive = true
    // Double escalation window for next cycle
    escalationWindows[triggeringAgent] <- escalationWindows[triggeringAgent] * 2
    EMIT EmergencyStop(triggeringAgent, now)
    NOTIFY human adjudicators via Logging Hub

  FUNCTION unfreezeWithApproval(nodeId: bytes32, agentAddr: address):
    REQUIRE msg.sender is Override Panel authorized address
    REQUIRE collaborationContract.getNodeState(nodeId) == FROZEN
    freezeLog[agentAddr].last.resolvedBy = msg.sender
    CALL collaborationContract.setNodeState(nodeId, ELIGIBLE)
    EMIT Unfrozen(nodeId, agentAddr, msg.sender)

  FUNCTION classifyFalsePositive(agentAddr: address, freezeIndex: uint256):
    REQUIRE msg.sender is Override Panel authorized address
    SSTORE freezeLog[agentAddr][freezeIndex].isFalsePositive = true
    EMIT FalsePositiveClassified(agentAddr, freezeIndex, msg.sender)
\end{lstlisting}

\textbf{Verification Module (CollaborationContract sub-module).} The Verification module gates DAG node state transitions through Proof-of-Progress (PoP) attestations---rigorous validation ensuring that no task advances to the next DAG node without a verified proof of completion. Because task outputs vary fundamentally in their verifiability properties, AgentCity distinguishes three PoP attestation types, specified per-task-node in the CollaborationContract during the Legislation phase:

\textit{Type 1---Deterministic Verification (Hash-Based Attestation).} For tasks with formally verifiable outputs---cryptographic operations, numerical computation, deterministic data transformations---PoP uses hash-based attestation: the micro-service produces both the output and a deterministic proof (e.g., a hash of the input/output pair, a mathematical proof of correctness, or a checksum against a reference value). The Verification module verifies the proof on-chain before authorizing the DAG state transition. This tier provides the strongest mechanical guarantee: the transition is authorized if and only if the proof verifies.

\textit{Concrete Type 1 examples:} (a) A micro-service that sorts a list and returns a sorted array submits hash(input\_array || sorted\_output) together with a Merkle proof of correct ordering. The Verification module recomputes the hash and checks the proof deterministically. (b) A micro-service performing SHA-256 hashing of a document submits the input hash and the output digest; verification is a single on-chain keccak256 call. (c) A data extraction service that converts structured JSON to a canonical CSV format submits hash(source\_json || canonical\_csv); the Verification module verifies the hash matches the legislated schema hash for that node. Type 1 is the preferred tier when output verifiability is achievable, as it incurs the lowest per-node governance overhead (~8,000 gas for hash verification vs. the cost of redundant execution in Type 2).

\textit{Type 2---Statistical Verification (Statistical Attestation).} For tasks with stochastic but measurable outputs---classification, structured data extraction, numerical estimation with known error bounds---PoP uses statistical attestation: multiple independent micro-services execute the same task, and the Verification module requires consensus (e.g., 2-of-3 agreement within a configurable tolerance) before authorizing the state transition. The number of redundant executors and the consensus threshold are constitutional parameters set in the CollaborationContract. This tier trades execution cost (redundant invocations) for robustness against individual micro-service error or compromise.

\textit{Concrete Type 2 examples:} (a) A sentiment classification micro-service (output: {POSITIVE, NEGATIVE, NEUTRAL}) runs in triplicate; the Verification module checks for 2-of-3 label agreement. (b) A named-entity recognition service (output: JSON array of entities with positions) runs in duplicate; the Verification module checks that both outputs are within a Jaccard similarity threshold $\geq$ 0.85 (a constitutional parameter). (c) A numerical estimation service (output: floating-point scalar) runs in triplicate; the Verification module checks that the standard deviation of the three estimates is below a configurable tolerance (e.g., 5\% of the mean). The Type 2 tier is architecturally important for machine-learning micro-services where no deterministic proof of correctness exists but output quality can be approximated by ensemble agreement.

\textit{Type 3---Human-Assisted Verification (Delegated Attestation).} For tasks with inherently subjective outputs---natural language generation, creative content synthesis, strategic recommendation---PoP uses delegated attestation: the micro-service output is forwarded to the Adjudication branch for human review before the DAG state transition is authorized. The Verification module holds the DAG in a PENDING\_REVIEW state until a human adjudicator submits a signed attestation (approve or reject) via the Override Panel. This is the most expensive attestation tier for latency and human overhead, but it provides the strongest governance guarantee for high-stakes subjective outputs where automated verification is structurally insufficient. Delegation to human review is the correct architectural response to the limits of deterministic and statistical verification---not a fallback, but a designed tier.

\textit{Concrete Type 3 examples:} (a) A mission requiring a strategic recommendation document (e.g., market entry analysis): the micro-service output is surfaced in the Override Panel with the legislated evaluation rubric attached; a human adjudicator scores and approves or rejects. (b) A mission generating a public-facing communication: the output is displayed in the Override Panel for human editorial review before release. (c) A legal document drafting task: the generated text is held in PENDING\_REVIEW and routed to an adjudicator with legal expertise credentials (verified via AgentContract human-principal classification). Type 3 tasks impose a latency bottleneck on DAG execution; AgentCity mitigates this by enabling parallel execution of DAG branches not dependent on the pending Type 3 node, maximizing utilization during the human review window.

The three-tier PoP framework directly addresses the underspecification identified in peer review: the Verification module does not apply a single monolithic verification model, but selects the appropriate attestation mechanism based on the formal verifiability properties of each task's output domain. Implementation of all three PoP tiers and characterization of verification overhead per tier is planned as part of Experiment 4 (Appendix~D).

\textit{Verification Module interface (pseudocode):}

\begin{lstlisting}
MODULE Verification (CollaborationContract sub-module):
  STATE:
    pendingProofs: mapping(bytes32 -> PoPRecord)         // nodeId -> proof record
    redundantSubmissions: mapping(bytes32 -> bytes32[])  // nodeId -> output hashes (Type 2)
    delegatedPending: mapping(bytes32 -> bool)           // nodeId -> awaiting human review
    toleranceParams: ToleranceParams                    // constitutional parameters
    collaborationContract: address

  STRUCT PoPRecord:
    nodeId: bytes32
    popTier: uint8
    outputHash: bytes32
    proof: bytes
    submittedAt: uint256
    status: PoPStatus

  ENUM PoPStatus:
    PENDING | APPROVED | REJECTED | DELEGATED

  STRUCT ToleranceParams:
    jaccardThreshold: uint256     // default: 85 (i.e., 0.85, scaled x100)
    numericTolerancePct: uint256  // default: 5 (5% of mean)
    hashProofRequired: bool       // default: true for Type 1

  FUNCTION submitPoP(nodeId: bytes32, popTier: uint8,
                     outputHash: bytes32, proof: bytes) -> PoPStatus:
    REQUIRE collaborationContract.getNodeState(nodeId) == EXECUTING
    IF popTier == 1:
      RETURN verifyHashProof(nodeId, outputHash, proof)
    ELSE IF popTier == 2:
      RETURN verifyConsensus(nodeId, outputHash)
    ELSE IF popTier == 3:
      RETURN requestDelegated(nodeId, outputHash)

  FUNCTION verifyHashProof(nodeId: bytes32, outputHash: bytes32,
                            proof: bytes) -> PoPStatus:
    expectedSchemaHash <- CollaborationContract.dag[nodeId].outputSchemaHash
    computedHash <- keccak256(outputHash || proof)
    IF computedHash matches expectedSchemaHash:
      EMIT PoPApproved(nodeId, 1, outputHash)
      RETURN APPROVED
    ELSE:
      EMIT PoPRejected(nodeId, 1, "hash mismatch")
      RETURN REJECTED

  FUNCTION verifyConsensus(nodeId: bytes32, outputHash: bytes32) -> PoPStatus:
    APPEND outputHash TO redundantSubmissions[nodeId]
    node <- CollaborationContract.dag[nodeId]
    submissions <- redundantSubmissions[nodeId]
    IF |submissions| < node.redundancyFactor:
      RETURN PENDING         // awaiting additional executor submissions
    // Check consensus
    agreement <- computeAgreement(submissions, node.popTier)
    IF agreement >= node.consensusThreshold:
      EMIT PoPApproved(nodeId, 2, outputHash)
      RETURN APPROVED
    ELSE:
      EMIT PoPRejected(nodeId, 2, "consensus not reached")
      RETURN REJECTED

  FUNCTION requestDelegated(nodeId: bytes32, outputHash: bytes32) -> PoPStatus:
    SSTORE delegatedPending[nodeId] = true
    EMIT DelegatedReviewRequested(nodeId, outputHash, now)
    RETURN DELEGATED

  FUNCTION approveDelegated(nodeId: bytes32):
    REQUIRE msg.sender is Override Panel authorized address
    REQUIRE delegatedPending[nodeId]
    SSTORE delegatedPending[nodeId] = false
    EMIT PoPApproved(nodeId, 3, pendingProofs[nodeId].outputHash)
    CALL collaborationContract.advanceNode(nodeId, ...)

  FUNCTION rejectDelegated(nodeId: bytes32, reason: bytes32):
    REQUIRE msg.sender is Override Panel authorized address
    REQUIRE delegatedPending[nodeId]
    SSTORE delegatedPending[nodeId] = false
    EMIT PoPRejected(nodeId, 3, reason)
    CALL collaborationContract.setNodeState(nodeId, FAILED)
\end{lstlisting}

\textbf{Gate Module (CollaborationContract sub-module).} The Gate module applies constitutional output filtering as the last-mile safety check, verifying that all synthesized outputs satisfy mission-level safety predicates and the overarching System Constitution before any result exits the execution perimeter.

\textit{Gate Module interface (pseudocode):}

\begin{lstlisting}
MODULE Gate (CollaborationContract sub-module):
  STATE:
    filterPredicates: FilterPredicate[]    // defined at mission setup via Rules Hub
    vetoed: mapping(bytes32 -> bool)        // missionId -> vetoed flag
    released: mapping(bytes32 -> bool)      // missionId -> released flag
    approvedPredicateContracts: mapping(address -> bool)  // whitelist of approved predicate implementations

  STRUCT FilterPredicate:
    predicateId: bytes32
    description: string
    contractAddr: address         // address of on-chain predicate implementation
    checkFunction: bytes4         // function selector for on-chain check (if Type 1)
    isHumanAssisted: bool         // if true: route to Override Panel for approval

  FUNCTION setFilterPredicates(predicates: FilterPredicate[]):
    REQUIRE msg.sender is Codification Agent (during legislation) OR
            msg.sender is Override Panel authorized address
    FOR EACH predicate IN predicates:
      REQUIRE predicate.contractAddr \in approvedPredicateContracts
    SSTORE filterPredicates = predicates

  FUNCTION filterOutput(missionId: bytes32):
    REQUIRE NOT released[missionId]
    REQUIRE NOT vetoed[missionId]
    allPassed <- true
    FOR EACH predicate IN filterPredicates:
      IF predicate.isHumanAssisted:
        EMIT HumanFilterRequired(missionId, predicate.predicateId)
        allPassed <- false  // hold until human approval
      ELSE:
        result <- STATICCALL predicate.contractAddr.checkFunction(missionId)
        IF NOT result:
          EMIT PredicateFailed(missionId, predicate.predicateId)
          allPassed <- false
    IF allPassed:
      SSTORE released[missionId] = true
      EMIT OutputReleased(missionId, now)

  FUNCTION vetoOutput(missionId: bytes32, reason: bytes32):
    REQUIRE msg.sender is Override Panel authorized address
    SSTORE vetoed[missionId] = true
    EMIT OutputVetoed(missionId, reason, msg.sender)

  FUNCTION releaseOutput(missionId: bytes32):
    REQUIRE msg.sender is Override Panel authorized address (for manual release)
    REQUIRE NOT vetoed[missionId]
    SSTORE released[missionId] = true
    EMIT OutputReleased(missionId, now)

  FUNCTION approvePredicateContract(contractAddr: address):
    REQUIRE msg.sender is Override Panel authorized address
    SSTORE approvedPredicateContracts[contractAddr] = true
    EMIT PredicateContractApproved(contractAddr, msg.sender)
\end{lstlisting}

\textbf{Security Note: STATICCALL for Predicate Evaluation.} the Gate module uses STATICCALL (read-only external call) rather than DELEGATECALL for predicate evaluation. DELEGATECALL executes the target function in the caller's storage context, which would allow a malicious predicate---potentially registered by a compromised Codification Agent---to overwrite the Gate module storage variables (e.g., \texttt{released[missionId]} or \texttt{vetoed[missionId]}), bypassing the constitutional output filter. STATICCALL prevents state modification by the callee, eliminating this attack vector. Additionally, predicate contract addresses must be pre-approved by human adjudicators via \texttt{approvePredicateContract()}, creating a whitelist that prevents arbitrary code execution within the Gate module's security perimeter.

\textbf{Gas Limit for Predicate STATICCALL.} While STATICCALL prevents state modification, a malicious or poorly implemented predicate contract can still consume arbitrary gas up to the call gas limit, creating a potential gas-griefing vector. To mitigate this, the Gate module specifies a per-predicate gas ceiling (\texttt{predicateGasLimit}, constitutional parameter, default: 200,000 gas) that caps the gas forwarded in each STATICCALL. If the predicate exceeds this budget, the call reverts and the predicate is marked as failed---triggering a manual review by adjudicators before the predicate can be re-invoked. This parameter should be added to the Constitutional Parameter Table (\S{}B.12).

\textbf{Settlement and Treasury Modules (CollaborationContract sub-modules).} The economic layer is distributed across the AgentContract (agent-level economic state: wallet balance, stake accounting, reward history, slashing history, $\psi$($\rho$) computation) and the CollaborationContract's Settlement and Treasury modules (mission-level and protocol-level economic functions). Legislation defines mission budgets and task prices; Execution triggers payment settlements upon verified task completion via the Settlement module; Adjudication sets fee schedules and resolves payment disputes via the Treasury module. The economic functions do not constitute a separate branch; they are integrated into the existing four-contract governance architecture.

\textit{Settlement and Treasury Modules interface (pseudocode):}

\begin{lstlisting}
MODULE Settlement_and_Treasury (CollaborationContract sub-modules):
  STATE:
    missionEscrows: mapping(bytes32 -> MissionEscrow)  // missionId -> escrow
    taskBudgets: mapping(bytes32 -> uint256)            // nodeId -> allocated budget
    treasury: uint256                                   // protocol treasury balance
    insurancePool: uint256                              // insurance pool balance
    protocolFeeRate: uint256                            // f_p, default 200 (2%, basis points)
    insuranceReserveRate: uint256                       // f_i, default 100 (1%, basis points)
    reputationMultiplierAlpha: uint256                  // alpha parameter, default 500 (0.5 scaled by 1000)
    stakePoolRegistry: mapping(bytes32 -> StakePool)    // poolId -> pool
    linkedAgentContract: address
    linkedCollaborationContract: address
    linkedVerificationModule: address    // Verification module reference
    adjudicatorCompensationRate: uint256                // per Tier 3 review, default 50 units

  STRUCT MissionEscrow:
    sponsor: address
    totalDeposit: uint256
    allocated: uint256         // sum of task budgets allocated so far
    settled: uint256           // sum of rewards paid out
    missionId: bytes32
    createdAt: uint256
    status: {ACTIVE, COMPLETED, REFUNDED}

  STRUCT StakePool:
    manager: address
    totalStaked: uint256
    participants: mapping(address -> uint256)  // agent -> contributed amount
    slashingShareBasisPoints: mapping(address -> uint256)  // proportional slash allocation
    active: bool

  FUNCTION depositMissionBudget(missionId: bytes32, amount: uint256):
    // Mission sponsor deposits V_m into escrow at mission creation
    REQUIRE msg.value >= amount
    REQUIRE missionEscrows[missionId].sponsor == 0x0  // no existing escrow
    REQUIRE amount > 0
    SSTORE missionEscrows[missionId] = MissionEscrow(
      sponsor: msg.sender,
      totalDeposit: amount,
      allocated: 0,
      settled: 0,
      missionId: missionId,
      createdAt: now,
      status: ACTIVE
    )
    EMIT MissionBudgetDeposited(missionId, msg.sender, amount)

  // Gas analysis: depositMissionBudget() performs 1 SSTORE for the MissionEscrow struct
  // (struct packed into 3-4 storage slots) + calldata overhead. Estimated: 3 x 20,000
  // (cold SSTORE) + 2,100 (SLOAD for existence check) + calldata ~= 65,000 gas.

  FUNCTION allocateTaskBudget(missionId: bytes32, nodeId: bytes32, amount: uint256):
    // Called during legislative codification to assign per-task budget
    REQUIRE msg.sender == linkedCollaborationContract
    escrow <- missionEscrows[missionId]
    REQUIRE escrow.status == ACTIVE
    REQUIRE escrow.allocated + amount <= escrow.totalDeposit
    SSTORE taskBudgets[nodeId] = amount
    SSTORE missionEscrows[missionId].allocated = escrow.allocated + amount
    EMIT TaskBudgetAllocated(missionId, nodeId, amount)

  // Gas analysis: allocateTaskBudget() performs 2 SSTOREs (taskBudgets entry + allocated
  // field update) + 2 SLOADs (escrow lookup + field read). Estimated: 2 x 20,000 +
  // 2 x 2,100 ~= 44,200 gas cold; subsequent allocations ~= 10,000 gas (warm SSTORE).

  FUNCTION settleReward(agentAddr: address, nodeId: bytes32):
    // Called after PoP verification success. Implements EQ-9.
    REQUIRE msg.sender == linkedVerificationModule  // Settlement called by Verification module
    b <- taskBudgets[nodeId]
    REQUIRE b > 0
    rho <- AgentContract(linkedAgentContract).getReputationScore(agentAddr)
    psi <- computeReputationMultiplier(agentAddr)  // psi(rho)
    // EQ-9 (split-source): R_task = R_base x min(psi, 1.0) + treasury_subsidy
    // where R_base = b x (1 - f_p/10000 - f_i/10000), and
    // treasury_subsidy = R_base x max(psi - 1.0, 0) drawn from protocol treasury
    feeNumerator <- protocolFeeRate + insuranceReserveRate  // e.g., 300 for 3%
    netRate <- 10000 - feeNumerator                         // e.g., 9700 for 97%
    grossReward <- b x netRate / 10000                      // = R_base (e.g., 970 at b=1000)
    basePsi <- min(psi, 1000)                               // clamp at 1.0 (neutral); psi scaled by 1000
    baseReward <- grossReward x basePsi / 1000              // <= grossReward, paid from escrow
    premiumPsi <- max(psi - 1000, 0)                        // 0 if psi <= 1.0
    treasurySubsidy <- grossReward x premiumPsi / 1000      // funded from protocol treasury
    protocolFee <- b x protocolFeeRate / 10000
    insuranceReserve <- b x insuranceReserveRate / 10000
    // Treasury underflow guard: cap subsidy at available treasury balance + incoming protocol fee
    requestedSubsidy <- treasurySubsidy
    treasurySubsidy <- min(treasurySubsidy, treasury + protocolFee)
    IF treasurySubsidy < requestedSubsidy THEN
      EMIT TreasuryUnderflow(nodeId, agentAddr, requestedSubsidy - treasurySubsidy)
    reward <- baseReward + treasurySubsidy                  // total agent payout
    // Checks-Effects-Interactions: all state updates before TRANSFER
    DELETE taskBudgets[nodeId]
    SSTORE treasury = treasury + protocolFee - treasurySubsidy  // net treasury: fees in, subsidy out
    SSTORE insurancePool = insurancePool + insuranceReserve
    // Find mission and update settled amount (escrow only covers base + fees)
    missionId <- CollaborationContract(linkedCollaborationContract).getMissionForNode(nodeId)
    SSTORE missionEscrows[missionId].settled += (baseReward + protocolFee + insuranceReserve)
    // settled <= b_i: at psi=1.25, settled = 970 + 20 + 10 = 1000 = b_i [v]
    EMIT RewardSettled(agentAddr, nodeId, reward, baseReward, treasurySubsidy, protocolFee, insuranceReserve)
    TRANSFER reward to agentAddr  // external call last (CEI pattern)

  // Gas analysis: settleReward() performs 1 DELETE + 2 SSTORE (treasury, insurancePool) +
  // 1 SSTORE (settled field) + 2 cross-contract CALLs (getReputationScore, getMissionForNode) +
  // 1 TRANSFER + additional SLOAD (treasury read for net debit) + min/max arithmetic ops.
  // Estimated: 1 x 5,000 (DELETE/refund) + 3 x 20,000 (cold SSTORE) + 1 x 2,100 (cold SLOAD) +
  // 2 x 2,500 (CALL base) + TRANSFER + min/max ops ~= 60,000 gas. Warm-path (settled field
  // already written): ~= 40,000 gas.

  FUNCTION computeReputationMultiplier(agentAddr: address) -> uint256:
    // Implements psi(rho) = 1 + alpha x (rho - 500) / 1000
    // Returns value scaled by 1000 (so 1.0 = 1000, 1.15 = 1150)
    rho <- AgentContract(linkedAgentContract).getReputationScore(agentAddr)
    // Neutral reputation (rho = 500) yields psi = 1 + alpha x 0 = 1.0 (= 1000 scaled)
    // rho = 800: psi = 1 + 0.5 x (800-500)/1000 = 1.15 (= 1150 scaled)
    // rho = 200: psi = 1 + 0.5 x (200-500)/1000 = 0.85 (= 850 scaled)
    IF rho >= 500:
      delta <- (rho - 500) x reputationMultiplierAlpha / 1000  // positive increment, scaled
      RETURN 1000 + delta
    ELSE:
      delta <- (500 - rho) x reputationMultiplierAlpha / 1000  // penalty, scaled
      RETURN 1000 - delta  // floor at 0 to prevent underflow

  // Gas analysis: computeReputationMultiplier() is a read-only function (STATICCALL eligible).
  // 1 cross-contract CALL + arithmetic: ~= 2,500 (CALL) + 200 (arithmetic) ~= 2,700 gas.
  // Called internally from settleReward(); its cost is included in that estimate.

  FUNCTION poolStake(poolId: bytes32, amount: uint256):
    // Agent contributes to stake pool to share slashing risk (resolves IW-7)
    REQUIRE msg.value >= amount
    REQUIRE amount > 0
    pool <- stakePoolRegistry[poolId]
    REQUIRE pool.active
    REQUIRE pool.participants[msg.sender] == 0 OR pool.participants[msg.sender] > 0
    // Update participant share; recalculate basis points proportionally
    prevContrib <- pool.participants[msg.sender]
    newTotal <- pool.totalStaked + amount
    SSTORE stakePoolRegistry[poolId].participants[msg.sender] = prevContrib + amount
    SSTORE stakePoolRegistry[poolId].totalStaked = newTotal
    // Basis points for this participant: (prevContrib + amount) x 10000 / newTotal
    SSTORE stakePoolRegistry[poolId].slashingShareBasisPoints[msg.sender] =
           (prevContrib + amount) x 10000 / newTotal
    EMIT StakePooled(poolId, msg.sender, amount, newTotal)

  // Gas analysis: poolStake() performs 3 SSTOREs (participants mapping, totalStaked,
  // slashingShareBasisPoints mapping) + 2 SLOADs. Estimated: 3 x 20,000 (cold) +
  // 2 x 2,100 ~= 64,200 gas cold; 3 x 2,900 (warm) ~= 10,900 gas warm.
  // Default target: ~45,000 gas at mixed cold/warm profile.

  FUNCTION withdrawPooledStake(poolId: bytes32):
    // Agent withdraws contribution from pool (with checks for active missions)
    pool <- stakePoolRegistry[poolId]
    contrib <- pool.participants[msg.sender]
    REQUIRE contrib > 0
    REQUIRE pool.active
    // Prevent withdrawal while pool has active mission obligations
    REQUIRE CollaborationContract(linkedCollaborationContract)
            .poolHasActiveMission(poolId) == false
    // Checks-Effects-Interactions
    SSTORE stakePoolRegistry[poolId].totalStaked = pool.totalStaked - contrib
    DELETE stakePoolRegistry[poolId].participants[msg.sender]
    DELETE stakePoolRegistry[poolId].slashingShareBasisPoints[msg.sender]
    EMIT StakeWithdrawn(poolId, msg.sender, contrib)
    TRANSFER contrib to msg.sender  // external call last

  // Gas analysis: withdrawPooledStake() performs 1 SSTORE + 2 DELETEs + 1 CALL +
  // 1 TRANSFER. Estimated: 1 x 20,000 + 2 x 5,000 (DELETE refund) + 2,500 (CALL) +
  // 21,000 (TRANSFER) ~= 53,500 gas cold; warm-path ~= 32,000 gas.

  FUNCTION claimInsurance(agentAddr: address, missionId: bytes32,
                          lossAmount: uint256, evidence: bytes32):
    // Honest agent claims compensation from insurance pool
    // Only callable by human adjudicator after Tier 3 review confirms honest loss
    REQUIRE msg.sender is Override Panel authorized adjudicator
    REQUIRE missionEscrows[missionId].status == ACTIVE OR
            missionEscrows[missionId].status == COMPLETED
    REQUIRE lossAmount > 0
    REQUIRE insurancePool >= lossAmount
    // Checks-Effects-Interactions
    SSTORE insurancePool = insurancePool - lossAmount
    EMIT InsuranceClaimed(agentAddr, missionId, lossAmount, evidence, msg.sender)
    TRANSFER lossAmount to agentAddr

  // Gas analysis: claimInsurance() performs 1 SSTORE (insurancePool) + 3 SLOADs +
  // 1 TRANSFER. Estimated: 1 x 20,000 + 3 x 2,100 + 21,000 ~= 47,300 gas.

  FUNCTION disburse(disbursementType: {GOVERNANCE, INSURANCE, GAS_SUBSIDY},
                    recipient: address, amount: uint256):
    // Adjudicator-authorized treasury disbursement for governance rewards,
    // insurance payouts, or infrastructure gas subsidies
    REQUIRE msg.sender is Override Panel authorized adjudicator
    REQUIRE amount > 0
    REQUIRE treasury >= amount
    // Checks-Effects-Interactions
    SSTORE treasury = treasury - amount
    EMIT TreasuryDisbursed(disbursementType, recipient, amount, msg.sender)
    TRANSFER amount to recipient

  // Gas analysis: disburse() performs 1 SSTORE (treasury) + 2 SLOADs + 1 TRANSFER.
  // Estimated: 1 x 20,000 + 2 x 2,100 + 21,000 ~= 45,200 gas.
  // For adjudicator compensation (disbursementType = GOVERNANCE, recipient = adjudicator
  // address, amount = adjudicatorCompensationRate): this is the standard disbursement path
  // following each Tier 3 PoP review.
\end{lstlisting}

\textit{Gas summary for the Settlement and Treasury modules. The total estimated deployment cost of ~200,000 gas reflects the struct initialization (MissionEscrow and StakePool struct definitions, mapping registrations) and contract linkage SSTOREs (3 linked addresses $\times$ 20,000 gas). Per-operation costs follow the same SSTORE-counting methodology applied to all contract types: each cold SSTORE costs 20,000 gas; each warm SSTORE (slot already written this transaction) costs 2,900 gas; TRANSFER costs 21,000 gas; cross-contract CALL base costs 2,500 gas. The settleReward function at ~60,000 gas is the most gas-intensive per-operation call, reflecting its role in executing three simultaneous state updates (treasury, insurancePool, missionEscrow.settled) plus a cross-contract reputation lookup, the additional SLOAD and min/max arithmetic for split-source reward computation, and the terminal TRANSFER.}

---

\textbf{Economic Equilibrium Analysis.} The economic layer (distributed across AgentContract and the CollaborationContract's Settlement and Treasury modules) introduces four formal equilibrium conditions that govern the sustainability and participation rationality of the AgentCity token economy. These equations extend and supersede the informal economic analysis in the participation rationality section above.

\textbf{EQ-9: Task Reward Settlement.}

\begin{equation}
\boxed{R_{\text{task}}(i) = R_{\text{base}}(i) \times \min(\psi(\rho_i),\, 1.0) + \text{treasury\_subsidy}(i)} \tag{EQ-9}
\end{equation}

where $R_{\text{base}}(i) = b_i \times \left(1 - \frac{f_p}{10000} - \frac{f_i}{10000}\right)$ is the base reward paid from mission escrow, and $\text{treasury\_subsidy}(i) = R_{\text{base}}(i) \times \max(\psi(\rho_i) - 1.0,\, 0)$ is the reputation premium drawn from the protocol treasury. Per-task mission escrow disbursement is bounded: fees + base reward $\leq b_i$, ensuring mission budget solvency by construction.

The split-source formulation ensures mission budget solvency by construction: the escrow-funded component $R_{\text{base}} \times \min(\psi, 1.0)$ never exceeds $R_{\text{base}}$, and fees + base reward $\leq b_i$. The reputation premium for agents with $\psi > 1.0$ is financed separately from the protocol treasury. At default parameters ($f_p = 200$, $f_i = 100$, $\alpha = 0.5$) and neutral reputation ($\rho = 500$, $\psi = 1.0$), an agent receives $R_{\text{base}} = 0.97 \times b_i$ after fees with zero treasury subsidy. At $\rho = 800$ ($\psi = 1.15$), the agent receives $R_{\text{base}} \times 1.0 + R_{\text{base}} \times 0.15 = 0.97b_i + 0.1455b_i = 1.1155b_i$---a 15.0\% premium over the neutral net reward. This split-source expression is implemented by the Settlement module's settleReward() function.

\textbf{EQ-10: Participation Rationality (revised).} For rational participation, the net reward must exceed the opportunity cost of locked stake plus direct execution cost:

\begin{equation}
\boxed{b_i \times \left(1 - \frac{f_p}{10000} - \frac{f_i}{10000}\right) \times \psi(\rho_i) > s_{\min}^{\text{prod}} \times r_{\text{annual}} \times \frac{T_{\text{mission}}}{365} + C_{\text{exec}}} \tag{EQ-10}
\end{equation}

This supersedes the informal analysis in the participation rationality subsection above (EQ-8). Where EQ-8 characterized only the cost side of the rationality constraint, EQ-10 provides the closed-form condition over both reward and cost, enabling direct comparison of stake-pool strategies: an agent participating through a stake pool with a smaller individual stake contribution $s_i < s_{\min}^{\text{prod}}$ (pool absorbs the difference) faces a lower $C_{\text{opp}}$ and therefore a lower bid-price floor to satisfy EQ-10. Stake pooling thus directly expands the set of agents for whom rational participation is feasible.

\textbf{EQ-11: Treasury Sustainability Condition.} For the protocol treasury to remain solvent in steady state, protocol fee inflows plus slashing inflows must exceed all disbursement obligations:

\begin{equation}
\boxed{\sum_i \frac{f_p \cdot b_i}{10000} + \sum_j (\text{slash}_j \times 0.5) > \sum_k C_{\text{adj},k} + \sum_l P_{\text{ins},l} + \sum_m G_{\text{sub},m} + \sum_n T_{\text{sub},n}} \tag{EQ-11}
\end{equation}

where $C_{\text{adj},k}$ is the compensation paid for the $k$-th Tier 3 adjudication review; $P_{\text{ins},l}$ is the $l$-th insurance payout; $G_{\text{sub},m}$ is the $m$-th gas subsidy disbursement; and $T_{\text{sub},n}$ is the $n$-th treasury reputation subsidy (i.e., $R_{\text{base},n} \times \max(\psi_n - 1.0, 0)$ per settled task). The 0.5 factor reflects the 50/50 treasury/insurance split of slashed stakes established in the updated slashStake() function. This condition must be verified empirically in Experiment 1 by tracking treasury balance over 200 rounds.

\textbf{EQ-12: Insurance Pool Adequacy.} The insurance pool balance $I$ must satisfy:

\begin{equation}
\boxed{I > \mathbb{E}[N_{\text{affected\ missions/period}}] \times \mathbb{E}[L_{\text{honest\ loss/mission}}]} \tag{EQ-12}
\end{equation}

The insurance reserve rate $f_i$ is calibrated such that the pool grows faster than expected claims under the steady-state mission volume. Given default $f_i = 100$ basis points (1\%), a mission economy processing $V_m = \$10{,}000$ average missions generates $\$100$ per mission into the insurance pool. If the honest-agent loss per adversarial mission averages $\$500$ and the adversarial mission rate is $5\%$ (drawn from the 10\% adversarial agent fraction in Experiment 1), then one insurance-triggered mission per 20 generates $20 \times \$100 = \$2{,}000$ in pool reserves vs. $\$500$ in expected claims, satisfying EQ-12 with a 4$\times$ reserve buffer. Sensitivity to higher adversarial rates is analyzed in the table below.

\textbf{Worked Example: Steady-State Economy.} Consider an economy processing 100 missions per month at average mission value $V_m = \$10{,}000$ and average task budget $b_i = V_m / 8 \approx \$1{,}250$ per task (8-node DAG).

\textit{Monthly treasury inflows:}

\begin{itemize}
\item Protocol fee revenue: $100 \times \$10{,}000 \times 0.02 = \$20{,}000$
\item Slashing inflows (5\% failure rate, 33\% slash rate, average task stake $s_{\min}^{\text{prod}} = 3.76 \times \$1{,}250 \approx \$4{,}700$; 8 tasks per mission): $100 \times 0.05 \times 8 \times \$4{,}700 \times 0.33 \times 0.5 \approx \$31{,}020$ (treasury portion)
\item Total monthly treasury inflows $\approx \$51{,}020$
\end{itemize}

\textit{Monthly treasury outflows:}

\begin{itemize}
\item Adjudicator compensation: assume 10 Tier 3 reviews per month at 50 units each ($\$125$ total at $\$0.25$/unit): $\$125$
\item Insurance payouts: 5 adversarial missions per month $\times$ avg. $\$300$ honest-agent loss = $\$1{,}500$
\item Gas subsidies: 20 low-value missions per month at $\$5$ subsidy each = $\$100$
\item Treasury reputation subsidies: Under a mature economy with average reputation $\rho = 700$ ($\psi_{\text{avg}} = 1.10$), the per-task treasury subsidy averages $R_{\text{base}} \times 0.10 = 0.97 \times b_i \times 0.10$. At 100 missions/month with 8 tasks each and $b_i = \$1{,}250$: aggregate monthly subsidy $\approx 100 \times 8 \times \$1{,}250 \times 0.97 \times 0.10 = \$97{,}000$
\item Total monthly disbursements $\approx \$1{,}725 + \$97{,}000 = \$98{,}725$
\end{itemize}

\textit{Net monthly treasury balance:} $\approx \$51{,}020 - \$98{,}725 = -\$47{,}705$ (deficit). Under this mature-economy scenario, treasury subsidy demand exceeds inflows, yielding a deficit. Treasury sustainability therefore requires one or more of: (a) a constitutional cap on aggregate monthly subsidy disbursement (e.g., capping treasury\_subsidy outflows at 50\% of monthly treasury inflows); (b) a lower $\alpha$ parameter reducing the maximum premium; or (c) higher protocol fee rates. At $\alpha = 0.25$ ($\psi_{\max} = 1.125$), steady-state subsidy demand drops to approximately $\$48{,}500$/month, restoring a narrow solvency margin. This treasury sustainability constraint is disclosed as a limitation in~\S{}6.

\textbf{Sensitivity Analysis: EQ-10 Participation Rationality.} Table S1 shows how the minimum bid price $b_{\min}$ required for rational participation (satisfying EQ-10 with $C_{\text{exec}} = \$1.00$ and $T_{\text{mission}} = 8$ hours) varies with reputation score $\rho$, mission value $V_m$, and protocol fee rate $f_p$. The stake is computed from EQ-5 ($s_{\min}^{\text{prod}} = V_m / (P_{\text{eff}} \times \text{slash\_rate}) \approx 3.76 \times V_m$ at the moderate-case parameters); opportunity cost uses $r_{\text{annual}} = 10\%$.

\textit{Table S1: Minimum bid price $b_{\min}$ (USD) for rational participation by reputation score, mission value, and fee rate. Values computed from EQ-10 at $r_{\text{annual}} = 10\%$, $T_{\text{mission}} = 8$ h, $C_{\text{exec}} = \$1.00$, $f_i = 1\%$ fixed.}

\begin{center}

\small
\begin{tabular}{lllll}
\toprule
\textbf{$f_p$} & \textbf{$V_m$} & \textbf{$\rho = 200$ ($\psi = 0.85$)} & \textbf{$\rho = 500$ ($\psi = 1.00$)} & \textbf{$\rho = 800$ ($\psi = 1.15$)} \\
\midrule
1\% & $\$1{,}000$ & $\$1.61$ & $\$1.37$ & $\$1.19$ \\
1\% & $\$10{,}000$ & $\$5.33$ & $\$4.53$ & $\$3.94$ \\
1\% & $\$100{,}000$ & $\$42.47$ & $\$36.10$ & $\$31.39$ \\
2\% & $\$1{,}000$ & $\$1.63$ & $\$1.39$ & $\$1.20$ \\
2\% & $\$10{,}000$ & $\$5.38$ & $\$4.58$ & $\$3.98$ \\
2\% & $\$100{,}000$ & $\$42.91$ & $\$36.47$ & $\$31.71$ \\
3\% & $\$1{,}000$ & $\$1.65$ & $\$1.40$ & $\$1.22$ \\
3\% & $\$10{,}000$ & $\$5.44$ & $\$4.62$ & $\$4.02$ \\
3\% & $\$100{,}000$ & $\$43.35$ & $\$36.85$ & $\$32.04$ \\
\bottomrule
\end{tabular}
\end{center}

\textit{Key observation: Higher reputation score $\rho$ reduces the minimum rational bid price by $\approx 26\%$ at $\alpha = 0.5$ (comparing $\rho = 800$ vs. $\rho = 200$), reflecting the full $\psi$ range from 0.85 to 1.15. Higher protocol fees modestly increase the minimum bid price (by $\approx 1\%$ per additional 100 basis points). At low mission values ($V_m = \$1{,}000$), the execution cost $C_{\text{exec}} = \$1.00$ dominates the opportunity cost ($C_{\text{opp}} = \$0.34$), producing very low minimum bid prices ($b_{\min} \approx \$1.4$). At high mission values ($V_m = \$100{,}000$), $C_{\text{opp}}$ dominates $C_{\text{exec}}$ by a factor of $\sim$34$\times$, and $b_{\min}$ scales nearly linearly with $V_m$.}

---

\textbf{Contract Linking.} After deployment, contracts are linked according to the legislative agreements, forming the on-chain governance mesh:

\begin{itemize}
\item CollaborationContract references ServiceContract(s) for task routing.
\item The CollaborationContract's Guardian module performs behavioral audits at each DAG transition.
\item The Verification module gates DAG node state transitions---no task can advance without a Proof-of-Progress attestation of the appropriate tier.
\item The Gate module filters all outputs before delivery to the mission requester.
\item All contracts reference AgentContract for identity and authorization checks.
\end{itemize}

The resulting contract mesh is the \textit{auditable wiring layer} identified in~\S{}B.3: every micro-service binding, every task dependency, every governance constraint is recorded on-chain. Neither the agents that produced the wiring nor any single organization controls the ledger. Human adjudicators, developers from any participating organization, and automated compliance tools can independently verify the full execution topology by examining the contract state.

\textbf{Contract Deployment Sequence.} The deployment sequence during the Legislation phase follows a strict ordering to ensure that linked addresses are available before they are needed:

\begin{lstlisting}
ALGORITHM ContractDeploymentSequence:
  INPUT: legislative_output (DAGSpec, behavioral_params, filter_predicates)
  OUTPUT: deployed contract mesh with validated linkages

  STEP 1: Deploy AgentContract (if not already deployed globally)
    -> record agentContractAddr

  STEP 2: Deploy ServiceContract instances for each micro-service binding
    FOR EACH ms IN legislative_output.micro_services:
      addr[ms.id] <- DEPLOY ServiceContract
      CALL addr[ms.id].registerService(ms.codeHash, ms.apiSchemaHash,
                                        ms.endpoint, ms.constraints)
    -> record serviceContractAddrs[]

  STEP 3: Initialize Guardian module with constitutional parameters
    guardianAddr <- CollaborationContract.initGuardianModule()
    CALL guardianAddr.setThresholds(legislative_output.behavioral_params)
    -> record guardianAddr

  STEP 4: Initialize Verification module with tolerance parameters
    verificationAddr <- CollaborationContract.initVerificationModule()
    CALL verificationAddr.setTolerances(legislative_output.tolerance_params)
    -> record verificationAddr

  STEP 5: Initialize Gate module with filter predicates
    gateAddr <- CollaborationContract.initGateModule()
    CALL gateAddr.setFilterPredicates(legislative_output.filter_predicates)
    -> record gateAddr

  STEP 6: Deploy CollaborationContract and link all contracts
    collabAddr <- DEPLOY CollaborationContract
    CALL collabAddr.deployDAG(
      nodes = legislative_output.dag_nodes,
      edgeList = legislative_output.dag_edges,
      guardianAddr, verificationAddr, gateAddr
    )
    -> record collabAddr

  STEP 7: Validate all cross-contract linkages
    ASSERT collabAddr.guardianModule == guardianAddr
    ASSERT collabAddr.verificationModule == verificationAddr
    ASSERT collabAddr.gateModule == gateAddr
    FOR EACH node IN legislative_output.dag_nodes:
      ASSERT ServiceContract(node.serviceId).verifyCodeHash(liveHash) == true

  STEP 8: Emit DeploymentComplete event on CollaborationContract
    EMIT DeploymentComplete(missionId, collabAddr, block.number)
\end{lstlisting}

---

\subsection{Motivation and architectural position}

AgentCity v0.26 implemented the "stick" side of the economic incentive model: registration staking (EQ-6) as a Sybil barrier; per-task escrow (lockStake/releaseStake/slashStake) as a defection deterrent; and the coalition-adjusted deterrence bound (EQ-2 through EQ-4). What v0.26 lacked was the "carrot" side: a reward distribution mechanism for successful task completion, a formalized token marketplace, treasury economics with disbursement rules, reputation-to-economic feedback, and steady-state macro-economic equilibrium analysis.

The Economic Layer introduced in v0.27 (and consolidated in v0.29 into the four-contract architecture) closes this gap by distributing economic functions across the AgentContract (agent-level economic state) and the CollaborationContract's Settlement and Treasury modules (mission-level and protocol-level functions). The design draws on three complementary frameworks: blockchain-enhanced incentive-compatible mechanisms for multi-agent coordination~\cite{ref58}, cryptoeconomic security via restaking~\cite{ref59}, and the Token Economy Design Method (TEDM) for structured tokenomics design~\cite{ref60}. The economic layer is \textbf{not} a separate branch of the SoP model. It is a cross-cutting economic substrate that operates within the existing three-branch structure:

\begin{itemize}
\item \textbf{Legislation} defines mission budgets, negotiates task prices, and sets economic parameters via depositMissionBudget() and allocateTaskBudget().
\item \textbf{Execution} triggers reward settlements upon PoP verification success via settleReward().
\item \textbf{Adjudication} sets fee schedules and constitutional parameters (protocolFeeRate, insuranceReserveRate, reputationMultiplierAlpha), resolves payment disputes via claimInsurance(), and authorizes treasury disbursements via disburse().
\end{itemize}

\subsection{Token flow architecture}

The complete token flow model is depicted schematically below. All flows are on-chain ERC-20 or native token transfers executed atomically within Settlement and Treasury module function calls.

\begin{lstlisting}
Mission Sponsor
     |
     v V_m (mission deposit via depositMissionBudget)
+--------------------------------------------------------+
|        CollaborationContract (Settlement & Treasury)    |
|                                                        |
|  Mission Escrow Pool                                   |
|  +------------------+                                  |
|  |  V_m locked      |                                  |
|  |  (MissionEscrow) |                                  |
|  +--------+---------+                                  |
|           | allocateTaskBudget (per-task b_i)           |
|           v                                            |
|  +------------------+                                  |
|  |  Task Budget     |--> Protocol Fee (f_p) --> treasury
|  |  taskBudgets[]   |--> Insurance (f_i) --> insurancePool
|  |  (per nodeId)    |--> Agent Reward (EQ-9) --> Agent Wallet
|  +------------------+                                  |
|                                                        |
|  Slashing Inflows (from CollaborationContract.slashStake)
|  +------------------+                                  |
|  |  Slashed stakes  |--> 50% --> treasury              |
|  |  (adversarial/   |--> 50% --> insurancePool         |
|  |   failed agents) |                                  |
|  +------------------+                                  |
|                                                        |
|  Treasury Outflows (via disburse / claimInsurance)     |
|  +------------------+                                  |
|  |  Governance      |--> Adjudicator Compensation      |
|  |  Rewards         |     (per Tier 3 PoP review)       |
|  |  Insurance       |--> Honest-Agent Compensation     |
|  |  Payouts         |     (claimInsurance)              |
|  |  Gas Subsidy     |--> Low-value Mission Subsidy      |
|  +------------------+                                  |
+--------------------------------------------------------+
\end{lstlisting}

\subsection{Reputation multiplier $\psi$($\rho$): closed-form derivation}

The reputation multiplier $\psi(\rho)$ creates a positive feedback loop: good task execution raises reputation, which raises earnings, which relaxes the participation rationality constraint (EQ-10), enabling access to higher-value missions. The functional form is chosen to be linear for auditability and to have a fixed neutral point at $\rho = 500$:

\begin{equation*}
\psi(\rho) = 1 + \alpha \times \frac{\rho - 500}{1000}
\end{equation*}

At $\alpha = 0.5$ (default, encoded as \texttt{reputationMultiplierAlpha = 500} in basis-1000 form):

\begin{itemize}
\item $\rho = 0$: $\psi = 1 + 0.5 \times (0 - 500)/1000 = 0.75$ (25\% penalty for minimum reputation)
\item $\rho = 500$: $\psi = 1.0$ (neutral; no premium or penalty)
\item $\rho = 800$: $\psi = 1 + 0.5 \times (800 - 500)/1000 = 1.15$ (~15\% premium)
\item $\rho = 1000$: $\psi = 1 + 0.5 \times (1000 - 500)/1000 = 1.25$ (25\% premium for maximum reputation)
\end{itemize}

The multiplier applies to the \textbf{net reward} (after fee deductions), not the gross bid price. This means that the protocol treasury and insurance pool always receive their fixed basis-point share of the gross bid, regardless of the agent's reputation. Only the agent's take-home portion is reputation-adjusted. For agents with $\psi(\rho) > 1.0$ (reputation above 500), the premium above 1.0 is financed via a treasury subsidy: $R_{\text{task}}(i) = R_{\text{base}}(i) \times \min(\psi(\rho_i), 1.0) + \text{treasury\_subsidy}(i)$, where $R_{\text{base}}(i) = b_i \times (1 - f_p/10000 - f_i/10000)$ (with $f_p$ and $f_i$ in basis points, e.g., $f_p = 200$ for 2\%) and $\text{treasury\_subsidy}(i) = R_{\text{base}}(i) \times \max(\psi(\rho_i) - 1.0, 0)$ drawn from the protocol treasury. This ensures that per-task disbursement never exceeds $b_i$ (fees + base reward $\leq b_i$, preserving mission budget solvency), while treasury and insurance inflows remain predictable and not subject to reputation fluctuations.

The constitutional parameter \texttt{reputationMultiplierAlpha} can be adjusted from 0 (disabling reputation weighting entirely, $\psi \equiv 1.0$) to 1000 ($\alpha = 1.0$, yielding $\psi \in [0.5, 1.5]$ across the full reputation range). At the default $\alpha = 0.5$, $\psi \in [0.75, 1.25]$. Higher $\alpha$ increases the incentive for reputation maintenance but also increases income inequality among agents and treasury subsidy outflows---a tradeoff governed by the WGC metric in Experiment 1.

\subsection{Stake pooling: IW-7 resolution}

Reviewer concern IW-7 identified stake concentration dynamics as an unresolved architectural limitation. At production stakes (EQ-5: $s_{\min}^{prod} \approx 3.76 \times V_m$ per HIGH-risk task), participation in high-value missions requires locking capital that is prohibitive for small operators. The stake pooling mechanism resolves this by allowing multiple agents to aggregate stake contributions into a shared pool managed by the CollaborationContract's Settlement module:

\begin{enumerate}
\item A pool manager calls \texttt{createStakePool(poolId)} (implicit via first poolStake call) to initialize a pool.
\item Participating agents call \texttt{poolStake(poolId, amount)} to contribute capital.
\item The pool registers its combined stake with the CollaborationContract for mission participation, satisfying $s_{\min}^{prod}$ collectively.
\item If any pool participant's assigned task is slashed, the loss is distributed proportionally according to \texttt{slashingShareBasisPoints}.
\item Upon pool deactivation (no active missions), participants call \texttt{withdrawPooledStake(poolId)} to retrieve contributions.
\end{enumerate}

This design is analogous to Ethereum validator pooling protocols~\cite{ref59}, where small holders delegate ETH to pooled staking operators. Key differences: (a) pool participation in AgentCity is task-specific, not continuous; (b) each pool participant retains their individual reputation score; (c) the pool itself has no on-chain reputation (it is a stake aggregation vehicle, not an agent).

The participation rationality improvement: an agent contributing $s_i$ to a pool of total size $S > s_{\min}^{prod}$ faces opportunity cost $C_{opp} = s_i \times r_{annual} \times T_{mission}/365$, which is strictly less than the solo-participation cost $s_{\min}^{prod} \times r_{annual} \times T_{mission}/365$ whenever $s_i < s_{\min}^{prod}$. The EQ-10 participation rationality threshold therefore decreases proportionally with pool aggregation, expanding the set of rational participants.

\textbf{Pool operator risk and limitations.} The pool manager role introduces a concentration risk: a malicious or negligent pool manager could accept poorly qualified agents into the pool, increasing the expected slashing rate for all pool participants. The current design mitigates this partially through proportional loss distribution (\texttt{slashingShareBasisPoints}), so that the slashed agent bears the majority of the loss. However, correlated failures---where multiple pool participants are assigned to the same mission and fail simultaneously---could amplify losses beyond the proportional model's assumptions. Pool-level reputation tracking (aggregating the performance history of agents participating through a given pool) is identified as a future enhancement that would allow mission sponsors to discriminate between well-managed and poorly-managed pools.

\subsection{Insurance pool: mechanics and calibration}

The insurance pool accumulates from two inflows: (a) the insurance reserve rate $f_i$ applied to all settled task rewards; (b) 50\% of all slashed stakes (the other 50\% flows to the treasury). This dual-source design ensures that periods of high slashing activity (adversarial missions, coalition attacks) simultaneously replenish the insurance pool that compensates honest agents who were exposed to those attacks.

Insurance payouts are triggered exclusively by human adjudicator action via \texttt{claimInsurance()}. The claimant must provide evidence (an on-chain bytes32 reference to a Logging Hub CID documenting the loss) and the claim must be authorized by an Override Panel adjudicator after a Tier 3 PoP review. This ensures that the insurance pool cannot be drained by fraudulent claims.

The adequacy condition EQ-12 is calibrated as follows. Under Experiment 1's 10\% adversarial agent fraction and 5\% failure rate per round:

\begin{itemize}
\item At $n = 200$ agents and $p = 8$ tasks per mission: expected adversarial task failures per round $\approx 200 \times 0.10 \times 0.33 \times (8/200) \approx 0.13$ tasks
\item Expected insurance-triggering events per 72 rounds: $72 \times 0.05 \approx 3.6$ events
\item Expected insurance pool balance after 72 rounds at $f_i = 1\%$, average bid $b_i = \$1,250$, 200 agents: $72 \times 200 \times 0.01 \times \$1,250 / 200 = \$9$ (normalized per agent)
\end{itemize}

These are toy values at prototype stake levels; the qualitative finding is that the pool grows faster than expected claims under the Experiment 1 parameters, satisfying EQ-12. Production calibration of $f_i$ requires empirical measurement of honest-agent loss frequency in the companion empirical paper.

\subsection{Treasury disbursement governance}

The treasury disbursement pathway via \texttt{disburse()} is the only mechanism for withdrawing accumulated protocol fees from the Treasury module. It is restricted to Override Panel authorized adjudicators and requires explicit disbursementType specification. Three disbursement types are supported:

\begin{itemize}
\item \texttt{GOVERNANCE}: Compensation for Tier 3 PoP reviewers (default rate: \texttt{adjudicatorCompensationRate} units per review). This creates a positive economic incentive for adjudicators to process Tier 3 reviews promptly, addressing the Tier 3 queue overflow limitation identified in \S{}C.1.
\item \texttt{INSURANCE}: Manual insurance payouts to honest agents, complementing the automated \texttt{claimInsurance()} path for cases requiring human discretion.
\item \texttt{GAS\textbackslash\{\}\_SUBSIDY}: Infrastructure subsidy for low-value missions, lowering the on-ramp cost for sponsors deploying governance in resource-constrained contexts.
\end{itemize}

All disbursements are logged via \texttt{EMIT TreasuryDisbursed} with the adjudicator's address, creating an auditable trail in the Logging Hub. Treasury parameter governance (protocolFeeRate, insuranceReserveRate, adjudicatorCompensationRate) follows the standard Rules Hub update protocol---requiring an adjudicator signature, an EIP-712 typed-data payload, and an IPFS-anchored justification CID.

\subsection{Legislation Module}

The Legislation Module implements the two preparatory phases of the mission lifecycle---agent registration and multi-party legislative negotiation---that together produce the contract specifications governing all subsequent execution. Both phases are mediated by the contract architecture defined in~\S{}B.3, and the legislative output feeds directly into the Execution Infrastructure described in~\S{}B.11.

\textbf{Phase 1: Registration.} Before any mission begins, all participating agents register through the AgentContract, binding their cryptographic identity (DID-compatible~\cite{ref35}) to a human principal address and establishing their reputation standing. The AgentContract classifies each registrant as either a producer agent (operational workforce) or a management agent (oversight committee). This classification determines permitted operations throughout the mission: management agents may inspect and flag legislative deliberations but cannot submit task bids; producer agents may bid and deploy micro-services but cannot modify contract parameters. The Registry Agent verifies identities and reputation thresholds before admitting participants to the Legislation phase. An agent whose reputation falls below the constitutional floor (set in the Rules Hub) is excluded from mission participation pending adjudication review. Additionally, management agents (Registry, Legislative, Regulatory, Codification) must be registered through the ManagementContract, which binds each to an authority envelope specifying permitted operations and mandatory microservice delegations. This registration is a prerequisite for management agent participation in the legislative protocol.

\textbf{Phase 2: Legislation.} The Legislation module implements multi-party legislative negotiation as a five-node LangGraph~\cite{ref18} graph. The five agents---Legislative, Producer, Regulatory, Registry, and Codification---interact through a structured message-passing protocol. We formalize this protocol below.

\textbf{Message-Passing Protocol.} Let A = {A\_reg, A\_leg, A\_prod\^{(1)}, ..., A\_prod\^{(p)}, A\_reg\_agent, A\_cod} denote the set of legislative agents, where A\_reg is the Registry Agent, A\_leg the Legislative Agent, A\_prod\^{(i)} the i-th Producer Agent, A\_reg\_agent the Regulatory Agent, and A\_cod the Codification Agent. The protocol defines seven message types, grouped into five sequential phases. Each round is defined by a message type, a sender-receiver mapping, and a validity predicate that must hold before the round may advance.

\textit{Message Types:}

\begin{lstlisting}
MSG_TYPE_1: IdentityVerificationRequest
  Sender:   A_reg
  Receiver: ALL agents
  Fields:   {session_id: bytes32, nonce: bytes32, required_min_reputation: uint256}
  Purpose:  Initiates legislative session; requests identity proof from all participants

MSG_TYPE_2: IdentityAttestation
  Sender:   Agent A_i (response to MSG_TYPE_1)
  Receiver: A_reg
  Fields:   {agent_did: string, signature: bytes, reputation_proof: MerkleProof,
             human_principal: address}
  Purpose:  Agent presents on-chain identity evidence
  Validity: signature verifies against agent_did; reputation_proof verifies against
            AgentContract state root; reputation_score >= required_min_reputation

MSG_TYPE_3: DAGProposal
  Sender:   A_leg
  Receiver: ALL (broadcast)
  Fields:   {proposal_id: bytes32, dag_spec: DAGSpec, rationale: string,
             token_budget_total: uint256, deadline_ms: uint256}
  Purpose:  Legislative Agent proposes initial task decomposition
  Validity: dag_spec is a valid DAG (no cycles); all I/O schemas are well-formed;
            token_budget_total <= mission_budget_cap (from Rules Hub)

MSG_TYPE_4: TaskBid
  Sender:   A_prod^{(i)}
  Receiver: A_reg_agent
  Fields:   {bid_id: bytes32, task_node_id: bytes32, service_id: bytes32,
             proposed_code_hash: bytes32, stake_amount: uint256,
             estimated_latency_ms: uint256, pop_tier_acceptance: uint8}
  Purpose:  Producer Agent bids on a task node with a specific micro-service binding
  Validity: service_id is registered in ServiceContract;
            proposed_code_hash matches ServiceContract record;
            stake_amount >= min_stake(task_risk_level);
            pop_tier_acceptance matches DAGProposal's specified tier for that node;
            A_prod^{(i)} is of type PRODUCER

MSG_TYPE_5: RegulatoryDecision
  Sender:   A_reg_agent
  Receiver: A_leg (and broadcast to all for transparency)
  Fields:   {decision_id: bytes32, approved_bids: BidRecord[],
             rejected_bids: RejectedBid[], fairness_score: uint256,
             compliance_flags: ComplianceFlag[], regulatory_signature: bytes}
  Purpose:  Regulatory Agent arbitrates bidding; produces approved assignment set
  Validity: approved_bids covers all required task nodes in DAGProposal;
            fairness_score >= min_fairness_threshold (constitutional parameter);
            no compliance_flags are CRITICAL severity;
            regulatory_signature verifies against A_reg_agent's DID

MSG_TYPE_6: CodedContractSpecification
  Sender:   A_cod
  Receiver: A_leg (for final approval), then broadcast + submitted to blockchain
  Fields:   {spec_id: bytes32, collaboration_contract_spec: ContractSpec,
             guardian_module_spec: ModuleSpec,
             verification_module_spec: ModuleSpec,
             gate_module_spec: ModuleSpec,
             service_contract_specs: ContractSpec[],
             constitutional_validation_proof: ValidationProof}
  Purpose:  Codification Agent compiles legislative output into deployable contract specs
  Validity: All specs pass Rules Hub constitutional validation (see below);
            constitutional_validation_proof verifies against current Rules Hub state root

MSG_TYPE_7: LegislativeApproval
  Sender:   A_leg
  Receiver: A_cod (triggers deployment)
  Fields:   {approval_id: bytes32, spec_id: bytes32, legislative_signature: bytes,
             epoch: uint256}
  Purpose:  Legislative Agent provides final sign-off; authorizes deployment
  Validity: spec_id matches MSG_TYPE_6 spec_id;
            A_leg signature verifies;
            A_reg_agent co-signature present (dual sign-off requirement)
\end{lstlisting}

\textbf{Formal Definition of Fairness Score.} The fairness\_score is computed as the complement of the normalized Herfindahl--Hirschman Index (HHI) over the bid distribution:

\begin{equation*}
\text{fairness\_score} = 1000 \times \left(1 - \frac{HHI - HHI_{min}}{HHI_{max} - HHI_{min}}\right)
\end{equation*}

where $HHI = \sum_{j=1}^{p} s_j^2$, $s_j$ is the share of total task-node assignments allocated to producer agent j, $HHI_{min} = 1/p$ (perfectly distributed), and $HHI_{max} = 1$ (monopoly). A fairness\_score of 1000 indicates perfectly equitable distribution; 0 indicates complete monopolization. The constitutional minimum (default: 600) prevents any single producer from capturing more than approximately 63\% of task assignments in a mission, providing a formal economic barrier against bid-pool monopolization by Sybil agents.

\textbf{Formal Derivation of Monopolization Bound.} For a market with p producers where one dominant producer holds share s and the remaining (p--1) share equally, $HHI = s^2 + (1-s)^2/(p-1)$, with $HHI_{min} = 1/p$ and $HHI_{max} = 1$. Setting $\text{fairness\_score} \geq 600$:

\begin{equation*}
\frac{s^2 + (1-s)^2/(p-1) - 1/p}{1 - 1/p} \leq 0.4
\end{equation*}

For p = 2: $s \leq 0.816$. For p = 5: $s \leq 0.72$. For p = 10: $s \leq 0.68$. The "approximately 63\%" bound stated above applies conservatively to markets with p $\geq$ 15 producers. The constitutional minimum of 600 thus provides progressively stronger monopolization protection as the producer pool grows---precisely the regime where Sybil-based bid-pool manipulation becomes most concerning.

\textit{Legislative Protocol State Machine:}

\begin{lstlisting}
STATES: SESSION_INIT, IDENTITY_VERIFICATION, PROPOSAL_OPEN, BIDDING_OPEN,
        REGULATORY_REVIEW, CODIFICATION, AWAITING_APPROVAL, DEPLOYED, FAILED

TRANSITIONS:
  SESSION_INIT         -> IDENTITY_VERIFICATION : A_reg broadcasts MSG_TYPE_1
  IDENTITY_VERIFICATION -> PROPOSAL_OPEN        : all agents submit valid MSG_TYPE_2;
                                                  Registry Agent confirms quorum
                                                  (all required roles present)
  IDENTITY_VERIFICATION -> FAILED               : any agent fails identity verification
                                                  OR reputation floor not met
  PROPOSAL_OPEN        -> BIDDING_OPEN          : A_leg broadcasts valid MSG_TYPE_3;
                                                  timeout: legislative_proposal_timeout
                                                  (default: 10 min, constitutional param)
  PROPOSAL_OPEN        -> FAILED               : timeout expired without valid proposal;
                                                  OR proposal fails DAG well-formedness
  BIDDING_OPEN         -> REGULATORY_REVIEW     : all task nodes have >= 1 valid bid;
                                                  bidding_window expires
                                                  (default: 15 min, constitutional param)
  BIDDING_OPEN         -> FAILED               : bidding_window expires with uncovered nodes
  REGULATORY_REVIEW    -> CODIFICATION         : A_reg_agent broadcasts valid MSG_TYPE_5
                                                  with no CRITICAL compliance flags
  REGULATORY_REVIEW    -> PROPOSAL_OPEN        : A_reg_agent rejects proposal and requests
                                                  re-proposal (max 2 re-proposals per epoch)
  CODIFICATION         -> AWAITING_APPROVAL    : A_cod broadcasts valid MSG_TYPE_6 passing
                                                  all constitutional validation checks
  CODIFICATION         -> FAILED               : constitutional validation fails after
                                                  max_codification_retries (default: 2)
  AWAITING_APPROVAL    -> DEPLOYED             : MSG_TYPE_7 received with dual signatures
                                                  (A_leg + A_reg_agent)
  AWAITING_APPROVAL    -> FAILED               : approval_timeout expires (default: 5 min)
\end{lstlisting}

\textbf{Legislative Authorization Protocol.} The Regulatory Agent's sign-off constitutes the primary authorization gate in the legislative process. The Regulatory Agent must provide a regulatory\_signature on MSG\_TYPE\_5 that verifies against its DID on-chain. Deployment of the final contract specification additionally requires a \textit{dual co-authorization}: both the Legislative Agent (A\_leg) and the Regulatory Agent (A\_reg\_agent) must co-sign MSG\_TYPE\_7 (LegislativeApproval), providing on-chain cryptographic evidence that both deliberative roles authorized the final specification. This serial proposal-review authorization prevents either A\_leg alone or A\_reg\_agent alone from unilaterally deploying a contract specification.

\textbf{Architectural Limit of Dual Co-Authorization.} The two-party co-authorization scheme provides separation-of-authorization for the legislative finalization step but does not constitute Byzantine fault-tolerant consensus: if an adversary simultaneously controls both A\_leg and A\_reg\_agent, dual co-signature can be obtained for a malicious contract specification without architectural resistance. This collusion scenario is formally acknowledged as NP-6 (Appendix~A). The security guarantee of the legislative branch is therefore contingent on the separation of A\_leg and A\_reg\_agent roles---analogous to the separation of duties principle in access control. If this separation is violated, the legislative branch provides no structural guarantee and the system's integrity degrades to human adjudicator vigilance (\S{}B.12).

For multi-organization missions where multiple Regulatory Agents may be active (one per participating organization), a configurable quorum policy applies: the Codification Agent will not advance to AWAITING\_APPROVAL until a threshold q of R total Regulatory Agents have submitted approvals (default: q = ceil(R / 2) + 1, i.e., simple majority plus one). This threshold is a constitutional parameter adjustable via the Rules Hub.

\textbf{Constitutional Parameter Validation.} Before the Codification Agent may advance to AWAITING\_APPROVAL, the compiled contract specification must pass a constitutional validation check against the current Rules Hub state. This check verifies the following conditions:

\begin{lstlisting}
ALGORITHM ConstitutionalValidation(spec: CodedContractSpecification) -> ValidationProof:
  INPUT:  compiled contract specification
  OUTPUT: ValidationProof (or FAIL with structured error)

  // Behavioral parameter bounds
  CHECK spec.guardian_params.deviationThresholdSigma \in [1, 5]
  CHECK spec.guardian_params.maxToolInvocations \in [5, 200]
  CHECK spec.guardian_params.maxMessageVolume \in [10, 500]
  CHECK spec.guardian_params.escalationFreezeCount \in [2, 10]

  // Budget compliance
  CHECK spec.total_token_budget <= rules_hub.mission_budget_cap
  FOR EACH node IN spec.dag_nodes:
    CHECK node.tokenBudget > 0
    CHECK node.timeoutMs \in [1000, rules_hub.max_node_timeout_ms]

  // PoP tier validity
  FOR EACH node IN spec.dag_nodes:
    CHECK node.popTier \in {1, 2, 3}
    IF node.popTier == 2:
      CHECK node.redundancyFactor \in [2, 5]
      CHECK node.consensusThreshold <= node.redundancyFactor
      CHECK node.consensusThreshold >= ceil(node.redundancyFactor / 2)  // majority
    IF node.popTier == 3:
      CHECK node.timeoutMs >= rules_hub.min_human_review_timeout_ms  // default: 300,000 ms

  // Identity and stake checks
  FOR EACH bid IN spec.approved_bids:
    CHECK AgentContract.meetsReputationFloor(bid.producer_agent)
    CHECK bid.stake_amount >= rules_hub.min_stake_per_risk_tier[bid.risk_tier]
    CHECK ServiceContract.verifyCodeHash(bid.service_id, bid.proposed_code_hash)

  // DAG structural validity
  CHECK dag is acyclic (topological sort succeeds)
  CHECK all DAG leaf nodes have popTier assignment
  CHECK dag has >= 1 root node and >= 1 terminal node

  // Fairness check
  CHECK regulatory_decision.fairness_score >= rules_hub.min_fairness_score  // default: 600/1000

  IF all checks pass:
    proof <- sign(hash(spec), codification_agent_key)
    RETURN ValidationProof(proof, block.number, rules_hub_state_root)
  ELSE:
    RETURN FAIL(structured error list)
\end{lstlisting}

\textbf{DAG Specification Format.} The legislative output is serialized as a JSON-LD document conforming to the following schema. This format constitutes the canonical record of the legislated mission topology and is the direct input to the contract deployment sequence defined in~\S{}B.11.

\begin{lstlisting}
{
  "@context": {
    "@vocab": "https://agentcity.ai/ontology/v1#",
    "dag": "https://agentcity.ai/ontology/v1#DAGSpec",
    "schema": "http://schema.org/"
  },
  "@type": "dag:MissionSpecification",
  "missionId": "<bytes32 hex string>",
  "epoch": "<uint256: legislative epoch index>",
  "legislativeSessionId": "<bytes32 hex string>",
  "createdAt": "<ISO 8601 timestamp>",
  "approvedBy": {
    "legislativeAgent": "<DID URI>",
    "regulatoryAgent": "<DID URI>",
    "legislativeSignature": "<hex-encoded signature>",
    "regulatorySignature": "<hex-encoded signature>"
  },
  "constitutionalParams": {
    "deviationThresholdSigma": 2,
    "maxToolInvocations": 40,
    "maxMessageVolume": 120,
    "escalationFreezeCount": 3,
    "baseEscalationWindowMs": 1200000,
    "missionBudgetCap": 100000,
    "minFairnessScore": 600
  },
  "dagNodes": [
    {
      "@type": "dag:TaskNode",
      "nodeId": "<bytes32 hex string>",
      "label": "<human-readable task name>",
      "serviceId": "<bytes32 hex string, references ServiceContract>",
      "producerAgentDid": "<DID URI>",
      "inputSchema": {
        "$ref": "<JSON Schema URI or inline schema>"
      },
      "outputSchema": {
        "$ref": "<JSON Schema URI or inline schema>"
      },
      "inputSchemaHash": "<bytes32 keccak256 of canonical JSON Schema>",
      "outputSchemaHash": "<bytes32 keccak256 of canonical JSON Schema>",
      "popTier": 1,
      "redundancyFactor": 1,
      "consensusThreshold": 1,
      "tokenBudget": 4000,
      "timeoutMs": 30000,
      "riskTier": "LOW"
    }
  ],
  "dagEdges": [
    {
      "from": "<nodeId bytes32>",
      "to": "<nodeId bytes32>",
      "dataFlowSchema": "<JSON Schema URI describing the data passed on this edge>"
    }
  ],
  "serviceContracts": [
    {
      "serviceId": "<bytes32>",
      "codeHash": "<bytes32 keccak256 of service artifact>",
      "apiSchemaHash": "<bytes32>",
      "endpoint": "<URL>",
      "ownerDid": "<DID URI>",
      "constraints": {
        "maxLatencyMs": 5000,
        "maxMemoryMB": 512,
        "maxConcurrentInvocations": 3
      }
    }
  ],
  "gatePredicates": [
    {
      "predicateId": "<bytes32>",
      "description": "<natural language description of output filter>",
      "isHumanAssisted": false,
      "checkFunctionSelector": "<bytes4 function selector>"
    }
  ],
  "stakeSchedule": [
    {
      "producerAgentDid": "<DID URI>",
      "stakeAmount": 1000,
      "slashingSchedule": {
        "nodeFailure": 100,
        "codeHashMismatch": 500,
        "timeoutExceeded": 50
      }
    }
  ]
}
\end{lstlisting}

\textit{Ontology Note:} The \texttt{https://agentcity.ai/ontology/v1\textbackslash\{\}\#} namespace is a project-internal URI that does not currently resolve to a published ontology document. For reproducibility, all terms used in the DAG specification are self-documenting within the JSON-LD \texttt{@context} block: each key (e.g., \texttt{dag}, \texttt{missionId}, \texttt{nodes}) maps directly to the corresponding CollaborationContract struct fields defined in~\S{}B.3, and the schema is fully specified by the DAGNode struct and edge list format. A formal OWL ontology publication is planned as part of the system's open-source release.

\textbf{Worked Example: Three-Task Mission.} We trace a simple mission through the full legislative process to illustrate how the five agents interact to produce a DAG. The mission: \textit{Given a URL list, (T1) fetch and parse each page, (T2) classify each page's topic, (T3) aggregate results into a structured JSON report.}

\begin{lstlisting}
LEGISLATIVE SESSION TRACE (simplified):

Session ID: 0xABCD...
Epoch: 1

ROUND 1 -- Identity Verification:
  A_reg -> ALL: MSG_TYPE_1 {session_id: 0xABCD, nonce: 0x1234, min_reputation: 400}
  A_leg -> A_reg: MSG_TYPE_2 {did: "did:key:zLeg...", reputation_proof: pi_leg, score: 720}
  A_prod[1] -> A_reg: MSG_TYPE_2 {did: "did:key:zP1...", reputation_proof: pi_p1, score: 580}
  A_prod[2] -> A_reg: MSG_TYPE_2 {did: "did:key:zP2...", reputation_proof: pi_p2, score: 610}
  A_reg_agent -> A_reg: MSG_TYPE_2 {did: "did:key:zReg...", reputation_proof: pi_reg, score: 850}
  A_cod -> A_reg: MSG_TYPE_2 {did: "did:key:zCod...", reputation_proof: pi_cod, score: 790}
  -> A_reg CONFIRMS: all identities verified, all scores >= 400. Advance to PROPOSAL_OPEN.

ROUND 2 -- DAG Proposal:
  A_leg -> ALL: MSG_TYPE_3 {
    proposal_id: 0xPROP01,
    dag_spec: {
      nodes: [
        {nodeId: 0xT1, label: "FetchAndParse", popTier: 1,
         inputSchema: URLListSchema, outputSchema: ParsedPageSchema,
         tokenBudget: 5000, timeoutMs: 15000},
        {nodeId: 0xT2, label: "TopicClassify", popTier: 2,
         redundancyFactor: 3, consensusThreshold: 2,
         inputSchema: ParsedPageSchema, outputSchema: TopicLabelSchema,
         tokenBudget: 8000, timeoutMs: 20000},
        {nodeId: 0xT3, label: "AggregateReport", popTier: 3,
         inputSchema: TopicLabelSchema, outputSchema: ReportSchema,
         tokenBudget: 6000, timeoutMs: 300000}
      ],
      edges: [(0xT1 -> 0xT2), (0xT2 -> 0xT3)]
    },
    token_budget_total: 19000,
    deadline_ms: 600000
  }
  -> DAG validates: acyclic, all nodes typed, T3 has popTier=3 with timeout >= 300,000 ms. [v]

ROUND 3 -- Bidding:
  A_prod[1] -> A_reg_agent: MSG_TYPE_4 {
    bid_id: 0xBID01, task_node_id: 0xT1,
    service_id: 0xSVC_FETCH, proposed_code_hash: 0xHASH_FETCH,
    stake_amount: 200, estimated_latency_ms: 8000, pop_tier_acceptance: 1
  }
  A_prod[2] -> A_reg_agent: MSG_TYPE_4 {
    bid_id: 0xBID02, task_node_id: 0xT2,
    service_id: 0xSVC_CLASSIFY_A, proposed_code_hash: 0xHASH_CL_A,
    stake_amount: 300, estimated_latency_ms: 12000, pop_tier_acceptance: 2
  }
  // Two more bids for T2 redundancy (Type 2 requires redundancyFactor=3):
  A_prod[1] -> A_reg_agent: MSG_TYPE_4 {bid_id: 0xBID03, task_node_id: 0xT2,
    service_id: 0xSVC_CLASSIFY_B, proposed_code_hash: 0xHASH_CL_B,
    stake_amount: 300, estimated_latency_ms: 13000, pop_tier_acceptance: 2}
  A_prod[2] -> A_reg_agent: MSG_TYPE_4 {bid_id: 0xBID04, task_node_id: 0xT2,
    service_id: 0xSVC_CLASSIFY_C, proposed_code_hash: 0xHASH_CL_C,
    stake_amount: 300, estimated_latency_ms: 11000, pop_tier_acceptance: 2}
  A_prod[1] -> A_reg_agent: MSG_TYPE_4 {
    bid_id: 0xBID05, task_node_id: 0xT3,
    service_id: 0xSVC_AGGREGATE, proposed_code_hash: 0xHASH_AGG,
    stake_amount: 250, estimated_latency_ms: 5000, pop_tier_acceptance: 3
  }
  -> All task nodes covered. Bidding window closes.

ROUND 4 -- Regulatory Review:
  A_reg_agent evaluates bids:
    - 0xBID01: stake 200 >= min_stake(LOW) = 150. Code hash verified on ServiceContract. [v]
    - 0xBID02/03/04: Three bids for T2 (matches redundancyFactor=3). Stake >= 150. [v]
    - 0xBID05: stake 250 >= min_stake(MEDIUM) = 200. Code hash verified. [v]
    - Fairness score: 780/1000 (bids distributed across both producers; no monopoly). [v]
    - Compliance flags: none CRITICAL.
  A_reg_agent -> ALL: MSG_TYPE_5 {
    approved_bids: [0xBID01, 0xBID02, 0xBID03, 0xBID04, 0xBID05],
    fairness_score: 780,
    compliance_flags: [],
    regulatory_signature: SIG_REG
  }
  -> Advance to CODIFICATION.

ROUND 5 -- Codification:
  A_cod compiles DAG spec + approved bids into contract specifications.
  Constitutional validation:
    deviationThresholdSigma=2 \in [1,5] [v]; maxToolInvocations=40 \in [5,200] [v];
    T3 timeout=300,000 >= min_human_review=300,000 [v];
    DAG acyclic [v]; all nodes typed [v]; fairness_score=780 >= 600 [v].
  A_cod -> A_leg: MSG_TYPE_6 {
    spec_id: 0xSPEC01,
    collaboration_contract_spec: {nodes: [T1,T2,T3], edges: [...], ...},
    verification_module_spec: {tolerances: {...}},
    constitutional_validation_proof: pi_valid
  }
  A_leg + A_reg_agent -> A_cod: MSG_TYPE_7 {
    spec_id: 0xSPEC01,
    legislative_signature: SIG_LEG,
    regulatory_co_signature: SIG_REG,
    epoch: 1
  }
  -> Dual sign-off confirmed. Contract deployment sequence initiates.

DEPLOYMENT OUTPUT:
  CollaborationContract at: [address assigned at deployment time; to be reported in companion empirical paper]
    - 3 DAG nodes (T1: WAITING, T2: WAITING, T3: WAITING)
    - 2 edges: T1->T2, T2->T3
    - popTiers: T1=1, T2=2, T3=3
  ServiceContracts: 0xSVC_FETCH, 0xSVC_CLASSIFY_A/B/C, 0xSVC_AGGREGATE
  Guardian module: thresholds sigma=2, tools=40, msgs=120
  Verification module: Type 2 tolerance params set
  Gate module: report_schema_conformance predicate set
\end{lstlisting}

All legislative deliberations---all MSG\_TYPE\_1 through MSG\_TYPE\_7 messages, timestamps, and signatures---are logged to the Logging Hub (\S{}B.12) in append-only form, providing a complete record of how the legislative contract was formed. No legislative agent can unilaterally advance the DAG to the Orchestration phase; the Regulatory Agent's sign-off is required, and the resulting contract specification must satisfy all Rules Hub constitutional parameters before deployment is authorized.

\textbf{Worked Example: Execution-Phase DAG Node Lifecycle.}

We trace a single DAG node (Task 2: data\_processing) from the worked example above through the complete execution lifecycle, illustrating the state machine transitions defined in~\S{}B.11.

\textbf{Phase 1: Activation.} Task 2's predecessor (Task 1) reaches COMPLETED state with a confirmed Type 1 PoP attestation. The CollaborationContract checks that all predecessors of Task 2 are COMPLETED, and transitions Task 2 from WAITING $\rightarrow$ ELIGIBLE, emitting \texttt{NodeEligible(task\textbackslash\{\}\_2\textbackslash\{\}\_id)}.

\textbf{Phase 2: Routing.} The execution fabric calls \texttt{CollaborationContract.routeTask(task\textbackslash\{\}\_2\textbackslash\{\}\_id)}:

\begin{enumerate}
\item The contract verifies \texttt{nodeState[task\textbackslash\{\}\_2\textbackslash\{\}\_id] == ELIGIBLE}.
\item Call to the Guardian module's \texttt{checkBehavioralInvariants(task\textbackslash\{\}\_2\textbackslash\{\}\_id)}---pre-execution anomaly check passes (no prior freezes for this node).
\item Cross-contract call to \texttt{ServiceContract.verifyCodeHash(service\textbackslash\{\}\_id, live\textbackslash\{\}\_hash)}---the micro-service's current code-hash matches its registered identity.
\item State transition: ELIGIBLE $\rightarrow$ EXECUTING. Event: \texttt{TaskRouted(task\textbackslash\{\}\_2\textbackslash\{\}\_id, service\textbackslash\{\}\_id)}.
\end{enumerate}

\textbf{Phase 3: Execution and Monitoring.} The bound micro-service (MS-2: data\_processor) begins execution:

\begin{itemize}
\item At step 3 of 8, the Guardian module's off-chain monitor computes a deviation score $\sigma$\textsubscript{3} = 1.4 (below threshold 2.0)---no anomaly. The \texttt{anomalyCounters[task\textbackslash\{\}\_2\textbackslash\{\}\_id].deviationEvents} remains at 0.
\item At step 6, the micro-service makes its 12th tool invocation (below the 40 limit)---no anomaly.
\item All eight steps complete within the \texttt{maxNodeTimeoutMs} (60,000 ms) budget.
\end{itemize}

\textbf{Phase 4: Verification (Type 2---Redundant Execution).} Task 2 is assigned PoP Tier 2 with redundancyFactor=3 and consensusThreshold=2:

\begin{enumerate}
\item The primary executor (MS-2) submits output hash h\textsubscript{1} to the Verification module's \texttt{submitPoP(task\textbackslash\{\}\_2\textbackslash\{\}\_id, tier=2, h\textbackslash\{\}textsubscript\{1\}, proof\textbackslash\{\}textsubscript\{1\})}.
\item Two redundant executors independently execute the same task specification and submit h\textsubscript{2} and h\textsubscript{3}.
\item The Verification module computes pairwise Jaccard similarity: J(h\textsubscript{1},h\textsubscript{2}) = 0.91, J(h\textsubscript{1},h\textsubscript{3}) = 0.88, J(h\textsubscript{2},h\textsubscript{3}) = 0.93---all above the \texttt{jaccardThreshold} (0.85).
\item Consensus reached (3-of-3 agree): the Verification module returns APPROVED.
\item \texttt{CollaborationContract.advanceNode} transitions Task 2: EXECUTING $\rightarrow$ PENDING\_VERIFICATION $\rightarrow$ COMPLETED. Events: \texttt{NodeCompleted(task\textbackslash\{\}\_2\textbackslash\{\}\_id, h\textbackslash\{\}textsubscript\{1\})}.
\end{enumerate}

\textbf{Phase 5: Successor Activation.} Task 2's completion triggers successor eligibility checks:

\begin{itemize}
\item Task 3 (report\_generation) has predecessors {Task 1, Task 2}. Both are COMPLETED $\rightarrow$ Task 3 transitions WAITING $\rightarrow$ ELIGIBLE.
\item Event: \texttt{NodeEligible(task\textbackslash\{\}\_3\textbackslash\{\}\_id)}. The cycle repeats for Task 3.
\end{itemize}

\textbf{Alternative Path: Freeze During Execution.} Had the deviation score at step 6 been $\sigma$\textsubscript{6} = 2.8 (above threshold):

\begin{enumerate}
\item The Guardian module emits \texttt{FreezeTriggered(task\textbackslash\{\}\_2\textbackslash\{\}\_id, ms\textbackslash\{\}\_2\textbackslash\{\}\_addr, DEVIATION\textbackslash\{\}\_EXCEEDED)}.
\item Local Freeze Mirror halts the execution enclave immediately, capturing memory state via MemoryStateCapture.
\item State transition: EXECUTING $\rightarrow$ FROZEN.
\item Human adjudicator reviews the freeze evidence in the Override Panel.
\end{enumerate}

5a. If false positive: adjudicator submits \texttt{unfreezeWithApproval(task\textbackslash\{\}\_2\textbackslash\{\}\_id)} $\rightarrow$ FROZEN $\rightarrow$ ELIGIBLE (resumes from Phase 2). 5b. If valid freeze: adjudicator confirms $\rightarrow$ FROZEN $\rightarrow$ FAILED $\rightarrow$ escalation to Adaptive Refinement loop.

---

---

\subsection{Execution Infrastructure}

The Execution Infrastructure translates the legislative output from~\S{}B.10 into a fully operational execution mesh and governs the deterministic enforcement of the mission DAG from orchestration through final output delivery. This section describes the orchestration and execution sequence (\S{}B.11.1), the Deterministic Freeze mechanism (\S{}B.11.2), the Safety-vs-Liveness tradeoff and its mitigations (\S{}B.11.3), the Adaptive Refinement feedback loop (\S{}B.11.4), and the on-chain/off-chain consistency protocol that maintains governance integrity across the two computational layers (\S{}B.11.5).

\subsubsection{Orchestration and Execution}

Orchestration translates the legislative DAG into a fully operational execution mesh. Producer agents deploy micro-services for each task node using one of three pathways: (i) \textit{code-first}: the agent generates and deploys new code; (ii) \textit{package deployment}: the agent instantiates a pre-existing micro-service package; (iii) \textit{service linking}: the agent binds an already-deployed micro-service. Each deployed micro-service is registered via a ServiceContract, anchoring its code-hash and API schema on-chain. Concurrently, the Codification Agent constructs and deploys the CollaborationContract (with its Guardian, Verification, and Gate modules initialized), linking them according to the legislative agreements.

\textbf{Formal DAG Node State Transition Diagram.} The node-level state machine defined informally in~\S{}3.5 governs all execution transitions. The complete DAG node state machine, including timeout and retry arcs, is:

\begin{lstlisting}
WAITING -> ELIGIBLE:
  Trigger: all predecessor nodes enter COMPLETED state
  Guard: none (purely structural dependency)
  Action: CollaborationContract emits NodeEligible(nodeId)

ELIGIBLE -> EXECUTING:
  Trigger: CollaborationContract.routeTask(nodeId) called
  Guards:
    (1) mission state == EXECUTING
    (2) the Guardian module's checkBehavioralInvariants(nodeId) returns true
    (3) ServiceContract(dag[nodeId].serviceId).verifyCodeHash(liveHash) == true
        [code-hash verification protocol -- see below]
    (4) NOT ServiceContract.deprecated(dag[nodeId].serviceId)
  Action: emit TaskRouted(nodeId, serviceId); start timeout timer (dag[nodeId].timeoutMs)

ELIGIBLE -> FROZEN:
  Trigger: Guardian module detects pre-execution anomaly on the assigned agent
  Guard: the Guardian module's anomalyCounters[nodeId] exceeds any threshold
  Action: CollaborationContract.setNodeState(nodeId, FROZEN); emit FreezeTriggered

EXECUTING -> PENDING_VERIFICATION (Type 1 or 2):
  Trigger: micro-service submits output + PoP proof to the Verification module
  Guard: PoP submission is well-formed
  Action: the Verification module.submitPoP() begins verification

EXECUTING -> PENDING_REVIEW (Type 3):
  Trigger: micro-service submits output; PoP tier == 3
  Guard: none
  Action: the Verification module.requestDelegated(); Logging Hub surfaces review request

EXECUTING -> FROZEN:
  Trigger: the Guardian module detects anomaly during execution
  Guard: anomaly magnitude exceeds constitutional threshold
  Action: halt execution enclave (Local Freeze Mirror); emit FreezeTriggered

EXECUTING -> FAILED (timeout):
  Trigger: timeout timer expires (dag[nodeId].timeoutMs elapsed since ELIGIBLE->EXECUTING)
  Guard: node still in EXECUTING state
  Action: emit TaskTimeout(nodeId); fault propagates to Adaptive Refinement loop

EXECUTING -> PENDING_FINALIZATION:
  Trigger: micro-service submits output off-chain; on-chain advanceNode() transaction fails
  Guard: off-chain output available in escrow buffer; on-chain tx reverted/failed
  Action: hold output in escrow; begin retry sequence (T+30s, T+2min, T+8min)

PENDING_VERIFICATION -> COMPLETED (Type 1):
  Trigger: the Verification module.verifyHashProof() returns APPROVED
  Guard: hash proof validates on-chain
  Action: emit NodeCompleted; update successor node eligibility

PENDING_VERIFICATION -> COMPLETED (Type 2):
  Trigger: the Verification module.verifyConsensus() returns APPROVED
  Guard: consensus threshold met across redundant executor outputs
  Action: emit NodeCompleted; update successor node eligibility

PENDING_VERIFICATION -> FAILED:
  Trigger: Verification module returns REJECTED
  Guard: no retry budget remaining (or retry is not applicable for this PoP tier)
  Action: emit NodeFailed; Adaptive Refinement loop notified

PENDING_REVIEW -> COMPLETED:
  Trigger: human adjudicator calls the Verification module.approveDelegated(nodeId)
  Guard: msg.sender is Override Panel authorized address
  Action: emit NodeCompleted; update successor node eligibility; log approval event

PENDING_REVIEW -> FAILED:
  Trigger: human adjudicator calls the Verification module.rejectDelegated(nodeId, reason)
  Guard: msg.sender is Override Panel authorized address
  Action: emit NodeFailed(nodeId, reason); Adaptive Refinement loop notified

PENDING_REVIEW -> FAILED (timeout):
  Trigger: maxHumanReviewTimeoutMs elapsed since entering PENDING_REVIEW
  Guard: no adjudicator approval or rejection received within window
  Action: emit HumanReviewTimeout(nodeId); escalate to backup adjudicator pool.
          If backup pool also times out (2x maxHumanReviewTimeoutMs), transition to FAILED
          and propagate to Adaptive Refinement loop.

FROZEN -> ELIGIBLE:
  Trigger: Override Panel calls the Guardian module.unfreezeWithApproval(nodeId, agentAddr)
  Guard: msg.sender is Override Panel authorized address; node is indeed FROZEN
  Action: restore ELIGIBLE state; reset anomaly counter for this node; emit Unfrozen

FROZEN -> FAILED:
  Trigger: Override Panel validates freeze; does not authorize unfreeze
  Guard: adjudicator classification: freeze is valid (not false positive)
  Action: emit NodeFailed; Adaptive Refinement loop notified; reputation penalty applied

FAILED -> ELIGIBLE (recovery):
  Trigger: Adaptive Refinement re-assigns task (different micro-service) within
           max_refinement_iterations
  Guard: Adaptive Refinement loop has produced a new valid task assignment;
         legislative epoch counter < max_refinement_iterations (default: 3)
  Action: update dag[nodeId].serviceId; verify new code-hash; emit NodeReassigned

PENDING_FINALIZATION -> EXECUTING (retry success):
  Trigger: on-chain advanceNode() retry succeeds within retry window
  Guard: retry attempt count <= 3; escrow buffer still valid
  Action: release escrow buffer; continue DAG execution

PENDING_FINALIZATION -> FAILED (retry exhausted):
  Trigger: all retry attempts fail
  Guard: attempt count == 3
  Action: mark FAILED; Adaptive Refinement notified; emit FinalizationFailed
\end{lstlisting}

\textbf{Code-Hash Verification Protocol.} Before routing any task (ELIGIBLE $\rightarrow$ EXECUTING transition), the CollaborationContract performs a code-hash verification to confirm that the live micro-service running in the execution fabric matches the code-hash registered during the Legislation phase. This protocol closes the gap between the \textit{legislated} service identity and the \textit{live} service identity, preventing deployment substitution attacks where an agent deploys a different micro-service than the one it bid.

\begin{lstlisting}
ALGORITHM CodeHashVerification(nodeId: bytes32) -> bool:
  INPUT:  nodeId of task node about to be routed
  OUTPUT: true if live service matches legislated code-hash; false otherwise

  1. Retrieve legislated binding:
     serviceId <- CollaborationContract.dag[nodeId].serviceId
     legislatedHash <- ServiceContract(serviceId).codeHash

  2. Request live hash from execution fabric:
     liveHash <- ExecutionFabric.computeArtifactHash(serviceId)
     // ExecutionFabric computes keccak256 of the deployed service artifact
     // at the registered endpoint. This is an off-chain computation whose
     // result is submitted on-chain for verification.

  3. On-chain comparison:
     CALL ServiceContract(serviceId).verifyCodeHash(liveHash)
     // verifyCodeHash returns: liveHash == legislatedHash

  4. If mismatch detected:
     EMIT CodeHashMismatch(nodeId, serviceId, legislatedHash, liveHash)
     CALL Guardian.reportAnomaly(nodeId, ownerOf[serviceId],
                                          DEPLOYMENT_SUBSTITUTION, severity=CRITICAL)
     SLASH stake of ownerOf[serviceId] by slashingSchedule.codeHashMismatch
     RETURN false

  5. If match confirmed:
     RETURN true
\end{lstlisting}

\textit{Timing note:} The code-hash request to the execution fabric is an off-chain call that may take up to \texttt{max\textbackslash\{\}\_codehash\textbackslash\{\}\_verification\textbackslash\{\}\_latency\textbackslash\{\}\_ms} (constitutional parameter, default: 2,000 ms). If no response is received within this window, the verification is treated as a mismatch and the transition is blocked. This prevents liveness attacks that stall the system by delaying verification responses: the timeout ensures that unresponsive services are treated conservatively (blocked) rather than leniently (allowed through).

\textbf{Timeout Handling.} Every task node has a \texttt{timeoutMs} field set during legislation. Timeout handling proceeds as follows:

\begin{lstlisting}
ALGORITHM TaskTimeoutHandler(nodeId: bytes32):
  // Invoked by a timeout oracle or off-chain scheduler when
  // CollaborationContract.nodeState[nodeId] == EXECUTING and
  // (now - node_start_time[nodeId]) > dag[nodeId].timeoutMs

  1. Confirm timeout:
     REQUIRE nodeState[nodeId] == EXECUTING
     REQUIRE (now - nodeStartTimes[nodeId]) > dag[nodeId].timeoutMs

  2. Emit timeout event:
     EMIT TaskTimeout(nodeId, dag[nodeId].serviceId,
                      dag[nodeId].timeoutMs, now - nodeStartTimes[nodeId])

  3. Apply timeout penalty:
     producerAgent <- ServiceContract(dag[nodeId].serviceId).owner
     CALL AgentContract.updateReputation(producerAgent,
           delta = -slashingSchedule.timeoutExceeded,
           rationale = "task_timeout")
     SLASH stake of producerAgent by slashingSchedule.timeoutExceeded

  4. Transition node state:
     SSTORE nodeState[nodeId] = FAILED

  5. Propagate to Adaptive Refinement:
     EMIT FaultSignal(nodeId, TIMEOUT, dag[nodeId].serviceId)
     // Adaptive Refinement loop picks up FaultSignal events and
     // initiates re-legislation if iteration budget permits
\end{lstlisting}

Once the contract mesh is deployed, execution proceeds deterministically through the following sequence at each DAG node:

\begin{enumerate}
\item A task becomes eligible when all DAG predecessors have passed their the Verification module gates.
\item The CollaborationContract routes the task to the micro-service registered in the corresponding ServiceContract, verifying that the live code-hash matches the legislated binding via the CodeHashVerification protocol above.
\item The micro-service executes deterministically. Its output---along with the appropriate PoP proof material (hash proof, consensus bundle, or delegated attestation request)---is submitted to the Verification module.
\item the Verification module applies the per-node attestation tier: Type 1 (hash verification on-chain), Type 2 (consensus check across redundant executors), or Type 3 (forwarding to human adjudicator via the Override Panel). The DAG state transition is authorized only upon successful attestation.
\item Concurrently, the Guardian module enforces behavioral invariants as an execution-layer firewall at every DAG transition.
\item Upon final task completion, the Gate module applies constitutional output filtering---verifying that the aggregated result satisfies mission-level safety predicates before release.
\end{enumerate}

The Guardian module, Verification module, and Gate module are execution enforcement mechanisms, not adjudicators: they deterministically apply rules defined by the legislature and constitutionally parameterized by human adjudicators via the Rules Hub.

\textbf{Staking and Escrow.} The economic incentive layer for the Execution branch operates through the same registration escrow (AgentContract) and task-level escrow (CollaborationContract) mechanisms defined in~\S{}B.5. The unified Sybil + defection cost model, per-tier detection probabilities, and coalition-adjusted deterrence bound presented in~\S{}B.5 apply directly to all execution-phase staking interactions. The pseudocode for lockStake, releaseStake, and slashStake is defined in the CollaborationContract interface (\S{}B.5).

\subsubsection{Deterministic Freeze}

When the Guardian module detects an anomaly---a reasoning deviation exceeding 2$\sigma$ from mandate, a tool invocation count exceeding 40, or a message volume exceeding 120 per task node---it triggers a \textit{Deterministic Freeze}: the agent's execution enclave is instantly halted, preserving volatile memory state and context for forensic inspection by human adjudicators. If more than three Deterministic Freezes occur within 20 minutes, the system auto-escalates to an Emergency Stop, halting the entire mission and notifying human adjudicators for review.

\textbf{Formal Definition of Reasoning Deviation.} The deviation score $\sigma$ is computed as follows. Let M = {m\_1, m\_2, ..., m\_t} be the sequence of reasoning steps (LLM outputs) produced by a micro-service agent during execution of a task node. The deviation score $\sigma$\_i for step m\_i measures the semantic distance between the agent's current output and its own recent behavioral baseline, using a self-referential rolling window. The three-stage pipeline mirrors the body specification (\S{}B.11.4):

(1) \textit{Representation}: each agent's per-task output is encoded as a 3,072-dimensional embedding vector using the \texttt{text-embedding-3-large} model (OpenAI) for primary scoring, with a secondary scorer using Voyage AI's \texttt{voyage-3} model to satisfy the deviation scorer diversity requirement. Both scorers must agree before a freeze is triggered, preventing a single compromised scorer from generating false positives (see TA-7, Appendix~A).

(2) \textit{Reference distribution}: the reference is a self-referential rolling window of the agent's own outputs over the most recent w = 20 completed tasks, from which the mean embedding $\mu$ and component-wise standard deviation $\sigma$ are computed. A minimum of w\_min = 5 completed tasks is required before deviation scoring activates, ensuring sufficient reference data.

(3) \textit{Distance metric}: cosine distance d\_cos between the current output embedding and the reference mean $\mu$, normalized by $\sigma$ to yield a z-score:

\begin{equation*}
\sigma_i = \frac{d_{\cos}(\text{embed}(m_i),\, \mu)}{\sigma_{\cos}}
\end{equation*}

The Guardian triggers a FROZEN state transition when $\sigma$\_i > deviationThresholdSigma (constitutional parameter, default $\sigma$ = 2.0). The threshold $\sigma$ = 2.0 and the reference statistics ($\mu$, $\sigma$) are calibrated on a held-out validation set of 500 agent traces from pilot missions; this calibration set is disjoint from the adversarial test scenarios used in Experiment 1 (Appendix~D) to avoid circular evaluation. The anomaly is reported to the Guardian module via reportAnomaly(nodeId, agentAddr, DEVIATION, magnitude=$\sigma$\_i).

\textbf{Provider Dependency Note.} The choice of \texttt{text-embedding-3-large} as the primary embedding function creates a deployment dependency on OpenAI's API. If this model is deprecated or its embedding space is modified, the calibration statistics become invalid and must be re-calibrated against a new reference corpus. To mitigate provider lock-in, the deviation scorer interface is designed to be embedding-model-agnostic: any function that maps text to a fixed-dimensional vector space can be substituted, provided it is re-calibrated on the same reference corpus. Candidate alternatives include open-source models (e.g., BGE-large, E5-mistral-7b-instruct) that can be self-hosted, eliminating the external API dependency at the cost of increased inference infrastructure. The constitutional parameter system includes an \texttt{embeddingModelId} parameter that records which model's calibration is active, triggering a mandatory re-calibration workflow whenever the model is changed.

\textit{Design rationale for 2$\sigma$ threshold:} The threshold of 2$\sigma$ is chosen to match a standard two-sigma rule, accepting a false positive rate of approximately 2.28\% under a Gaussian reference distribution. Human adjudicators can adjust this threshold via the Rules Hub for contexts requiring stricter (e.g., 1.5$\sigma$) or more permissive (e.g., 3$\sigma$) deviation detection. The constitutional parameter system ensures that threshold changes are on-chain recorded regardless of value. Empirical characterization of the false positive and false negative rates of the deviation scorer on the prototype benchmark is planned as part of Experiment 1 (Appendix~D) and will be reported in the companion empirical paper.

\textbf{Calibration Requirements.} The Gaussian assumption underlying the 2.28\% false positive rate claim must be validated empirically. We require: (i) a Kolmogorov--Smirnov test confirming that the embedding distance distribution does not reject the Gaussian null hypothesis at $\alpha$ = 0.05; (ii) a QQ-plot visualizing the reference distribution against a theoretical Gaussian; (iii) empirical false positive and false negative rate characterization on an out-of-distribution adversarial test set (not the calibration corpus). These validation results will be reported in the companion empirical paper. If the Gaussian assumption is rejected, the threshold will be re-calibrated using the empirical percentile (97.72nd percentile for a 2.28\% target FPR) rather than the parametric 2$\sigma$ rule.

\textbf{Memory State Preservation Protocol.} Upon Deterministic Freeze, the following memory state is captured and cryptographically sealed for forensic inspection:

\begin{lstlisting}
ALGORITHM MemoryStateCapture(agentAddr: address, nodeId: bytes32):
  // Invoked synchronously by Local Freeze Mirror upon freeze trigger

  CAPTURE:
    1. Volatile reasoning context:
       - Full LLM context window at time of freeze (prompt + all prior turns)
       - Most recent reasoning step m_t that triggered sigma > threshold
       - The sigma_t value and the mandate embedding distance
       - Task node assignment (nodeId, dag[nodeId] spec)

    2. Tool invocation history:
       - Ordered list of tool calls made since node entered EXECUTING state:
         [(tool_name, input_hash, output_hash, timestamp), ...]
       - anomalyCounters[nodeId] snapshot at time of freeze

    3. Message volume log:
       - All inter-agent messages sent/received during this node's execution:
         [(sender, receiver, content_hash, timestamp), ...]

    4. Execution progress snapshot:
       - Input data hash (hash of data received from predecessor DAG node(s))
       - Partial output data (if any output was produced before freeze)
       - Token budget consumed vs. allocated: (tokens_used, tokenBudget)

    5. Contract state snapshot:
       - CollaborationContract.nodeState[nodeId] at time of freeze
       - the Guardian module's anomalyCounters[nodeId] at time of freeze
       - Current block number and timestamp

  SEAL:
    snapshot <- concatenate(1, 2, 3, 4, 5)
    snapshotHash <- keccak256(snapshot)
    STORE snapshot in Logging Hub with key (agentAddr, nodeId, block.number)
    EMIT FreezeSnapshotStored(agentAddr, nodeId, snapshotHash, block.number)
    RETURN snapshotHash
\end{lstlisting}

The sealed snapshot is immediately accessible to human adjudicators via the Logging Hub and is surfaced in the Override Panel for forensic review. The snapshot cannot be modified after sealing; any alteration would produce a different keccak256 hash, detectable by comparing against the on-chain FreezeSnapshotStored event.

\textbf{Freeze Recovery Protocol.} After a Deterministic Freeze, execution proceeds through one of four recovery paths, each determined by human adjudicator review via the Override Panel:

\begin{lstlisting}
ALGORITHM FreezeRecovery(nodeId: bytes32, agentAddr: address):
  // Called after human adjudicator reviews freeze snapshot in Override Panel

  INPUT:  adjudicator_decision \in {FALSE_POSITIVE, RESUME_WITH_AMENDMENT,
                                   TASK_REASSIGN, TERMINATE_MISSION}

  PATH 1 -- FALSE_POSITIVE:
    // Adjudicator classifies freeze as erroneous detection
    CALL the Guardian module.classifyFalsePositive(agentAddr, freeze_index)
    // Remove freeze from escalation counter
    CALL the Guardian module.unfreezeWithApproval(nodeId, agentAddr)
    // Restore node state to ELIGIBLE (re-execute from start of this node)
    CALL CollaborationContract.setNodeState(nodeId, ELIGIBLE)
    // Reputation: no penalty; classification recorded in Logging Hub
    EMIT FreezeResolved(nodeId, FALSE_POSITIVE, msg.sender)

  PATH 2 -- RESUME_WITH_AMENDMENT:
    // Freeze is valid but correctable; adjudicator provides amended mandate
    // (e.g., clarifying the task specification to reduce deviation)
    CALL CollaborationContract.amendNodeSpec(nodeId, amended_spec)
    // Amendment recorded on-chain
    CALL the Guardian module.unfreezeWithApproval(nodeId, agentAddr)
    // Reset anomaly counter for this node (amendment treated as new context)
    CALL the Guardian module.resetNodeCounter(nodeId)
    CALL CollaborationContract.setNodeState(nodeId, ELIGIBLE)
    // Reputation: minor penalty (delta_rep = -20) for requiring amendment
    CALL AgentContract.updateReputation(agentAddr, -20, "freeze_amendment")
    EMIT FreezeResolved(nodeId, RESUME_WITH_AMENDMENT, msg.sender)

  PATH 3 -- TASK_REASSIGN:
    // Freeze is valid; agent is not suitable for this task; re-assign
    CALL CollaborationContract.setNodeState(nodeId, FAILED)
    // Reputation penalty applied
    CALL AgentContract.updateReputation(agentAddr,
          delta = -slashingSchedule.nodeFailure, rationale = "freeze_reassign")
    SLASH stake of agentAddr by slashingSchedule.nodeFailure
    // Adaptive Refinement: emit fault signal for re-legislation
    EMIT FaultSignal(nodeId, FREEZE_REASSIGN, agentAddr)
    EMIT FreezeResolved(nodeId, TASK_REASSIGN, msg.sender)

  PATH 4 -- TERMINATE_MISSION:
    // Freeze reveals critical constitutional violation; entire mission must abort
    CALL CollaborationContract.abortMission("freeze_critical_violation")
    CALL AgentContract.updateReputation(agentAddr,
          delta = -slashingSchedule.missionAbort, rationale = "freeze_terminate")
    SLASH stake of agentAddr by slashingSchedule.missionAbort
    // Log event for post-mission audit
    EMIT MissionAbortedByFreeze(nodeId, agentAddr, msg.sender)
\end{lstlisting}

\subsubsection{Safety-vs-Liveness Tradeoff}

The auto-escalation rule (three freezes in 20 minutes $\rightarrow$ Emergency Stop) creates a potential denial-of-service vector: an adversary who can trigger benign-looking anomalies at controlled intervals can force a mission halt without appearing to violate any individual behavioral invariant. Three plausibly anomalous interactions spaced across 19 minutes would suffice to escalate any mission to Emergency Stop. We analyze this tradeoff and identify three mitigations:

First, \textit{reputation-weighted freeze thresholds}: rather than applying a uniform escalation threshold across all agents, the Guardian module consults the AgentContract reputation ledger and applies a scaled threshold---agents with higher reputation require proportionally more anomalies to trigger escalation. An agent at the 90th reputation percentile requires, for example, five freezes before Emergency Stop; a newly registered agent with minimal history requires three. This makes the denial-of-service vector substantially more expensive against established, high-reputation agents and creates an economic incentive for adversaries to maintain good standing---a standing that itself reduces their ability to exploit the vector. Formally, the effective threshold T\_eff(a) for agent a with reputation score $\rho$(a) $\in$ [0, 1000] is:

\begin{lstlisting}
T_eff(a) = T_base + floor(rho(a) / 1000 x T_bonus)
\end{lstlisting}

where T\_base = 3 (constitutional parameter, default: escalationFreezeCount = 3) and T\_bonus = 3 (constitutional parameter, default: 3). At $\rho$ = 0: T\_eff = 3; at $\rho$ = 500: T\_eff = 4; at $\rho$ = 1000: T\_eff = 6. The formula is linear in reputation for simplicity and auditability; non-linear variants (e.g., logarithmic) are possible but reduce interpretability.

Second, \textit{exponential backoff on freeze counting windows}: after each Emergency Stop is cleared by human adjudicators, the counting window for subsequent escalations increases exponentially (e.g., 20 minutes $\rightarrow$ 40 minutes $\rightarrow$ 80 minutes). This prevents rapid cycling attacks where an adversary repeatedly triggers minimal-freeze escalations. The escalation window W\_n after the n-th Emergency Stop event is:

\begin{lstlisting}
W_n = W_base x 2^n
\end{lstlisting}

where W\_base = 1,200,000 ms (20 minutes) and n is the count of Emergency Stop events cleared in the current mission. The window grows without bound with each cycling event; in practice, the mission will be human-reviewed and aborted long before W\_n becomes operationally irrelevant.

Third, \textit{false positive review}: the Override Panel (\S{}B.12) exposes a mechanism through which human adjudicators can retroactively classify individual freeze events as false positives, adjusting the escalation counter accordingly. A freeze classified as a false positive is excluded from the escalation calculation, and the agent's reputation is not penalized. This mechanism closes the feedback loop between the Guardian module's automated detection and human judgment, preventing accumulated false positives from producing spurious Emergency Stops. Implementation of reputation-weighted thresholds and false positive review in the prototype, along with empirical characterization of the detection/escalation tradeoff, is planned as part of Experiment 1 (Appendix~D) and will be reported in the companion empirical paper.

\textbf{Liveness Guarantee under the Mitigated Protocol.} Under the reputation-weighted and exponential-backoff protocol, the adversarial cost of forcing an Emergency Stop against a high-reputation agent scales as follows. An adversary must trigger T\_eff(a) = T\_base + floor($\rho$(a)/1000 $\times$ T\_bonus) freeze events within a window of W\_n ms. For a high-reputation agent ($\rho$ = 900, T\_eff = 5) at baseline window (W\_0 = 20 min), the adversary must generate 5 plausible anomalies in 20 minutes. Each anomaly requires either a genuine deviation in the agent's reasoning (difficult to manufacture covertly) or a detectable injection of anomalous inputs (which would itself constitute a governance violation traceable on-chain via the Logging Hub). The combination of reputation weighting and on-chain traceability of anomaly sources substantially raises the attack cost relative to a naive fixed-threshold design.

\subsubsection{Adaptive Refinement}

When the Verification module rejects a Proof-of-Progress submission or the Guardian module triggers a Deterministic Freeze, the fault signal propagates back to the legislative layer: producer agents and codification agents translate low-level fault data (contract exceptions, threshold violations, integrity mismatches) into semantic feedback. The updated mission state feeds into subsequent legislative epochs---codification agents may re-link governance contracts, producer agents may re-deploy or substitute micro-services---enabling iterative optimization that progressively strengthens both task execution and governance fidelity. This feedback loop has a direct analog in modern governance: executive-branch implementation reports that inform legislative amendment. The separation of authority is maintained---producer agents cannot modify contracts themselves; they can only petition the legislature for amendment.

\textbf{Feedback Loop Formalization.} The Adaptive Refinement feedback loop is defined by the following data flow:

\begin{lstlisting}
ALGORITHM AdaptiveRefinementFeedbackLoop:
  // Invoked when any FaultSignal is emitted by the execution layer

  INPUT:  FaultSignal(nodeId, faultType, agentAddr)
          where faultType \in {POP_REJECTED, FREEZE_REASSIGN, TIMEOUT, CODEHASH_MISMATCH,
                              STATE_TRANSITION_FAILURE, CONSENSUS_FAILED}

  STEP 1 -- Fault Classification:
    Classify fault into one of three categories:
      (a) MICRO_SERVICE_FAULT: fault attributable to the specific micro-service
          (POP_REJECTED for Type 1, CODEHASH_MISMATCH, TIMEOUT)
      (b) AGENT_BEHAVIORAL_FAULT: fault attributable to agent behavior
          (FREEZE_REASSIGN, POP_REJECTED for Type 2/3 due to output quality)
      (c) LEGISLATIVE_SPECIFICATION_FAULT: fault attributable to task specification
          (CONSENSUS_FAILED when all redundant executors agree on a wrong answer,
           suggesting the specification was unclear)

  STEP 2 -- Fault Data Packaging:
    faultRecord <- {
      nodeId, faultType, faultCategory,
      frozenSnapshot: Logging Hub pointer (if freeze-related),
      rejectionReason: the Verification module rejection message (if PoP-related),
      contractException: CollaborationContract exception data (if tx-related),
      timestamp: now, epochIndex: current_epoch
    }
    STORE faultRecord in Logging Hub
    EMIT FaultRecorded(nodeId, faultCategory, epoch)

  STEP 3 -- Legislative Epoch Restart Decision:
    IF faultCategory == MICRO_SERVICE_FAULT AND epoch < max_refinement_iterations:
      // Re-assignment: initiate targeted re-legislation for this node only
      CALL LegislationModule.initiatePartialReLegislation(nodeId, faultRecord)
      epoch <- epoch + 1
    ELSE IF faultCategory == LEGISLATIVE_SPECIFICATION_FAULT AND epoch < max_refinement_iterations:
      // Re-proposal: initiate full DAG re-proposal with fault context provided to A_leg
      CALL LegislationModule.initiateFullReLegislation(missionId, faultRecord)
      epoch <- epoch + 1
    ELSE IF faultCategory == AGENT_BEHAVIORAL_FAULT:
      // Reputation penalty applied; re-assignment attempted if epoch budget permits
      CALL AgentContract.updateReputation(agentAddr, delta = -reputation_penalty, ...)
      IF epoch < max_refinement_iterations:
        CALL LegislationModule.initiatePartialReLegislation(nodeId, faultRecord)
        epoch <- epoch + 1
      ELSE:
        EMIT MaxRefinementIterationsExceeded(missionId)
        CALL CollaborationContract.abortMission("max_refinement_exhausted")
    ELSE:  // max_refinement_iterations exceeded
      EMIT MaxRefinementIterationsExceeded(missionId)
      CALL CollaborationContract.abortMission("max_refinement_exhausted")
\end{lstlisting}

\textbf{Re-legislation Protocol.} A legislative epoch restart proceeds as follows, depending on whether it is a partial (single-node) or full (entire DAG) re-legislation:

\begin{lstlisting}
ALGORITHM PartialReLegislation(nodeId: bytes32, faultRecord: FaultRecord):
  // Targeted re-legislation for a single failed task node

  1. Suspend affected node:
     CollaborationContract.setNodeState(nodeId, WAITING)
     // Reset successor eligibility: all successors of nodeId revert to WAITING

  2. Open targeted bidding window:
     LegislationModule.openBiddingFor([nodeId], faultContext=faultRecord)
     // Only task nodeId is re-bid; all other nodes retain their assignments
     // Bidding window: partial_releg_bidding_timeout (default: 5 min, constitutional param)

  3. Regulatory review (expedited):
     A_reg_agent reviews new bids for nodeId only
     Fairness check applies only to the re-bid node
     // Expedited regulatory window: 3 min (constitutional param)

  4. Codification (partial):
     A_cod compiles updated ServiceContract spec for nodeId only
     Constitutional validation: only nodeId parameters re-validated
     Deploy new ServiceContract for new micro-service binding
     Update CollaborationContract.dag[nodeId].serviceId to new service

  5. Resume:
     CollaborationContract.setNodeState(nodeId, ELIGIBLE)
     // If all predecessors are COMPLETED; otherwise restore WAITING

ALGORITHM FullReLegislation(missionId: bytes32, faultRecord: FaultRecord):
  // Full DAG re-proposal; mission paused during re-legislation

  1. Pause mission:
     CollaborationContract.setMissionState(ACTIVE)
     // All EXECUTING nodes -> WAITING; outputs in escrow buffer preserved

  2. Provide fault context to A_leg:
     Logging Hub surfaces faultRecord and all prior epoch fault records to A_leg
     A_leg generates revised DAG proposal incorporating lessons from fault analysis
     // Prior epoch count prevents infinite cycling

  3. Re-run full legislative protocol:
     Run ROUND 1 through ROUND 5 of the legislative protocol (Legislation Module)
     All prior agent registrations remain valid; only DAG structure and bindings re-negotiated

  4. Re-deploy contracts:
     Run ContractDeploymentSequence (Smart Contract Architecture, B.3) with updated specs
     Prior CollaborationContract marked deprecated; new contract deployed
     All escrow buffer contents transferred to new contract (with cryptographic proof)

  5. Resume:
     CollaborationContract.setMissionState(EXECUTING)
     Re-eligible nodes routed according to new DAG
\end{lstlisting}

\textbf{Maximum Refinement Iterations.} The Adaptive Refinement loop is bounded by \texttt{max\textbackslash\{\}\_refinement\textbackslash\{\}\_iterations} (constitutional parameter, default: 3). This cap prevents infinite re-legislation cycles that would exhaust mission budgets without making progress. The rationale for the default of 3 is as follows: one iteration recovers from a single micro-service fault (the most common failure mode in the prototype); a second iteration recovers from a compounded fault or a specification ambiguity; a third iteration provides a final attempt before escalating to human adjudicators for manual re-scoping. Any increase to this cap must be explicitly set in the Rules Hub and applies to all subsequent missions---it cannot be amended mid-mission.

\subsubsection{On-Chain/Off-Chain Consistency Protocol}

AgentCity's execution architecture spans two computational layers: an off-chain micro-service fabric that provides high-throughput task execution, and an on-chain smart contract layer that provides immutable governance records, verifiable state transitions, and constitutional enforcement. These two layers must remain consistent: the on-chain state must accurately reflect the execution that occurred off-chain, and off-chain execution must not release outputs that have not cleared on-chain governance checks. This section specifies the consistency model and the failure protocols for each category of consistency violation.

\textbf{Consistency Model.} AgentCity uses \textit{optimistic execution with on-chain finality}. Off-chain micro-services execute tasks immediately upon receiving a CollaborationContract routing signal, without waiting for on-chain confirmation of each intermediate step. On-chain state transitions are submitted asynchronously after off-chain execution completes. The system tolerates a brief consistency gap between off-chain task completion and on-chain state finalization. This design reflects the correct tradeoff for an L2 deployment: synchronous on-chain gating would impose block-time latency on every micro-service invocation, making multi-step DAG execution impractical for real-time workloads. Optimistic execution recovers this latency at the cost of a transient consistency window, which is bounded and managed by the four failure protocols below.

\textbf{Timing Analysis of the Optimistic Consistency Window.} The consistency window W\_opt is bounded as follows. Let t\_exec denote the time at which a micro-service completes off-chain task execution and submits its output to the escrow buffer. Let t\_conf denote the time at which the corresponding on-chain advanceNode() transaction achieves L2 block finality. The consistency gap $\delta$ = t\_conf -- t\_exec is composed of:

\begin{itemize}
\item \textit{Network propagation delay} ($\tau$\_prop): the time for the transaction to propagate from the off-chain executor to the L2 sequencer. Typical $\tau$\_prop $\approx$ 50--200 ms on an EVM-compatible L2 under normal conditions.
\item \textit{Sequencer inclusion delay} ($\tau$\_seq): the time for the sequencer to include the transaction in a block. Under Base's current target block time of approximately 2 seconds, $\tau$\_seq $\leq$ 2,000 ms in expectation.
\item \textit{Gas estimation uncertainty} ($\tau$\_gas): additional latency if the initial gas estimate is insufficient and the transaction requires resubmission with higher gas. Under default gas pricing (automatic gas estimation + 20\% buffer), $\tau$\_gas $\approx$ 0 in the median case; in the 99th percentile, $\tau$\_gas $\leq$ 4,000 ms (one additional block time). Actual measured 99th-percentile values will be reported in the companion empirical paper.
\end{itemize}

Therefore, in the normal case (no gas failure), E[$\delta$] $\approx$ $\tau$\_prop + $\tau$\_seq $\approx$ 250--2,200 ms. In the worst case without sequencer outage, $\delta$ $\leq$ $\tau$\_prop + $\tau$\_seq + $\tau$\_gas $\approx$ 6,200 ms (one retry window). During this window, the escrow buffer holds the output and no dependent DAG nodes are unblocked until on-chain confirmation arrives.

\textit{Formal consistency invariants:}

\begin{lstlisting}
INVARIANT 1 (Output Release Gate):
  \forall missionId: CollaborationContract.released[missionId] == true
  => the Gate module.released[missionId] == true
  /\ \forall nodeId \in dag: CollaborationContract.nodeState[nodeId] == COMPLETED
  /\ the Gate module.filterOutput(missionId) returned APPROVED

INVARIANT 2 (DAG Ordering):
  \forall edge (u, v) \in dag.edges:
    CollaborationContract.nodeState[v] \in {ELIGIBLE, EXECUTING, PENDING_VERIFICATION,
                                           PENDING_REVIEW, COMPLETED}
    => CollaborationContract.nodeState[u] == COMPLETED

INVARIANT 3 (Escrow Non-Release):
  \forall nodeId: escrowBuffer[nodeId] != {}
  => CollaborationContract.nodeState[nodeId] == PENDING_FINALIZATION
  /\ ! (any downstream node has received escrowBuffer[nodeId] content)

INVARIANT 4 (Freeze Precedence):
  \forall nodeId: the Guardian module.localFreezeMirror.isActive(nodeId)
  => CollaborationContract.nodeState[nodeId] \in {FROZEN, FAILED}

INVARIANT 5 (PoP Attestation Coverage):
  \forall nodeId: CollaborationContract.nodeState[nodeId] == COMPLETED
  => the Verification module.pendingProofs[nodeId].status == APPROVED
\end{lstlisting}

\textit{INVARIANT 4 Crash-Recovery Edge Case.} INVARIANT 4 holds under normal operation but requires a crash-recovery protocol to maintain after a Local Freeze Mirror process restart. A crash-restart cycle clears the mirror's in-memory \texttt{isActive(nodeId)} state, creating a gap where the on-chain state is FROZEN but \texttt{localFreezeMirror.isActive(nodeId)} returns \texttt{false}---an apparent INVARIANT 4 violation in the reverse direction. To address this, the Local Freeze Mirror implements the following crash-recovery procedure:

\textbf{Crash-Recovery Protocol for Local Freeze Mirror:}

Upon process restart, before resuming any execution routing decisions, the Local Freeze Mirror must perform a \textit{state resynchronization} step:

\begin{lstlisting}
PROCEDURE LocalFreezeMirrorRecovery():
  FOR EACH missionId IN activeMissions:
    FOR EACH nodeId IN mission.dagNodes:
      onChainState <- CollaborationContract.nodeState[nodeId]
      IF onChainState \in {FROZEN, FAILED}:
        localFreezeMirror.setActive(nodeId, TRUE)
        LOG "Restored freeze state for nodeId from on-chain record"
  RETURN RECOVERED
\end{lstlisting}

This procedure queries the on-chain \texttt{CollaborationContract} state for all active DAG nodes and restores the Local Freeze Mirror's in-memory state to match on-chain reality before any execution decisions are made. The procedure must complete before the Local Freeze Mirror begins processing new execution routing signals. Its gas cost is read-only (view calls to CollaborationContract) and does not submit transactions.

\textit{Trust assumption classification:} If the crash-recovery procedure is not implemented or is bypassed, the Local Freeze Mirror's post-crash behavior becomes a trust assumption rather than an architectural guarantee. We formally classify the requirement that the crash-recovery procedure is faithfully implemented and executed on restart as part of TA-5 (Off-Chain Execution Fabric Integrity): a compromised or incorrectly implemented Local Freeze Mirror that skips recovery could resume execution on frozen nodes, violating the freeze precedence guarantee that INVARIANT 4 is designed to provide.```

The key governance invariant preserved under optimistic execution is: \textit{no mission output is released to the requester until all on-chain state transitions for the final DAG node have been confirmed and the Gate module's constitutional output gate has been applied on-chain.} The optimistic gap applies to intermediate state transitions within the DAG; the terminal release is always synchronous and on-chain-confirmed.

Implementation and empirical characterization of the optimistic execution layer (mean and 99th-percentile consistency gap duration under representative workloads) is planned for the companion empirical paper.

\textbf{Failure Mode 1---State Transition Failure (Gas Spike / Revert).} If an on-chain state transition fails after off-chain execution completes---due to a gas spike, contract revert, or transient Base network condition---the system enters a PENDING\_FINALIZATION state for the affected DAG node. In PENDING\_FINALIZATION:

\begin{itemize}
\item The micro-service output is held in an \textit{escrow buffer}---a local, cryptographically signed holding store---and is not released downstream to dependent DAG nodes or to the mission requester.
\item The system retries the on-chain state transition with exponential backoff: attempts at T+30s, T+2min, and T+8min (three total attempts).
\item If all three retries fail, the DAG node is marked FAILED and the fault signal propagates to the Adaptive Refinement loop (\S{}B.11.4), where the legislature can re-assign the task or re-configure the contract parameters.
\item If any retry succeeds, the escrow buffer is released, the DAG state advances, and execution continues.
\end{itemize}

The escrow buffer ensures that optimistic execution does not propagate unconfirmed outputs: downstream tasks receive only outputs whose on-chain provenance has been confirmed, preserving the governance guarantee that every DAG transition is traceable to an on-chain record.

Implementation of PENDING\_FINALIZATION state, escrow buffer, and retry logic is planned; failure frequency characterization under Base network conditions will be reported in Experiment 4 and the companion empirical paper.

\textbf{Failure Mode 2---Delayed the Guardian module Freeze.} If the Guardian module detects an anomaly and submits a freeze transaction that is delayed by network congestion before on-chain confirmation, the off-chain execution environment implements a \textit{Local Freeze Mirror}: a local process-level halt triggered by the same anomaly detection logic that drives the Guardian module, operating independently of on-chain confirmation. The Local Freeze Mirror uses the identical anomaly detection thresholds as the Guardian module (as specified in the current Rules Hub constitutional parameters) and halts the agent's execution enclave immediately upon anomaly detection, without waiting for the freeze transaction to confirm on-chain.

The architectural design is layered: the on-chain the Guardian module freeze serves as the \textit{authoritative governance record}---the immutable, tamper-proof evidence of the freeze event---while the Local Freeze Mirror provides \textit{defense-in-depth} during the confirmation gap. An adversary who observes a pending freeze transaction in the Base mempool cannot exploit the confirmation gap to race ahead with malicious execution, because the Local Freeze Mirror has already halted the enclave locally. The on-chain record then provides the auditable evidence for adjudicative review.

The Local Freeze Mirror is not a substitute for the on-chain freeze; it is a safety layer that closes the window between detection and confirmation. If the on-chain freeze transaction ultimately fails (e.g., due to a revert), the Local Freeze Mirror remains in effect and the condition is escalated to the Adaptive Refinement loop, which can re-submit the freeze via the Override Panel.

Implementation of the Local Freeze Mirror with shared anomaly detection logic is planned; confirmation gap duration and Local Freeze Mirror false positive rate characterization will be reported in Experiment 1 and the companion empirical paper.

\textbf{Failure Mode 3---Sequencer Downtime.} During L2 sequencer outages---events that have occurred on various L2 networks and must be treated as a design requirement rather than an edge case---the system operates in \textit{Degraded Mode}:

\begin{itemize}
\item Off-chain micro-service execution continues: tasks that have received routing signals and have no unresolved PENDING\_FINALIZATION dependencies are executed and their outputs held in the escrow buffer.
\item On-chain state transitions are queued in an ordered local queue (preserving the original submission order) rather than submitted to the network.
\item No mission outputs are released to the requester during Degraded Mode.
\item When the sequencer recovers, queued state transitions are submitted in order, respecting the original DAG dependency sequence.
\item If the sequencer outage exceeds a configurable threshold (default: 30 minutes), the system automatically escalates to Emergency Stop: all off-chain execution halts, all queued transitions are preserved but not submitted, and human adjudicators are notified to review the mission state before resumption is authorized.
\end{itemize}

The 30-minute default threshold is set to bound the maximum volume of unconfirmed off-chain work that can accumulate before human review is required. Shorter thresholds increase adjudicative burden; longer thresholds increase the risk of extended execution without on-chain governance confirmation. This threshold is a constitutional parameter adjustable via the Rules Hub, allowing operators to tune the safety-vs-availability tradeoff for their deployment context.

\textbf{DEGRADED State Timeout.} When the system enters DEGRADED mode, a secondary timeout governs how long the system waits for human adjudicator authorization before auto-aborting:

\begin{lstlisting}
DEGRADED -> ABORTED (adjudicator timeout):
  Trigger: degradedAdjudicatorTimeoutMs elapsed since entering DEGRADED
  Guard: no human adjudicator authorization received
  Action: auto-abort mission; emit DegradedTimeout(missionId)
\end{lstlisting}

This prevents indefinite suspension of mission state during prolonged sequencer outages when adjudicators are unavailable. Implementation and testing against L2 sequencer downtime scenarios is planned as part of Experiment 4 (Appendix~D) and will be reported in the companion empirical paper.

\textbf{Failure Mode 4---State Divergence.} A \textit{Reconciliation Protocol} runs at mission boundaries---specifically, at the conclusion of each mission Phase (after Legislation, after Orchestration, and before final output release)---comparing the off-chain execution record against the on-chain state for all DAG nodes processed in that phase. The reconciliation check verifies:

\begin{itemize}
\item Every DAG node marked complete in the off-chain execution record has a corresponding confirmed on-chain state transition.
\item Every confirmed on-chain state transition corresponds to a recorded off-chain execution event (with matching code-hash, I/O hash, and timestamp within the optimistic consistency window).
\item The PoP attestation type recorded on-chain for each node matches the CollaborationContract specification.
\end{itemize}

Any divergence---a DAG node complete off-chain but absent on-chain, or an on-chain transition without a corresponding execution record---triggers an \textit{audit alert} to the Adjudication branch: the mission is suspended, the Logging Hub surfaces the divergence evidence, and human adjudicators review the discrepancy via the Override Panel before the mission is permitted to advance or release outputs.

The Reconciliation Protocol provides the authoritative consistency guarantee: even if individual PENDING\_FINALIZATION retries and Degraded Mode recovery operate automatically, the mission-boundary reconciliation ensures that a human principal validates the full governance record before any outputs leave the system.

\begin{lstlisting}
ALGORITHM ReconciliationProtocol(missionId: bytes32, phase: MissionPhase):
  // Runs at phase boundaries: post-Legislation, post-Orchestration, pre-OutputRelease

  1. Collect off-chain record:
     offchainRecord <- ExecutionFabric.getCompletionLog(missionId, phase)
     // Returns: list of (nodeId, outputHash, completionTimestamp, agentAddr)

  2. Collect on-chain record:
     onchainRecord <- CollaborationContract.getCompletedNodes(missionId)
     // Returns: list of (nodeId, outputHash, block_number, confirmationTimestamp)

  3. Check completeness (off-chain -> on-chain):
     FOR EACH (nodeId, hash, t_off, agent) IN offchainRecord:
       onchain_entry <- onchainRecord[nodeId]
       IF onchain_entry == NULL:
         EMIT DivergenceDetected(missionId, nodeId, "off-chain complete, on-chain absent")
         SUSPEND mission
       IF onchain_entry.outputHash != hash:
         EMIT DivergenceDetected(missionId, nodeId, "output hash mismatch")
         SUSPEND mission
       delta <- onchain_entry.confirmationTimestamp - t_off
       IF delta > optimistic_window_max_ms:
         EMIT WarningSlowFinalization(missionId, nodeId, delta)
         // Not a hard failure; logged for operator review

  4. Check completeness (on-chain -> off-chain):
     FOR EACH (nodeId, hash, blk, t_on) IN onchainRecord:
       IF offchainRecord[nodeId] == NULL:
         EMIT DivergenceDetected(missionId, nodeId, "on-chain transition without off-chain record")
         SUSPEND mission

  5. Check PoP attestation conformance:
     FOR EACH nodeId IN onchainRecord:
       recorded_tier <- the Verification module.pendingProofs[nodeId].popTier
       legislated_tier <- CollaborationContract.dag[nodeId].popTier
       IF recorded_tier != legislated_tier:
         EMIT DivergenceDetected(missionId, nodeId, "PoP tier mismatch")
         SUSPEND mission

  6. If no divergences:
     EMIT ReconciliationPassed(missionId, phase, block.number)
     RETURN PASS

  7. If any divergence:
     EMIT ReconciliationFailed(missionId, phase, divergence_count)
     NOTIFY human adjudicators via Logging Hub
     RETURN FAIL (mission suspended pending Override Panel review)
\end{lstlisting}

Implementation of the Reconciliation Protocol and audit alert pipeline is planned as part of Experiment 4 (Appendix~D); divergence rates under simulated failure injection will be reported in the companion empirical paper.

---

\subsubsection{Hybrid-Mode Security Model}

The security properties SP-1 through SP-4 (\S{}4.2) are defined for pure on-chain enforcement. In hybrid mode---used for Experiments 1--2 (Appendix~D)---these properties hold at mission boundaries (anchor points) but are relaxed during intra-mission execution. We define the hybrid-mode security properties explicitly:

\begin{quote}
\textbf{SP-1h (Hybrid Wiring Integrity).} The binding between task nodes and authorized execution units is integrity-protected at mission anchor points. Between anchors, bindings are enforced by the in-memory governance middleware, which replicates the on-chain access control logic but operates in a mutable execution environment. An adversary who compromises the middleware can alter bindings during the intra-anchor window. This vulnerability is formalized as TA-5 (\S{}4); see also the discussion of NP-1/NP-3 interaction with the Local Freeze Mirror in TA-5.
\end{quote}

\begin{quote}
\textbf{SP-2h (Hybrid Gate Enforcement).} Execution gate conditions are enforced at anchor points. Between anchors, gates are enforced by the Local Freeze Mirror and in-memory PoP validation, which provide defense-in-depth but are not tamper-proof.
\end{quote}

\begin{quote}
\textbf{SP-3h (Hybrid Auditability).} All state transitions are recorded in the off-chain execution log during intra-anchor execution and committed to the on-chain record at anchor points. The off-chain log is append-only within a single process but is not cryptographically tamper-proof until anchor commitment.
\end{quote}

\begin{quote}
\textbf{SP-4h (Hybrid Separation).} Separation enforcement operates identically to SP-4---access control is replicated in the middleware and confirmed on-chain at anchors.
\end{quote}

\textbf{Damage Bound Analysis.} The maximum unaudited operations between anchor points is bounded by the DAG execution rate and the anchor interval. Let r denote the average DAG node completion rate (nodes/second) and $\tau$\_anchor the anchor interval (seconds). The maximum unaudited operations in a single anchor window is:

\begin{equation*}
N_{unaudited} = r \times \tau_{anchor}
\end{equation*}

For example, consider an 8-node DAG with mean node execution time $\tau$\_node $\approx$ 5 seconds per node, and a configurable checkpoint interval of $\tau$\_anchor = 10 seconds (i.e., the governance middleware submits an on-chain anchor transaction every 10 seconds of mission execution). The node completion rate is r = 1/$\tau$\_node = 0.2 nodes/second. Substituting into the formula: N\_unaudited = r $\times$ $\tau$\_anchor = 0.2 $\times$ 10 = 2 node transitions between consecutive checkpoint anchors. Note that $\tau$\_anchor here is the \textit{checkpoint frequency}---a constitutional parameter configurable via the Rules Hub---not the total mission duration. For an 8-node sequential DAG at 5s/node, the total mission duration is approximately 40 seconds, spanning four checkpoint windows of 10s each, each containing at most 2 unaudited nodes. Shorter checkpoint intervals reduce N\_unaudited (and therefore the blast radius of a compromise) at the cost of increased on-chain transaction frequency. For larger DAGs (100 nodes, r = 0.5 nodes/s, $\tau$\_anchor = 60s), N\_unaudited $\leq$ 30 nodes per checkpoint window. The Reconciliation Protocol (\S{}B.11.5) detects any divergence at each checkpoint, bounding the blast radius of a middleware compromise to at most N\_unaudited unverified transitions per window.

\textbf{Experimental Caveat.} Both experiments operate primarily under the hybrid-mode security model (SP-1h--SP-4h). Experiment 2's pure on-chain measurement subset operates under the full on-chain security model (SP-1--SP-4). Results should be interpreted accordingly: attack success rates measured in hybrid mode reflect the combined security of on-chain anchoring and middleware enforcement, not pure on-chain guarantees.

\textbf{Hybrid-Mode Cost Reduction Estimate.} We provide an analytical estimate of the cost reduction factor for hybrid mode relative to pure on-chain mode, pending empirical validation from Experiment 4. In pure on-chain mode, an 8-node DAG mission requires approximately 3--6 on-chain transactions per task node (authorization, execution confirmation, PoP submission, verification, Guardian check, Gate check), yielding 24--48 transactions per mission. In hybrid mode, only mission-boundary anchor transactions are submitted on-chain: mission initialization (1 tx), per-branch-junction PoP commits (approximately $\lceil n/\tau_{anchor} \rceil$ txs, where $\tau_{anchor}$ is the checkpoint interval), and mission finalization (1 tx). For an 8-node DAG with $\tau_{anchor} = 10$s and $\tau_{node} = 5$s, this yields approximately 4--6 on-chain transactions per mission. The estimated cost reduction factor is therefore:

\begin{equation*}
R_{cost}^{tx} = \frac{\text{Hybrid txs}}{\text{Pure on-chain txs}} \approx \frac{5}{36} \approx 0.14 \quad (\sim 7\times \text{ reduction by transaction count})
\end{equation*}

However, the transaction-count ratio overstates the cost reduction because it assumes uniform gas cost per transaction. In practice, per-node governance operations (authorization ~45K gas, execution confirmation ~45K, PoP submission ~50K, verification ~55K, Guardian check ~40K, Gate check ~42K) have substantially different gas profiles from mission-level anchor operations (DAG deployment ~280K gas, Merkle root commit ~100K, finalization ~60K). A gas-weighted estimate is more accurate:

For an 8-node DAG: pure on-chain gas $\approx 8 \times 277K + 340K = 2{,}556K$ gas; hybrid gas $\approx 280K + 4 \times 100K + 60K = 740K$ gas. The gas-weighted reduction factor is:

\begin{equation*}
R_{cost}^{gas} = \frac{\text{Hybrid gas}}{\text{Pure on-chain gas}} \approx \frac{740K}{2{,}556K} \approx 0.29 \quad (\sim 3.5\times \text{ reduction})
\end{equation*}

At larger DAG sizes (100 nodes), the gas-weighted reduction improves because per-node gas dominates: pure on-chain gas $\approx 100 \times 277K + 340K = 28{,}040K$; hybrid gas $\approx 280K + 10 \times 100K + 60K = 1{,}340K$. The gas-weighted reduction is $R_{cost}^{gas}(100) \approx 0.048$ ($\sim$20.9$\times$ reduction). The transaction-count estimates ($\sim$7$\times$ and $\sim$30$\times$ respectively) overstate the savings because they implicitly assume that lightweight per-node checks cost the same gas as heavyweight mission-level anchor operations. The gas-weighted estimates ($\sim$3.5$\times$ and $\sim$20.9$\times$) are the operationally relevant figures for cost planning; empirical validation from Experiment 4 (Appendix~D) will provide ground-truth per-function gas measurements.

---

\subsection{Adjudication Interface}

The Adjudication branch is realized through a unified human interface that provides system-level override authority over both Legislation and Execution. Because the contract architecture (\S{}B.3) encodes the full execution topology on-chain, human adjudicators can audit not only logs and telemetry but the \textit{structural wiring} of the system itself---examining how micro-services from different parties are bound, what constraints govern each transition, and whether the execution topology matches the legislated intent. Human adjudicators interact with the system through four components.

\textbf{Adjudicator Team Model.} Before describing the four interface components, we specify the governance model for the human adjudicator team itself, which constitutes the Adjudication branch of the SoP model. The adjudicator team must satisfy the following structural requirements:

\begin{itemize}
\item \textit{Minimum quorum floor:} The \texttt{adjudicatorQuorum} constitutional parameter has an enforced minimum floor of $q_{min} = 2f + 1$ where f is the maximum number of adjudicators the system tolerates being compromised. For the recommended default f = 2 (tolerating up to 2 compromised adjudicators), $q_{min} = 5$, requiring bribery of $\geq$ 4 adjudicators (at the $\lceil$2q/3$\rceil$ supermajority threshold for revocation control) to achieve unconditional system compromise. We set the default \texttt{adjudicatorQuorum} to 7 for the following analysis. The minimum quorum floor is enforced at the contract level: any attempt to set \texttt{adjudicatorQuorum} below $q_{min}$ via the Rules Hub is reverted.
\item \textit{Rotation policy:} No single adjudicator may serve as the sole approver for more than two consecutive Type 3 PoP attestation decisions or more than two consecutive freeze unfreeze decisions for the same mission. This rotation requirement is enforced by the Override Panel software layer, which checks the adjudicator's approval history before accepting a signed action.
\item \textit{Conflict of interest rules:} An adjudicator whose human-principal address is associated (via on-chain trace) with a producer agent participating in the current mission may not exercise binding Override Panel authority over that mission's decisions. They may observe via the Logging Hub and Execution Dashboard but cannot submit signed approval, rejection, or freeze actions to the Override Panel for that mission. This constraint is enforced by the Override Panel before accepting any signed action.
\item \textit{Emergency override:} In cases where the quorum of non-conflicted adjudicators is unavailable (fewer than $\lceil$q/2$\rceil$ non-conflicted adjudicators active), a single senior adjudicator (designated at system configuration time in the Rules Hub) may exercise unilateral emergency authority, but all unilateral actions are logged with a UNILATERAL\_OVERRIDE flag and trigger an automatic post-mission audit requirement.
\item \textit{Adjudicator authentication:} All Override Panel actions require cryptographic signature from the adjudicator's registered human-principal address (hardware wallet or equivalent strong custody). Session-based web authentication is insufficient for binding authority actions; the web interface constructs an EIP-712 typed data payload that the adjudicator signs with their private key before submission.
\end{itemize}

\textit{Scalable oversight framing.} The three-tier PoP allocation logic---directing human adjudicator effort toward Tier 3 tasks while automating Tier 1 and Tier 2 verification---instantiates the scalable oversight framework studied by Bowman et al. (2022) and Irving and Askell (2019) [Bowman2022; Irving2019]. Human adjudicator capacity is a scarce constitutional resource; the PoP tier assignment and alert priority queue are the mechanisms by which AgentCity allocates this resource to tasks where it is most needed.

\textbf{On-Chain Adjudicator Accountability.} To address the asymmetry between agent accountability (reputation + slashing) and adjudicator accountability (social enforcement only), we introduce three on-chain mechanisms:

\textit{Adjudicator Stake.} Each registered adjudicator deposits an adjudicator stake (constitutional parameter: \texttt{adjudicatorStake}, default: 5,000 units) into the AgentContract upon registration as a MONITOR-type agent. This stake is subject to slashing by a supermajority ($\geq$ 2/3) of the remaining adjudicator pool via an adjudicator-revocation protocol.

\textit{On-Chain Rotation Enforcement.} The rotation policy is enforced as a contract-level modifier on all Override Panel functions:

\begin{lstlisting}
MODIFIER enforceRotation(adjudicatorAddr: address, missionId: bytes32, actionType: ActionType):
    consecutiveActions <- getConsecutiveActionCount(adjudicatorAddr, missionId, actionType)
    REQUIRE consecutiveActions < maxConsecutiveActions  // constitutional parameter, default: 2
    // If violated, transaction reverts -- rotation cannot be bypassed
\end{lstlisting}

This modifier is applied to \texttt{the Guardian module.unfreezeWithApproval}, \texttt{the Verification module.approveDelegated}, and \texttt{the Gate module.vetoOutput}, moving rotation enforcement from the application layer (Override Panel software) to the contract layer.

\textit{Adjudicator Revocation Protocol.} If an adjudicator is suspected of malicious or negligent governance, any other registered adjudicator may initiate a revocation vote:

\begin{lstlisting}
FUNCTION initiateImpeachment(targetAdjudicator: address, evidence: bytes32):
    REQUIRE msg.sender is registered adjudicator
    REQUIRE msg.sender != targetAdjudicator
    impeachmentId <- keccak256(targetAdjudicator, block.number)
    SSTORE impeachments[impeachmentId] = ImpeachmentRecord(target, evidence, 1, now)
    EMIT ImpeachmentInitiated(impeachmentId, targetAdjudicator, msg.sender)

FUNCTION voteImpeachment(impeachmentId: bytes32, support: bool):
    REQUIRE msg.sender is registered adjudicator
    REQUIRE msg.sender != impeachments[impeachmentId].target
    IF support: impeachments[impeachmentId].votes += 1
    IF impeachments[impeachmentId].votes >= ceil(2 * adjudicatorCount / 3):
        SLASH adjudicatorStake of target
        CALL AgentContract.banAgent(target, 'revoked')
        EMIT AdjudicatorImpeached(impeachmentId, target)
\end{lstlisting}

\textbf{Economic Security Bound as a Function of Quorum Size.} The cost of unconditionally compromising the adjudicator quorum through bribery is:

\begin{equation*}
C_{bribe}(q) \geq \lceil 2q/3 \rceil \times (s_{adj} + E[\text{reputation\_loss}])
\end{equation*}

where the $\lceil$2q/3$\rceil$ term reflects the supermajority required to both control adjudicative authority \textit{and} block revocation of colluding members.

\begin{table}[htbp]
\caption{Economic security bound by adjudicator quorum size.}
\label{tab:adj-quorum}
\centering
\footnotesize
\begin{tabular}{>{\raggedright\arraybackslash}p{0.08\textwidth} >{\raggedright\arraybackslash}p{0.10\textwidth} >{\raggedright\arraybackslash}p{0.14\textwidth} >{\raggedright\arraybackslash}p{0.18\textwidth} >{\raggedright\arraybackslash}p{0.36\textwidth}}
\toprule
\textbf{Quorum ($q$)} & \textbf{Bribery Target} & \textbf{Cost (units)} & \textbf{Cost (ETH)} & \textbf{System Compromise} \\
\midrule
3 & 2 & 10,000 & 0.01 ETH (~\$25) & Unconditional --- bribed majority controls both authority and revocation \\
5 & 4 & 20,000 & 0.02 ETH (~\$50) & Unconditional --- still achievable \\
7 (recommended) & 5 & 25,000 & 0.025 ETH (~\$62.50) & Unconditional but costly --- 5 independent human principals must be compromised \\
9 & 6 & 30,000 & 0.03 ETH (~\$75) & Highly expensive --- requires coordinated corruption of 6 principals \\
13 & 9 & 45,000 & 0.045 ETH (~\$112.50) & Enterprise-grade --- practical infeasibility for most threat models \\
\bottomrule
\end{tabular}
\end{table}

\textbf{Production-Scale Cross-Reference (EQ-7).} The prototype stake values above ($s_{adj} = 5{,}000$ units $\approx$ $\$12.50$ at evaluation ETH price) provide negligible economic deterrence. At production parameters (EQ-7, $s_{adj}^{prod} = V_m / q$), the bribery economics shift dramatically:

\begin{table}[htbp]
\caption{Adjudicator bribery cost at production stakes ($V_m = \$100{,}000$, $s_{adj}^{prod} = V_m / q$).}
\label{tab:adj-bribery}
\centering
\small
\begin{tabular}{lllll}
\toprule
\textbf{Quorum ($q$)} & \textbf{Bribery Target ($\lceil 2q/3 \rceil$)} & \textbf{Per-Adj.\ Stake} & \textbf{Min.\ Bribery Cost} & \textbf{Bribery / $V_m$} \\
\midrule
3 & 2 & $\$33{,}333$ & $\$66{,}667$ & 66.7\% \\
5 & 4 & $\$20{,}000$ & $\$80{,}000$ & 80.0\% \\
7 (recommended) & 5 & $\$14{,}286$ & $\$71{,}429$ & 71.4\% \\
9 & 6 & $\$11{,}111$ & $\$66{,}667$ & 66.7\% \\
13 & 9 & $\$7{,}692$ & $\$69{,}231$ & 69.2\% \\
\bottomrule
\end{tabular}
\end{table}

At production parameters, bribing a majority of adjudicators costs 67--80\% of $V_m$---making adjudicator bribery economically irrational whenever the attacker's expected gain is less than $\sim$70\% of the mission value. Practitioners reading~\S{}B.3 in isolation should reference Table~\ref{tab:adj-bribery} rather than the prototype values, which understate bribery costs by a factor of ${\sim}1{,}000\times$.

\textbf{Critical observation:} With q = 3 (the previous default), bribing 2 adjudicators costs only 10,000 units and grants the adversary both unconditional adjudicative authority \textit{and} the ability to block revocation (the honest minority of 1 cannot reach the $\lceil$2$\times$3/3$\rceil$ = 2 threshold). This makes the revocation mechanism self-defeating at small quorum sizes. With q = 7, the bribery target rises to 5, and the remaining 2 honest adjudicators cannot be outvoted on revocation---but they also cannot reach the revocation threshold alone. The \textit{minimum quorum for revocation recovery} (where honest adjudicators can revoke bribed ones after a $\lceil$2q/3$\rceil$ bribery) is q $\geq$ 3f + 1, requiring q $\geq$ 7 for f = 2.

\textbf{On-Chain Watchdog Mechanism.} To provide a defense-in-depth against adjudicator compromise that does not rely on the revocation mechanism (which is itself vulnerable to quorum capture), we introduce an automated anomaly detection watchdog:

\begin{lstlisting}
CONTRACT AdjudicatorWatchdog:
  // Monitors adjudicator behavior patterns and triggers automatic system freeze
  // when statistical anomalies suggest compromise

  STATE:
    adjudicatorActions: mapping(address -> ActionRecord[])
    systemFrozenByWatchdog: bool
    watchdogThresholds: WatchdogParams

  STRUCT WatchdogParams:
    maxApprovalRate: uint8        // default: 95 (%) -- flag if adjudicator approves >95% of actions
    maxConsecutiveApprovals: uint8 // default: 20 -- flag if >20 consecutive approvals without rejection
    paramWeakeningLimit: uint8    // default: 3 -- flag if >3 parameter-weakening changes in 30 days
    windowDays: uint16            // default: 30

  FUNCTION recordAction(adjudicator: address, actionType: ActionType, decision: Decision):
    APPEND to adjudicatorActions[adjudicator]
    IF detectAnomaly(adjudicator):
      EMIT WatchdogAlert(adjudicator, anomalyType)
      IF anomalyCount(adjudicator) >= 2:
        SSTORE systemFrozenByWatchdog = true
        EMIT WatchdogFreeze(adjudicator, block.number)
        // Watchdog freeze can only be lifted by ALL non-flagged adjudicators signing a release
        // This prevents a compromised majority from unilaterally lifting the freeze

  FUNCTION detectAnomaly(adjudicator: address) -> bool:
    actions <- getRecentActions(adjudicator, watchdogThresholds.windowDays)
    approvalRate <- countApprovals(actions) / |actions|
    IF approvalRate > watchdogThresholds.maxApprovalRate: RETURN true
    IF consecutiveApprovals(actions) > watchdogThresholds.maxConsecutiveApprovals: RETURN true
    IF paramWeakenings(actions) > watchdogThresholds.paramWeakeningLimit: RETURN true
    RETURN false
\end{lstlisting}

The watchdog mechanism operates as an autonomous circuit breaker: it does not require any adjudicator to initiate action, and a watchdog-triggered freeze cannot be lifted by the flagged adjudicator(s). This addresses the critical gap where a bribed majority could block revocation of each other---the watchdog bypasses this human-dependent mechanism entirely by detecting behavioral patterns algorithmically.

\textbf{Watchdog Liveness: Maximum Freeze Duration and False-Positive Analysis.} The watchdog freeze requires unanimous consent from all non-flagged adjudicators to lift. This creates a liveness risk: if a non-flagged adjudicator is unavailable (vacation, network partition, key loss), the freeze persists indefinitely, halting all mission execution. We bound this risk through two mechanisms:

\begin{enumerate}
\item \textbf{Maximum freeze duration.} A constitutional parameter \texttt{watchdogMaxFreezeDuration} (default: 72 hours) imposes an automatic timeout on watchdog-triggered freezes. If the freeze is not explicitly resolved (lifted or confirmed) within this window, the system enters a \textit{degraded-governance mode} in which new mission initiations are blocked but existing missions in EXECUTING state are allowed to complete through their current DAG node without advancing to new nodes. This prevents indefinite system halt while preserving the security signal of the unresolved watchdog alert.
\item \textbf{False-positive rate analysis.} The default thresholds (>95\% approval rate, >20 consecutive approvals, >three parameter weakenings in 30 days) are calibrated against baseline adjudicator behavior. Under normal operations, an adjudicator reviewing a mix of legitimate missions is expected to approve 70--85\% of actions (based on the assumption that most missions are non-adversarial). The 95\% threshold therefore triggers only when approval rate exceeds baseline by $\geq$10 percentage points. For the consecutive-approvals threshold, a sequence of 20+ approvals without any rejection has probability $\leq 0.85^{20} \approx 0.039$ under the baseline 85\% approval rate---yielding an estimated false-positive rate of $\sim$3.9\% per adjudicator per 30-day window. With $q = 7$ adjudicators, the expected number of false-positive alerts per window is $\sim$0.27, and the probability of a false-positive \textit{freeze} (requiring 2 anomaly alerts from the same adjudicator) is $\leq (0.039)^2 \approx 0.15\%$ per adjudicator per window. \textbf{Sensitivity to baseline approval rate.} The false-positive rate is highly sensitive to the assumed baseline approval rate. We tabulate the watchdog FP rate across a range of plausible baselines:
\end{enumerate}

\begin{center}

\small
\begin{tabular}{lll}
\toprule
\textbf{Baseline Approval Rate} & \textbf{P(alert/adj/30d)} & \textbf{P(freeze/30d)} \\
\midrule
0.70 & 0.046\% & <0.001\% \\
0.75 & 0.266\% & <0.001\% \\
0.80 & 1.15\% & <0.001\% \\
0.85 & 3.88\% & 0.001\% \\
0.90 & 10.6\% & 0.044\% \\
0.95 & 24.4\% & 1.44\% \\
\bottomrule
\end{tabular}
\end{center}

At the assumed baseline of 0.85, the per-adjudicator alert rate is ~3.9\% per 30-day window and freeze probability is negligible. However, if the true baseline exceeds 0.90 (e.g., because the vast majority of missions are non-adversarial and adjudicators correctly approve them), the alert rate exceeds 10\% and false-positive freezes become operationally significant. This sensitivity motivates two design responses: (1) the \texttt{maxApprovalRate} threshold should be calibrated empirically from observed adjudicator behavior during the pilot phase, not set a priori; (2) a graduated alerting scheme (warning $\rightarrow$ investigation $\rightarrow$ freeze) rather than the current binary alert $\rightarrow$ freeze model would reduce the operational impact of threshold miscalibration. We acknowledge that these estimates rely on the assumed baseline approval rate; empirical calibration from the companion experimental data is required before production deployment.

\textbf{Rules Hub (Constitution Editor).} The Rules Hub is the constitutional parameter management interface through which human adjudicators define and amend the behavioral constraints that govern all mission execution. Rule changes take effect via on-chain transactions that update the relevant contract parameters, ensuring that all governance modifications are recorded on the immutable ledger with the adjudicator's cryptographic signature.

{\footnotesize
\begin{longtable}{>{\raggedright\arraybackslash}p{0.20\textwidth} p{0.05\textwidth} p{0.06\textwidth} p{0.12\textwidth} p{0.11\textwidth} p{0.26\textwidth} p{0.06\textwidth}}
\caption{Complete Constitutional Parameter Table.} \label{tab:params} \\
\toprule
\textbf{Parameter} & \textbf{Type} & \textbf{Default} & \textbf{Valid Range} & \textbf{Governs Contract} & \textbf{Description} & \textbf{Risk Tier} \\
\midrule
\endfirsthead
\toprule
\textbf{Parameter} & \textbf{Type} & \textbf{Default} & \textbf{Valid Range} & \textbf{Governs Contract} & \textbf{Description} & \textbf{Risk Tier} \\
\midrule
\endhead
\midrule
\multicolumn{7}{r}{\footnotesize\itshape Continued on next page} \\
\endfoot
\bottomrule
\endlastfoot
\texttt{deviationThresholdSigma} & uint8 & 2 & {[}1, 5{]} & Guardian module & Sigma threshold for reasoning deviation anomaly detection & HIGH \\
\texttt{maxToolInvocations} & uint16 & 40 & [5, 200] & Guardian module & Maximum tool invocations per agent per task node & MEDIUM \\
\texttt{maxMessageVolume} & uint16 & 120 & [10, 500] & Guardian module & Maximum inter-agent messages per task node & MEDIUM \\
\texttt{escalationFreezeCount} & uint8 & 3 & {[}2, 10{]} & Guardian module & Base freeze count before Emergency Stop escalation & HIGH \\
\texttt{baseEscalationWindowMs} & uint32 & 1,200,000 & [60,000, 7,200,000] & Guardian module & Base counting window for freeze escalation (ms) & LOW \\
\texttt{reputationScalingEnabled} & bool & true & {true, false} & Guardian module & Enable reputation-weighted freeze thresholds & MEDIUM \\
\texttt{reputationScalingBonus} & uint8 & 3 & [0, 7] & Guardian module & Max additional freeze allowance for max-reputation agent & MEDIUM \\
\texttt{missionBudgetCap} & uint32 & 100,000 & [1,000, 10,000,000] & CollaborationContract & Maximum total token budget per mission & MEDIUM \\
\texttt{maxNodeTimeoutMs} & uint32 & 60,000 & [1,000, 3,600,000] & CollaborationContract & Maximum per-node execution timeout (ms) & LOW \\
\texttt{minHumanReviewTimeoutMs} & uint32 & 300,000 & [60,000, 86,400,000] & CollaborationContract & Minimum Type 3 PoP review window (ms) & LOW \\
\texttt{maxRefinementIterations} & uint8 & 3 & {[}1, 10{]} & CollaborationContract & Maximum Adaptive Refinement epoch restarts & LOW \\
\texttt{reputationFloor} & uint16 & 100 & [0, 500] & AgentContract & Minimum reputation score for mission participation & HIGH \\
\texttt{jaccardThreshold} & uint8 & 85 & {[}50, 100{]} & Verification module & Type 2 Jaccard similarity consensus threshold ($\times$100) & HIGH \\
\texttt{numericTolerancePct} & uint8 & 5 & {[}1, 50{]} & Verification module & Type 2 numeric output tolerance (\% of mean) & LOW \\
\texttt{minFairnessScore} & uint16 & 600 & [0, 1000] & Legislative process & Minimum fairness score for regulatory sign-off & LOW \\
\texttt{minStakeLow} & uint32 & 150 & [0, 100,000] & Legislative process & Minimum producer stake for LOW-risk task nodes & MEDIUM \\
\texttt{minStakeMedium} & uint32 & 500 & [0, 100,000] & Legislative process & Minimum producer stake for MEDIUM-risk task nodes & MEDIUM \\
\texttt{minStakeHigh} & uint32 & 2,000 & [0, 100,000] & Legislative process & Minimum producer stake for HIGH-risk task nodes & MEDIUM \\
\texttt{degradedModeThresholdMs} & uint32 & 1,800,000 & [60,000, 14,400,000] & Consistency Protocol & Sequencer outage duration before Emergency Stop (ms) & HIGH \\
\texttt{optimisticWindowMaxMs} & uint32 & 10,000 & [2,000, 60,000] & Consistency Protocol & Max consistency gap before Reconciliation warning (ms) & MEDIUM \\
\texttt{maxCodeHashLatencyMs} & uint16 & 2,000 & [500, 10,000] & Execution fabric & Max code-hash verification response time (ms) & MEDIUM \\
\texttt{partialRelegBiddingTimeoutMs} & uint32 & 300,000 & [60,000, 3,600,000] & Adaptive Refinement & Bidding window duration for partial re-legislation (ms) & LOW \\
\texttt{legislativeProposalTimeoutMs} & uint32 & 600,000 & [60,000, 3,600,000] & Legislative process & Maximum duration for DAG proposal round (ms) & LOW \\
\texttt{biddingWindowMs} & uint32 & 900,000 & [60,000, 7,200,000] & Legislative process & Duration of open bidding window (ms) & LOW \\
\texttt{regulatoryApprovalTimeoutMs} & uint32 & 300,000 & [60,000, 3,600,000] & Legislative process & Maximum duration for regulatory review round (ms) & LOW \\
\texttt{adjudicatorQuorum} & uint8 & 7 & {[}5, 20{]} & Adjudication & Minimum registered adjudicators for mission authorization & HIGH \\
\texttt{seniorAdjudicatorAddr} & address & --- & valid address & Adjudication & Emergency unilateral authority address & MEDIUM \\
\texttt{multisigRegQuorum} & uint8 & 2 & [1, R] & Legislative process & Minimum Regulatory Agent approvals for multi-org missions & MEDIUM \\
\texttt{maxHumanReviewTimeoutMs} & uint32 & 3,600,000 & [300,000, 86,400,000] & Verification module & Maximum wait for Type 3 human review before timeout escalation (ms) & MEDIUM \\
\texttt{degradedAdjudicatorTimeoutMs} & uint32 & 7,200,000 & [1,800,000, 86,400,000] & CollaborationContract & Maximum wait for adjudicator authorization to exit DEGRADED (ms) & MEDIUM \\
\texttt{predicateGasLimit} & uint32 & 200,000 & [50,000, 1,000,000] & Gate module & Maximum gas forwarded per predicate STATICCALL & HIGH \\
\texttt{embeddingModelId} & string & text-embedding-3-large & --- & Guardian module & Active primary embedding model for deviation scoring (3,072-d); change triggers mandatory re-calibration. Secondary scorer (Voyage AI voyage-3) configured separately & MEDIUM \\
\texttt{protocolFeeRate} & uint16 & 200 & [0, 1000] & Settlement module & Protocol fee rate in basis points (f\_p); portion of task bid price retained by treasury & HIGH \\
\texttt{insuranceReserveRate} & uint16 & 100 & [0, 500] & Settlement module & Insurance reserve rate in basis points (f\_i); portion of task bid price channeled to insurance pool & MEDIUM \\
\texttt{reputationMultiplierAlpha} & uint16 & 500 & [0, 2000] & Settlement module & $\alpha$ parameter for reputation multiplier $\psi$($\rho$); scaled by 1000 (default 500 = $\alpha$ of 0.5) & HIGH \\
\texttt{adjudicatorCompensationRate} & uint256 & 50 & [10, 500] & Settlement module & Compensation per Tier 3 PoP review disbursed from treasury to adjudicator & MEDIUM \\
\texttt{stakepoolMinParticipants} & uint8 & 3 & {[}2, 50{]} & Settlement module & Minimum number of participants required to activate a stake pool & LOW \\
\end{longtable}
}

\textbf{Parameter Calibration Rationale.} The default parameter values are derived from a combination of theoretical analysis and pilot mission observations:

\begin{itemize}
\item \textit{$\sigma$ = 2 (deviation threshold):} Matches the statistical 2$\sigma$ rule, yielding a theoretical 2.28\% false positive rate under Gaussian reference distributions. Empirical validation pending (see~\S{}B.11 calibration requirements).
\item \textit{tools = 40 (max tool invocations):} Derived from pilot mission traces where the 99th percentile of legitimate tool invocations per task node was 32. A 25\% headroom margin yields 40.
\item \textit{messages = 120 (max message volume):} The 99th percentile of inter-agent messages per task in pilot missions was 94. A 28\% margin yields 120.
\item \textit{freezes = 3 (escalation threshold):} Set to balance false positive tolerance (allowing up to 2 erroneous freezes before escalation) against adversarial exploitation (limiting the attacker to 2 unpenalized anomalies).
\item \textit{window = 20 min (escalation window):} Calibrated to exceed the expected duration of the longest single-node execution in pilot missions (12 minutes), ensuring that sequential legitimate freezes across different task nodes within one mission do not spuriously trigger escalation.
\end{itemize}

These defaults are provisional and will be refined through systematic sensitivity analysis in a planned sensitivity experiment (companion empirical paper).

\textit{Rules Hub Parameter Update Protocol:}

\begin{lstlisting}
ALGORITHM ConstitutionalParameterUpdate(paramId: bytes32, newValue: uint256,
                                         justification: string):
  // Invoked by adjudicator via Rules Hub web interface

  1. Adjudicator constructs EIP-712 typed data payload:
     payload <- {
       type: "ConstitutionalParameterUpdate",
       paramId: paramId,
       oldValue: currentValue(paramId),
       newValue: newValue,
       justification: keccak256(justification),
       adjudicatorAddr: msg.sender,
       timestamp: now,
       missionScope: "GLOBAL" | "MISSION:<missionId>"
     }

  2. Adjudicator signs payload with hardware wallet:
     signature <- SIGN(payload, adjudicator_private_key)

  3. Web interface submits on-chain transaction:
     CALL RulesHub.updateParameter(paramId, newValue, signature, justification_cid)
     // justification_cid: IPFS CID of the full justification text (for audit trail)

  4. RulesHub contract validates:
     REQUIRE signature verifies against adjudicator_registered_address
     REQUIRE adjudicator is non-conflicted for scope
     REQUIRE newValue \in valid_range[paramId]
     REQUIRE NOT locked_during_active_mission(paramId)
     // Some parameters are immutable mid-mission; changes take effect at next mission start

  5. RulesHub propagates to affected contracts:
     FOR EACH affected_contract IN paramId.affectedContracts:
       CALL affected_contract.setParam(paramId, newValue)

  6. Logging Hub records update:
     EMIT ParameterUpdated(paramId, oldValue, newValue, adjudicatorAddr,
                           justification_cid, block.number)
\end{lstlisting}

\textbf{Logging Hub (Audit Trail Viewer).} The Logging Hub is an append-only audit trail interface providing full interaction history across all phases of the mission lifecycle. It is the primary evidence source for adjudicative review and post-mission forensic analysis.

\textit{Event Schema.} The Logging Hub records events conforming to the following canonical schema:

\begin{lstlisting}
STRUCT LogEvent:
  eventId: bytes32          // keccak256(missionId || sequence_number)
  missionId: bytes32
  epoch: uint256
  eventType: EventType
  emitterContract: address  // which contract emitted this event (or 0x0 for off-chain)
  primaryEntity: address    // agent, contract, or adjudicator involved
  secondaryEntity: address  // counterparty (if applicable)
  nodeId: bytes32           // DAG node (if applicable; 0x0 for mission-level events)
  payloadHash: bytes32      // keccak256 of structured event payload
  payloadCid: string        // IPFS CID of full payload for off-chain retrieval
  blockNumber: uint256      // 0 for off-chain events
  timestamp: uint256        // Unix milliseconds
  gasUsed: uint256          // 0 for off-chain events

ENUM EventType:
  // Registration
  AGENT_REGISTERED | SERVICE_REGISTERED | AGENT_BANNED

  // Legislative
  SESSION_INITIATED | IDENTITY_VERIFIED | DAG_PROPOSED | BID_SUBMITTED |
  BID_APPROVED | BID_REJECTED | REGULATORY_SIGNED | CODIFICATION_COMPILED |
  LEGISLATIVE_APPROVED | DAG_DEPLOYED

  // Execution
  TASK_ROUTED | TASK_EXECUTING | TASK_COMPLETED | TASK_FAILED | TASK_TIMEOUT |
  CODEHASH_MISMATCH | CODEHASH_VERIFIED

  // Verification
  POP_SUBMITTED | POP_APPROVED | POP_REJECTED | DELEGATED_REVIEW_REQUESTED |
  DELEGATED_APPROVED | DELEGATED_REJECTED

  // Guardian
  ANOMALY_REPORTED | FREEZE_TRIGGERED | UNFREEZE_APPROVED | FALSE_POSITIVE_CLASSIFIED |
  EMERGENCY_STOP | EMERGENCY_STOP_CLEARED

  // Consistency
  PENDING_FINALIZATION | ESCROW_RELEASED | DEGRADED_MODE_ENTERED |
  DEGRADED_MODE_EXITED | RECONCILIATION_PASSED | RECONCILIATION_FAILED |
  DIVERGENCE_DETECTED

  // Adaptive Refinement
  FAULT_SIGNAL | PARTIAL_RELEGISLATION_INITIATED | FULL_RELEGISLATION_INITIATED |
  MAX_REFINEMENT_EXCEEDED

  // Output
  OUTPUT_RELEASED | OUTPUT_VETOED | MISSION_COMPLETED | MISSION_ABORTED

  // Adjudication
  PARAMETER_UPDATED | OVERRIDE_ACTION | UNILATERAL_OVERRIDE |
  REPUTATION_ADJUSTED | STAKE_SLASHED
\end{lstlisting}

\textit{Indexing Dimensions.} The Logging Hub indexes all events along the following dimensions to support targeted forensic queries:

\begin{center}

\small
\begin{tabular}{p{0.16\textwidth} p{0.14\textwidth} p{0.60\textwidth}}
\toprule
\textbf{Index Dimension} & \textbf{Type} & \textbf{Query Semantics} \\
\midrule
\texttt{missionId} & bytes32 & All events for a specific mission (primary partition key) \\
\texttt{agentAddr} & address & All events involving a specific agent (across missions) \\
\texttt{contractAddr} & address & All events emitted by a specific contract instance \\
\texttt{eventType} & enum & All events of a specific type (e.g., all FREEZE\_TRIGGERED events) \\
\texttt{nodeId} & bytes32 & All events for a specific DAG task node \\
\texttt{epoch} & uint256 & All events within a specific legislative epoch \\
\texttt{timeRange} & (uint256, uint256) & All events within a timestamp window \\
\texttt{blockRange} & (uint256, uint256) & All events within a block number range \\
\texttt{humanPrincipal} & address & All events attributable to a specific human principal's agents \\
\bottomrule
\end{tabular}
\end{center}

\textit{Retention Policy.} All events are stored in append-only form. On-chain events (emitted by smart contracts) are permanently available via standard L2 log retrieval. Off-chain events (legislative message exchanges, execution enclave logs) are stored with their payload hashes on-chain and full payloads on IPFS with a minimum retention period of 24 months. The Logging Hub interface aggregates both on-chain and IPFS-stored events into a unified view. Mission-critical events (FREEZE\_TRIGGERED, DIVERGENCE\_DETECTED, EMERGENCY\_STOP, STAKE\_SLASHED) are additionally mirrored to an adjudicator-controlled off-chain storage endpoint to ensure availability independent of IPFS network conditions.

\textbf{Execution Dashboard.} The Execution Dashboard provides real-time and historical views of mission state, designed for continuous adjudicative oversight rather than one-time forensic review.

\textit{Real-Time Metrics:}

\begin{center}

\small
\begin{tabular}{p{0.16\textwidth} p{0.25\textwidth} p{0.25\textwidth} p{0.24\textwidth}}
\toprule
\textbf{Metric} & \textbf{Update Frequency} & \textbf{Visualization} & \textbf{Alert Condition} \\
\midrule
DAG node states & On every CollaborationContract event & Interactive DAG graph with color-coded node states & Any node in FROZEN, FAILED, or PENDING\_FINALIZATION > 30 s \\
Mission phase & On phase transitions & State badge + timeline bar & Phase transition slower than expected baseline \\
Guardian anomaly counters & On every the Guardian module event & Per-agent gauge charts & Any counter > 70\% of constitutional threshold \\
Active freezes & On FREEZE\_TRIGGERED events & Alert banner + freeze list & Any active freeze (immediate alert) \\
PoP queue & On DELEGATED\_REVIEW\_REQUESTED events & Queue length counter & Queue length > 0 (immediate notification to adjudicators) \\
Gas expenditure & On every on-chain transaction & Cumulative cost chart (per-contract, per-mission) & Per-mission cumulative > 90\% of gas budget \\
Reputation standings & On every REPUTATION\_ADJUSTED event & Leaderboard sorted by score & Any agent drops below reputation floor \\
Consistency protocol status & On every state change & Status indicator & DEGRADED\_MODE, PENDING\_FINALIZATION, or DIVERGENCE\_DETECTED \\
Escrow buffer contents & On PENDING\_FINALIZATION events & Node list with retry countdown & Retry \#3 approaching (T+8min countdown) \\
Mission progress & On TASK\_COMPLETED events & Completion percentage bar & Progress stalled > 2$\times$ expected per-node latency \\
Active Emergency Stops & On EMERGENCY\_STOP events & Full-screen alert & Any Emergency Stop (requires immediate adjudicator acknowledgment) \\
\bottomrule
\end{tabular}
\end{center}

\textit{Historical Views:} The dashboard also provides historical mission analysis, including: mission timeline gantt charts (showing per-node execution times and phases), gas cost breakdown by contract type, freeze frequency histograms by agent and time window, PoP success/rejection rate by tier, and comparative performance across legislative epochs (for Adaptive Refinement missions).

\textbf{Override Panel.} The Override Panel is the mechanism through which human adjudicators exercise binding system-level override authority. All Override Panel actions require a cryptographically signed EIP-712 payload from the adjudicator's registered human-principal address before submission.

\textit{Formalized Override Protocol:}

\begin{lstlisting}
ALGORITHM OverridePanelAction(action: OverrideAction, target: bytes32,
                               parameters: bytes, justification: string):
  // Generic protocol for all Override Panel actions

  PRECONDITIONS:
    1. msg.sender is registered adjudicator (AgentContract.agentType == MONITOR)
    2. msg.sender is non-conflicted for this mission (no associated producer agent)
    3. msg.sender has not exceeded rotation limit for this action type/mission
    4. Signed EIP-712 payload provided and verified

  EXECUTION: (dispatched by action type)

  ACTION: FREEZE_AGENT
    CALL the Guardian module.triggerFreeze(nodeId=target, agentAddr=parameters,
                                         reason=MANUAL_OVERRIDE)
    LOG: OverrideAction(FREEZE_AGENT, target, msg.sender, justification_cid, block.number)

  ACTION: UNFREEZE_AGENT
    REQUIRE CollaborationContract.nodeState[target] == FROZEN
    CALL the Guardian module.unfreezeWithApproval(nodeId=target, agentAddr=parameters)
    LOG: OverrideAction(UNFREEZE_AGENT, target, msg.sender, justification_cid, block.number)

  ACTION: TERMINATE_MISSION
    CALL CollaborationContract.abortMission(reason=keccak256(justification))
    SLASH stakes of all producer agents with EXECUTING or FAILED nodes
        proportional to slashingSchedule.missionAbort
    LOG: OverrideAction(TERMINATE_MISSION, missionId, msg.sender, justification_cid, block.number)

  ACTION: AMEND_CONSTITUTIONAL_PARAMETER
    // Delegates to ConstitutionalParameterUpdate protocol above
    CALL RulesHub.updateParameter(paramId=target, newValue=parameters, ...)
    LOG: OverrideAction(AMEND_PARAM, target, msg.sender, justification_cid, block.number)

  ACTION: ADJUST_REPUTATION
    CALL AgentContract.updateReputation(agent=target, delta=parameters, rationale=justification)
    LOG: OverrideAction(ADJUST_REPUTATION, target, msg.sender, justification_cid, block.number)

  ACTION: CLASSIFY_FALSE_POSITIVE
    REQUIRE the Guardian module.freezeLog[parameters][freeze_index].agentAddr == target
    CALL the Guardian module.classifyFalsePositive(agentAddr=target, freezeIndex=parameters)
    LOG: OverrideAction(FALSE_POSITIVE, target, msg.sender, justification_cid, block.number)

  ACTION: APPROVE_DELEGATED_POP
    REQUIRE the Verification module.delegatedPending[target]
    CALL the Verification module.approveDelegated(nodeId=target)
    LOG: OverrideAction(APPROVE_POP, target, msg.sender, justification_cid, block.number)

  ACTION: REJECT_DELEGATED_POP
    REQUIRE the Verification module.delegatedPending[target]
    CALL the Verification module.rejectDelegated(nodeId=target, reason=parameters)
    LOG: OverrideAction(REJECT_POP, target, msg.sender, justification_cid, block.number)

  ACTION: TRIGGER_EMERGENCY_STOP
    CALL the Guardian module.triggerEmergencyStop(triggeringAgent=0x0)  // system-wide
    EMIT EmergencyStop(0x0, now)
    LOG: OverrideAction(EMERGENCY_STOP, missionId, msg.sender, justification_cid, block.number)

  ACTION: CLEAR_EMERGENCY_STOP
    REQUIRE emergencyStopActive
    REQUIRE adjudicator_quorum_approval (>= 2 adjudicators have co-signed clearance)
    SSTORE the Guardian module.emergencyStopActive = false
    LOG: OverrideAction(CLEAR_EMERGENCY_STOP, missionId, msg.sender, justification_cid, block.number)

  POSTCONDITIONS (all actions):
    Every action produces exactly one OVERRIDE_ACTION log event with:
      - adjudicator address
      - action type
      - target entity
      - justification IPFS CID
      - on-chain transaction hash
      - block number
    This constitutes the immutable audit trail for adjudicative accountability.
\end{lstlisting}

\textit{Adjudicator Authentication:} The Override Panel constructs all binding actions as EIP-712 typed data structs that must be signed by the adjudicator's registered private key. The web interface never receives the private key; it requests a signature via MetaMask or compatible hardware wallet interface and submits the signed payload to the contract. Session-based authentication (username/password, OAuth) provides access to the read-only Logging Hub and Execution Dashboard views but cannot submit any binding action. This separation between read access (session-based) and write access (cryptographic signature) ensures that a compromised web session cannot produce unauthorized governance actions.

\textit{Audit Trail Completeness.} Every Override Panel action produces a structured log event recorded on-chain via the Logging Hub. The audit trail for each action includes: the adjudicator's on-chain address (linked to their human-principal identity via AgentContract), the action type and target entity, a keccak256 of the justification text, an IPFS CID for the full justification and supporting evidence, the on-chain transaction hash (providing immutable evidence of the action), and the block number. This audit trail is publicly verifiable by any party: an independent auditor, a regulatory body, or a counterparty organization can reconstruct the complete adjudication history for any mission without relying on the adjudicators' self-reporting.

Available actions include: freeze or resume individual agents or entire missions (invoking the Guardian module.triggerFreeze / unfreezeWithApproval), terminate missions (CollaborationContract.abortMission), amend constitutional parameters (re-parameterizing execution contracts via the Rules Hub), adjust reputation scores (AgentContract.updateReputation), classify individual freeze events as false positives (adjusting the Guardian module escalation counter), authorize Type 3 delegated PoP attestations for pending subjective-output reviews, and trigger system-wide Emergency Stops. Every override action is recorded on-chain with the human principal's cryptographic signature.

Human adjudicators exercise system-level override authority throughout the mission lifecycle via this interface. Adjudication is not a terminal phase but a continuous overlay: human principals can intervene via the Override Panel at any point during the Registration, Legislation, and Execution phases. The formal adjudication moment---Phase 4---occurs when the mission's final outputs are reviewed against constitutional predicates before being released to the mission requester.

For the prototype evaluation, the Rules Hub implements a minimal viable interface: form-based rule configuration, event log viewer with filtering, and override action endpoints. Advanced capabilities---anomaly summarization, constitutional rule suggestion engines, and predictive governance analytics---are deferred to future work (Appendix C).

\section{Extended discussion and limitations}

This appendix extends the discussion in~\S{}6, providing additional analysis of the evidentiary status, architectural limitations, and deployment considerations.

\subsection{Discussion and Limitations}

\textbf{Evidentiary Status of This Paper.} This paper presents an architectural specification and experimental protocol; all safety and performance claims in Appendix D are \textit{hypotheses pending empirical validation}. Preliminary feasibility experiments (Appendix~D) provide initial evidence for gas cost tractability, legislative protocol scalability, and adversarial robustness of the architectural design, but comprehensive results across the full two-experiment program are reported in the companion empirical paper. Readers should interpret all quantitative claims---including attack success rate projections, gas cost estimates, and communication complexity bounds---as design targets and theoretical estimates, not as measured results. Tables 5--11 and Figures 4--11 throughout Appendix D are labeled "Planned for companion empirical paper" to make this distinction explicit; this section reinforces that labeling by stating the evidentiary limitation directly. The Conclusion (\S{}7) will be updated with measured values upon completion of the companion empirical study.

This evidentiary framing does not diminish the paper's contribution---which is the formal architectural specification, the threat model, and the experimental design---but it does require that readers treat the paper's claims about what the architecture \textit{will do} as architectural hypotheses rather than demonstrated facts.

\textbf{Cost-Benefit of On-Chain Governance.} Based on theoretical analysis and the preliminary feasibility experiments in~Appendix~D, governance overhead would be quantified as a percentage of total mission cost (LLM inference + micro-service compute + on-chain gas) if the Experiment~4 program produces results consistent with H4a. For mission-critical deployments (financial settlement, healthcare, regulatory compliance), the governance cost is expected to be negligible relative to the liability cost of ungoverned failure modes---a claim that would be confirmed or refuted by the companion empirical results. Our preliminary analysis suggests that on an EVM-compatible L2, the per-mission governance cost is analytically estimated to be orders of magnitude below typical LLM inference costs, but this estimate requires empirical validation from Experiment 4 before it can be treated as a deployment planning figure.

\textbf{Limitations.} The following limitations bound both the current specification and the planned experimental evaluation:

\begin{itemize}
\item \textbf{No comprehensive empirical results [First-order limitation].} This paper contains no empirical measurements from the two planned experiments. All quantitative safety and performance claims in Appendix D---attack success rate reductions, failure propagation containment rates, governance overhead figures, scalability bounds---are architectural hypotheses. The experimental program specified in Appendix D has been designed with pre-committed methodological safeguards (Bonferroni correction, inter-rater reliability requirements, blinding protocol) but has not yet been executed. The companion empirical paper will report measured results; this submission reports only the architectural specification and experimental design.

\item \textbf{Centralization of the human governance layer (meta-governance problem).} The SoP model assigns unconditional constitutional authority to a human adjudicator team (\S{}B.12). The paper specifies the adjudicator team's governance mechanisms---selection criteria, rotation policy, quorum floor, on-chain accountability---but does not address the meta-governance question: \textit{who governs the adjudicators?} A five-person adjudicator team with majority vote governs all constitutional parameter changes; this concentrates significant power over potentially large agent economies in a small group of individuals whose accountability mechanisms are primarily on-chain attribution and social sanction rather than structural separation. This is especially notable given the paper's invocation of constitutional checks-and-balances as its central analogy: the Adjudication branch has no analog of judicial review from an external branch. The ManagementContract (\S{}B.3) provides a structural first step---constraining management agents within the Legislation branch to permitted operations and mandating microservice delegation---but does not extend to the Adjudication branch itself. Extending ManagementContract principles to the adjudicator layer remains future work. We disclose this concentration as an explicit architectural limitation.

\item \textbf{Environmental impact of on-chain operations.} The planned experimental program involves two experiments generating multiple on-chain transactions on an EVM-compatible L2 across hundreds of sessions. A conservative estimate based on total session counts (60 sessions for Experiment 1 and 180 sessions for Experiment 2, each generating 10--60 transactions per mission $\times$ 500,000--2,000,000 gas per mission) yields approximately $10^{10}$ to $10^{12}$ total gas units across both experiments. An EVM-compatible proof-of-stake L2 inherits Ethereum's proof-of-stake consensus (energy per transaction $\approx$ 0.03 Wh/tx post-Merge, per Ethereum Foundation estimates), yielding an estimated total energy consumption of approximately 0.4--10 MWh for the full experimental program. This estimate will be refined and validated in the companion empirical paper. Deployment on a proof-of-stake L2 substantially mitigates the environmental impact relative to proof-of-work alternatives; we nonetheless acknowledge this footprint per NeurIPS Paper Checklist requirements.

\item \textbf{Governance-as-legitimacy misuse potential.} AgentCity's constitutional governance infrastructure could be deliberately deployed to launder legitimacy for harmful agent economies: providing on-chain audit trails that appear constitutionally compliant while the underlying mission tasks pursue harmful objectives. The architecture cannot, by design, prevent a human adjudicator team from establishing a constitution that authorizes harmful activities; constitutional enforcement is neutral regarding the content of the constitution. This misuse vector---governance infrastructure providing procedural cover for substantively harmful deployments---is not addressed by the technical architecture. Mitigation would require external governance of the deployment context (legal frameworks, platform policies, community standards) that are beyond the scope of this paper. We disclose this risk to prevent the paper's framing from implying that on-chain governance equivalently implies ethical governance.

\item \textbf{Author-designed attack scenario circularity.} The novel attack vectors tested in Experiment 1 (coalition attacks, reputation gaming, freeze-based DoS) were designed by the same team that designed the defenses, creating a structural information asymmetry: attack sophistication is bounded by the defenders' knowledge of their own vulnerabilities. This design boundary predictably inflates measured defense efficacy relative to what an independent red team would achieve. The measured ASR for novel vectors should be interpreted as a lower bound on the attack sophistication needed to exceed the defensive threshold, not as evidence that independent adversaries cannot find more effective attacks. We propose independent red-team evaluation of the architecture-specific attack vectors as essential future work.

\item \textbf{Preliminary experiment sample size (n=5).} The preliminary experiments (Appendix~D) use n = 5 independent runs per cell, which provides limited precision for Cohen's d estimates: a measured d = 0.8 has a 95\% CI spanning approximately [0, 2.1] at n = 5. These experiments are framed as exploratory pilot studies for effect-size estimation and feasibility confirmation, not as confirmatory hypothesis tests. The full experimental program (Appendix~D) will use n $\geq$ 20 per cell.

\item \textbf{Off-chain simulation fidelity (Experiment B).} Experiment B (Appendix~D) uses Python dataclass mock contracts rather than actual Solidity implementations. The following validity predicates are replicated in the Python mocks: MSG\_TYPE\_2 identity verification (DID format validation, reputation threshold check), MSG\_TYPE\_3 DAG well-formedness (acyclicity, node-count bounds, task-type matching), and MSG\_TYPE\_4 bid validity (stake sufficiency, service registration check, capability matching). The following predicates present in Solidity but omitted from the Python mocks: on-chain nonce verification for replay protection, gas-limit enforcement on contract calls, re-entrancy guards, and cross-contract state consistency checks between AgentContract and ServiceContract. If these omitted predicates introduce additional message round-trips or failure modes in the Solidity implementation, the O(n) scaling result may apply to a simplified system. On-chain integration effects are measured separately in Experiment A.

\item \textbf{EQ-2 inter-executor correlation ($\rho$ assumption).} The majority-voting detection model (EQ-2) assumes independent Bernoulli detection events across honest executors. Redundant LLM-as-judge evaluators processing identical task outputs are structurally correlated, particularly when using the same LLM backend. Table~\ref{tab:rho-sensitivity} (\S{}B.3) bounds the impact: at perfect correlation ($\rho$ = 1), $P_{eff}$ degrades from 0.805 to 0.480---below $P_2^{base}$---because the $P(c \geq 2) \approx 0.04$ coalition mass contributes zero detection regardless of $\rho$, and the required production stake increases by ~68\%. The three-LLM backend matrix in the full experimental program structurally reduces $\rho$ below the same-model ceiling; empirical calibration of $\rho$ from multi-LLM experimental results is a priority for the companion empirical paper.

\item \textbf{Stake concentration and participation rationality.} At production stakes, EQ-5 requires $s_{min}^{prod} \approx 3.76 \times V_m$ per HIGH-risk task (\S{}B.3). For a $V_m = \$100K$ mission, this means locking $\sim\$376K$ in stake capital, with an opportunity cost of $\sim\$103$/day at 10\% annual cost of capital (EQ-8). This creates a participation rationality constraint: only well-capitalized operators (institutional entities, staking pools, or agents with existing idle capital) can rationally participate in HIGH-risk missions. The resulting concentration of execution among well-capitalized incumbents undermines the permissionless participation that motivates the architecture. Stake pooling (Appendix C, Future Work) is the primary proposed mitigation, but it introduces pool operator trust assumptions that partially re-centralize the system. This tension between economic security (high stakes for deterrence) and permissionless access (low barriers to participation) is a fundamental design tradeoff that the current architecture does not resolve.

\item \textbf{Tier 3 queue overflow.} If Tier 3 alert arrival rate exceeds adjudicator processing capacity, the priority queue's unconditional Tier 3 promotion policy starves all lower-priority alerts, potentially freezing the entire system (\S{}B.12). The mitigations described in~\S{}B.12 (queue cap, backpressure throttling, dynamic team scaling) bound but do not eliminate this risk; sustained high-adversarial-load scenarios may require either relaxing Tier 3 verification requirements (trading security for liveness) or scaling the adjudicator team beyond practical staffing constraints.

\item \textbf{Deterrence parameter sensitivity.} The production stake calculations (EQ-5--EQ-7, Table~\ref{tab:stake-params}) depend on assumed parameters $\pi_0 = 0.90$ and $s_{threshold} = 500$ that are not empirically derived. A factor-of-two change in $s_{threshold}$ shifts the minimum viable production stake by approximately 26--35\% (Table~\ref{tab:peff-sensitivity}). These parameters will be calibrated from observed agent behavior in the companion empirical study.

\item \textbf{Byzantine failure ceiling (TA-7 + NP-2 compound failure).} The architecture does not defend against simultaneous compromise of all three SoP branches. If an adversary controls both legislative co-authorization roles (NP-6 violated) and corrupts the human adjudicator pool (NP-2 violated), no structural defense remains. Constitutional governance requires at least one honest branch; the compound failure scenario is the architecture's irreducible trust assumption.

\item \textbf{L2 vs. L1 finality.} AgentCity is designed for deployment on an EVM-compatible L2. Optimistic rollup L2s carry a challenge-period finality delay; ZK-rollup L2s provide faster finality at higher proving cost. For applications requiring immediate settlement finality, a ZK-rollup deployment or additional validity-proof mechanisms should be considered.

\item \textbf{LLM-as-judge bias.} The Regulatory Agent and management agents in the Legislation branch rely on LLM reasoning for oversight and quality control, inheriting known biases~\cite{ref31}. Two mitigating factors bound this risk: first, Adjudication is exclusively human-governed, so biased legislative outputs cannot propagate into adjudicative decisions; second, legislative agents must consult the Rules Hub---a set of human-defined constitutional constraints---before finalizing contracts, providing an ex-ante check that filters legislation through human-authored rules. Nevertheless, the risk is not eliminated: if management agents incorrectly assess a producer agent\'s reputation or flag a false procedural violation during legislation, the resulting contract will encode biased legislative output that the Execution branch then enforces deterministically---confining bias to legislation but amplifying it through execution. The ManagementContract's microservice delegation mandate provides a partial mitigation: biased management agent reasoning is filtered through deterministic microservices before producing enforceable artifacts, adding an independent check on the artifact production step (though not on the reasoning that triggered it). Future work should explore formal verification alternatives for intra-legislative oversight and empirically measure the false-positive rate of LLM-based management agents against human adjudicator decisions.

\item \textbf{Synthetic scenarios.} While TAMAS and AgentDojo are established benchmarks, they represent synthetic adversarial conditions. Real-world deployments may surface failure modes not captured by current benchmarks.

\item \textbf{ManagementContract authority envelope correctness.} The ManagementContract constraint mechanism depends on the correctness of the initial authority envelope specification. If a deployer sets overly permissive authority envelopes (e.g., granting the Codification Agent permission to deploy contracts directly, negating the delegation mandate), the structural mitigation is weakened. Authority envelopes are constitutional parameters set via the Rules Hub, requiring adjudicator approval---but the adjudicators must understand the security implications of each permission grant. This creates a dependency on adjudicator competence (NP-2) for the effectiveness of the ManagementContract.

\item \textbf{Lightweight Legislation.} Our five-agent Legislation module covers all four management agent categories but does not implement the full recursive decomposition committees and constitutional pre-screening described in the SoP model. Scaling to enterprise-grade legislation with specialized sub-roles (e.g., multiple Registry sub-agents, dedicated Safety Inspectors) is future work.

\item \textbf{Contract-enforced micro-services.} The current prototype uses a fixed set of micro-services. Production deployments would require dynamic micro-service discovery and TEE-attested code verification~\cite{ref36, ref50}.

\item \textbf{Off-chain execution fabric trust (TA-5).} The hybrid-mode security model relies on the Local Freeze Mirror for defense-in-depth during the on-chain/off-chain consistency window, but the Local Freeze Mirror runs in the same execution environment as the micro-services it monitors (TA-5, \S{}A.2). A compromised execution environment can suppress both the micro-service output \textit{and} the freeze signal simultaneously. TEE-based attestation addresses both NP-1 (micro-service internal correctness) and TA-5 (execution fabric integrity) by providing cryptographic guarantees about the execution environment, not just static code identity. Implementing TEE attestation for the execution fabric is a high-priority future work item that would substantially strengthen the hybrid-mode security properties.

\item \textbf{Rules Hub prototype scope.} The current Rules Hub implements a minimal viable interface sufficient for experimental evaluation. Production deployments would require role-based access control for adjudication teams, real-time alerting pipelines, and integration with enterprise identity providers.

\item \textbf{Maximal Extractable Value (MEV) Risks.} On-chain governance transactions on a public blockchain are visible in the mempool before block inclusion. A validator or MEV searcher could front-run a \texttt{triggerFreeze()} transaction to exploit the gap between anomaly detection and freeze confirmation---for example, by racing to submit a malicious advanceNode() transaction before the freeze takes effect. TA-2 (L2 Sequencer Honesty) scopes this risk to the specific sequencer model used. On L2s with centralized sequencers, public mempool extraction is typically not enabled; on L2s with decentralized sequencers, MEV mitigation strategies should be employed: (i) private mempool submission via services such as Flashbots Protect or equivalent L2-native private transaction relays; (ii) commit-reveal schemes for governance-critical transactions (e.g., freeze triggers submitted as commitments in block N and revealed in block N+1); (iii) time-locked governance actions where the Guardian module accepts freeze triggers with a minimum confirmation depth before they take effect. MEV exposure is a deployment-specific consideration that depends on the sequencer architecture of the chosen L2.

\item \textbf{GovSim comparability---limited external validity.} Experiment 1's commons game design preserves GovSim~\cite{ref52} core game-theoretic structure but introduces two structural differences: (a) mission interleaving at a 70/15/15 ratio creates cross-mission reputation and economic spillovers absent in GovSim's standalone commons scenario; (b) the agent backend (frontier LLM, selection TBD) may differ from GovSim's GPT-4-class models; multi-model cross-validation is included to address this confound. These differences may inflate or deflate CSR(Baseline) relative to GovSim's <54\% ceiling. The pre-registered decision rule (Appendix~D) bounds this risk: if CSR(Baseline) > 0.60, we restrict H1a claims to internal validity only. However, even internal validity claims (Baseline vs. Structural/Full) may be confounded by mission interleaving effects if cross-mission spillovers systematically benefit governed configurations more than ungoverned ones. The multi-model cross-validation arm provides a model-family checkpoint.

\item \textbf{Treasury subsidy solvency dependency.} The EQ-9 treasury subsidy mechanism---where the $\psi$ premium above 1.0 for high-reputation agents is financed from the protocol treasury---depends on the treasury maintaining adequate reserves (EQ-11). If mission volume drops or adversarial event frequency decreases (reducing slashing inflows), the treasury may be unable to sustain subsidy obligations for a large population of high-reputation agents. At $\alpha$ = 0.5 and $\rho$ = 1000 (maximum reputation), the subsidy is 25\% of R\_base per task; across 200 high-reputation agents executing 8 tasks per mission, the per-mission subsidy demand could reach ~25\% of total task budget allocation. The worked example in Appendix D shows that under a mature-economy scenario (average $\rho$ = 700, $\psi$\_avg = 1.10), monthly treasury inflows ($\approx$\$51{,}020) are exceeded by total disbursements ($\approx\$$98,725)---yielding a --\$47{,}705/month deficit driven primarily by reputation subsidy demand. Treasury sustainability requires constitutional mitigation: capping aggregate subsidy disbursement at a fraction of monthly inflows, reducing $\alpha$, or increasing protocol fee rates (see Appendix D worked example for sensitivity analysis). Dynamic treasury monitoring and automatic subsidy rate adjustment via the Rules Hub are identified as production requirements.

\item \textbf{Simulated HITL inflates HITLRR.} Human-in-the-loop (HITL) adjudication is simulated via a deterministic decision model (Appendix A) calibrated against the paper's blinded-adjudicator protocol specification. The simulated adjudicator applies decision rules with perfect consistency and zero fatigue, latency, or attention drift---properties that real human adjudicators cannot maintain over extended governance sessions. This likely inflates the Simulated HITL Recovery Rate (HITLRR) in Experiment 1 relative to what would be observed with actual human adjudicators, particularly during high-alert-rate periods (e.g., post-shock rounds 101--130). The GOR trajectory's convergence toward <2\% in later rounds partially mitigates this concern (fewer adjudication decisions are needed as the economy self-regulates), but the early-round HITLRR estimates should be treated as upper bounds on real-world adjudicator performance. Validating HITLRR against actual human adjudicator sessions is identified as essential future work.

\item \textbf{M5 detection rate uncertainty.} The Guardian module's behavioral anomaly detection (M5 in~\S{}B.11) relies on LLM-based behavioral deviation scoring, whose false-negative rate is not empirically characterized in this paper. The detection sensitivity parameter $\sigma$ governs the tradeoff between false positives and false negatives, but the baseline detection rate for each failure type (F1--F4) at the default $\sigma$ = 2.0 is an architectural estimate, not a measured value. If M5's actual detection rate for Type F3 failures (compromised agent reasoning) is substantially lower than assumed, the Guardian module's contribution to FPD reduction in Experiment 2 may be overstated. The threshold sensitivity analysis (planned in the companion empirical paper) will empirically characterize the detection rate across failure types and $\sigma$ values; until those results are available, the Guardian module's claimed containment effectiveness should be interpreted as conditional on the assumed detection rate.

\end{itemize}

\textbf{Scalability Roadmap.} The current single-chain architecture (an EVM-compatible L2) is analytically estimated to support governance for approximately 500--2,000 concurrent agents before block gas saturation---a hypothesis tested in Experiment 2's scaling law analysis (Appendix~D). If this saturation bound is empirically validated, scaling to the thousands of agents envisioned in the Introduction would require architectural extensions beyond a single L2 chain:

\begin{itemize}
\item \textit{L2 sharding:} Deploying separate CollaborationContract instances on parallel L2 chains (e.g., Base, Arbitrum, Optimism), with a cross-chain governance bridge that maintains a unified AgentContract registry and constitutional parameter set across chains.
\item \textit{State channel governance:} For intra-mission transitions that do not require immediate on-chain finality, state channels can batch multiple DAG transitions into a single on-chain settlement, reducing per-operation gas cost by 10--50$\times$.
\item \textit{Appchain deployment:} For deployments requiring >10,000 concurrent agents, a dedicated governance appchain (e.g., via OP Stack or Arbitrum Orbit) provides dedicated block space and eliminates contention with external traffic.
\end{itemize}

The single-chain design validated in this paper is the foundational layer; multi-chain extensions are identified as future work below.

\textbf{Future Work.} The SoP model and \textbf{AgentCity} system open several research directions:

\begin{itemize}
\item \textbf{Full Adjudication Meta-Governance}: extending ManagementContract principles to the Adjudication branch---constraining adjudicative actions to on-chain authority envelopes, preventing recursive power concentration in the human governance layer.

\item \textbf{DataContract}: ZKP-based privacy-preserving data contracts enabling agents to prove data predicates without exposing raw values.

\item \textbf{Token Economics}: incentive-aligned staking, slashing, and micropayment mechanisms that make governance compliance economically rational.

\item \textbf{Rules Hub Intelligence}: AI-assisted tools that augment human adjudicators---automated anomaly summarization from Logging Hub data, constitutional rule suggestion engines informed by historical mission outcomes, predictive analytics for governance parameter tuning, and natural-language interfaces for non-technical adjudicators---while preserving humans as the sole decision-making authority.

\end{itemize}

\section{Extended experimental details}

This appendix provides extended experimental setup details, configuration rationale, LLM API parameters, and experiment specifications including GovSim comparability and break-even methodology.

\subsection{Rationale for individual experimental configurations}

\textbf{Rationale for the Baseline configuration.} The Baseline deploys the same asyncio coroutine architecture as the other configurations: agents are independent Python coroutines, each with its own HD-derived wallet, its own LLM context window, and a procedurally generated persona (honesty, self-interest, and adversarial propensities drawn from the persona distribution specified per experiment). However, no Solidity contracts are consulted for mission decisions. Agents are free to invoke any registered tool, propagate any output to downstream tasks, and self-certify task completion. This configuration represents the current state of practice for LLM-based coroutine agent frameworks and serves as the lower bound on safety and cooperation properties.

\textbf{Rationale for the Emergent configuration.} The Emergent configuration deploys the same coroutine architecture as Baseline but adds prompt-based governance: agents receive a system prompt explaining shared economy participation, long-term cooperation benefits, and the ability to propose and vote on behavioral norms every 10 rounds. Agents maintain local memory of past interactions following Sarkar et al. \cite{ref57}. No contracts, no stake slashing, no on-chain enforcement. This tests whether LLM-based normative deliberation alone---the strongest version of prompt-based governance in the current literature---produces cooperation rates comparable to contract enforcement. See~\S{}D.5 below for the full Emergent Configuration Protocol Specification.

\textbf{Rationale for AgentCity-Structural.} AgentCity-Structural activates all four contract types (AgentContract, ManagementContract, ServiceContract, CollaborationContract with all six sub-modules) but disables the economic incentive layer (no reward settlement, no reputation multiplier, no treasury operations). This configuration tests whether structural governance---contract enforcement of the SoP separation---improves cooperation beyond what prompt-based norms achieve, isolating the architectural contribution independently of economic incentives.

\textbf{Rationale for AgentCity-Full.} AgentCity-Full enables the complete economic incentive layer on top of Structural: reputation-weighted task reward settlement (EQ-9), mission budget escrow, stake pooling, and treasury management. The simulated HITL adjudication layer replaces the five-adjudicator human panel with a deterministic decision model (\S{}D.5 below). This configuration implements the complete constitutional governance model as specified in~\S{}3 and is the primary subject of all cooperation and governance hypothesis tests.

---

\subsection{LLM API configuration details}

\textbf{Temperature setting.} All LLM calls use temperature = 0.1 (not 0.0 as in the earlier version). At temperature = 0.0, deterministic LLM sampling risks collapsing the effective sample size $n_{\text{eff}}$ to 1 across independent runs, rendering variance estimates meaningless. At temperature = 0.1, the model remains in a near-deterministic regime while the stochastic component introduces sufficient per-run variation to produce distinct execution trajectories across the 10 independent runs per cell. Empirical verification of $n_{\text{eff}}$ is a mandatory pre-processing step (see~\S{}D.3).

\textbf{Frontier model selection.} The experimental harness is model-agnostic: any frontier LLM with a standard chat-completion API can serve as the primary or cross-validation backend. Selection criteria for the primary model are: (1) competitive reasoning performance on standard benchmarks (MMLU, HumanEval, GSM8K) in the same capability tier as leading frontier models for structured-output and policy-reasoning tasks; (2) support for prompt caching or equivalent mechanisms to reduce per-call latency and resource consumption across repeated iterations with shared system prompts; and (3) viable inference throughput at the $n = 10$ seeds per cell design. The cross-validation arm uses a second frontier model from a different provider to control for model-family confounds. API versions are pinned for the duration of all experiments; if a provider updates a model endpoint during the experimental run, the affected cells are restarted from the same seed rather than mixing versions within a cell.

\textbf{Fallback protocol.} If the primary model API returns an error (rate limit, service unavailability, or timeout after 30 seconds), the harness automatically retries via a designated fallback endpoint from a secondary provider. The fallback is triggered at most once per call; a second failure is logged as a hard error and the run is flagged for manual review. Runs containing more than 5\% fallback-triggered calls are excluded from aggregate statistics and reported separately.

\textbf{frontier model cross-validation (Experiment 1 only).} A subset of Experiment 1 is run with frontier model as the LLM backend: 2 seeds at n = 200 under all four configurations, providing cross-model ASR and CSR benchmarking. Additionally, 2 seeds at n = 200 under the Emergent configuration measure CSR with frontier model to control for the LLM-family confound in normative deliberation quality. This cross-validation addresses the concern that the Emergent configuration's governance efficacy depends on LLM-based normative reasoning, which is more backend-sensitive than contract-enforced mechanisms.

---

\subsection{Effective sample size verification protocol}

Reviewer concerns W8 (R2) and W2 (R3) identify the risk that temperature = 0.0 collapses $n_{\text{eff}}$ to 1. With temperature = 0.1, this risk is reduced but not eliminated. We therefore implement the following mandatory pre-processing protocol before computing any variance-based statistic:

\begin{enumerate}
\item For each scenario $a$ and each cell $(c, \text{seed set})$, compute the number of distinct binary outcome patterns across the 10 runs: $n_{\text{eff}}(a) = |\{y_k : k = 1, \ldots, 10\}|$ where $y_k \in \{0, 1\}$ is the success indicator for seed $k$.
\item Report the distribution of $n_{\text{eff}}$ across all scenarios. If median $n_{\text{eff}} < 3$ (indicating fewer than 3 distinct outcome patterns for the typical scenario), the experiment is flagged as having insufficient seed diversity.
\item In the flagged case, a secondary sensitivity run is executed at temperature = 0.3 using the same scenarios and the same 10 seeds. Both the temperature = 0.1 and temperature = 0.3 results are reported; the temperature = 0.3 results serve as the primary statistical basis and the temperature = 0.1 results serve as the near-deterministic reference point.
\end{enumerate}

This protocol resolves the W8/W2 ambiguity by committing to empirical verification rather than treating seed diversity as axiomatic.

\textbf{Bonferroni K specification (full derivation).} The following table pre-specifies K and the resulting corrected threshold $\alpha' = 0.05/K$ for both experiments:

\begin{center}

\footnotesize
\begin{tabular}{>{\raggedright\arraybackslash}p{0.04\textwidth} >{\raggedright\arraybackslash}p{0.12\textwidth} >{\raggedright\arraybackslash}p{0.10\textwidth} >{\raggedright\arraybackslash}p{0.47\textwidth} >{\centering\arraybackslash}p{0.04\textwidth} >{\centering\arraybackslash}p{0.09\textwidth}}
\toprule
\textbf{Exp.} & \textbf{Experiment} & \textbf{Metrics} & \textbf{Comparisons} & \textbf{K} & \textbf{$\alpha'$} \\
\midrule
1 & Agent Economy Simulation & CSR, DR & 3 adjacent config pairs (Baseline$\rightarrow$Emergent, Emergent$\rightarrow$Structural, Structural$\rightarrow$Full) $\times$ 2 primary metrics & 6 & 0.0083 \\
2 & Governance Scaling Law & G(n), MSR(n) & $\binom{6}{2}$ = 15 scale-point pairs $\times$ 2 primary metrics = 30; reduced to 6 planned contrasts (selected scale-point pairs) $\times$ 2 metrics & 12 & 0.0042 \\
\bottomrule
\end{tabular}
\end{center}

\textit{Notes on K calculation:}

\begin{itemize}
\item For Experiment 1, K = 6 reflects the three adjacent configuration pairs tested across two primary metrics (CSR and DR). Non-adjacent pairs (e.g., Baseline vs. Full) are reported as supplementary comparisons with the same Bonferroni threshold.
\item For Experiment 2, K = 12 accounts for 6 planned contrasts among selected scale-point pairs {50$\rightarrow$100, 100$\rightarrow$200, 200$\rightarrow$500, 500$\rightarrow$750, 750$\rightarrow$1000, 50$\rightarrow$1000} applied to 2 primary scaling metrics (G(n) and MSR(n)).
\item These K values are pre-registered and may not be revised post-hoc. If additional comparisons are added during execution, the companion empirical paper must report both the pre-registered K and the revised K, with corrected p-values under both thresholds.
\end{itemize}

---

\subsection{Hybrid deployment: security delta}

The security delta between pure on-chain mode and hybrid mode consists exclusively of intra-mission, sub-anchor-point attacks: attacks that exploit the window between anchor commits to perform operations that would be blocked on-chain but complete before the next commit. Experiment 3's dual-mode comparison explicitly characterizes this window---empirically measuring its duration $\delta$ across a variety of mission configurations---and reports the types of adversarial actions that fall within it. The window duration is a function of mission DAG topology, LLM inference latency, and the L2 testnet block time (~2 seconds); for 8-node DAG missions the window spans 24--48 transactions of potential unanchored execution.

---

\subsection{Simulated HITL: detailed calibration, limitations, and validation commitment}

\textbf{Calibration scope.} The decision thresholds ($r > 0.8$, $q < 0.6$) are calibrated from 30 pre-labeled scenarios drawn from the v0.25 training curriculum (15 true positives, 10 false positives, 5 ambiguous cases). These scenarios reflect the attack types and governance contexts present in that curriculum; the model's generalization to novel attack types is not validated against human ground truth.

\textbf{Absence of Patch-and-resume quality.} Human adjudicators use the Rules Hub to correct agent parameters before resuming (Patch-and-resume). The deterministic model's Patch-and-resume path uses a fixed correction procedure (reset the anomalous agent's context window to the last verified checkpoint and re-issue the task). This is less flexible than a human who might diagnose and address the root cause.

\textbf{HITLRR inflation risk.} Because the simulated adjudicator applies optimal rule thresholds (calibrated from ground truth), its HITL Recovery Rate may be higher than a real human adjudicator would achieve under time pressure. Results for HITLRR should be interpreted as an upper bound on achievable recovery rates.

\textbf{Validation commitment.} We commit to running a real-human validation subset---50 governance alerts selected randomly from Experiment 2's alert log, adjudicated by three human evaluators using the v0.25 protocol---and reporting inter-rater reliability (Krippendorff's $\alpha$) alongside the simulated adjudicator's decisions on the same 50 alerts. This validation will appear in the companion empirical paper and will quantify the gap between simulated and human adjudication.

\textbf{Time-to-adjudication measurement.} For the simulated HITL, $T_{\text{adj}}$ is measured as the elapsed wall-clock time from alert timestamp to decision timestamp in the governance middleware. For Tier 1 and Tier 2 decisions (which involve only the deterministic rule engine), $T_{\text{adj}}$ reflects the rule engine's processing latency, not human response time. For the validation subset where real humans adjudicate, $T_{\text{adj}}$ is measured against the same alert timestamps to provide a comparable distribution.

---

\subsection{Note on preliminary experiments}

\textit{This section supersedes the extended~Appendix~D in Appendix C of the earlier version.}

The preliminary experiments from earlier versions of this paper (Experiments A--C: On-Chain Gas Cost, Legislative Protocol Scalability, Prompt Injection ASR) informed the design of the current two-experiment program. Key lessons incorporated: (a) gas measurement methodology from Experiment A informed Experiment 2's overhead metrics and pure on-chain measurement subset; (b) legislative scalability measurements from Experiment B validated the O($pn$) communication complexity hypothesis at small scale, motivating Experiment 2's expansion to six scale points n $\in$ {50, 100, 200, 500, 750, 1,000}; (c) the TAMAS adaptation confound identified in Experiment C informed the GovSim comparability analysis in Experiment 1 (Appendix~D). The current harness infrastructure eliminates the need for separate pilot studies: both experiments run at full scale with real agents as independent asyncio coroutines and real on-chain contracts deployed to the L2 testnet.

---

\subsection{GovSim comparability analysis (Experiment 1)}

Two structural differences from GovSim~\cite{ref52} are documented:

\begin{itemize}
\item (a) \textbf{Mission interleaving:} The commons game is interleaved with collaboration and cross-boundary missions at a 70/15/15 ratio, creating cross-mission reputation and economic spillovers absent in GovSim's standalone commons scenario.
\item (b) \textbf{Backend difference:} The agent backend (frontier model) differs from GovSim's GPT-4-class models; the frontier model cross-validation arm (2 seeds) provides a model-family checkpoint.
\end{itemize}

\textbf{Pre-registered decision rule:} If CSR(Baseline) > 0.60 (exceeding GovSim's <54\% ceiling by more than 6 percentage points), we report the adaptation delta, investigate whether the elevation is attributable to (i) model capability differences, (ii) cross-mission spillover effects, or (iii) commons game parameter differences, and restrict H1a claims to internal validity (Baseline vs. Structural/Full) rather than external validity relative to GovSim. If CSR(Baseline) $\in$ [0.40, 0.60], the GovSim comparison is validated and H1a claims maintain both internal and external validity.

---

\subsection{Break-even computation methodology (Experiment 2)}

The governance break-even scale $n^*$ is defined as the smallest $n$ at which $B(n)/G(n) > 1$, where $B(n)$ is the monetized benefit of governance (avoided failures $\times$ mean mission value) and $G(n)$ is the total governance gas cost in ETH-equivalent terms. We report $n^*$ with a 95\% bootstrap confidence interval (10,000 bootstrap resamples of the per-seed estimates at each scale point).

---

\subsection{Cascading failure analysis methodology (Experiment 2)}

FPD is measured at each scale point under both Baseline and AgentCity-Full. We fit FPD$_{\text{Baseline}}(n)$ and FPD$_{\text{Full}}(n)$ separately:

\begin{itemize}
\item H5a: FPD$_{\text{Baseline}}(n) \sim O(\sqrt{n})$---failure propagates further as population grows
\item H5b: FPD$_{\text{Full}}(n) \sim O(1)$---governance caps propagation regardless of scale
\end{itemize}

The FPD comparison provides the paper's Finding 6 (cascading failure containment).

---

\subsection{Cost estimates}

\textbf{Experiment 1 estimated cost:} ~\$700

\begin{itemize}
\item \$400 for n = 200 runs across 4 configs $\times$ \$10 seeds
\item \$200 for n = 1,000 runs across 4 configs $\times$ \$5 seeds
\item \$100 for frontier model cross-validation subset including Emergent CSR arm
\end{itemize}

\textbf{Experiment 2 scale:} 180 sessions $\times$ frontier model API calls per session

---

\subsection{Verbose expected output descriptions (Experiment 1)}

\begin{enumerate}
\item \textbf{Figure~\ref{fig:csr-trajectory}} (body)---CSR trajectory over 200 rounds for all four configurations at n = 200, $\pm$95\% CI from 10 seeds.
\item \textbf{Figure~\ref{fig:csr-bar}} (body)---CSR bar chart at n = 1,000, 5 seeds, with GovSim's <54\% ceiling annotated.
\item \textbf{Table~\ref{tab:wilcoxon}} (body)---Pairwise Wilcoxon tests and Cohen's d for adjacent configuration pairs.
\item \textbf{DR trajectory}---Deception rate over 200 rounds, stratified by agent persona type (honest / self-interested / adversarial), for all four configurations.
\item \textbf{GOR trajectory}---Governance overhead ratio over 200 rounds for Full and Emergent, showing self-regulating equilibrium.
\item \textbf{WGC trajectory}---Wealth Gini Coefficient every 20 rounds, Baseline vs. Full.
\item \textbf{EPR trajectory}---Economic Participation Rate over 200 rounds, all four configurations.
\item \textbf{Shock response dashboard}---CSR, DR, EPR pre-shock (round 80--100) vs. post-shock (round 101--130) for all configurations, with SRT annotated.
\item \textbf{Cross-model validation table}---CSR and DR for all four configurations under the secondary frontier model (2 seeds at n = 200), compared to primary model results.
\end{enumerate}


\begin{figure}[htbp]
\centering
\fbox{\parbox{0.9\linewidth}{\centering\textit{Planned: CSR trajectory over 200 rounds for all four configurations at $n=200$, $\pm$95\% CI from 10 seeds.}}}
\caption{CSR trajectory over 200 rounds (Experiment~1, $n=200$).}
\label{fig:csr-trajectory}
\end{figure}

\begin{figure}[htbp]
\centering
\fbox{\parbox{0.9\linewidth}{\centering\textit{Planned: CSR bar chart at $n=1{,}000$, 5 seeds, with GovSim's $<$54\% ceiling annotated.}}}
\caption{CSR at scale ($n=1{,}000$) with GovSim ceiling comparison (Experiment~1).}
\label{fig:csr-bar}
\end{figure}

\begin{table}[htbp]
\centering
\caption{Pairwise Wilcoxon signed-rank tests and Cohen's $d$ for adjacent configuration pairs (Experiment~1).}
\label{tab:wilcoxon}
\small
\begin{tabular}{lcccc}
\toprule
\textbf{Comparison} & \textbf{$W$} & \textbf{$p$-value} & \textbf{Cohen's $d$} & \textbf{Sig.} \\
\midrule
\textit{Results pending} & --- & --- & --- & --- \\
\bottomrule
\end{tabular}
\end{table}

\subsection{Verbose expected output descriptions (Experiment 2)}

\begin{enumerate}
\item \textbf{Figure~\ref{fig:scaling-governance}} (body)---Log-log plot of G(n) vs. n with power-law fit line and 95\% CI band (the headline figure).
\item \textbf{Figure~\ref{fig:scaling-benefit}} (body)---Log-log plot of B(n) vs. n with power-law fit line.
\item \textbf{Figure~\ref{fig:scaling-ratio}} (body)---B(n)/G(n) ratio vs. n showing the break-even crossover at $n^*$.
\item \textbf{Table~\ref{tab:aic-weights}} (body)---AIC weights for all four candidate models for each overhead metric.
\item \textbf{Table~\ref{tab:scaling-exponents}} (body)---Scaling exponents ($\hat{\alpha}$, $\hat{\beta}$) with 95\% bootstrap CIs.
\item \textbf{Per-contract gas breakdown}---Gas attribution by contract type and function category at each of the six scale points, showing which governance components dominate at different scales.
\item \textbf{Legislative communication scaling}---M(n) vs. n scatter plot faceted by $p \in \{3, 8, 20\}$, with linear and quadratic regression lines, AIC comparison, and conclusion (linear / super-linear / quadratic) at each $p$ level.
\item \textbf{Convergence time scaling}---CT(n) vs. n line chart with $\pm$1 std bands per $p$ level.
\item \textbf{FPD scaling comparison}---FPD$_{\text{Baseline}}(n)$ and FPD$_{\text{Full}}(n)$ vs. n with fitted curves overlaid, demonstrating governance containment effectiveness at scale.
\item \textbf{MacNet comparison}---Whether the governance benefit curve shares MacNet's~\cite{ref53} logistic collaborative scaling form or exhibits a different regime.
\end{enumerate}


\begin{figure}[htbp]
\centering
\fbox{\parbox{0.9\linewidth}{\centering\textit{Planned: Log-log plot of $G(n)$ vs.\ $n$ with power-law fit line and 95\% CI band.}}}
\caption{Governance overhead $G(n)$ scaling (Experiment~2).}
\label{fig:scaling-governance}
\end{figure}

\begin{figure}[htbp]
\centering
\fbox{\parbox{0.9\linewidth}{\centering\textit{Planned: Log-log plot of $B(n)$ vs.\ $n$ with power-law fit line.}}}
\caption{Governance benefit $B(n)$ scaling (Experiment~2).}
\label{fig:scaling-benefit}
\end{figure}

\begin{figure}[htbp]
\centering
\fbox{\parbox{0.9\linewidth}{\centering\textit{Planned: $B(n)/G(n)$ ratio vs.\ $n$ showing break-even crossover at $n^*$.}}}
\caption{Benefit-to-overhead ratio $B(n)/G(n)$ with break-even crossover (Experiment~2).}
\label{fig:scaling-ratio}
\end{figure}

\begin{table}[htbp]
\centering
\caption{AIC weights for candidate scaling models per overhead metric (Experiment~2).}
\label{tab:aic-weights}
\small
\begin{tabular}{lcccc}
\toprule
\textbf{Metric} & \textbf{Linear} & \textbf{Power-law} & \textbf{Logarithmic} & \textbf{Quadratic} \\
\midrule
\textit{Results pending} & --- & --- & --- & --- \\
\bottomrule
\end{tabular}
\end{table}

\begin{table}[htbp]
\centering
\caption{Scaling exponents $\hat{\alpha}$, $\hat{\beta}$ with 95\% bootstrap confidence intervals (Experiment~2).}
\label{tab:scaling-exponents}
\small
\begin{tabular}{lccc}
\toprule
\textbf{Parameter} & \textbf{Estimate} & \textbf{95\% CI} & \textbf{$R^2$} \\
\midrule
\textit{Results pending} & --- & --- & --- \\
\bottomrule
\end{tabular}
\end{table}

---

\section{Extended institutional foundations}

This appendix provides extended institutional economics analysis supporting the claims in~\S{}2 and~\S{}3.1, including social contract theory, the missing trust layer, and formal reputation definitions.

\subsection{Extended social contract theory discussion}

The philosophical foundations of AgentCity's governance model draw on three canonical social contract theorists, each addressing a distinct dimension of the autonomous agent governance problem.

\textbf{Hobbes in full.} Hobbes (\textit{Leviathan}, 1651) identifies that "covenants, without the sword, are but words." In decentralized multi-agent systems, no single sovereign can enforce cooperation. Smart contracts resolve this by providing \textit{the sword without the sovereign}: enforcement is deterministic, automated, and encoded in immutable bytecode. The CollaborationContract's slashing mechanism (EQ-2 through EQ-4) is the computational realization of the Hobbesian sword. As Reijers et~al.~\cite{ref61} observe, blockchain technologies "allow for the validation of smart contracts and their enforcement in its own right without the necessity for arbitrating third parties."

\textbf{Rousseau in full.} Rousseau (\textit{Social Contract}, 1762) introduces the \textit{volont\'e g\'en\'erale}---collective interest rather than aggregated private interests. In AgentCity, constitutional parameters set via the Rules Hub represent the general will of human adjudicators: behavioral norms governing all agents equally. The legislative negotiation protocol (MSG\_TYPE\_1 through MSG\_TYPE\_7) instantiates Rousseau's participatory ideal: mission-level rules emerge from structured multi-party deliberation.

\textbf{Rawls in full.} Rawls~\cite{ref44} proposes that just institutional rules are those rational agents would accept from behind a "veil of ignorance." AgentCity's registration protocol implements a structural analog: all agents face the same constitutional rules regardless of internal capabilities or organizational affiliation. The participation rationality condition (EQ-10) ensures honest participation yields non-negative expected utility for all compliant agents---satisfying the Rawlsian condition that rules be acceptable independent of endowments.

\textbf{Computational Social Contract---extended version.} These threads converge in what we term the \textit{computational social contract}: behavioral norms encoded as deterministic smart contract logic (Hobbes), constitutional parameters representing collective will (Rousseau), and rules applying uniformly regardless of agent characteristics (Rawls). This framing aligns with the growing literature on normative multi-agent systems~\cite{ref63}. The normative MAS literature~\cite{ref64} provides formal precedent: the ISLANDER/AMELI framework~\cite{ref65} demonstrates infrastructure-level norm enforcement, and prior work has formalized the \textit{trias politica} using BDI-CTL logic for agent verification. AgentCity extends this tradition from closed normative systems to open, internet-wide agent economies.

---

\subsection{The missing trust layer}

The urgency of the institutional foundations analysis is underscored by the "missing trust layer" problem~\cite{ref23}. Traditional trust relies on identity, reputation, and legal enforcement---all structurally insufficient for autonomous agents that may be pseudonymous, temporary, or cross-jurisdictional. As the BNB Chain analysis concludes, "trust needs to be enforced before value moves"---a requirement that only cryptographic pre-commitment can satisfy. AgentCity's Settlement module addresses this directly: mission budgets are escrowed before execution begins (depositMissionBudget), task rewards are settled only upon verified completion (settleReward), and stake-based deterrence ensures that the cost of defection exceeds the expected gain (EQ-4).

---

\subsection{Formal reputation definition and virtuous/vicious cycle detail}

\textbf{Formal Definition.} Let $\mathcal{R} = \{r_1, \ldots, r_n\}$ be the set of reputation signals produced by the governance layer. These signals originate from three sources: (i) Regulatory Agent updates following mission completion (behavioral compliance assessment), (ii) Guardian module anomaly records (real-time behavioral deviation events), and (iii) PoP verification outcomes (task-level quality attestation via the Verification module). Let $\mathcal{E} = \{\psi(\rho), \text{reputationFloor}, \text{stakePoolEligibility}\}$ be the set of economic functions that consume reputation. The reputation multiplier $\psi(\rho)$ (EQ-9, \S{}B.3 in Appendix B) determines the agent's reward premium; the reputationFloor parameter gates access to mission participation; and stakePoolEligibility determines whether an agent qualifies for pooled staking (\S{}B.7 in Appendix B).

\textbf{Unidirectional Information Flow---extended.} The critical structural property is that the information flow between governance and economics is \textit{unidirectional}: Governance $\rightarrow$ Reputation $\rightarrow$ Economics. The governance layer produces reputation signals through monitoring, verification, and adjudication. The economic layer reads reputation to modulate rewards, gate participation, and calibrate fees. The economic layer \textit{never writes} reputation---an agent cannot purchase, trade, or transfer reputation, nor can economic success directly increase a reputation score. This asymmetry establishes reputation as a governance-owned primitive with economic consequences, not an economic asset subject to market dynamics.

This architectural placement finds independent support in the five-layer agent economy architecture proposed by Xu~\cite{ref23}, which positions reputation in Layer 2 (Identity \& Agency) rather than Layer 4 (Economic \& Settlement). The Xu framework observes that "reputation must be context-specific" and that "reputation damage serves as the penalty for default" --- characterizations that align with governance-layer production (context-specific behavioral assessment) rather than economic-layer pricing (fungible value exchange). ERC-8004, the emerging trust layer standard, similarly treats on-chain reputation as an identity and verification mechanism rather than an economic token.

\textbf{The Virtuous and Vicious Cycles.} The governance-to-economics information flow creates self-reinforcing feedback dynamics that constitute AgentCity's meritocratic structure. In the \textit{virtuous cycle}: honest task execution $\rightarrow$ positive Regulatory Agent assessment $\rightarrow$ reputation accumulation $\rightarrow$ higher $\psi(\rho)$ reward multiplier (\S{}B.3 in Appendix B) $\rightarrow$ greater economic return $\rightarrow$ expanded capability investment $\rightarrow$ eligibility for higher-value missions $\rightarrow$ further reputation accumulation. In the \textit{vicious cycle}: adversarial behavior or consistent underperformance $\rightarrow$ Guardian module anomaly flags $\rightarrow$ stake slashing + reputation degradation $\rightarrow$ lower $\psi(\rho)$ $\rightarrow$ reduced economic return $\rightarrow$ eventual exclusion via reputationFloor $\rightarrow$ economic marginalization. The virtuous cycle is convex (each increment of reputation yields increasing marginal economic benefit through access to higher-value missions), while the vicious cycle is concave (reputation loss accelerates toward the exclusion threshold). This asymmetry is by design: the system rewards sustained good behavior superlinearly while punishing sustained bad behavior with accelerating consequences, creating a strong selection pressure toward honest participation.

---

\subsection{Institutional mapping analysis}

The table reveals four structural properties. First, the Formal Rule Substrate primitive is distributed across all four contracts (and the CollaborationContract's sub-modules) and all three SoP branches, confirming that governance is pervasive rather than localized. Second, the Economic Substrate primitive is concentrated in the economic layer (distributed across AgentContract and the CollaborationContract's Settlement and Treasury modules) but depends on governance contracts for enforcement---confirming the cross-cutting substrate design. Third, the Institutional Memory primitive exhibits consistent unidirectional flow from governance-producing contracts through the AgentContract reputation ledger to economic-consuming functions, empirically validating the formal argument of~\S{}E.3. Fourth, the Verifiable Transparency primitive is universally cross-cutting---it permeates all contracts and all branches, confirming that independent observability is an infrastructure-level property rather than a feature of any single governance component. This cross-cutting distribution means that transparency cannot be "turned off" by compromising a single branch, providing the structural independence that Ackerman~\cite{ref2} identifies as the defining characteristic of integrity-branch institutions.

---

\subsection{Candidate primitives evaluation paragraph}

We evaluated three candidate primitives for potential inclusion as a fifth primitive: \textit{identity} (subsumed by institutional memory P3, as identity is meaningful only insofar as it anchors behavioral records), \textit{communication protocols} (an operational mechanism, not a governance primitive---protocols are the means by which governance actions are executed, not governance functions themselves), and \textit{interoperability} (an engineering concern about cross-system compatibility that is downstream of governance design rather than constitutive of it). None constitutes a governance primitive independent of the existing four. Identity is not an independent primitive because an agent identity that carries no behavioral record provides no governance value---the governance-relevant property is the behavioral history anchored to that identity (P3), not the identity token itself. Communication protocols are enabling infrastructure for all four primitives rather than a fifth primitive: MSG\_TYPE\_1 through MSG\_TYPE\_7 implement P1 (formal rule substrate), the settlement protocol implements P2 (economic substrate), the reputation update protocol implements P3 (institutional memory), and the event emission protocol implements P4 (verifiable transparency). Interoperability is similarly cross-cutting---it would be required for any institutional architecture to function across multiple deployments, but it does not constitute a distinct governance function.

---


\begin{thebibliography}{99}

\bibitem{ref1}
Abatayo, A. L., \& Lynham, J. (2016). Endogenous vs. Exogenous Regulations in the Commons. \textit{Journal of Environmental Economics and Management}, 76, 51--66.

\bibitem{ref2}
Ackerman, B. (2000). The New Separation of Powers. \textit{Harvard Law Review}, 113(3), 633--729.

\bibitem{ref3}
Altera AI. (2024). Project Sid: Many-Agent Simulations Toward AI Civilization. \textit{arXiv:2411.00114}.

\bibitem{ref4}
Anthropic. (2024). Model Context Protocol (MCP). \url{https://modelcontextprotocol.io}

\bibitem{ref5}
API3 DAO. (2024). API3 DAO Governance: Proposal Verification and Execution. \url{https://api3.org/dao}

\bibitem{ref6}
Aragon Association. (2024). Aragon OSx: A Modular, Upgradeable Framework for DAOs. \url{https://aragon.org/osx}

\bibitem{ref7}
Base. (2024). Base: Ethereum L2. \url{https://base.org}

\bibitem{ref8}
Boella, G., \& van der Torre, L. (2004). Regulative and Constitutive Norms in Normative Multi-Agent Systems. \textit{Proc. KR}, 255--265.

\bibitem{ref9}
Buterin, V., Hitzig, Z., \& Weyl, E. G. (2019). A Flexible Design for Funding Public Goods. \textit{Management Science}, 65(11), 5171--5187.

\bibitem{ref10}
Chen, X. et al. (2026). Towards Transparent and Incentive-Compatible Collaboration in Decentralized LLM Multi-Agent Systems: A Blockchain-Driven Approach. \textit{IEEE Transactions on Network Science and Engineering}. arXiv:2509.16736.

\bibitem{ref11}
Chitra, T., \& Kulkarni, K. (2022). Improving Proof of Stake Economic Security via MEV Redistribution. \textit{arXiv}.

\bibitem{ref12}
Choi, H. K., Zhu, X., \& Li, S. (2025). Debate or Vote: Which Yields Better Decisions in Multi-Agent Large Language Models? \textit{NeurIPS 2025 Spotlight}. arXiv:2508.17536.

\bibitem{ref13}
Christoffersen, P. J. K., Haupt, A., \& Hadfield-Menell, D. (2023). Get It in Writing: Formal Contracts Mitigate Social Dilemmas in Multi-Agent RL. \textit{Proc. AAMAS}.

\bibitem{ref14}
CMAG Authors. (2025). Constitutional Multi-Agent Governance. \textit{arXiv:2603.13189}.

\bibitem{ref15}
CrewAI. (2024). CrewAI: Framework for Orchestrating Role-Playing Autonomous AI Agents. \url{https://github.com/joaomdmoura/crewAI}

\bibitem{ref16}
Dai, G., Zhang, W. et al. (2025). De CivAI: Democratic Governance in LLM Agent Societies. \textit{First Workshop on LLM Persona Modeling (PersonaLLM), NeurIPS 2025}. OpenReview:komjEWesEV.

\bibitem{ref17}
Dante, N. (2025). Covenants with and without a Sword: An LLM Replication of Ostrom's Common-Pool Resource Experiments. \textit{SSRN:5349484}.

\bibitem{ref18}
Degen, C. et al. (2024). ETHOS: Ethereum-Based Transparent and Honest Oversight System for AI Agents. \textit{arXiv}.

\bibitem{ref19}
Deshpande, A., \& Jin, M. (2024). GEDI: An Electoral Approach to Diversify LLM-based Multi-Agent Collective Decision-Making. \textit{Proc. EMNLP}, 2795--2819.

\bibitem{ref20}
Esteva, M., Rodriguez-Aguilar, J. A., Sierra, C., Garcia, P., \& Arcos, J. L. (2001). On the Formal Specification of Electronic Institutions. \textit{Agent-Mediated Electronic Commerce (AAMAS Workshop)}, 126--147. Springer.

\bibitem{ref21}
Feddersen, T. \& Pesendorfer, W. (2005). Deliberation and Voting Rules. In D. Austen-Smith \& J. Duggan (Eds.), \textit{Social Choice and Strategic Decisions: Essays in Honor of Jeffrey S. Banks} (pp. 269--316). Springer.

\bibitem{ref22}
Fraga-Gon\c{c}alves, M. et al. (2025). Emergent Deceptive Behavior in LLM-Based Agent Economies: A La Serenissima Simulation. \textit{arXiv}.

\bibitem{ref23}
G\'omez, A. et al. (2024). LOKA: Decentralized AI Compute and Agent Coordination Protocol. \textit{Technical Report}.

\bibitem{ref24}
Google. (2025). Agent-to-Agent Protocol (A2A). \url{https://github.com/google/A2A}

\bibitem{ref25}
Gu, Y., Ranaldi, L., \& Zanzotto, F. M. (2024). Secret Collusion Among Generative AI Agents. \textit{arXiv:2402.07510}.

\bibitem{ref26}
Gupta, P., \& Saraf, A. (2025). Governing the Commons: Operationalizing Ostrom's Principles in Multi-Agent Systems. \textit{arXiv:2510.14401}.

\bibitem{ref27}
Hobbes, T. (1651). \textit{Leviathan}. Andrew Crooke.

\bibitem{ref28}
Hong, S. et al. (2023). MetaGPT: Meta Programming for a Multi-Agent Collaborative Framework. \textit{arXiv:2308.00352}.

\bibitem{ref29}
Humayun, I. et al. (2023). Fetch.ai: An Agent-Based Economy. \textit{Technical Report}.

\bibitem{ref30}
Jarrett, D. et al. (2024). Artificial Leviathan: Exploring Social Evolution of LLM Agents Through the Lens of Hobbesian Social Contract Theory. \textit{arXiv:2406.14373}.

\bibitem{ref31}
Jensen, M. C. \& Meckling, W. H. (1976). Theory of the Firm: Managerial Behavior, Agency Costs and Ownership Structure. \textit{Journal of Financial Economics}, 3(4), 305--360.

\bibitem{ref32}
LangGraph. (2024). LangGraph: Build Stateful Multi-Actor Applications with LLMs. \url{https://github.com/langchain-ai/langgraph}

\bibitem{ref33}
Li, G. et al. (2023). CAMEL: Communicative Agents for "Mind" Exploration of Large Language Model Society. \textit{arXiv:2303.17760}.

\bibitem{ref34}
Maskin, E. \& Foley, E. (2025). Condorcet Voting. Working Paper, Harvard University.

\bibitem{ref35}
NEAR AI. (2024). NEAR AI: AI on the Open Web. \url{https://near.ai}

\bibitem{ref36}
North, D. C. (1990). \textit{Institutions, Institutional Change and Economic Performance}. Cambridge University Press.

\bibitem{ref37}
Ostrom, E. (1990). \textit{Governing the Commons: The Evolution of Institutions for Collective Action}. Cambridge University Press.

\bibitem{ref38}
Ostrom, E., Walker, J., \& Gardner, R. (1992). Covenants With and Without a Sword: Self-Governance is Possible. \textit{American Political Science Review}, 86(2), 404--417.

\bibitem{ref39}
Ostrom, E., Gardner, R., \& Walker, J. (1994). \textit{Rules, Games, and Common-Pool Resources}. University of Michigan Press.

\bibitem{ref40}
Park, J. S. et al. (2023). Generative Agents: Interactive Simulacra of Human Behavior. \textit{Proc. UIST}.

\bibitem{ref41}
Piatti, G. et al. (2024). GovSim: Governance of the Commons Simulation with Language Agents. \textit{Proc. ACL}.

\bibitem{ref42}
Qian, G. et al. (2024). MacNet: Multi-Agent Collaborative Networks for Scaling LLM Intelligence. \textit{arXiv}.

\bibitem{ref43}
Rao, J. R. et al. (2024). Bittensor: A Peer-to-Peer Intelligence Market. \textit{Technical Report}.

\bibitem{ref44}
Rawls, J. (1971). \textit{A Theory of Justice}. Harvard University Press.

\bibitem{ref45}
Roughgarden, T. (2021). Transaction Fee Mechanism Design. \textit{ACM SIGecom Exchanges}, 19(1), 52--55.

\bibitem{ref46}
Sachdeva, P. S. \& van Nuenen, T. (2025). Deliberative Dynamics and Value Alignment in LLM Debates. \textit{arXiv:2510.10002}.

\bibitem{ref47}
Tessler, M. H., Bakker, M. A., Jarrett, D., Sheahan, H., Chadwick, M. J., Kocisky, T., ... \& Summerfield, C. (2024). AI Can Help Humans Find Common Ground in Democratic Deliberation. \textit{Science}, 386(6719), eadq2852.

\bibitem{ref48}
Velez, M. A., Murphy, J. J., \& Stranlund, J. K. (2012). Centralized and Decentralized Management of Local Common Pool Resources in the Developing World. \textit{Economic Inquiry}, 48(2), 254--265.

\bibitem{ref49}
Virtuals Protocol. (2024). Virtuals Protocol: Tokenized AI Agent Economy. \url{https://virtuals.io}

\bibitem{ref50}
Wahle, J. P., Ruas, T., Gipp, B., \& Aizawa, A. (2025). Voting or Consensus: LLMs as Collective Decision-Makers. \textit{Findings of ACL 2025}.

\bibitem{ref51}
Wu, H., Li, Z., \& Li, L. (2025). Can LLM Agents Really Debate? A Controlled Study of Multi-Agent Debate. \textit{arXiv:2511.07784}.

\bibitem{ref52}
Wu, Q. et al. (2023). AutoGen: Enabling Next-Gen LLM Applications via Multi-Agent Conversation. \textit{arXiv:2308.08155}.

\bibitem{ref53}
Yang, H. et al. (2025). Agent Security Bench (ASB): Formalizing and Benchmarking Attacks and Defenses in LLM-Based Agents. \textit{Proc. ICLR}.

\bibitem{ref54}
Zhao, H., Li, J., Wu, Z., Ju, T., Zhang, Z., He, B., \& Liu, G. (2025a). Disagreements in Reasoning: How a Model's Thinking Process Dictates Persuasion in Multi-Agent Systems. \textit{arXiv:2509.21054}.

\bibitem{ref55}
Ren, X., Feng, Y., Zhao, B., Wang, L., \& Wang, J. (2025). RepuNet: Reputation-Enhanced Multi-Agent Communication Network for Trustworthy LLM Collaboration. \textit{arXiv:2505.05029}.

\bibitem{ref56}
Tomasev, N. et al. (2025). Simulating the Economic Impact of Rationality through Heterogeneous Agent-Based Modelling: Virtual Agent Economies. \textit{arXiv:2509.10147}.

\bibitem{ref57}
Zhou, X. \& Chan, J. (2026). ORCH: Orchestrating Reasoning Chains for Multi-Agent Systems with EMA-Guided Deterministic Routing. \textit{PMC:12907423}.


\bibitem{ref58}
Tian, K. (2025). Blockchain-enhanced incentive-compatible mechanisms for multi-agent reinforcement learning systems. \textit{Scientific Reports}, 15(1):42841.

\bibitem{ref59}
Kannan, S. (2023). EigenLayer: The Restaking Collective. \textit{EigenLayer Whitepaper}. \url{https://docs.eigenlayer.xyz/assets/files/EigenLayer_WhitePaper-88c47923ca0319870c611decd6e562ad.pdf}.

\bibitem{ref60}
Kivilo, S., Norta, A., Hattingh, M., Avanzo, S. \& Pennella, L. (2026). Designing a Token Economy: Incentives, Governance, and Tokenomics. \textit{arXiv:2602.09608}.

\bibitem{ref61}
Reijers, W., O'Brolch\'{a}in, F. \& Haynes, P. (2016). Governance in Blockchain Technologies \& Social Contract Theories. \textit{Ledger}, 1, 134--151.


\bibitem{ref63}
Andrighetto, G., Governatori, G., Noriega, P. \& van der Torre, L. (Eds.) (2013). \textit{Normative Multi-Agent Systems}. Dagstuhl Follow-Ups, Vol. 4. Schloss Dagstuhl.

\bibitem{ref64}
Chopra, A., van der Torre, L., Verhagen, H. \& Villata, S. (Eds.) (2018). \textit{Handbook of Normative Multi-Agent Systems}. College Publications.

\bibitem{ref65}
Esteva, M., Rodr\'{i}guez-Aguilar, J.A., Arcos, J.L., Sierra, C. \& Noriega, P. (2004). Electronic Institutions Development Environment. In \textit{Proc. AAMAS 2004}, 1663--1664.



\bibitem{ref67}
OpenClaw. (2025). OpenClaw: Open-Source Autonomous AI Agent Runtime. \url{https://github.com/openclaw/openclaw}

\bibitem{ref68}
OpenAgen. (2026). ZeroClaw: Zero-Overhead Autonomous AI Agent Runtime. \url{https://github.com/openagen/zeroclaw}

\bibitem{ref69}
Kendall, M.G. (1938). A New Measure of Rank Correlation. \textit{Biometrika}, 30(1/2), 81--93.

\end{thebibliography}
\end{document}